\documentclass[12pt]{article}
\pdfoutput=1
\usepackage[colorlinks,linkcolor=Blue,citecolor=Blue,bookmarks,bookmarksnumbered]{hyperref}
\usepackage[scaled=0.85]{helvet}
\usepackage{amsmath,amssymb,accents,mathrsfs,XoohmE}
\usepackage{graphicx,color}
\usepackage{booktabs}
\usepackage{multirow}
\usepackage{placeins}
\usepackage{amsmath}
\usepackage{subcaption}
\usepackage{XoohmE}

\definecolor{Green}  {rgb}{0.10,0.70,0.10} 
\definecolor{Orange} {rgb}{1.00,0.50,0.15} 
\definecolor{Red}    {rgb}{0.90,0.00,0.12} 
\definecolor{Purple} {rgb}{0.50,0.25,0.55} 
\definecolor{Turque} {rgb}{0.00,0.65,0.85} 
\definecolor{Blue}   {rgb}{0.00,0.00,1.00} 
\definecolor{Magenta}{rgb}{1.00,0.00,1.00} 
\definecolor{Gold}   {rgb}{1.00,0.75,0.25} 
\definecolor{Seaweed}{rgb}{0.01,0.24,0.09} 
\definecolor{Brown}  {rgb}{0.43,0.26,0.32} 
\definecolor{grey1}  {rgb}{0.20,0.20,0.20} 
\definecolor{grey2}  {rgb}{0.40,0.40,0.40} 
\definecolor{grey3}  {rgb}{0.60,0.60,0.60} 
\definecolor{grey4}  {rgb}{0.80,0.80,0.80} 
\definecolor{grey5}  {rgb}{0.90,0.90,0.90} 
\def\C#1#2{{\ifcase#1\or
             \color{Green}\or \color{Orange}\or \color{Red}\or
              \color{Purple}\or \color{Turque}\or \color{Blue}\or
               \color{Magenta}\or \color{Gold}\or \color{Seaweed}\or
                \color{Brown}\or\color{grey1}\or\color{grey2}\or
                 \color{grey3}\else\color{grey4}\fi#2}}

\definecolor{Slate} {rgb}{0.00,0.45,0.55}

\def\fracm#1#2{\hbox{\large{${\frac{{#1}}{{#2}}}$}}}

\def\be{\begin{equation}}
\def\ee{\end{equation}}
\newcommand{\bea}{\begin{eqnarray}}
\newcommand{\eea}{\end{eqnarray}}
\newcommand{\ena}{\end{eqnarray}}

\def\pp{{\mathchoice
          {
              \kern 1pt%
              \raise 1pt
              \vbox{\hrule width5pt height0.4pt depth0pt
                    \kern -2pt
                    \hbox{\kern 2.3pt
                          \vrule width0.4pt height6pt depth0pt
                          }
                    \kern -2pt
                    \hrule width5pt height0.4pt depth0pt}%
                    \kern 1pt
           }
            {
              \kern 1pt%
              \raise 1pt
              \vbox{\hrule width4.3pt height0.4pt depth0pt
                    \kern -1.8pt
                    \hbox{\kern 1.95pt
                          \vrule width0.4pt height5.4pt depth0pt
                          }
                    \kern -1.8pt
                    \hrule width4.3pt height0.4pt depth0pt}%
                    \kern 1pt
            }
            {
              \kern 0.5pt%
              \raise 1pt
              \vbox{\hrule width4.0pt height0.3pt depth0pt
                    \kern -1.9pt  
                    \hbox{\kern 1.85pt
                          \vrule width0.3pt height5.7pt depth0pt
                          }
                    \kern -1.9pt
                    \hrule width4.0pt height0.3pt depth0pt}%
                    \kern 0.5pt
            }
            {
              \kern 0.5pt%
              \raise 1pt
              \vbox{\hrule width3.6pt height0.3pt depth0pt
                    \kern -1.5pt
                    \hbox{\kern 1.65pt
                          \vrule width0.3pt height4.5pt depth0pt
                          }
                    \kern -1.5pt
                    \hrule width3.6pt height0.3pt depth0pt}%
                    \kern 0.5pt
            }
        }}

\def\mm{{\mathchoice
                       {
                             \kern 1pt
               \raise 1pt    \vbox{\hrule width5pt height0.4pt depth0pt
                                  \kern 2pt
                                  \hrule width5pt height0.4pt depth0pt}
                             \kern 1pt}
                       {
                            \kern 1pt
               \raise 1pt \vbox{\hrule width4.3pt height0.4pt depth0pt
                                  \kern 1.8pt
                                  \hrule width4.3pt height0.4pt depth0pt}
                             \kern 1pt}
                       {
                            \kern 0.5pt
               \raise 1pt
                            \vbox{\hrule width4.0pt height0.3pt depth0pt
                                  \kern 1.9pt
                                  \hrule width4.0pt height0.3pt depth0pt}
                            \kern 1pt}
                       {
                           \kern 0.5pt
             \raise 1pt  \vbox{\hrule width3.6pt height0.3pt depth0pt
                                  \kern 1.5pt
                                  \hrule width3.6pt height0.3pt depth0pt}
                           \kern 0.5pt}
                       }}

\def\ad{{\kern0.5pt
                   \alpha \kern-5.05pt \raise5.8pt\hbox{$\textstyle.$}\kern
0.5pt}}

\def\bd{{\kern0.5pt
                   \beta \kern-5.05pt \raise5.8pt\hbox{$\textstyle.$}\kern
0.5pt}}

\def\qd{{\kern0.5pt
                   q \kern-5.05pt \raise5.8pt\hbox{$\textstyle.$}\kern
0.5pt}}
\def\Dot#1{{\kern0.5pt
     {#1} \kern-5.05pt \raise5.8pt\hbox{$\textstyle.$}\kern
0.5pt}}

\catcode`@=11
\def\un#1{\relax\ifmmode\@@underline#1\else
        $\@@underline{\hbox{#1}}$\relax\fi}
\catcode`@=12

\def\a{\alpha}
\def\b{\beta}
\def\c{\chi}
\def\d{\delta}
\def\e{\epsilon}

\def\g{\gamma}

\def\i{\iota}
\def\j{\psi}
\def\k{\kappa}
\def\l{\lambda}
\def\m{\mu}

\def\o{\omega}

\def\q{\theta}
\def\r{\rho}
\def\s{\sigma}
\def\t{\tau}

\def\x{\xi}

\def\G{\Gamma}

\def\L{\Lambda}
\def\O{\Omega}
\def\P{\Pi}

\def\S{\Sigma}

\def\ca{{\cal A}}

\def\cf{{\cal F}}

\def\dslash{\not{\hbox{\kern-2pt $\partial$}}}
\def\Dslash{\not{\hbox{\kern-4pt $D$}}}
\def\pslash{\not{\hbox{\kern-2.3pt $p$}}}
 \newtoks\slashfraction
 \slashfraction={.13}
 \def\slash#1{\setbox0\hbox{$ #1 $}
 \setbox0\hbox to \the\slashfraction\wd0{\hss \box0}/\box0 }
 
\def\kcr{{\hbox{\ro \char'170}}}                
\def\ktl{{\hbox{\ro \char'170}}}        
\def\ktr{{\hbox{\ro \char'170}}}        
\def\kbl{{\hbox{\ro \char'170}}}        
\def\kbr{{\hbox{\ro \char'170}}}        

\def\plpl{\raise-2pt\hbox{$\raise3pt\hbox{$_+$}\hskip-6.67pt\raise0.0pt
\hbox{$^+$}\hskip 0.01pt$}}
\def\mimi{\raise-2pt\hbox{$\raise3pt\hbox{$_-$}\hskip-6.67pt\raise0.0pt
\hbox{$^-$}\hskip 0.01pt$}} 

\def\bo{{\raise.15ex\hbox{\large$\Box$}}}               
\def\pa{\partial}                                       
\def\TH{{\raise.2ex\hbox{$\displaystyle \bigodot$}\mskip-4.7mu \llap H \;}}
\def\face{{\raise.2ex\hbox{$\displaystyle \bigodot$}\mskip-2.2mu \llap {$\ddot
        \smile$}}}                                      

\def\dt#1{\on{\hbox{\bf .}}{#1}}                
\def\Dot#1{\dt{#1}}

\def\Bar#1{\overline{#1}}                       
\def\leftrightarrowfill{$\mathsurround=0pt \mathord\leftarrow \mkern-6mu
        \cleaders\hbox{$\mkern-2mu \mathord- \mkern-2mu$}\hfill
        \mkern-6mu \mathord\rightarrow$}
\def\dvec#1{\vbox{\ialign{##\crcr
        \leftrightarrowfill\crcr\noalign{\kern-1pt\nointerlineskip}
        $\hfil\displaystyle{#1}\hfil$\crcr}}}           
\def\dt#1{{\buildrel {\hbox{\LARGE .}} \over {#1}}}     

\def\fracm#1#2{\hbox{\large{${\frac{{#1}}{{#2}}}$}}}
\def\sfrac#1#2{{\vphantom1\smash{\lower.5ex\hbox{\small$#1$}}\over
        \vphantom1\smash{\raise.4ex\hbox{\small$#2$}}}} 
\def\bfrac#1#2{{\vphantom1\smash{\lower.5ex\hbox{$#1$}}\over
        \vphantom1\smash{\raise.3ex\hbox{$#2$}}}}       
\def\afrac#1#2{{\vphantom1\smash{\lower.5ex\hbox{$#1$}}\over#2}}    

\def\pa{\partial}

\def\ad{{\dot{\alpha}}}
\def\bd{{\dot{\beta}}}

 \font\rOpe=cmsy10                        
 \def\ktl{{\hbox{\rOpe\char'170}}}        
 \def\kbl{{\hbox{\rOpe\char'170}}}        
 \def\kcr{{\reflectbox{\rOpe\char'170}}}        
 \def\ktr{{\reflectbox{\rOpe\char'170}}}        
 \def\kbr{{\reflectbox{\rOpe\char'170}}}        
 \def\Border{\vbox{\hsize0pt
        \setlength{\unitlength}{1mm}
        \newcount\xco
        \newcount\yco
        \xco=-21
        \yco=12
        \begin{picture}(0,0)(-7.5,0)
        \put(\xco,\yco){$\ktl$}
        \advance\yco by-1
        {\loop
        \put(\xco,\yco){$\kcr$}
        \advance\yco by-2
        \ifnum\yco>-240
        \repeat
        \put(\xco,\yco){$\kbl$}}
        \xco=170
        \yco=12
        \put(\xco,\yco){$\ktr$}
        \advance\yco by-1
        {\loop
        \put(\xco,\yco){$\kcr$}
        \advance\yco by-2
        \ifnum\yco>-240
        \repeat
        \put(\xco,\yco){$\kbr$}}
        \put(-19.5,13){\scalebox{.6065}{%
         University of Maryland Center for String and Particle  Theory \&\ Physics Department%
        |University of Maryland Center for String and Particle  Theory \&\ Physics Department}}
        \put(-19.5,-241.5){\scalebox{.5835}{%
         ****University of Maryland * Center for String and
         Particle  Theory* Physics Department****University of Maryland *Center
        for String and Particle  Theory* Physics Department}}
        \end{picture}
        \par\vskip-8mm}}
\definecolor{UMred}{rgb}{.9,.05,.2}
\definecolor{HUblue}{rgb}{.0,.3,.7}

\definecolor{Red}    {rgb}{0.90,0.00,0.12} 
\definecolor{Blue}   {rgb}{0.00,0.00,1.00} 
\definecolor{Green}  {rgb}{0.10,0.70,0.10} 
\definecolor{Turque} {rgb}{0.00,0.65,0.85} 
\definecolor{Orange} {rgb}{1.00,0.50,0.15} 
\definecolor{Magenta}{rgb}{1.00,0.00,1.00} 
\definecolor{Gold}   {rgb}{1.00,0.75,0.25} 
\definecolor{Seaweed}{rgb}{0.01,0.24,0.09} 
\definecolor{Purple} {rgb}{0.50,0.25,0.55} 
\definecolor{Brown}  {rgb}{0.43,0.26,0.32} 
\definecolor{grey1}  {rgb}{0.20,0.20,0.20} 
\definecolor{grey2}  {rgb}{0.40,0.40,0.40} 
\definecolor{grey3}  {rgb}{0.60,0.60,0.60} 
\definecolor{grey4}  {rgb}{0.80,0.80,0.80} 
\definecolor{grey5}  {rgb}{0.90,0.90,0.90} 
\def\C#1#2{{\ifcase#1\or
             \color{Red}\or \color{Green}\or \color{Blue}\or\
              \color{Turque}\or \color{Orange}\or \color{Magenta}\or 
               \color{Gold}\or \color{Seaweed}\or \color{Purple}\or
                \color{Brown}\or\color{grey1}\or\color{grey2}\or
                 \color{grey3}\else\color{grey4}\fi#2}}

\definecolor{Slate} {rgb}{0.00,0.45,0.55}

\newdimen\parshift\parshift=\parindent
\catcode`@=11
 \long\def\@footnotetext#1{\insert\footins{\reset@font\footnotesize
           \interlinepenalty\interfootnotelinepenalty\splittopskip%
            \footnotesep\splitmaxdepth\dp\strutbox\floatingpenalty\@MM%
             \hsize\columnwidth\addtolength{\hsize}{-2\parindent}
              \@parboxrestore\protected@edef\@currentlabel%
              {\csname p@footnote\endcsname\@thefnmark}%
                \color@begingroup%
                 \@makefntext{\rule\z@\footnotesep\ignorespaces#1%
                  \@finalstrut\strutbox}%
                \color@endgroup}}
 \long\def\@makefntext#1{\hglue\parshift%
           \vbox{\noindent\baselineskip=11pt plus.5pt minus.5pt\hb@xt@0em{\hss\@makefnmark\kern1pt}#1}}
\catcode`@=12

\newskip\humongous \humongous=0pt plus 1000pt minus 1000pt
\def\caja{\mathsurround=0pt}
\def\eqalign#1{\,\vcenter{\openup2\jot \caja
        \ialign{\strut \hfil$\displaystyle{##}$&$
        \displaystyle{{}##}$\hfil\crcr#1\crcr}}\,}
\newif\ifdtup

\makeatletter
\def\section{\@startsection{section}{1}{\z@}
        {3ex plus-1ex minus-.2ex}{1pt plus1pt}{\large\sf\bfseries\boldmath}}
\def\subsection{\@startsection{subsection}{2}{\z@}
         {1.5ex plus-1ex minus-.2ex}{0.01pt plus1pt}{\sf\slshape}}
\def\subsubsection{\@startsection{subsubsection}{3}{\z@}
          {1.5ex plus-1ex minus-.2ex}{0.01pt plus0.2pt}{\sf\boldmath}}
\def\paragraph{\@startsection{paragraph}{4}{\z@}
           {.75ex \@plus.5ex \@minus.2ex}{-2mm}{\sf\bfseries\boldmath}}
\makeatother

\allowdisplaybreaks
\seceq
\numberwithin{figure}{section}

\usepackage{tikz}
\usepackage[vcentermath,enableskew]{youngtab}
\usepackage[centertableaux]{ytableau}
\let\TC=\textcolor
\definecolor{Hey}{rgb}{.9,.05,.4}
\definecolor{orange}{rgb}{1,.5,0}
\definecolor{plum}{rgb}{.4,0,.6}
\definecolor{R}{rgb}{1,0,0}
\definecolor{G}{rgb}{0,1,0}
\definecolor{B}{rgb}{0,0,1}
\long\def\CMTblu#1{\leavevmode\TC{blue}{\sf#1}}
\long\def\CMTgrn#1{\leavevmode\TC{green}{\sf#1}}
\long\def\CMTred#1{\leavevmode\TC{red}{\sf#1}}

\long\def\CMTR#1{\leavevmode\TC{R}{\sf#1}}

\long\def\CMTB#1{\leavevmode\TC{B}{\sf#1}}

\usepackage{lipsum}
\usepackage{listings}
\definecolor{MyDarkGreen}{rgb}{0.0,0.4,0.0} 
\begin{document}

\thispagestyle{empty}
\noindent{\small
\hfill{$~~$}  \\ 
{}
}
\begin{center}
{\large \bf
Superfield Component Decompositions and the Scan for  \vskip0.02in
Prepotential Supermultiplets in 10D Superspaces
}   \\   [8mm]
{\large {
S.\ James Gates, Jr.\footnote{sylvester$_-$gates@brown.edu}${}^{,a, b}$,
Yangrui Hu\footnote{yangrui$_-$hu@brown.edu}${}^{a,b}$, and
S.-N. Hazel Mak\footnote{sze$_-$ning$_-$mak@brown.edu}${}^{a,b}$
}}
\\*[6mm]
\emph{
\centering
$^{a}$Brown Theoretical Physics Center,
\\[1pt]
Box S, 340 Brook Street, Barus Hall,
Providence, RI 02912-1843, USA 
\\[10pt]
$^{b}$Department of Physics, Brown University,
\\[1pt]
Box 1843, 182 Hope Street, Barus \& Holley,
Providence, RI 02912, USA 
}
 \\*[40mm]
{ ABSTRACT}\\[5mm]
\parbox{142mm}{\parindent=2pc\indent\baselineskip=14pt plus1pt
The first complete and explicit SO(1,9) Lorentz descriptions of all component fields contained 
in the $\mathcal{N} = 1$, $\mathcal{N} = 2$A, and $\mathcal{N} = 2$B unconstrained scalar 10D 
superfields are presented.  These are made possible by a discovery of the dependence of the 
superfield component expansion on the branching rules of irreducible representations in one 
ordinary Lie algebra into one of its Lie subalgebras. Adinkra graphs for ten dimensional 
superspaces are defined for the first time, whose nodes depict spin bundle representations of 
SO(1,9).  A consequential deliverable of this advance is it provides the first explicit, in 
terms of component fields, examples of all the off-shell 10D Nordstr\"om SG theories relevant 
to string theory, without off-shell central charges that are reducible but with finite numbers 
of fields.  An analogue of Breitenlohner's approach is implemented 
to scan for superfields that contain graviton(s) and gravitino(s), which are the candidates for 
the superconformal prepotential superfields of 10D off-shell supergravity theories and Yang-Mills 
theories.}
 \end{center}
\vfill
\vfill
\clearpage
%


\newpage
\section{Introduction}

On the basis of superspace geometry, recently a study \cite{NordSG}
of the problem of describing scalar gravitation at the linearized level within the context of eleven and ten dimensional 
superspaces was completed.  The results of this effort showed with 
regards to supergeometical concepts there were no significant differences (nor more importantly 
obstructions) between this problem in the high dimensional superspaces than in a four 
dimensional superspace.  Though one important distinction was noted... there is a long 
standing lack of a general theory of irreducibility in the area of scalar superfield representations.  
Related to this is the problem of defining a superconformal multiplet in these regimes.  
We were thus re-energized to look at these problems.

One of the longest unsolved problems in the study of supersymmetry is the fact that {\it 
{an irreducible off-shell formulation containing a finite number of component fields}} for the
ten dimensional (along with the extended and eleven dimensional ones) supergravity multiplet has not been presented. This 
statement accurately describes the current state-of-the-art of the field well after thirty years 
since it began.  
An even simpler problem is to give a detailed component level presentation of a {\it {reducible off-shell formulation explicitly showing a finite number of component fields}}.  
It is the purpose of this work to report on new techniques for the second of these problems. 
Though more work will need to be done to find these.

In this work we have new developed techniques, both algorithmic and analytical, that allow the {\it {first complete and explicit SO(1,9) Lorentz descriptions of all component fields 
contained in 10D}}, $\mathcal{N}=1$, $\mathcal{N}=2A$, and $\mathcal{N}=2B$ {\it {unconstrained 
scalar superfields.}} They form the maximal reducible and relevant supermultiplets. For the 
case of the $\mathcal{N}=1$ theory, our results agree with the first report on this topic 
\cite{10DScLR} given by Bergshoeff and de Roo.  These authors took as their starting point 
an analysis of the ``supercurrent corresponding to the $\mathcal{N}=1$ supersymmetic Maxwell theory 
in ten dimensions.''  An approach of this type is tied to making {\it {an initial assumption 
about the dynamics of some supersymmetrical system}}, i.e. the choice of the energy-momentum 
tensor.

The previous sentence illustrates a peculiarity of the study of spacetime supersymmetry in
comparison to other symmetries.  Namely, almost all studies of the representations of spacetime 
supersymmetry take as a starting a set of assumptions related to some dynamic system.  This
is shared by no other symmetry known to these authors.  Thus, there is a possibility by
exploring avenues that depart from dynamical assumptions, a resolution to the puzzles may be
found.

We will present the maximal reducible relevant representations of supergravity in 10D related 
to unconstrained scalar superfields by ``tensoring'' them with different representations of 
the SO(1,9) Lorentz group and search for graviton(s) and gravitino(s).  To achieve this goal, 
we borrow Breitenlohner's approach \cite{B1}, which gave the first off-shell 4D, $\mathcal{N}
=1$ supergravity description. It utilized the known off-shell structure of the 4D, $\mathcal{N}=1$ 
vector supermultiplet $(v_{\un{a}}, \l_{\b}, d)$. Since the structure of off-shell supermultiplets 
in ten dimensions is poorly understood, we implement the approach of Breitenlohner by use of 
a 10D scalar superfield in place of the vector supermultiplet.  This will ensure an off-shell 
supersymmetry realization.

There are three guiding principles implied by the Breitenlohner appoaoch that
are implemented in our discussion, use:\vskip0.02in \noindent
$~~~~~~~$ (a.) no {\it {dynamical}} assumptions in
the derivation of representation  \newline
$~~~~~~~~$
$~~~~~$ theory results, \newline
$~~~~~~~~$ (b.) {\it {only}} conventional Wess-Zumino superspaces, and \newline
$~~~~~~~~$ (c.) superfields that admit an {\it {unconstrained}} {\it {prepotential}} formulation. 
\vskip0.02in \noindent
These are the principles that allowed for the complete off-shell superspace descriptions of
supergravity in 4D, $\cal N$ = 1 theories.  Our current investigation looks to extend these
beyond the lower dimensional context and into the domain of the 10D theories.  The operational steps to implement the paradigm of the approach include
\vskip0.02in \noindent
$~~~~~~~$ (a.) identify an appropriate scalar supermultipet free of {\it {dynamical}}
 \newline
$~~~~~~~~$
$~~~~~$ assumptions with regards to the
closure of the superalgebra,
 \newline
$~~~~~~~~$ (b.) tensor the scalar supermultiplet with representations of
the 
 \newline
$~~~~~~~~$
$~~~~~$ 
Lorentz algebra, and
  \newline
$~~~~~~~~$ (c.) among the results found by the
tensoring operation, search 
 \newline
$~~~~~~~~$
$~~~~~$ 
for components that align with the Lorentz representations of 
 \newline
$~~~~~~~~$
$~~~~~$ conformal graviton and gravitino.

As this approach is independent of any assumptions about dynamics of the component fields 
contained in supermultiplets, there should be an expectation of differences in comparison 
with any past results derived from an analysis where dynamics are built into a conceptual 
foundation.  

It is known that when a supersymmetrical theory is expressed in terms of prepotentials, the 
reasons for many unexpected enhancement of vanishing quantum corrections become obvious.  So 
we restrict our considerations to formulations that are consistent with prepotentials.  This 
further restricts our considerations to solely theories that demonstrate an absence of
 ``off-shell central charges'' \cite{CZ1,CZ2,CZ3,CZ4,CZ5}. 

The keys to our progress in this realm are higher dimensional adinkras (built upon
the well established foundation of one dimensional ones 
\cite{adnk1,adnkM1,YZ,adnkM2,codes1,codes2,codes3,adnkGEO1,adnkGEO2,adnk1dHoloR1,adnk1dHoloR2,adnk4dN2,HYMN})
and advanced computational tools.  These techniques are derived via properties of representation 
theory and thus are {\it {independent of assumptions about dynamics.}}

\begin{figure}[ht]
\centering
\includegraphics[width=1.0\textwidth]{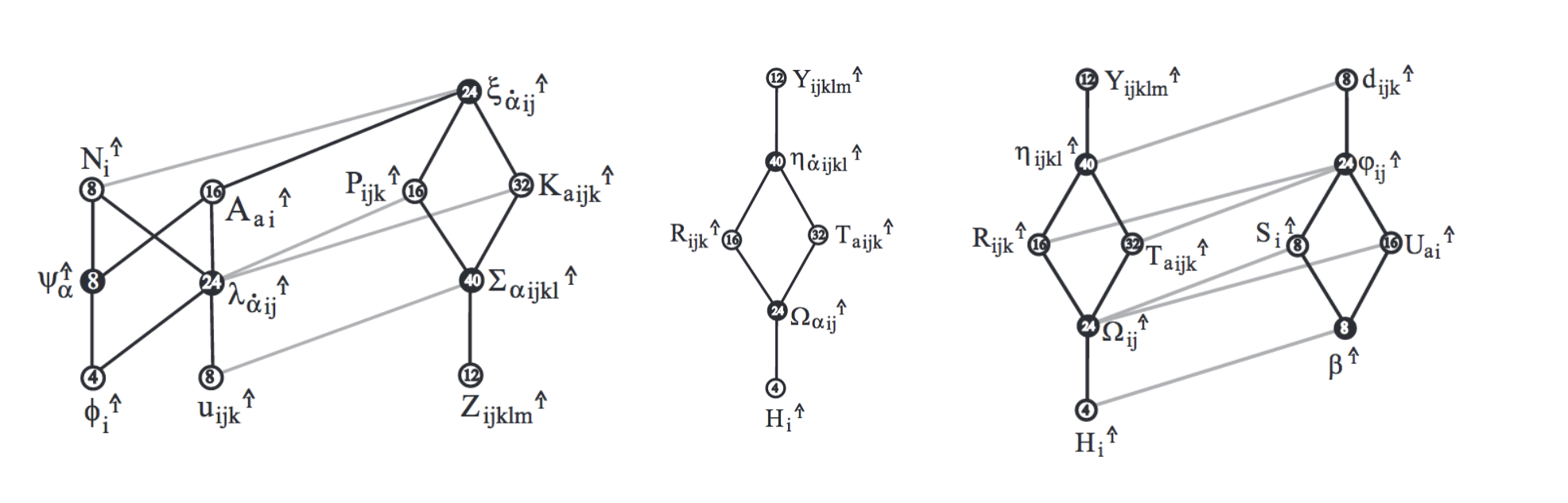}
\caption{Adinkra-like Representations of Three 4D, $\cal N$ = 2 Supermultiplets}
\label{TetMin}
\end{figure}

For the first of these keys, we have built upon efforts that began in the work of \cite{N2matter} 
wherein the first two images shown in Figure~\ref{TetMin} were presented.
More recently Faux \cite{cHyPRr} introduced the third image in Figure~\ref{TetMin}.  All the 
supermultiplets depicted are ``off-shell,'' i.e. the closure of the supersymmetry algebra does 
not require the use of the equation of motion for any component field.

The graphs from \cite{N2matter} are among the first in the literature where the individual nodes
of one dimensional adinkras are ``aggregated'' together into structures that carry non-trivial
representations of the Lorentz group.  Furthermore, the links in this class of diagrams are 
also ``aggregations'' of the links that appear in the one dimensional adinkras.  Both of these 
attributes pointed to the possibility of constructing adinkras in higher dimensions where both 
nodes and links can be regarded as aggregations of the one-dimensional concepts.  

The important principle established by the explicit demonstration of these graphs is that it 
is {\it a} {\it {priori}} possible to construct adinkras in greater than one dimensional 
contexts.  This is not, however, an ``organic'' constructive process.  Namely, these structures 
were obtained by {\it {first}} completing a component-level analytical construction of the 
supermultiplet.  An ``organic'' process would dispense with any component-level lead-in.  
Instead on the basis of a set of principles and tools, there should be a way to construct 
such adinkras in higher dimensions.  This is the primary goal of this current work.

Also on the analytical side, our efforts will build upon a class of works \cite{hst1,hst2,hst3} 
wherein the concept of the ``Tableaux calculus'' was discussed by Howe, Stelle, and Townsend.  
To our knowledge, these were the first works that explicitly discussed the use of Young Tableaux 
as applied to problems in the representation theory of spacetime supersymmetry.

The supermultiplets illustrated in Figure~\ref{TetMin} are {\it {constrained}}, i.e. 
subject to supercovariant derivative differential equations.  This is also the case of the 
works by Howe, Stelle, and Townsend.  In particular, these authors proposed to use Young  
Tableaux that are associated with the superspace covariant derivative operators acting on 
constrained supermultiplets.  The concept of associating the superspace covariant derivative 
operators with Tableaux images will be important in our work.  In our exploration, we also 
find it useful to utilize the more conventional class of Tableaux that are associated with bosonic Lorentz indices carried by fields.  We will call the former of these Young Tableaux 
``fermionic Young Tableaux'' (FYT) while distinguishing the latter by the term ``bosonic 
Young Tableaux'' (BYT).  An important distinction of our work is the supermultiplets are
unconstrained.

Our work is also a beneficiary of a report undertaken by N. Yamatsu \cite{so11table} who 
presented results for grand unified model-building based on finite-dimensional Lie algebras.  
In this work, the reader can find results for projection matrices and branching rules 
between Lie algebras and their subalgebras up to high ranks, as well as Dynkin labels 
and Weyl dimensional formulae of irreducible representations, and much more.  The results 
we have explicitly called out in the last sentence have proven to be particularly useful 
when applied to the SO(1,9) representation theory we consider.  It is of historical note,
that the work by Yamatsu falls among the path of earlier explorations \cite{SLnk,LttL,Mrn}.

The final enabling tool for our work is based on algorithmic efficiency and here the work 
by Feger and Kephart \cite{LieART} played an important role.  These authors developed and 
presented a Mathematica application - Lie Algebras and Representation Theory (LieART) which 
carries out computations typically occurring in the context of Lie algebras and representation 
theory.   This provides a robust algorithm that enables the study of weight systems.  One 
of its features that enhances its efficiency is the use of Dynkin labels for irreducible 
representations.  It is thus ideal for specializing to SO(1,9) representation theory. 

We organize this current paper in the manner described below. In Chapter two we discuss the 
numbers of independent bosonic and fermionic components in a scalar superfield. Chapter three is a transitional one where we review the $\theta-$expansion of a real scalar superfield 
in 4D, $\mathcal{N} = 1$ theory and introduce the higher dimensional adinkra technology. In 
Chapter four, we present the component decomposition result of the scalar superfield in 10D, $\mathcal{
N} = 1$ theory. Two different approaches that lead to the same result are presented. One we 
call the ``handicraft approach,\footnote{The term ``handicraft approach" comes for P.\ van 
Nieuwenhuizen who coined it to refer to a mathematical ``Rube Goldberg'' type of approach 
that nevertheless yields a correct result in the context of supersymmetry calculations. }" 
where we introduce Bosonic and Fermionic Young Tableaux. The other is the application 
of branching rules, the restrictions of representations from a Lie algebra $\mathfrak{g}$ to 
one of its subalgebra $\mathfrak{h}$. The complete ten dimensional $\mathcal{N} = 1$ adinkra 
is drawn. In Chapter five, we start from the well-studied off-shell description of 4D, $\mathcal{
N} = 1$ supergravity established by Breitenlohner and show how to apply his idea to carry 
out constructions of candidates for the prepotential and Yang-Mills supermultiplets in 10D, 
$\mathcal{N} = 1$ theory.  In Chapter six, we present the methodology and results of 
constructing the 10D, $\mathcal{N}=2$A scalar superfield from the 10D, $\mathcal{N}=1$ 
scalar superfield, as well as the discussion of the search of prepotential supermultiplet 
in Type IIA superspace. The ten dimensional IIA adinkra diagram is shown at low orders.  
However, its complete structure is given in the form of a list of the component field 
representations it contains.  These same steps are repeated in  chapter seven that gives 
the methodology and explicit decomposition  results as well as the discussion of constructing 
prepotential supermultiplet in Type IIB superspace. The ten dimensional IIB adinkra 
diagram is also shown at low orders.

We follow the presentation of our work with conclusions, three appendices, and the bibliography.
The first appendix gives detailed discussions about chiral and vector supermultiplets obtained 
from 4D, $\cal N$ = 1 unconstrained scalar superfield.  The second appendix contains tables 
of SO(1,9) representations drawn from the work of Yamatsu.  The third appendix presents the 
results of ``tensoring'' low order bosonic representations of SO(1,9) with the basic unconstrained 
10D, $\cal N$ = 1 scalar superfield.  Finally, the fourth appendix presents the results 
of ``tensoring'' low order fermionic representations of SO(1,9) with the basic unconstrained 
10D, $\cal N$ = 1 scalar superfield. 

\newpage
\section{Superfield Diophantine Considerations}

Via the use of simple toroidal compactification, one can count the numbers of 
independent bosonic and fermionic component fields that occur in a scalar 
superfield.  As a fermionic coordinate  cannot be squared, this means 
in the Grassmann coordinate expansion of a superfield, any one specific 
fermionic coordinate can only occur to the zeroth power or the first power. 
As component fields occur as coefficients of monomials in superspace Grassmann 
coordinates, counting the latter is the same as the former... as long as the 
superfield is not subject to any spinorial ``supercovariant derivative'' constraints.  

Each higher dimensional superspace with $D$ bosonic dimensions, 
for purposes of counting is equivalent to some value of $d$, where $d$ is the number 
of independent equivalent real one-dimensional fermionic coordinates on which the 
superfield depends. The total number of independent monomials is given by $2^d$. Next, to count the total number of bosonic components $n_B$ in the scalar superfield, we simply divide by a factor of two.  This same argument applies to the total number of fermionic components $n_F$ in the scalar superfield.  Thus we have $n_B$ = $n_F$ = $2^{d-1}$. A few cases are shown in the table below.
\begin{table}[h!]
\vspace{0.4cm}
\centering
\begin{tabular}{|c|c|c|c|c|} \hline
    $D$ & $d$ & $2^d$ & $n_B$ & $n_F$  \\ \hline \hline
    4 & 4 & 16 & 8 & 8  \\ \hline
    5 & 8 & 256 & 128 & 128  \\ \hline
    10 & 16 & 65,536 & 32,768 & 32,768  \\ \hline
    ~~11~~ & ~~32~~ & ~4,294,967,296~ & ~2,147,483,648~ & ~2,147,483,648~   \\ \hline
\end{tabular}
\caption{Number of independent components in {\em {unconstrained}} scalar superfields 
in $D$ dimensional \newline $~~~~~~{\,}~~~~~~$
spacetime \label{tab:count}}
\end{table}

While superfields easily provide a methodology for finding collections of component 
fields that are representations of spacetime supersymmetry, one thing that superfields 
do not yield so easily is a theory of the constraints that provides irreducible representations. 
For any constrained superfield, it must be the case that the number of component 
fields does {\em {not}} {\em {depend}} {\em {solely}} on the parameter $d$.  This is most certainly 
true for the maximally constrained and therefore minimal irreducible representations. 
This is illustrated by comparing the final two columns of the first three rows in 
Tables \ref{tab:count} and \ref{tab:constrained} that $n_{B}$ = $n_{F}$ $\ne$ $n_{
B(min)}$ = $n_{F(min)}$. 
\begin{table}[h]
\vspace{0.4cm}
\centering
\begin{tabular}{|c|c|c|c|c|}\hline
4D Minimal Off-Shell Supermultiplet & $d$ & $2^{d-1}$ & $n_{B(min)}$ 
& $n_{F(min)}$  \\ \hline \hline
$\cal N$ = 1 Chiral  & 4 & 8 & 4 & 4   \\ \hline
$\cal N$ = 2 Vector  & 8 & 128 & 8 & 8 \\ \hline
$\cal N$ = 4 SG  & ~~16~~ & ~32,768~ & 128  &  128  \\ \hline
\end{tabular}
\caption{Number of independent components in {\em {maximally constrained}} 
superfields \label{tab:constrained}}
\end{table}

In this paper, we will decompose the 10D unconstrained scalar superfields 
into component fields that transform under the Lorentz group SO(1,9).

\newpage
\section{4D Scalar Superfield Decomposition}\label{sec:4D}

The off-shell four dimensional vector supermultiplet is well understood now 
to be a part of the unconstrained superfield as given below. In 4D, $\mathcal{
N}=1$ real superspace, the spinor index (the Greek index) on $\theta^{\a}$ 
runs from 1 to 4, and we have 8 bosons and 8 fermions, as counted in Table 
\ref{tab:count}. The $\theta-$expansion of a scalar superfield is 
\begin{equation}
\label{equ:4D_superfield_expansion}
\begin{split}
    {\cal V}(x^{\un{a}},\theta^{\a}) ~=&~ f(x^{\un{a}}) ~+~ \theta^{\a} \, \psi_{\a}(x^{
    \un{a}}) ~+~ \theta^{\a}\theta^{\b}C_{\a\b} \, g(x^{\un{a}}) \\
    &+~ \theta^{\a}\theta^{\b}i(\g^{5})_{\a\b} \, h(x^{\un{a}}) ~+~ \theta^{\a}\theta^{
    \b} i(\g^{5}\g^{\un b})_{\a\b} \, v_{\un b}(x^{\un{a}})  \\
    &+~ \theta^{\a}\theta^{\b}\theta^{\g} C_{\a\b}C_{\g\d} \, \chi^{\d}(x^{\un{a}}) 
    ~+~ \theta^{\a}\theta^{\b} \theta^{\g}\theta^{\d} C_{\a\b}C_{\g\d} \, N(x^{\un{a}})
\end{split}
\end{equation}
where $f(x^{\un{a}})$, $g(x^{\un{a}})$, $h(x^{\un{a}})$, $v_{\un b}(x^{\un{a}})$, 
and $N(x^{\un{a}})$ are bosonic component fields with 8 d.o.f.\footnote{We use the notation
d.o.f to indicate ``degrees of freedom."} in total; $\psi_{
\a}(x^{\un{a}})$ and $\chi^{\d}(x^{\un{a}})$ are fermionic component fields with 
8 d.o.f. in total. We will discuss this $\theta-$expansion result from different perspectives.

\subsection{Group Theory Perspective}

From the perspective of group theory, we can translate the scalar superfield 
decomposition problem to the irreducible decomposition problem of representations in $\mathfrak{so}(4)$. 
First, we can write the general expression of the $\theta-$expansion of a superfield 
\begin{equation}\label{equ:4D_Gen_expansion}
\begin{split}
{\cal V} (x^{\un{a}},\theta^{\alpha}) ~=&~ v^{(0)} (x^{\un{a}}) ~+~ \theta^{\alpha} \, 
v^{(1)}_{\alpha} (x^{\un{a}}) ~+~ \theta^{\alpha}\theta^{\beta} \, v^{(2)}_{\alpha
\beta}(x^{\un{a}}) \\
&~+~ \theta^{\alpha}\theta^{\beta}\theta^{\gamma} \, v^{(3)}_{\alpha\beta\gamma} 
(x^{\un{a}}) ~+~ \theta^{\alpha}\theta^{\beta}\theta^{\gamma}\theta^{\delta} \, v^{(4)
}_{\alpha\beta\gamma\delta} (x^{\un{a}}) ~~~.
\end{split}
\end{equation}
Due to the antisymmetric property of the Grassmann coordinates $\theta^{\alpha}$, 
the quantities $\theta^{\alpha}$, $\theta^{\alpha}\theta^{\beta}$, $\theta^{\alpha}
\theta^{\beta}\theta^{\gamma}$, and $\theta^{\alpha}\theta^{\beta}\theta^{\gamma
}\theta^{\delta}$ have 4, 6, 4, and 1 degrees of freedom, respectively. They can be 
interpreted as representations of $\mathfrak{so}(4)$ with 4, 6, 4, and 1 dimensions. 
We use level-$n$ to denote the $\theta-$monomial with $n$ $\theta$s. The problem 
is reduced to do the irreducible decompositions of these representations and the 
results can be listed as
\begin{equation}\label{equ:4D}
{\cal V} ~=~ \begin{cases}
{~~}{\rm {Level}}-0 \,~~~~~~~~~~ \CMTB {\{ 1 \}} ~~~,  \\
{~~}{\rm {Level}}-1 \,~~~~~~~~~~ \CMTred {\{ 4 \}} ~~~,  \\
{~~}{\rm {Level}}-2 \,~~~~~~~~~~ \CMTB {\{ 1 \}} \oplus \CMTB {\{ 4 \}} \oplus 
\CMTB {\{ 1\}} ~~~, \\
{~~}{\rm {Level}}-3 \,~~~~~~~~~~ \CMTred {\{ 4 \}} ~~~, \\
{~~}{\rm {Level}}-4 \,~~~~~~~~~~ \CMTB {\{ 1 \}} ~~~.
\end{cases}
{~~~~~~~~~~~~~~~}
\end{equation}
Note that level-4 and 3 are conjugate to level-0 and 1, respectively, while level-2 
is self-conjugate.  The conjugates of $\CMTB {\{ 1 \}}$ and $\CMTred {\{ 4 \}}$ are still 
$\CMTB {\{ 1 \}}$ and $\CMTred {\{ 4 \}}$. Level-2 has two $\CMTB {\{ 1 \}}$'s\footnote{If one is exercising extra care, the $\CMTB {\{ 1 \}}$'s at level-2
are not identical.  In fact, one of these corresponds to a scalar field while the other 
corresponds to a pseudoscalar.  However, for our purposes, we can neglect this
difference.  This distinction is shown in Table \ref{tab:4D_Basis} where the ``count'' 
is the same for both, but their projections to $\g$-matrices are distinct.}. 
Although in the group theory context,  $\mathfrak{so}(4)$ only has one $\CMTB 
{\{ 1 \}}$, here two $\CMTB {\{ 1 \}}$'s represent two  different $\theta-$monomials 
as you will see shortly.

In order to distinguish between bosonic irreps and fermionic irreps, we color their 
dimensions: blue if bosonic and red if fermionic. In the rest of the paper, we will 
use these conventions. 

Recall that in 4D, $\mathcal{N} = 1$ real superspace, we can create the covariant 
gamma matrices which are $4\times4$ real matrices. The basis of the space of matrices
over these spinors is summarized in Table~\ref{tab:4D_Basis}. 
\begin{table}[htp!]
\centering
\begin{tabular}{|c|c|c|c|c|c|} \hline
Basis & $C_{\alpha\beta}$ & $(\gamma^{\underline{a}})_{\alpha\beta}$ & $
(\gamma^{[2]})_{\alpha\beta}$ & $i(\gamma^5)_{\alpha\beta}$ & $i(\gamma^5
\gamma^{\underline{a}})_{\alpha\beta}$ \\ \hline
Sym/Antisym & A & S & S & A & A  \\ \hline
Count       & 1 & 4 & 6 & 1 & 4  \\ \hline
\end{tabular}
\caption{Summary of 4D Basis: all of these are $4~\times~4$ real matrices\label{tab:4D_Basis}}
\end{table}

\noindent
It yields analytical expressions of the $\theta-$monomials below:
\begin{itemize}
\item Level-0: no $\theta$ 
\item Level-1: $\CMTred {\{ 4 \}}~=~\theta^{\a}$
\item Level-2: $\CMTB {\{ 1 \}}~=~\theta^{\a}\theta^{\b}C_{\alpha\beta} $, $\CMTB {\{ 
4 \}}~=~\theta^{\a}\theta^{\b}i(\gamma^5\gamma^{\underline{a}})_{\alpha\beta} $, 
$\CMTB {\{ 1 \}}~=~\theta^{\a}\theta^{\b}i(\gamma^5)_{\alpha\beta} $
\item Level-3: $\CMTred {\{ 4 \}}~=~ \theta^{\a}\theta^{\b}\theta^{\g}C_{\alpha
\beta}C_{\gamma\delta}$
\item Level-4: $\CMTB {\{ 1 \}}~=~ \theta^{\a}\theta^{\b}\theta^{\g}\theta^{\d}C_{\alpha
\beta}C_{\gamma\delta}$
\end{itemize}
Thus, Equation~(\ref{equ:4D_Gen_expansion}) can be rewritten as
\begin{equation}\label{equ:4D_expansion}
\begin{split}
    {\cal V} (x^{\un{a}},\theta^{\alpha}) ~=&~ v^{(0)} (x^{\un{a}}) ~+~ \theta^{\alpha} \, v^{(1)}_{\alpha} (x^{\un{a}}) ~+~ \theta^{\alpha}\theta^{\beta} \, \Big[~ C_{\alpha\beta} \, v^{(2)}_1(x^{\un{a}}) ~+~ i(\gamma^5)_{\alpha\beta} \, v^{(2)}_2(x^{\un{a}}) ~+~ i(\gamma^5\gamma^{\underline{b}})_{\alpha\beta} \, v^{(2)}_{\un b}(x^{\un{a}}) ~\Big]\\
    &~+~ \theta^{\alpha}\theta^{\beta}\theta^{\gamma}C_{\alpha\beta}C_{\gamma\delta} \, v^{(3)\delta} (x^{\un{a}}) ~+~ \theta^{\alpha}\theta^{\beta}\theta^{\gamma} \theta^{\delta}C_{\alpha\beta}C_{\gamma\delta} \, v^{(4)}(x^{\un{a}}) ~~~.
\end{split}
\end{equation}
Note that applying the following replacements
\begin{equation}
\begin{split}
& v^{(0)} (x^{\un{a}})~\rightarrow~f(x^{\un{a}}) ~~,~~ 
v^{(1)}_{\alpha} (x^{\un{a}}) ~\rightarrow~ \psi_{\alpha}(x^{\un{a}}) ~~,~~
v^{(2)}_{1} (x^{\un{a}}) ~\rightarrow~ g(x^{\un{a}}) ~~,~~ 
v^{(2)}_{2} (x^{\un{a}}) ~\rightarrow~ h(x^{\un{a}}) ~~,~~ \\
& v^{(2)}_{\un{b}} (x^{\un{a}}) ~\rightarrow~ v_{\un{b}}(x^{\un{a}}) ~~,~~
v^{(3)\delta} (x^{\un{a}}) ~\rightarrow~ \chi^{\delta}(x^{\un{a}}) ~~,~~
v^{(4)} (x^{\un{a}}) ~\rightarrow~ N(x^{\un{a}}) ~~,
\end{split}
\end{equation}
will reproduce the result in Equation~(\ref{equ:4D_superfield_expansion}).

\subsection{Graph Theory Perspective: Adinkra}

From the perspective of graph theory, particularly adinkra diagrams, we can define a four dimensional adinkra based on Equation~(\ref{equ:4D}).
First of all, the adinkra diagram carries information about component fields rather than $\theta-$monomials, so we need to translate Equation~(\ref{equ:4D}) to field variable language. 
Consider a variable with one upstairs spinor index $\chi^{\alpha}$ and assign irrep $\CMTred{\{4\}}$ to this field. What is the irrep corresponding to the variable with one downstairs spinor index $\chi_{\beta}$? Since 
\begin{equation}
    \chi_{\beta} ~=~ \chi^{\alpha}C_{\alpha\beta}  ~~~,
\end{equation}
where $C_{\alpha\beta}$ is the spinor metric, the irrep of $\chi_{\beta}$ is still $\CMTred{\{4\}}$. Generally speaking, in 4D, $\cal N$ = 1 real notation, the irreps corresponding to component fields are the same as their $\theta-$monomials. 

Then we can use open nodes to denote bosonic component fields and put the dimensions of their corresponding irreps in the centers of the open nodes. For fermionic component fields, we use closed nodes with a similar convention for dimensionality.  The level number represents the height assignment and it increases with height. Black edges connect nodes in adjacent levels and represent supersymmetric transformation operations on the component fields.
We can also interpret the adinkra using the idea in \cite{Adinkra_Prepotential} (i.e. adinkra nodes are the $\theta$ $\to$ 0 limit of superfields), as illustrated in the following table.
\begin{table}[h!]
\centering
\begin{tabular}{|c|c|c|c|} \hline
    Level & Adinkra nodes & Component fields & Irrep(s) in $\mathfrak{so}(4)$  \\ \hline
    0 & ${\cal V}|$ & $f(x^{\un{a}})$ & $\CMTB{\{1\}}$   \\ \hline
    1 & ${\rm D}_{\a}{\cal V}|$ & $\psi_{\a}(x^{\un{a}})$ &  $\CMTred{\{4\}}$  \\ \hline
    2 & ${\rm D}_{[\a}{\rm D}_{\b]}{\cal V}|$ & $g(x^{\un{a}})$, $h(x^{\un{a}})$, $v_{\un{b}}(x^{\un{a}})$ & $\CMTB{\{1\}}$, $\CMTB{\{1\}}$, $\CMTB{\{4\}}$  \\ \hline
    3 & ${\rm D}_{[\a}{\rm D}_{\b}{\rm D}_{\g]}{\cal V}|$ & $\chi^{\d}(x^{\un{a}})$ & $\CMTred{\{4\}}$  \\ \hline
    4 & ${\rm D}_{[\a}{\rm D}_{\b}{\rm D}_{\g}{\rm D}_{\d]}{\cal V}|$ & $N(x^{\un{a}})$ & $\CMTB{\{1\}}$   \\ \hline
\end{tabular}
\caption{Explicit Relations between Adinkra Nodes, Component Fields, and Irreps}
\end{table}

\begin{figure}[htp!]
\centering
\includegraphics[width=0.4\textwidth]{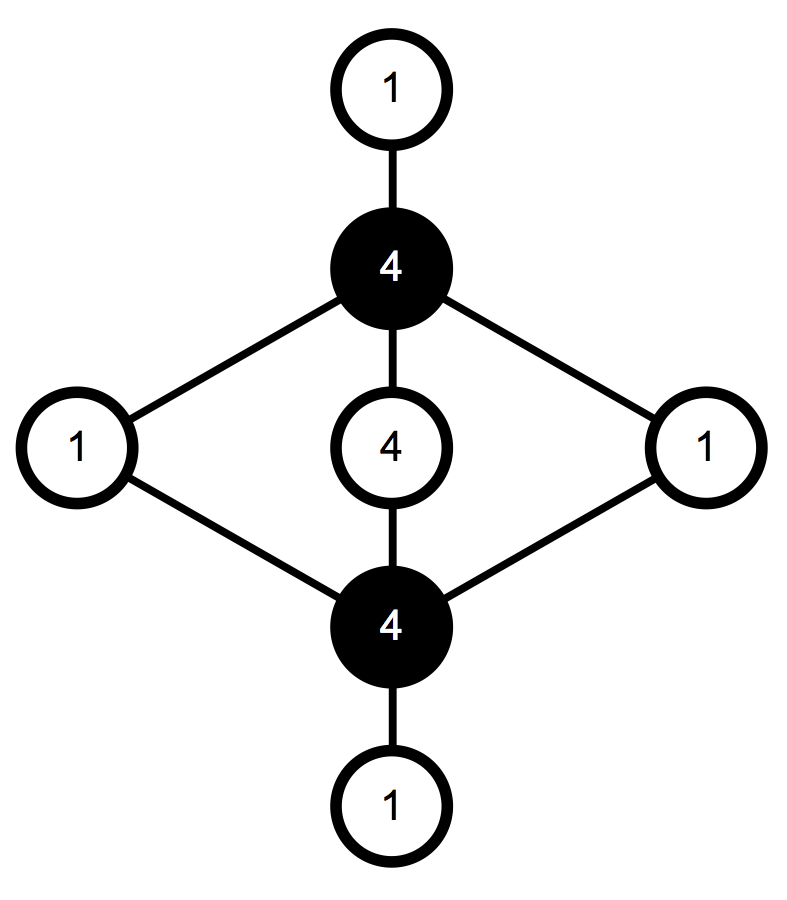}
\caption{Adinkra Diagram for 4D, $\mathcal{N} = 1$\label{Fig:4D}}
\end{figure}

Since the physical d.o.f. of $\theta^{\alpha}$ is $-1/2$, the height assignment describes the the physical d.o.f. of component fields as well. 
The corresponding adinkra diagram is Figure~\ref{Fig:4D}.  The graph shows
the ``$f$'' bosonic component field at the lowest level, the ``$\psi$'' fermionic field at level one, the ``$g$,'' ``$h$,'' and ``$v_{\un{b}}$'' bosonic component fields  at level two, the ``$\chi$'' fermionic field at level three, and finally the ``$N$'' bosonic component field at level four. 

\begin{figure}[htp!]
\centering
\includegraphics[width=0.85\textwidth]{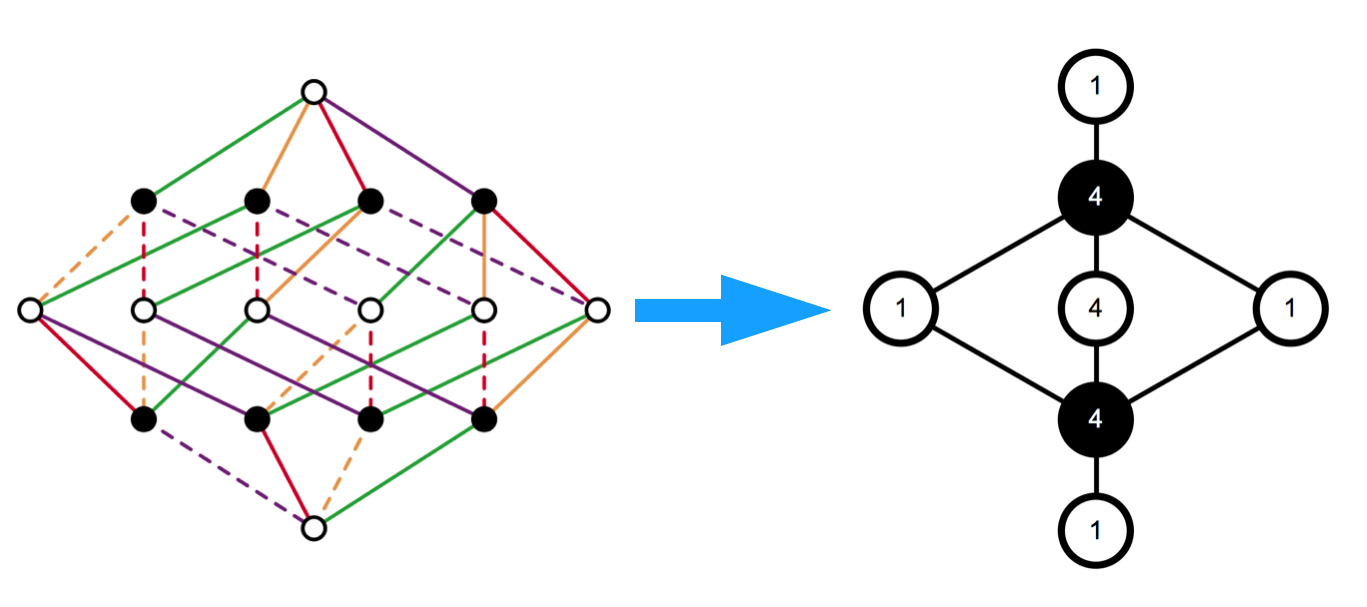}
\caption{From 1D, $\mathcal{N}=4$ Adinkra to 4D, $\mathcal{N}=1$ Adinkra\label{Fig:1Dto4D}}
\end{figure}

Since the 4D, $\cal N$ = 1 theory can be truncated to a 1D, $\cal N$ = 4 theory, we can discuss the relation between the 4D, $\cal N$ = 1 adinkra and the 1D, $\cal N$ = 4 adinkra as shown in Figure~\ref{Fig:1Dto4D}. The 1D, $\cal N$ = 4 maximal supermultiplet and the corresponding adinkras are well-studied. The graph shown to the left graph in Figure~\ref{Fig:1Dto4D} illustrates the 1D $\cal N$ = 4 system. 


By comparison, we see that starting from the 1D, $\cal N$ = 4 adinkra, if we aggregate a set of bosons or fermions into a single node, then a set of corresponding links will be merged (we use black links to replace them), and thus emerges the 4D, $\cal N$ = 1 adinkra.  Of course, to reach this goal we must decide to enforce some rules.  Four and only four black nodes are merged at the first and third levels. At the second level, the only merging choices are either one
or four nodes as permitted in an ``aggregated'' node.
From this procedure, although we have a large number of different 1D, ${\cal N}=4$ adinkras, they all collapse into the same 4D, ${\cal N} = 1$ version.  It is useful to also recall that the aggregation of nodes leads to ``dimensional enhancement'' \cite{Enhnc1,Enhnc2,Enhnc3} that allows the adinkra nodes
to carry representations of the four dimensional Lorentz group.


\begin{figure}[t]
\centering
\includegraphics[width=0.7\textwidth]{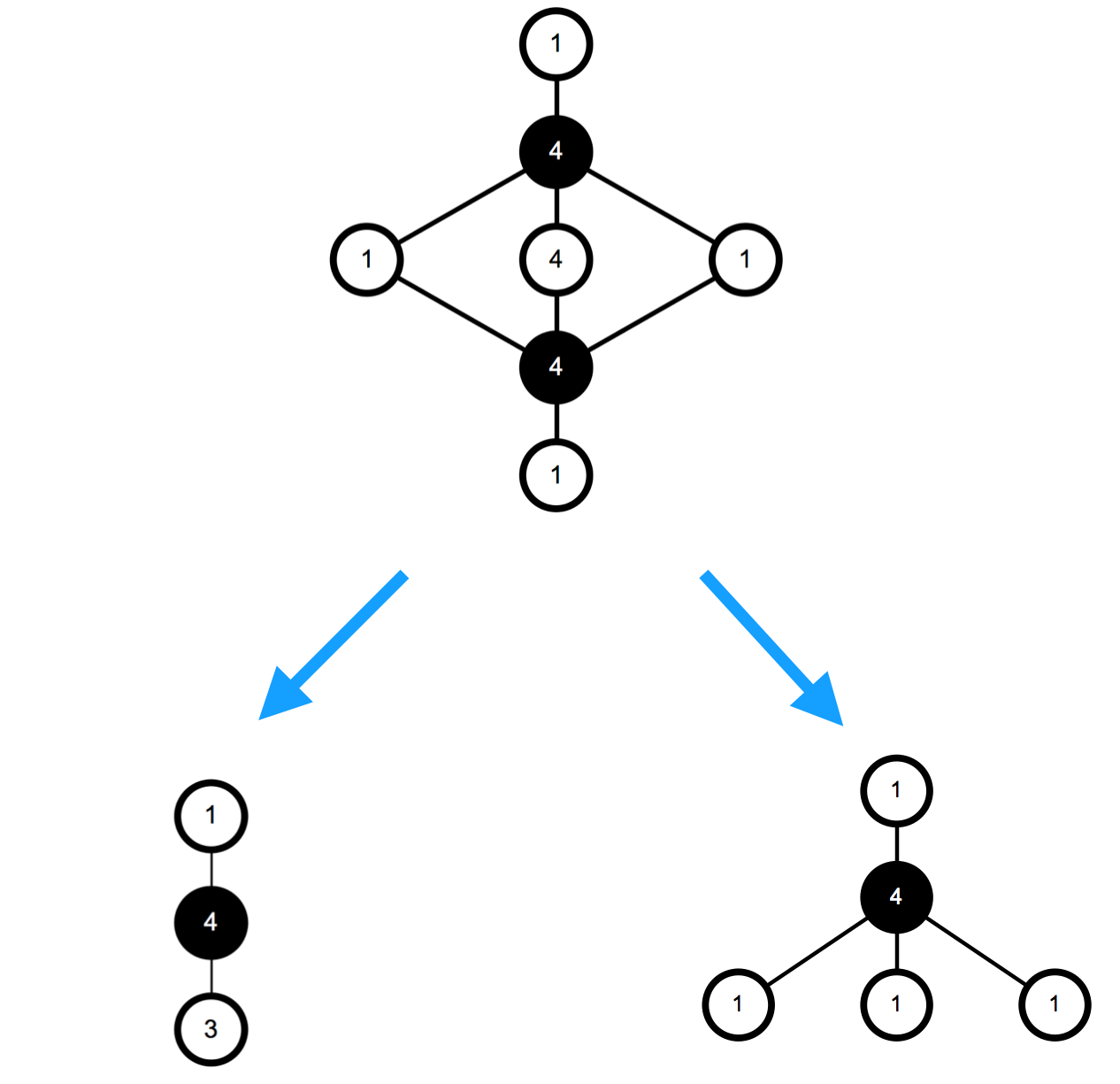}
\caption{From 4D, $\mathcal{N}=1$ Reducible Adinkra to Irreducible Adinkras (Vector and Chiral Supermultiplets)\label{Fig:4DtoVSCS}}
\end{figure}

Before we leave this section, there is another relevant comment.
Figure~\ref{Fig:4D} provides a representation of the data needed to construct the component field description of the 4D, ${\cal N }=1$ unconstrained superfield $\cal V$.  However, this superfield is well known to provide a {\it {reducible}} representation of 4D, ${\cal N }=1$ supersymmetry.  The {\it {irreducible representations}} contained in $\cal V$ are the two adinkra diagrams on the bottom of Figure~\ref{Fig:4DtoVSCS}.

It is useful to know one of the nodes at the lowest level of the bottom right graph in Figure~\ref{Fig:4DtoVSCS} (i.e. chiral supermultiplet) must be regarded as the spacetime integral of a spin-0 field.  The origin of this field was from the initial starting point as being one of the aggregated part of the ``4'' at the middle level of the $\cal V$ adinkra. 
The bottom left graph in Figure~\ref{Fig:4DtoVSCS} has nodes associated with the component fields of the vector supermultiplet where the gauge field is restricted to the Coulomb gauge. 
Refer to Appendix \ref{appen:4DCSVS} for details.

In 4D, there is a comprehensive understanding of how to start with a reducible representation such as the real scalar superfield $\cal V$ and ``break'' it apart into its irreducible components. However, the extension of such a procedure is totally unknown for the cases of 10D superfields.

For a general representation of spacetime supersymmetry, there is currently no understanding of how to carry out the process which has been accomplished in the realm of the 4D, $\cal N$ = 1 supersymmetry.  Resolving this class of problems in the realm of supersymmetric representation theory is a primary motivation for the adinkra approach to the study of superfields.

If we borrow the language of genetics, adinkras play the role of genomes for superfields.  The problem of finding the irreducible off-shell representations of all superfields is equivalent to a genome-editing problem. We are still without the mathematical analog of a
Cas9\footnote{Cas9 (CRISPR associated protein 9) 
is a protein which plays a vital role in the immunological defense of certain bacteria against DNA viruses, and its main function is to cut DNA and therefore alter a cell's genome. 
See for the Wikipedia - Cas9, https://en.wikipedia.org/wiki/Cas9.
} capable of beginning with a {\it {reducible}} graph, that is a viable off-shell representation of spacetime supersymmetry, and ending with off-spring {\it {irreducible}} graphs that also provides viable off-shell representations of spacetime supersymmetry.  An example of
the content of the preceding sentence is shown in Figure (4.3) above.

\newpage
\section{10D Scalar Superfield Decomposition}

In the case of the 10D, $\cal N$ = 1 theory, the number of independent Grassmann coordinates is $2^{10/2-1} = 16$ due to the Majorana-Weyl condition. Then the superspace has coordinates $(x^{\un{a}},\theta^{\a})$, where $\un{a} = 0, 1, \ldots, 9$ and $\a = 1, \ldots, 16$. Hence, the $\theta$-expansion of the ten dimensional scalar superfield begins at Level-0 and continues to Level-16, where Level-$n$ corresponds to the order $\mathcal{O}(\theta^{n})$. The unconstrained real scalar superfield ${\cal V}$ contains $2^{16-1}$ bosonic and $2^{16-1}$ fermionic components, as counted in Table \ref{tab:count}. Let's express the 10D, $\mathcal{N}=1$ scalar superfield as
\begin{equation}
    {\cal V} (x^{\un{a}}, \, \theta^{\a}) ~=~ \varphi^{(0)} (x^{\un{a}}) ~+~ \theta^{\a} \, \varphi^{(1)}_{\a} (x^{\un{a}}) ~+~ \theta^{\a} \theta^{\b} \, \varphi^{(2)}_{\a\b}  (x^{\un{a}}) ~+~ \dots  ~~~.  \label{equ:10DIsuperfield}
\end{equation}
We can decompose $\theta$-monomials $\theta^{[\a_{1}} \cdots \theta^{\a_{n}]}$ into a direct sum of irreducible representations of Lorentz group SO(1,9). With the antisymmetric property of Grassmann coordinates, we have
\begin{equation}
{\cal V} ~=~ \begin{cases}
{~~}{\rm {Level}}-0 \,~~~~~~~~~~ \{ 1 \} ~~~,  \\
{~~}{\rm {Level}}-1 \,~~~~~~~~~~ \{ 16 \} ~~~,  \\
{~~}{\rm {Level}}-2 \,~~~~~~~~~~ \{ 16 \} \, \wedge \, \{ 16 \} ~~~,  \\
{~~}{\rm {Level}}-3 \,~~~~~~~~~~ \{ 16 \} \, \wedge \, \{ 16 \} \, \wedge \, \{ 16 \} ~~~,  \\
{~~~~~~}  {~~~~} \vdots  {~~~~~~~~~\,~~~~~~} \vdots \\
{~~}{\rm {Level}}-n \,~~~~~~~~~ \underbrace{\{ 16 \} \, \wedge \, ~\dots~ \, \wedge \, \{ 16 \}}_\text{$n$ times}  ~~~,  \\
{~~~~~~}  {~~~~} \vdots  {~~~~~~~~~\,~~~~~~} \vdots \\
{~~}{\rm {Level}}-16 ~~~~~~~~~ \{ 1 \} ~~~.
\end{cases}
{~~~~~~~~~~~~~~~}
\end{equation}
All even levels are bosonic representations, while all odd levels are fermionic representations. Note that in a 16-dimensional Grassmann space, the Hodge-dual of a $p$-form is a $(16-p)$-form. Therefore, level-$(16-n)$ is the dual of level-$n$ for $n = 0, \ldots, 8$, and they have the same dimensions. By simple use of the values of the function ``16 choose n,'' these dimensions are found to be the ones that follow,
\begin{equation}
{\cal V} ~=~ \begin{cases}
{~~}{\rm {Level}}-0 \,~~~~~~~~~~ 1 ~~~,  \\
{~~}{\rm {Level}}-1 \,~~~~~~~~~~ 16 ~~~,  \\
{~~}{\rm {Level}}-2 \,~~~~~~~~~~   120~~~,  \\
{~~}{\rm {Level}}-3 \,~~~~~~~~~~ 560  ~~~,  \\
{~~~~~~}  {~~~~} \vdots  {~~~~~~~~~\,~~~~~~} \vdots \\
{~~}{\rm {Level}}-n \,~~~~~~~~~ \frac {16!}{n! (16 - n)!}  ~~~,  \\
{~~~~~~}  {~~~~} \vdots  {~~~~~~~~~\,~~~~~~} \vdots \\
{~~}{\rm {Level}}-16 ~~~~~~~~~  1 ~~~.
\end{cases}
{~~~~~~~~~~~~~~~}
\end{equation}

The $\theta-$monomials contained within the 10D scalar superfield ${\cal V}$ can be specified by the irreducible representations of $\mathfrak{so}(1,9)$ as follows. In Appendix \ref{appen:so10}, the irreps with small dimensions are listed. 
\begin{equation}\label{equ:10DTypeI}
{\cal V} ~=~ \begin{cases}
{~~}{\rm {Level}}-0 \,~~~~~~~~~~ \CMTB {\{ 1 \}} ~~~,  \\
{~~}{\rm {Level}}-1 \,~~~~~~~~~~ \CMTred {\{ 16 \}} ~~~,  \\
{~~}{\rm {Level}}-2 \,~~~~~~~~~~ \CMTB {\{120\}}  ~~~, \\
{~~}{\rm {Level}}-3 \,~~~~~~~~~~ \CMTred {\{ \overline{560} \}}   ~~~, \\
{~~}{\rm {Level}}-4 \,~~~~~~~~~~ \CMTB {\{770\}} \oplus \CMTB {\{1050\}} ~~~,  \\
{~~}{\rm {Level}}-5 \,~~~~~~~~~~ \CMTred {\{672\}} \oplus \CMTred {\{3696\}} ~~~,  \\
{~~}{\rm {Level}}-6 \,~~~~~~~~~~ \CMTB {\{3696'\}} \oplus \CMTB {\{4312\}} ~~~,  \\
{~~}{\rm {Level}}-7 \,~~~~~~~~~~ \CMTred {\{\overline{2640}\}} \oplus \CMTred {\{\overline{8800}\}} ~~~,  \\
{~~}{\rm {Level}}-8 \,~~~~~~~~~~ \CMTB {\{660\}} \oplus \CMTB {\{4125\}} \oplus \CMTB {\{8085\}} ~~~,  \\
{~~}{\rm {Level}}-9 \,~~~~~~~~~~ \CMTred {\{2640\}} \oplus \CMTred {\{8800\}} ~~~, \\
{~~}{\rm {Level}}-10 ~~~~~~~~~ \CMTB {\{\overline{3696'}\}} \oplus \CMTB {\{4312\}} ~~~,  \\
{~~}{\rm {Level}}-11 ~~~~~~~~~ \CMTred {\{\overline{672}\}} \oplus \CMTred {\{\overline{3696}\}} ~~~,  \\
{~~}{\rm {Level}}-12 ~~~~~~~~~ \CMTB {\{770\}} \oplus \CMTB {\{\overline{1050}\}} ~~~,  \\
{~~}{\rm {Level}}-13 ~~~~~~~~~ \CMTred {\{ 560 \}}   ~~~, \\
{~~}{\rm {Level}}-14 ~~~~~~~~~ \CMTB {\{ 120 \}}  ~~~, \\
{~~}{\rm {Level}}-15 ~~~~~~~~~ \CMTred {\{ \overline{16} \}} ~~~,  \\
{~~}{\rm {Level}}-16 ~~~~~~~~~ \CMTB {\{ 1 \}} ~~~.
\end{cases}
\end{equation}
Note that Level-16 to Level-9 are the conjugates of Level-0 to Level-7 respectively, and Level-8 is self-conjugate. That's the meaning of the duality mentioned above.  It is also clear that the dimensions given by the binomial
coefficient $\frac {16!}{n! (16 - n)!} $ ``align'' with the order of irreducible SO(1,9) representations for the first four levels (and the corresponding dual levels) of the 10D, $\cal N$ = 1 scalar superfield $\cal V$.  When the dimension of the wedge product described by the binomial coefficient 
does not align with the dimension of a SO(1,9) irreducible representation, it must be the case that a judiciously 
chosen sum of
irreducible SO(1,9) representations is equal to $\frac {16!}{n! (16 - n)!} $ at the $n$-th level.

Stated in the form of an equation the content of the last paragraph takes the form
\begin{equation} \label{equ:BIN1}
\frac {16!}{n! (16 - n)!} ~=~
\sum_{{\cal R}} \, b_{ \CMTB {\{  {\cal R}  \}  } }    \,  d_{ \CMTB {\{  {\cal R}  \}  } }    ~~~,
\end{equation}
which is satisfied at any even level of the superfield for some choice of coefficients $b_{ \CMTB {\{  {\cal R}  \}  } }  $ and  
representations ${ \CMTB {\{  {\cal R}  \}  }  }$.  Furthermore in this equation $ d_{ \CMTB {\{  {\cal R}  \}  } } $
denotes the dimensionality of the representation ${ \CMTB {\{  {\cal R}  \}  }  }$.

One may wonder why ``bar'' representations occur in 10D, $\mathcal{N}=1$ theory with the absence of dotted spinor indices. Let us turn to the $\CMTred {\{ 16 \}}$ and $\CMTred {\{ \overline{16} \}}$ representations. If we define $\chi^{\a}$ as the $\CMTred {\{ 16 \}}$ representation, then it follows that $\chi^{\dot{\a}}$ is the $\CMTred {\{ \overline{16}\}}$ representation. They both have upper spinor indices. However, it is important to keep in mind that the spinor metric for 10D takes the form $C_{\a\dot{\b}}$ and this means that it is possible to define another spinor that has a subscript index via the equation
\begin{equation}
    \chi_{\a} ~=~ \chi^{\dot{\b}} C_{\a\dot{\b}}  ~~~,
\end{equation}
where obviously $\chi_{\a}$ has a subscript undotted spinor index. So if we define $\chi^{\a}$ with a superscript as the $\CMTred {\{ 16 \}}$ representation, then it follows that $\chi{}_{\a}$ with a subscript is the $\CMTred {\{ \overline{16} \}}$ representation!

This provides a way to understand the meaning of the ``bar'' representations shown in the studies of the ten dimensional superfield. This also tells us the irreps corresponding to the component fields are nothing but the conjugate of the irreps corresponding to the $\theta-$monomials.

We can also rewrite the scalar superfield decomposition result in Equation (\ref{equ:10DTypeI}) in Dynkin labels as follows.
\begin{equation}
{\cal V} ~=~ \begin{cases}
{~~}{\rm {Level}}-0 \,~~~~~~~~~~ \CMTB {(00000)} ~~~,  \\
{~~}{\rm {Level}}-1 \,~~~~~~~~~~ \CMTred {(00001)} ~~~,  \\
{~~}{\rm {Level}}-2 \,~~~~~~~~~~ \CMTB {(00100)}  ~~~, \\
{~~}{\rm {Level}}-3 \,~~~~~~~~~~ \CMTred {(01010)}   ~~~, \\
{~~}{\rm {Level}}-4 \,~~~~~~~~~~ \CMTB {(02000)} \oplus \CMTB {(10020)} ~~~,  \\
{~~}{\rm {Level}}-5 \,~~~~~~~~~~ \CMTred {(00030)} \oplus \CMTred {(11010)} ~~~,  \\
{~~}{\rm {Level}}-6 \,~~~~~~~~~~ \CMTB {(01020)} \oplus \CMTB {(20100)} ~~~,  \\
{~~}{\rm {Level}}-7 \,~~~~~~~~~~ \CMTred {(30001)} \oplus \CMTred {(10110)} ~~~,  \\
{~~}{\rm {Level}}-8 \,~~~~~~~~~~ \CMTB {(40000)} \oplus \CMTB {(00200)} \oplus \CMTB {(20011)} ~~~,  \\
{~~}{\rm {Level}}-9 \,~~~~~~~~~~ \CMTred {(30010)} \oplus \CMTred {(10101)} ~~~, \\
{~~}{\rm {Level}}-10 ~~~~~~~~~ \CMTB {(01002)} \oplus \CMTB {(20100)} ~~~,  \\
{~~}{\rm {Level}}-11 ~~~~~~~~~ \CMTred {(00003)} \oplus \CMTred {(11001)} ~~~,  \\
{~~}{\rm {Level}}-12 ~~~~~~~~~ \CMTB {(02000)} \oplus \CMTB {(10002)} ~~~,  \\
{~~}{\rm {Level}}-13 ~~~~~~~~~ \CMTred {(01001)}   ~~~, \\
{~~}{\rm {Level}}-14 ~~~~~~~~~ \CMTB {(00100)}  ~~~, \\
{~~}{\rm {Level}}-15 ~~~~~~~~~ \CMTred {(00010)} ~~~,  \\
{~~}{\rm {Level}}-16 ~~~~~~~~~ \CMTB {(00000)} ~~~.
\end{cases}
\end{equation}

In the following subsections, we will present two different approaches to obtain the result in Equation (\ref{equ:10DTypeI}). We will also draw the 10D, $\mathcal{N} = 1$ adinkra diagram as we did in the 4D, $\mathcal{N} = 1$ case. 

In the discussion above, it was noted that the problem of finding a judicious choice of irreducible representations of the Lorentz group appropriate for 
level-$n$ of the superfield is at the crux of identifying what component fields actually appear at the level-$n$.  In the following discussions we will show two ways to carry out this process.

\subsection{Handicraft Approach: Fermionic Young Tableaux}

In this section, we will utilize the fermionic Young Tableau (FYT) and its application to obtaining the irreducible Lorentz decompositions of the component fields appearing in the $\theta$-expansion of the ten dimensional scalar superfield\footnote{As we noted in our introduction, the use of Young Tableaux for supersymmetric representation theory was pioneered in the literature \cite{hst1,hst2,hst3} some years ago. }. Since in superspace, there are not only spacetime coordinates but also Grassmann coordinates, we introduce the fermionic Young Tableau as an extension of the normal (bosonic) Young Tableau. In order to distinguish the bosonic Young Tableaux from the fermionic Young Tableaux, we use different colored boxes: Young Tableaux with blue boxes are bosonic and the ones with red boxes are fermionic. Namely, when calculating the dimension of a representation associated with a Young Tableau, we put ``10" into the first box if it is bosonic and ``16" if it is fermionic in 10D. 

One can start with the quadratic level. In $\mathfrak{so}(10)$, we can use Young Tableaux to denote \emph{reducible} representations. The rules of tensor product of two Young Tableaux are still valid. Thus, we have
\begin{equation}
    \CMTred {\yng(1)} ~\otimes~ \CMTred {\yng(1)} ~=~ \CMTred {\yng(2)} ~\oplus~ \CMTred {\yng(1,1)} ~~~,
\end{equation}
where entries in $\CMTred {\yng(2)}$ are completely symmetric in a corresponding set of spinor indices and entries in $\CMTred {\yng(1,1)}$ are completely antisymmetric in these same spinor indices. Therefore, the dimensions of these two representations are 136 and 120 respectively. Moreover, $\CMTred {\yng(2)}$ and $\CMTred {\yng(1,1)}$ are the only Young Tableaux that contain two boxes. By using the Mathematica package LieART \cite{LieART}, one obtains\footnote{Of course simply ways can be used to find these results also.} the following results about tensor product decomposition in $\mathfrak{so}(10)$:
\begin{equation}
    \CMTred {\{16\}} ~\otimes~ \CMTred {\{16\}} ~=~ \CMTB {\{10\}} ~\oplus~ \CMTB {\{120\}} ~\oplus~ \CMTB {\{\overline{126}\}}  ~~~. 
\end{equation}
Then we know the decompositions of $\CMTred {\yng(2)}$ and $\CMTred {\yng(1,1)}$
\begin{align}
    \CMTred {\yng(2)} ~=&~ 
    \CMTB {\{10\}} ~\oplus~ \CMTB {\{\overline{126}\}} 
    ~~~,\label{equ:quadratic-sym} \\
    \CMTred {\yng(1,1)} ~=&~ 
    \CMTB {\{120\}} 
    ~~~, \label{equ:quadratic-anti}
\end{align}
entirely from dimensionality.  This exercise also teaches the lesson that
\begin{equation}
    \CMTred {\{16\}} ~\wedge~ \CMTred {\{16\}} ~=~  \CMTB {\{120\}}   ~~~. 
\end{equation}

Note that the sigma matrices with five vector indices satisfy the self-dual / anti-self-dual identities
\begin{equation}
\begin{aligned}
    (\s_{[5]})_{\a\b} ~=&~ \frac{1}{5!} \e_{[5]}{}^{[\bar{5}]}(\s_{[\bar{5}]})_{\a\b} 
    \,~~~~~~,&~~~ 
    (\s_{[5]})^{\Dot{\a}\Dot{\b}} ~=&~ \frac{1}{5!} \e_{[5]}{}^{[\bar{5}]}(\s_{[\bar{5}]})^{\Dot{\a}\Dot{\b}}  \,~~~~~~, \\
    (\s_{[5]})^{\a\b} ~=&~ -\frac{1}{5!} \e_{[5]}{}^{[\bar{5}]}(\s_{[\bar{5}]})^{\a\b} 
    ~~~,&~~~
    (\s_{[5]})_{\Dot{\a}\Dot{\b}} ~=&~ -\frac{1}{5!} \e_{[5]}{}^{[\bar{5}]}(\s_{[\bar{5}]})_{\Dot{\a}\Dot{\b}}  ~~~.
\end{aligned} \label{equ:5formdual}
\end{equation}
Thus, the degrees of freedom is halved. We denote the 5-forms satisfying the self-dual condition
as $\CMTB{\{\overline{126}\}}$, and 
the 5-forms satisfying the anti-self-dual condition
as $\CMTB{\{126\}}$. Then we have $\CMTB{\yng(1,1,1,1,1)} = \frac{10\times 9\times 8\times 7\times 6}{5!} = \{252\} = \CMTB{\{126\}} \oplus \CMTB{\{\overline{126}\}}$.
This discussion reveals that the reducible $\{ 252 \}$ may be graphically reduced according to
\begin{equation}
\CMTB{\yng(1,1,1,1,1)} ~ = ~ {\CMTB{\yng(1,1,1,1,1)}}_{\,-}  ~ \oplus ~
~{\CMTB{\yng(1,1,1,1,1)}}_{\,+} ~~~,
\end{equation}
as an image.  The level-2 result that we need for the scalar superfield is $\CMTred{\{16\}} \wedge \CMTred{\{16\}}$, which is the totally antisymmetric piece $\CMTred {\yng(1,1)}$.

Next we go to the cubic level. The tensor product of three $\CMTred{ \{16\} }$ is
\begin{equation}
\begin{split}
    \CMTred {\yng(1)} ~\otimes~ \CMTred {\yng(1)} ~\otimes~ \CMTred {\yng(1)} 
    ~=&~ \Big[~ \CMTred {\yng(1)} ~\otimes~ \CMTred {\yng(1)} ~\Big] ~\otimes~ \CMTred {\yng(1)}  \\
    ~=&~ \CMTred {\yng(3)}  ~\oplus~ 2 ~ \CMTred {\yng(2,1)} ~\oplus~ \CMTred {\yng(1,1,1)}  ~~~.
\end{split}
\end{equation}Therefore we obtain
\begin{align}
    \CMTred {\yng(3)}  ~\oplus~ \CMTred {\yng(2,1)} ~=&~ \CMTred {\yng(2)}~\otimes~ \CMTred {\yng(1)} ~=~ \CMTred{ \{ \overline{16} \} } ~\oplus~ 2 \CMTred{ \{ \overline{144} \} } ~\oplus~ \CMTred{ \{ \overline{672} \} } ~\oplus~ \CMTred{ \{ \overline{1200} \} }   ~~~, \label{equ:cubicAB3} \\
    \CMTred {\yng(2,1)} ~\oplus~ \CMTred {\yng(1,1,1)} ~=&~ \CMTred {\yng(1,1)}~\otimes~ \CMTred {\yng(1)} ~=~ \CMTred{ \{ \overline{16} \} } ~\oplus~ \CMTred{ \{ \overline{144} \} } ~\oplus~ \CMTred{ \{ \overline{560} \} } ~\oplus~ \CMTred{ \{ \overline{1200} \} }  ~~~. \label{equ:cubicBC3}
\end{align}
To solve for $\CMTred {\yng(1,1,1)}$, first note that
\begin{equation}
    \CMTred {\yng(1,1,1)} ~\supseteq~ \CMTred {\yng(2,1)} ~\oplus~ \CMTred {\yng(1,1,1)} ~-~ \CMTred {\yng(3)}  ~\oplus~ \CMTred {\yng(2,1)} ~=~ \CMTred{ \{ \overline{560} \} }  ~~~,
\end{equation}
where ``$-$'' means complement including duplicates if we treat each direct sum of irreps as a set of irreps. Now note that the dimension of $\CMTred {\yng(1,1,1)}$ is exactly 560. Therefore, we can solve for all the decompositions in cubic level,
\begin{align}
    \CMTred {\yng(3)} ~=&~ \CMTred{ \{ \overline{144} \} } ~\oplus~ \CMTred{ \{ \overline{672} \} } ~~~, \\
    \CMTred {\yng(2,1)} ~=&~ \CMTred{ \{ \overline{16} \} } ~\oplus~ \CMTred{ \{ \overline{144} \} } ~\oplus~ \CMTred{ \{ \overline{1200} \} } ~~~, \\ 
    \CMTred {\yng(1,1,1)} ~=&~ \CMTred {\{\overline{560}\}} ~~~. \label{equ:cubic-anti}
\end{align}

When we look at the quartic level, we find
\begin{equation}
\begin{split}
    \CMTred {\yng(1)} ~\otimes~ \CMTred {\yng(1)} ~\otimes~ \CMTred {\yng(1)} ~\otimes~ \CMTred {\yng(1)} 
    ~=&~ \Big[~ \CMTred {\yng(1)} ~\otimes~ \CMTred {\yng(1)} ~\otimes~ \CMTred {\yng(1)} ~\Big] ~\otimes~ \CMTred {\yng(1)}  \\
    ~=&~ \Big[~ \CMTred {\yng(1)} ~\otimes~ \CMTred {\yng(1)} ~\Big] ~\otimes~ \Big[~ \CMTred {\yng(1)} ~\otimes~ \CMTred {\yng(1)} ~\Big]  \\
    ~=&~ \CMTred {\yng(4)}  ~\oplus~ 3 ~ \CMTred {\yng(3,1)} ~\oplus~ 2 ~ \CMTred {\yng(2,2)} ~\oplus~ 3 ~ \CMTred {\yng(2,1,1)} ~\oplus~ \CMTred {\yng(1,1,1,1)}  ~~~.
\end{split} \label{equ:quartic16}
\end{equation}
By the hook dimension formula, the tensors with four spinorial indices above have dimensions 3876, 9180, 5440, 7140 and 1820 respectively.  It is more useful to express these in terms of tensors with vector indices. By using the second line of Equation (\ref{equ:quartic16}), from Equations (\ref{equ:quadratic-sym}) and (\ref{equ:quadratic-anti}), we find
\begin{equation}
\begin{split}
    \CMTred {\yng(1)} ~\otimes~ \CMTred {\yng(1)} ~\otimes~ \CMTred {\yng(1)} ~\otimes~ \CMTred {\yng(1)} 
     ~=&~ \CMTB{ \yng(2) } ~\oplus~ \CMTB{ \yng(1,1) } ~\oplus~ 3~\CMTB{ \yng(2,1,1,1,1) } ~\oplus~ 3~\CMTB{ \yng(1,1,1,1,1,1) } ~\oplus~\CMTB{\bf  \cdot } ~\oplus~ \CMTB{ \yng(2,2,2,2,2) } ~\oplus~ \CMTB{ \yng(2,2,2,2,1,1) } \\
     &~\oplus~\CMTB{ \yng(2,2,2,1,1,1,1) } ~\oplus~ \CMTB{ \yng(2,2,1,1,1,1,1,1) } ~\oplus~ \CMTB{ \yng(2,1,1,1,1,1,1,1,1) } ~\oplus~ 2~ \CMTB{ \yng(1,1,1,1) } ~\oplus~ 2~\CMTB{ \yng(2,1,1) } ~\oplus~ 2~\CMTB{ \yng(1,1,1,1,1,1,1,1) }\\
     &~\oplus~ 2~\CMTB{ \yng(2,1,1,1,1,1,1) } ~\oplus~ 2~\CMTB{ \yng(2,2,1,1,1,1) } ~\oplus~ 2~\CMTB{ \yng(2,2,2,1,1) } ~\oplus~\CMTB{ \yng(2,2,2) } ~\oplus~ \CMTB{ \yng(2,2,1,1) } 
\end{split}
\label{equ:fourvectorindices}
\end{equation}
However, the r.h.s. of Equation~(\ref{equ:quartic16}) and (\ref{equ:fourvectorindices}) are not irreducible representations! By LieART, we obtain the irreducible decomposition 
\begin{equation}
\begin{split}
    \CMTred {\yng(1)} ~\otimes~ \CMTred {\yng(1)} ~\otimes~ \CMTred {\yng(1)} ~\otimes~ \CMTred {\yng(1)}  ~=&~ (2) \CMTB {\{1\}} ~\oplus~ (6) \CMTB {\{45\}} ~\oplus~ (3) \CMTB {\{54\}} ~\oplus~ (8) \CMTB {\{210\}} ~\oplus~ \CMTB {\{770\}} \\
    & ~\oplus~ (6) \CMTB {\{945\}} ~\oplus~ \CMTB {\{1,050\}} ~\oplus~ (6) \CMTB {\{\overline{1,050}\}} ~\oplus~ \CMTB {\{\overline{2,772}\}} \\
    & ~\oplus~ (2) \CMTB {\{4,125\}} ~\oplus~ (3) \CMTB {\{5,940\}} ~\oplus~ (3) \CMTB {\{\overline{6,930}\}}
\end{split} \label{equ:quarticirrep}
\end{equation}
In subsequent work, we will develop graphical techniques to get from Equation~(\ref{equ:quartic16}) and (\ref{equ:fourvectorindices}) to Equation (\ref{equ:quarticirrep}).

Similar to Equations~(\ref{equ:cubicAB3}) and (\ref{equ:cubicBC3}), we have 4 independent equations with 5 different Fermionic Young Tableaux with 4 boxes. 
Note that in level-$n$, the number of Young Tableaux with $n$-boxes is the number of integer partition of $n$, $p(n)$. In level-5, 6, 7 and 8, there are 7, 11, 15 and 22 different types of Young Tableaux respectively. 
The number of independent equations in level-$n$ is $p(n)-1$.
This method becomes increasingly tedious. We won't go through the details of the subsequent levels. Although the systems of equations always seem underconstrained, in all levels the restrictions from dimensionality are able to nail down all the solutions in the 10D, $\mathcal{N}=1$ case, and give us back the scalar superfield decomposition in Equation (\ref{equ:10DTypeI}). Thus we call this as a handicraft approach.

\newpage
\subsection{Branching Rules for $\mathfrak{su}(16)\supset \mathfrak{so}(10)$}

At level-$n$ of the 10D, $\mathcal{N}=1$ scalar superfield, we want to decompose the totally antisymmetric product $\CMTred{\{16\}}^{\wedge n}$ to $\mathfrak{so}(10)$ irreps. We are essentially looking at all the one-column Young Tableaux with 16 filled at the first box,
\begin{equation}
    \CMTred{\cdot} ~~~,~~~ {\CMTred{\begin{ytableau} 16 \end{ytableau} }}~~~,~~~ 
     {\CMTred{\begin{ytableau} 16 \\ 15 \end{ytableau} }}~~~,~~~ \cdots ~~~,~~~  {\CMTred{\begin{ytableau} 16 \\ \vdots \\ 1 \end{ytableau} }} ~~~.
\end{equation}
They have dimensions $\begin{pmatrix} 16 \\ n \end{pmatrix}$, $n = 0, 1, \ldots, 16$.
One natural way is to interpret these Young Tableaux as $\mathfrak{su}(16)$ irreps (while they do not correspond to irreps when interpreted in the $\mathfrak{so}(10)$ context), and consider how they branch into $\mathfrak{so}(10)$ irreps. We summarize the relevant branching rules in the following table. They give us the decomposition in Equation (\ref{equ:10DTypeI}).

\begingroup
\def\arraystretch{1.8}
\begin{table}[h!]
\centering
\begin{tabular}{|c|c|c|} \hline
    Level & Irrep in $\mathfrak{su}(16)$ & Irrep(s) in $\mathfrak{so}(10)$ \\ \hline \hline
    0 & $\{1\}$ & $\CMTB{\{1\}}$  \\ \hline
    1 & $\{16\}$ & $\CMTred{\{16\}}$  \\ \hline
    2 & $\{120\}$ & $\CMTB{\{120\}}$  \\ \hline
    3 & $\{560\}$ & $\CMTred{\{\overline{560}\}}$  \\ \hline
    4 & $\{1,820\}$ & $\CMTB{\{770\}} \oplus \CMTB{\{1,050\}}$  \\ \hline
    5 & $\{4,368\}$ & $\CMTred{\{672\}} \oplus \CMTred{\{3,696\}}$  \\ \hline
    6 & $\{8,008\}$ & $\CMTB{\{3,696'\}} \oplus \CMTB{\{4,312\}}$  \\ \hline
    7 & $\{11,440\}$ & $\CMTred{\{\overline{2,640}\}} \oplus \CMTred{\{\overline{8,800}\}}$  \\ \hline
    8 & $\{12,870\}$ & $\CMTB{\{660\}} \oplus \CMTB{\{4,125\}} \oplus \CMTB{\{8,085\}}$  \\ \hline
\end{tabular}
\caption{Summary of $\mathfrak{su}(16)$ to $\mathfrak{so}(10)$ branching rules for level-0 to 8 in 10D, $\mathcal{N}=1$ scalar superfield}
\end{table}
\endgroup

{\em {To our knowledge, the expansion of component fields of a off-shell superfield arising\\ $~~~~~$ by using the branching rules of one ordinary Lie algebra over one of its Lie subalge-\\ $~~~~~$ bra has never been noted in any of the prior literature concerning superfields or super-\\ $~~~~~$ symmetry. However, in the work of \cite{CURT}, one can find the same
   technique pioneered \\ $~~~~~$ by application of this idea to the on-shell component fields in the twelve dimensional\\ $~~~~~$
   context.
}} \vskip0.01in
\noindent
This is one of the major discoveries we are reporting in this research paper.  It provides a clean, precise, and new way to define the component fields in superfields.  It is very satisfying also from the point of view of our use of the ``handicraft'' techniques discussed in the last section.  The two methods yield the same conclusions in all levels.  However, the method in this section is both labor-saving and in a sense more mathematically rigorous.

\subsection{10D, $\mathcal{N} = 1$ Adinkra Diagram}

\begin{figure}[h!]
\centering
\begin{minipage}{0.46\textwidth}
    \centering
    \includegraphics[width=0.88\textwidth]{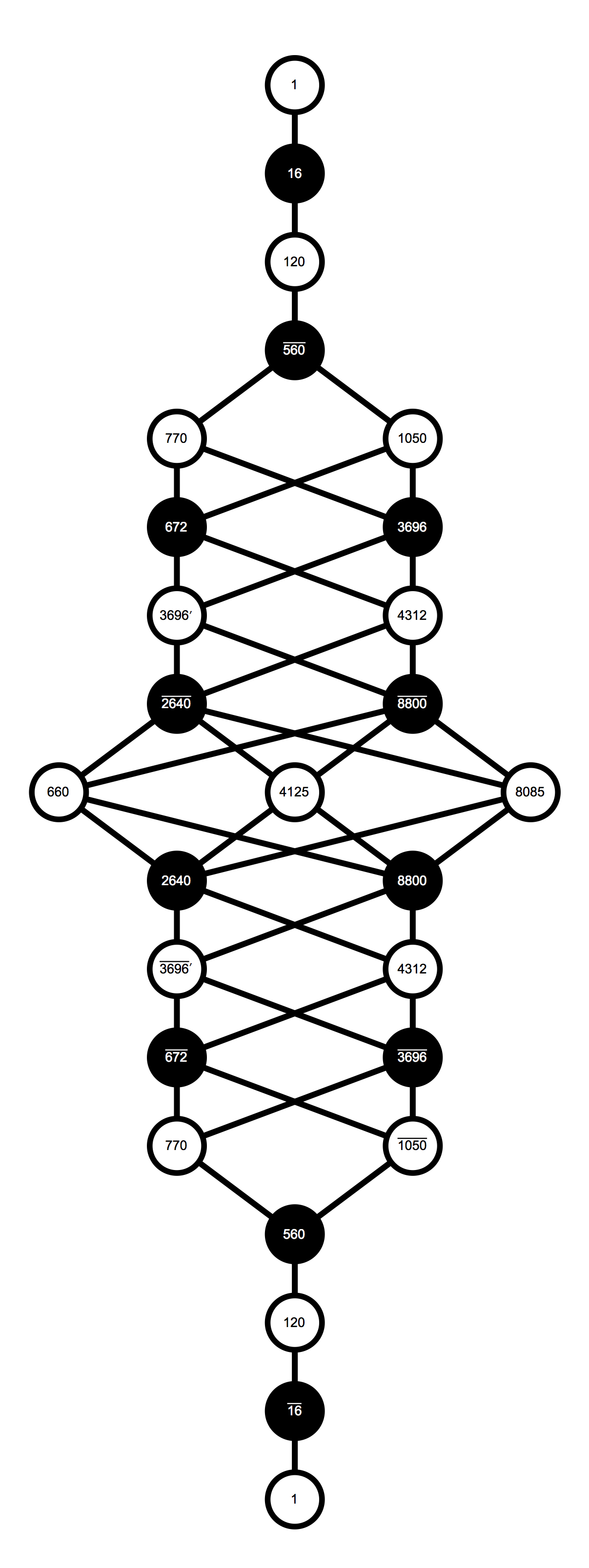}
    \caption{Adinkra with Dimensionality\\
    $~~~~~~~{\,}~~~~~~~~$ Indicated In Nodes}
    \label{Fig:10DTypeI}
\end{minipage}
\begin{minipage}{0.46\textwidth}
    \centering
    \includegraphics[width=0.88\textwidth]{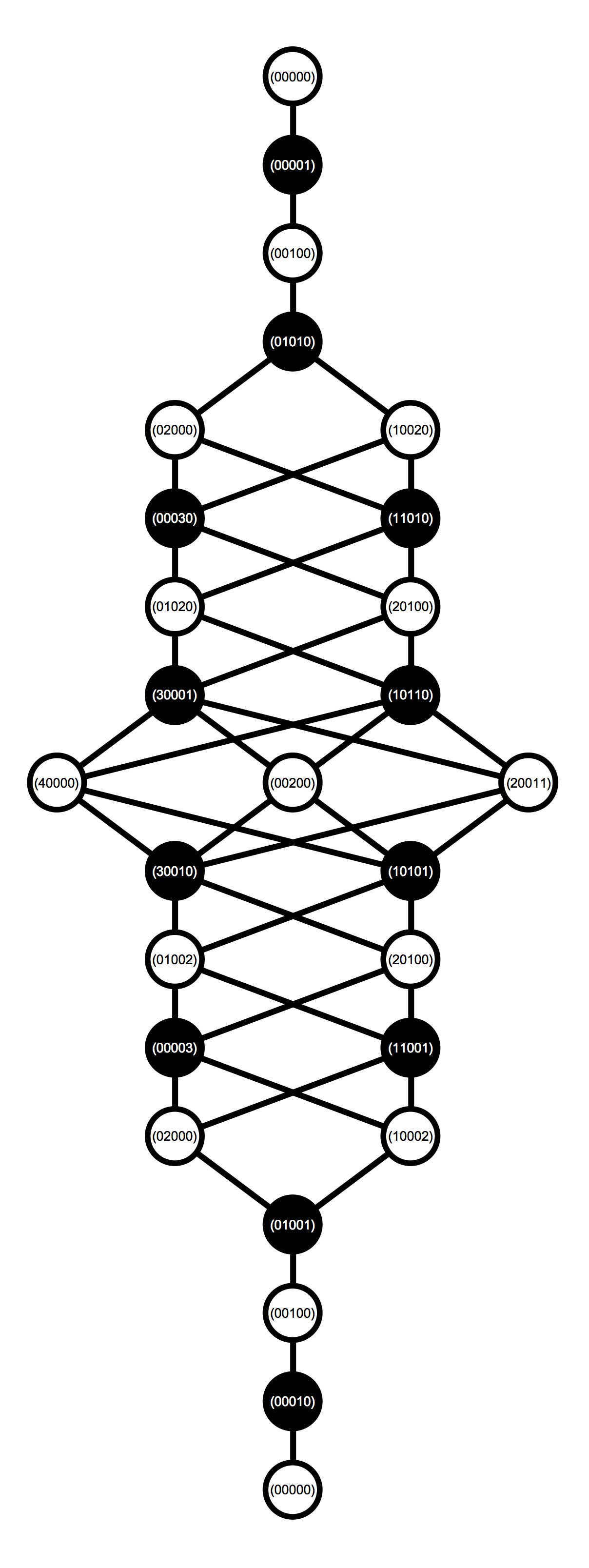}
    \caption{Adinkra with Dynkin Labels \\
    $~~~~~~~{\,}~~~~~~~~$ Indicated In Nodes}
    \label{Fig:10DTypeI_dynkin}
\end{minipage}
\end{figure}

In a manner similar to how we drew the 4D, $\mathcal{N} = 1$ adinkra, we are now in position to explicitly demonstrate the 10D, $\mathcal{N} = 1$ adinkra by the same process. The adinkra with irrep dimensionality shown in each node is drawn in Figure \ref{Fig:10DTypeI}, while the one with the corresponding Dynkin labels appears in Figure \ref{Fig:10DTypeI_dynkin}.


With this result, we have achieved two advances we believe to be significant.

(1.)
The graphs shown in Figure \ref{Fig:10DTypeI} and Figure \ref{Fig:10DTypeI_dynkin} mark the first appearance \newline 
\indent
$~~~$ ${}\,~~$of ten dimensional adinkras in the literature.

(2.)
The graphs shown in Figure \ref{Fig:10DTypeI} and Figure \ref{Fig:10DTypeI_dynkin} mark an independent con-\newline 
\indent
$~~~$ ${}\,~~$firmation of the component-level Lorentz representation
content of a 10D, 
\newline 
\indent
$~~~$ ${}\,~~$ $\cal N$ = 1 scalar superfield given by Bergshoeff and de Roo
\cite{10DScLR}.

Considering the case of 4D, $\cal N$ = 1 supersymmetry, the graphs shown in Figure \ref{Fig:10DTypeI} and Figure \ref{Fig:10DTypeI_dynkin} for the
10D, $\cal N$ = 1 scalar superfield should be considered as the equivalents to the graph shown in Figure~\ref{Fig:4D} for the 4D, $\cal N$ = 1 scalar superfield.  The obvious difference in ``heights'' was to be expected.  However, a surprising feature is the maximum width of the 10D, $\cal N$ = 1 scalar superfield adinkra is exactly the same as that of the 4D, $\cal N$ = 1 scalar superfield, namely only three bosonic nodes. 
Another surprising feature of the 10D, $\cal N$ = 1 scalar superfield is its economy.
It only contains 15 independent bosonic field representations and 
12 independent fermionic field representations.

A great and obvious challenge here is to discover the analog of the rules that govern the splitting that was described as to how 4D, $\cal N$ = 1 chiral (irreducible) and 4D, $\cal N$ = 1 vector (irreducible) adinkras can be obtained from a 4D, $\cal N$ = 1 real scalar (reducible) adinkra.

\subsection{The 10D, $\mathcal{N} = 1$ Adinkra \& Nordstr\" om SG Components}

In our previous work of \cite{NordSG}, we found that the construction of 10D, $\mathcal{N} = 1$ supergravitation at the level of superfields and
in terms of linearized supergeometry is very direct and without obstructions.  It involved the introduction of a 10D, $\mathcal{N} = 1$ scalar superfield $\Psi$ that appears in linearized frame operators in the forms
\begin{flalign}  {~~~~~~~~~~~~~~~~~~~~~~~~~~~~~~}
{\rm E}{}_{\alpha} &= {\rm D}_{\alpha}+ \fracm 12 \Psi {\rm D}_{\alpha}  ~~~, &&\\
{\rm E}{}_{\un{a}} &= \pa_{\un{a}} + \Psi\pa_{\un{a}} - i \fracm 25    
(\sigma_{\un{a}})^{\alpha\beta}
({\rm D}_{\alpha}\Psi){\rm D}_{\beta}  ~~~.
\end{flalign}
Taking the $\theta$ $\to$ 0 limit to the second of these equations implies that
\begin{flalign}  {~~~}
{\rm E}{}_{\un{a}}{\Big |} &= {\Big [ } ~ 1 ~+~ \Psi~ {\Big ] } {\Big |}
\d{}_{\un {a}}{}^{\un{m}} \pa{}_{\un {m}} ~+~ {\Big [}
- i \fracm 25  (\sigma_{\un{a}})^{\alpha\beta}
({\rm D}_{\alpha}\Psi) {\Big ]} {\Big |} \, {\rm D}_{\beta} ~~~,
\end{flalign}
where ${\Big |}$ is a notation used to indicate the taking of
the limit.  The left hand side of this equation can be defined as 
\begin{flalign}  {~~~}
{\rm E}{}_{\un{a}}{\Big |} &= {\rm e}{}_{\un {a}}{}^{\un{m}} \pa{}_{\un {m}}
~+~ \tilde{\psi}_{\un {a}}{}^{\b}\, {\rm D}_{\beta}
~~~,
\end{flalign}
with ${\rm e}{}_{\un {a}}{}^{\un{m}} $ describing
the 10D graviton (expressed as a frame field) and
$\tilde{\psi}_{\un {a}}{}^{\b}$
describing the 10D gravitino. 
Combining the last two equations yields
\begin{flalign} 
\label{equ:Nord_fram_gravitino}
{\rm e}{}_{\un {a}}{}^{\un{m}} ~=~
{\Big [ } ~ 1 ~+~ \Psi~ {\Big ] } {\Big |}
\d{}_{\un {a}}{}^{\un{m}}
~~~,~~~
\tilde{\psi}_{\un {a}}{}^{\b} ~=~ 
 {\Big [}
- i \fracm 25  (\sigma_{\un{a}})^{\alpha\beta}
({\rm D}_{\alpha}\Psi) {\Big ]} {\Big |}
~~~.
\end{flalign}
In Nordstr\" om theory, only the non-conformal spin-$0$ part of graviton described
by a scalar component field and the non-conformal spin-$\frac{1}{2}$ part of the gravitino
\begin{equation}
\psi_{\b} ~\equiv~ (\s^{\un{a}})_{\b\g} \tilde{\psi}_{\un{a}}{}^{\g} ~~~, 
\label{equ:nonconfgravitino}
\end{equation}
which is described
by a $\CMTred{ \{\Bar {16} \}}$ component field show up.  Both of these
are appropriate for a Nordstr\" om-type theory.

Two of the results in the work of \cite{NordSG} take
the respective forms
\begin{equation}
    T_{\a\un{b}}{}^{\g} ~=~ -~ \frac{3}{10} \d_{\a}{}^{\g} (\pa_{\un{b}}\Psi) ~+~ \frac{3}{10} (\s_{\un{b}}{}^{\un{c}})_{\a}{}^{\g} (\pa_{\un{c}}\Psi) ~+~ i \frac{1}{160} \left[ -(\s^{[2]})_{\a}{}^{\g}(\s_{\un{b}[2]})^{\b\d} ~+~ \frac{1}{3} (\s_{\un{b}[3]})_{\a}{}^{\g}(\s^{[3]})^{\b\d} \right]  G_{\b\d} ~~~,
\end{equation}
and
\begin{equation}
    R_{\a\b}{}^{\un{d}\un{e}} ~=~ -~ i \frac{6}{5}(\s^{[\un{d}})_{\a\b} (\partial^{\un{e}]}\Psi) ~-~ \frac{1}{80} \left[ \frac{1}{3!} (\s^{\un{d}\un{e}[3]})_{\a\b} (\s_{[3]}){}^{\g\d} ~+~ (\s^{\un{a}})_{\a\b} (\s_{\un{a}}{}^{\un{d}\un{e}})^{\g\d} \right] G_{\g\d}  ~~~.
\end{equation}
where $G_{\a\b} = ( [ {\rm D}_{\a}, {\rm D}_{\b} ] \Psi )$.
In the $\theta$ $\to$ 0 limit, the first two terms of each of these equations involve the spacetime derivative of the scalar graviton in the theory.  The final two terms in each of these equations involve the quantity $G_{\a\b} = ( [ {\rm D}_{\a}, {\rm D}_{\b} ] \Psi )$ corresponding to the Level-2 $\CMTB{ \{ {120}  \}}$ component field node shown in either of the adinkras shown in Figure \ref{Fig:10DTypeI} or Figure \ref{Fig:10DTypeI_dynkin}.  It is the lowest dimensional auxiliary field of Nordstr\" om SG.

Application of a spinorial derivative to $G_{\a\b}$ (i.e. $ {\rm D}_{[{\g}_1} {\rm D}_{\a} {\rm D}_{\b]} \Psi$) yields the Level-3 $\CMTred{ \{ {560}  \}}$ fermionic component field node. Application of two spinorial derivatives to $G_{\a\b}$ (i.e. $ {\rm D}_{[ {\g}_1} {\rm D}_{{\g}_2 } {\rm D}_{\a} {\rm D}_{\b]} \Psi$) yields the Level-4 $\CMTB{ \{ {770}  \}}$ and $\CMTB{ \{\Bar {1050}  \}}$ bosonic component field nodes. Application of three spinorial derivatives to $G_{\a\b}$ (i.e. $ {\rm D}_{[ {\g}_1} {\rm D}_{{\g}_2} {\rm D}_{{\g}_3 } {\rm D}_{\a} {\rm D}_{\b]} \Psi$) yields the Level-5 $\CMTR{ \{\Bar {672}  \}}$ and $\CMTR{ \{\Bar {3696}  \}}$ fermionic component field nodes. Application of four spinorial derivatives to $G_{\a\b}$ (i.e. $ {\rm D}_{[ {\g}_1} {\rm D}_{{\g}_2} {\rm D}_{{\g}_3} {\rm D}_{{\g}_4 } {\rm D}_{\a} {\rm D}_{\b]} \Psi$) yields the Level-6 $\CMTB{ \{\Bar {3696}  \}}$ and $\CMTB{ \{{4312}  \}}$ bosonic component field nodes. Application of five spinorial derivatives to $G_{\a\b}$ (i.e. $ {\rm D}_{[ {\g}_1} {\rm D}_{{\g}_2} {\rm D}_{{\g}_3}  {\rm D}_{{\g}_4} {\rm D}_{{\g}_5 } {\rm D}_{\a} {\rm D}_{\b]} \Psi$) yields the Level-7 $\CMTR{ \{\Bar {2640}  \}}$ and $\CMTR{ \{\Bar {8800}  \}}$ fermionic component field nodes.  Application of six spinorial derivatives to $G_{\a\b}$ (i.e. $ {\rm D}_{[ {\g}_1} {\rm D}_{{\g}_2} {\rm D}_{{\g}_3}  {\rm D}_{{\g}_4} {\rm D}_{{\g}_5}  {\rm D}_{{\g}_6 } {\rm D}_{\a} {\rm D}_{\b]} \Psi$) yields the Level-8 $\CMTB{ \{\Bar {660}  \}}$, $\CMTB{ \{\Bar {4125}  \}}$, and $\CMTB{ \{\Bar {8085}  \}}$ bosonic component field nodes.

From the form of the adinkras, we know that application of more than six fermionic derivatives to $G_{\a\b}$ (or eight to $\Psi$) only leads to the dual representations occurring.  So for example seven
fermionic derivatives applied to $G_{\a\b}$ (or nine to $\Psi$) must lead to the dual representations of those that occurred with the application of five fermionic derivatives (or seven to $\Psi$).  Thus, all the component fields of 10D, $\cal N$ = 1 Nordstr\" om SG are obtained from (\ref{equ:Nord_fram_gravitino}) as well as applying all possible spinorial derivatives to the superspace covariant supergravity field strength $G_{\a\b}$ followed by taking the $\theta$ $\to$ 0 limit.
 
Thus, the knowledge of the 10D, $\cal N$ = 1  scalar superfield adinkras provides an avenue to the complete component field description of the corresponding reducible Nordstr\" om supergravity theory derived from superspace geometry \cite{NordSG}.
In particular, the knowledge that we know explicitly what representations can appear 
above the $\CMTB{\{ 120 \} }$ determines (up to some constants) the form of the
solution of the superspace Bianchi Identities for the Nordstr\" om theory.  When
this is explicitly carried out, it opens the pathway to begin an exploration of 
dynamics in this context.  However, this route to dynamics is similar in spirit to
that of 10D, $\cal N$ = 2B SG \cite{SG2Bsp8c1,SG2Bsp8c2,SG2Bsp8c3} where there is
no action, but the equations for the dynamical fields in the supermultiplet are derived from analysis of the superspace Bianchi Identities alone.  There is no
a priori action required.

This is work for the future.

\newpage
\section{10D Breitenlohner Approach}
In this chapter, we will apply the Breitenlohner approach to construct candidates for the 10D Type I superconformal multiplet.  
The principle of this approach is to attach bosonic and fermionic indices on the 10D scalar superfield, and search for the traceless graviton and the traceless gravitino. 

Recall that the first off-shell description of 4D, $\cal N$ = 1 supergravity
was actually carried out by Breitenlohner \cite{B1}, who took an approach equivalent to starting
with the component fields of the Wess-Zumino gauge 4D, $\cal N$ = 1 vector supermultiplet 
$(v{}_{\un a}, \, \l_{\b}, {\rm d})$ together with their familiar SUSY transformation laws
\be \eqalign{
{\rm D}_{\a} \, v{}_{\un a} ~&=~  (\g_{\un a}){}_{\a} {}^{\b} \,  \l_{\b}  ~~~, ~~~ {~~~ 
~~~~~~~~~~~~~~~~~~~~~~~~~~~~~~~~} \cr
{\rm D}_{\a} \l_{\b} ~&=~   - \,i \, \fracm 14 ( [\, \g^{\un a}\, , \,  \g^{\un b} 
\,]){}_{\a\b} \, (\,  \pa_{\un a}  \, v{}_{\un b}    ~-~  \pa_{\un b} \, v{}_{\un a}  \, )
~+~  (\g^5){}_{\a\b} \,    {\rm d} ~~, {~~~~~~} \cr
{\rm D}_{\a} \, {\rm d} ~&=~  i \, (\g^5 \g^{\un a} ){}_{\a} {}^{\b} \, 
\,  \pa_{\un a} \l_{\b}  ~~~, \cr
} \label{V1}
\ee
followed by choosing as the gauge group the spacetime translations, SUSY generators, and the
spin angular momentum generators as well as allowing additional internal symmetries.  For the 
spacetime translation, this requires a series of replacements of the fields according to
\be  \eqalign{
 v{}_{\un a} ~& \to ~  h{}_{\un{a} \, \un{b}} ~~~,   ~~~
  \l_{\b} ~ \to ~ \psi{}_{\un{b} \, \b}  \,~~~,   ~~~
 {\rm d} ~ \to ~ A{}_{\un b}  ~~~~~, 
}  \label{V2} \ee
(in the notation in \cite{B1}, $A{}_{\un b} $ is $B^5{}_{\m}$). Because the vector supermultiplet
is off-shell (up to WZ gauge transformations), the resulting supergravity theory
is off-shell and includes a redundant set of auxiliary component fields,  i.\ e.\ this is
not an irreducible description of supergravity.  But as seen from (\ref{V2}), the
supergravity fields are all present, and together with the remaining component fields
a complete superspace geometry can be constructed.

The key of this process is that if you look at Equation~(\ref{equ:4D}), there is a $\CMTB{\{4\}}$ irrep (which is a vector gauge field) in level-2. When you consider the expansion of the superfield with one vector index $H^{\un a}$ which is also called the prepotential, you will find 
\begin{equation}
    \CMTB{\{4\}} ~\otimes~ \CMTB{\{4\}} ~=~ \CMTB{\{1\}} ~\oplus~ \CMTB{\{3\}} ~\oplus~ \CMTB{\{\overline{3}\}} ~\oplus~ \CMTB{\{9\}} ~~~, \label{equ:4Dbreitenlohner}
\end{equation}
where $\CMTB{\{9\}}$ is the traceless graviton, as the degree of freedom is $9 = \frac{4\times 5}{2} - 1$. 

In 10D, $\mathcal{N}=1$ theory, we want to search for the traceless graviton $h_{\un{a}\un{b}}$ and the traceless gravitino $\psi_{\un{a}}{}^{\b}$. Let us first figure out which to which irreducible representations they correspond. The graviton has two symmetric vector indices, which corresponds to $\frac{10\times 11}{2} = 55$, and can be decomposed into the traceless part and the trace,
\begin{equation}
\begin{split}
    \tilde{h}_{\un{a}\un{b}} ~~=&~~ h_{\un{a}\un{b}} ~+~ \eta_{\un{a}\un{b}} \, h   ~~~, \\
    \{55\} ~=&~ \CMTB{\{54\}} ~\oplus~ \CMTB{\{1\}}  ~~~.
\end{split}
\end{equation}
where both $\CMTB{\{54\}}$ and $\CMTB{\{1\}}$ are SO(1,9) irreps, while $\{55\}$ is not. In the following, we will denote the traceless graviton irrep to be $\CMTgrn{\bf \{54\}}$, where the green color is added just to highlight this irrep. To determine the irreducible representations corresponding to the gravitino $\tilde{\psi}_{\un{a}}{}^{\b}$, first note that it has one vector index which corresponds to $\CMTB{\{10\}}$ and one {\em upper} spinor index which corresponds to $\CMTred {\{ 16 \}}$. We can then decompose the gravitino into its traceless part and its $\s$-trace part by the tensor product of these two irreps,
\begin{equation}
\begin{split}
    \tilde{\psi}_{\un{a}}{}^{\b}  ~~~~~~=&~~  \psi_{\un{a}}{}^{\b} ~+~ \fracm{1}{10} (\s_{\un{a}})^{\b\g} \psi_{\g}    ~~~, \\
    \CMTB{ \{10\} } \otimes \CMTred {\{ 16 \}} ~=&~ \CMTred {\{ \overline{144} \}} ~\oplus~ \CMTred {\{ \overline{16} \}}  ~~~,
\end{split}
\end{equation}
where the non-conformal spin-$\frac{1}{2}$ $\s$-trace part $\psi_{\g}$ is defined in Equation (\ref{equ:nonconfgravitino}). The traceless part $\psi_{\un{a}}{}^{\b}$ of the gravitino is conformal, and this $\CMTred {\{ \overline{144} \}}$ is what we should search for in the superconformal prepotential multiplet.

Now, if we hope to follow the exact same method in Equation (\ref{equ:4Dbreitenlohner}), we would immediately confront the problem of the lacking of $\CMTB{\{10\}}$ in the expansion in Equation~(\ref{equ:10DTypeI}). Therefore, we cannot introduce the 10D vector superfield as the prepotential and construct the off-shell conformal supergravity supermultiplet in the 10D, $\mathcal{N} = 1$ case right away. Instead, we attach other combinations of bosonic and fermionic indices onto the scalar superfields and search for traceless gravitons $\CMTgrn{\{54\}}$ and gravitinos  $\CMTred{\{\overline{144}\}}$ directly.

Given the Lorentz component content of a scalar superfield, how do we know the components of the superfields with various bosonic (or fermionic) indices? Let's write out the scalar superfield $\theta-$expansion in Equation (\ref{equ:10DIsuperfield}) again,
\begin{equation}
    {\cal V} ~=~ \varphi^{(0)} ~+~ \theta^{\a} \, \varphi^{(1)}_{\a} ~+~ \theta^{\a} \theta^{\b} \, \varphi^{(2)}_{\a\b} ~+~ \dots  ~~~.  
\end{equation}
Attaching indices onto the scalar superfield would mean
\begin{equation}
    {\cal V}_{[\text{indices}]} ~=~ \varphi^{(0)}_{[\text{indices}]} ~+~ \theta^{\a} \, \varphi^{(1)}_{\a \, [\text{indices}]} ~+~ \theta^{\a} \theta^{\b} \, \varphi^{(2)}_{\a\b \, [\text{indices}]} ~+~ \dots  ~~~.  
\end{equation}
where [indices] could be any combination of bosonic and/or fermionic indices, and it would correspond to a (sum of) bosonic or fermionic irreducible representation(s) in SO(1,9). If we let the level-$n$ $\theta-$monomial decompositions of 10D, $\mathcal{N}=1$ scalar superfield be $\ell_{n}$ for $n = 0, 1, \dots, 16$, the level-$n$ component field content would be $\overline{\ell_{n}}$. For simplicity, consider only the [indices] that corresponds to one and only one irreducible representation \{irrep\}. Then the level-$n$ component field content of ${\cal V}_{[\text{indices}]}$ would be $\overline{\ell_{n}} \,\otimes\,$\{irrep\}. We denote the {\em component} content of the entire superfield by the expression ${\cal V} \,\otimes\,$\{irrep\}, which we will use throughout the following sections.

In the following two sections, we will study the expansions of some bosonic and fermionic superfields respectively to see whether we can identify possible off-shell supergravity supermultiplet(s). This is the meaning of the term ``scanning.''  Once
more we emphasize that our arguments are totally group theoretical and therefore
do not rely on dynamical assumptions.

\newpage
\subsection{Bosonic Superfields}

We will start by attaching bosonic indices on the 10D, $\cal N$ = 1 scalar superfield, which is equivalent to tensoring the corresponding bosonic irreps to the component decomposition of the scalar superfield.
Here we explicitly study these bosonic irreps: $\CMTB{\{10\}}$, $\CMTB{\{45\}}$, $\CMTB{\{54\}}$, $\CMTB{\{120\}}$, $\CMTB{\{126\}}$, $\CMTB{\{\overline{126}\}}$, $\CMTB{\{210\}}$, $\CMTB{\{210'\}}$, $\CMTB{\{320\}}$, $\CMTB{\{660\}}$ and $\CMTB{\{770\}}$. 
The summary of their expansions is listed in Table~\ref{Tab:tensoring_bosonic}. The third column shows which tensor product contributes to $\CMTgrn{\bf\{54\}}$, and the decomposition results show that each one only provides one $\CMTgrn{\bf\{54\}}$.

\begingroup
\def\arraystretch{1.8}
\begin{table}[h!]
\centering
\begin{tabular}{|c|c|c|} \hline
    Dynkin Label & Irrep & Is there $\CMTgrn{\bf\{54\}}$? \\ \hline \hline
    $\CMTB{(10000)}$ & $\CMTB{\{10\}}$ & no \\ \hline
    $\CMTB{(01000)}$ & $\CMTB{\{45\}}$ & no \\ \hline
    \multirow{3}{*}{ $\CMTB{(20000)}$ } & \multirow{3}{*}{ $\CMTB{\{54\}}$ } & level-0: $\CMTB{\{1\}} \otimes \CMTB{\{54\}}$ \\
     & & level-4: $\CMTB{\{770\}} \otimes\CMTB{\{54\}}$ \\
     & & level-8: $\CMTB{\{660\}} \otimes\CMTB{\{54\}}$ \\ \hline
    \multirow{2}{*}{ $\CMTB{(00100)}$ } & \multirow{2}{*}{ $\CMTB{\{120\}}$ } & level-2: $\CMTB{\{120\}} \otimes \CMTB{\{120\}}$ \\
     & & level-6: $\CMTB{\{4312\}} \otimes \CMTB{\{120\}}$ \\ \hline
     $\CMTB{(00020)}$  & $\CMTB{\{126\}}$ & no \\ \hline
    $\CMTB{(00002)}$ & $\CMTB{\{\overline{126}\}}$ & no \\ \hline
    \multirow{2}{*}{ $\CMTB{(00011)}$ } & \multirow{2}{*}{ $\CMTB{\{210\}}$ } & level-4: $\CMTB{\{\overline{1050}\}} \otimes \CMTB{\{210\}}$ \\
     & & level-8: $\CMTB{\{8085\}} \otimes \CMTB{\{210\}}$ \\ \hline
    $\CMTB{(30000)}$ & $\CMTB{\{210'\}}$ & no \\ \hline
    \multirow{2}{*}{ $\CMTB{(11000)}$ } & \multirow{2}{*}{ $\CMTB{\{320\}}$ } & level-2: $\CMTB{\{120\}} \otimes \CMTB{\{320\}}$ \\
     & & level-6: $\CMTB{\{4312\}} \otimes \CMTB{\{320\}}$ \\ \hline
    $\CMTB{(40000)}$ & $\CMTB{\{660\}}$ & level-8: $\CMTB{\{660\}} \otimes \CMTB{\{660\}}$ \\ \hline
    \multirow{2}{*}{ $\CMTB{(02000)}$ } & \multirow{2}{*}{ $\CMTB{\{770\}}$ } & level-4: $\CMTB{\{770\}} \otimes \CMTB{\{770\}}$ \\
     & & level-8: $\CMTB{\{4125\}} \otimes \CMTB{\{770\}}$ \\ \hline
\end{tabular}
\caption{Summary of tensoring bosonic irreps onto the scalar superfield\label{Tab:tensoring_bosonic}}
\end{table}
\endgroup

Of course, the result for ${\cal V} \otimes\CMTB{\{54\}} $ is not surprising and can be removed as a candidate for a conformal 10D, $\cal N$ = 1 supergravity prepotential.  All of the other entries in the table do provide candidate prepotentials.  It is noticeable that several of these candidates ( $ {\cal V} \otimes \CMTB{\{210\}} $,  ${\cal V} \otimes \CMTB{\{660\}} $,  and ${\cal V} \otimes\CMTB{\{770\}} $) allow conformal graviton embeddings at the eighth level.
If Wess-Zumino gauges exist to eliminate lower order bosons, these imply ``short'' supergravity multiplets.

Now, one can look into one of those bosonic superfields with traceless gravitons $\CMTgrn{\bf\{54\}}$ in detail as an example. The irreps shown here are {\em component} fields. The reader can find other bosonic superfields with traceless gravitons in Appendix~\ref{appen:bosonic_superfield}. 
\begingroup
\small
\begin{equation}\label{equ:Psitimes120}
{\cal V} \otimes \CMTB{\{120\}} ~=~ \begin{cases}
{~~}{\rm {Level}}-0 \,~~~~~~~ \CMTB{\{120\}} ~~~,  \\
{~~}{\rm {Level}}-1 \,~~~~~~~ \CMTred{\{16\}} \oplus \CMTred{\{144\}} \oplus \CMTred{\{560\}} \oplus \CMTred{\{1200\}} ~~~,  \\
{~~}{\rm {Level}}-2 \,~~~~~~~ \CMTB{\{1\}} \oplus \CMTB{\{45\}} \oplus \CMTgrn{\bf \{54\}} \oplus (2) \CMTB{\{210\}} \oplus \CMTB{\{770\}} \oplus \CMTB{\{945\}} \oplus \CMTB{\{1050\}}  \\
~~~~~~~~~~~~~~~~~~~~~~~ \oplus \CMTB{\{\overline{1050}\}} \oplus \CMTB{\{4125\}} \oplus \CMTB{\{5940\}}  ~~~, \\
{~~}{\rm {Level}}-3 \,~~~~~~~ \CMTred{\{\overline{16}\}} \oplus (2) \CMTred{\{\overline{144}\}} \oplus (2) \CMTred{\{\overline{560}\}} \oplus \CMTred{\{\overline{672}\}} \oplus \CMTred{\{\overline{720}\}} \oplus (2) \CMTred{\{\overline{1200}\}} \oplus \CMTred{\{\overline{1440}\}} \\
~~~~~~~~~~~~~~~~~~~~~~~ \oplus (2) \CMTred{\{\overline{3696}\}} \oplus \CMTred{\{\overline{8064}\}} \oplus \CMTred{\{\overline{8800}\}} \oplus \CMTred{\{\overline{11088}\}} \oplus \CMTred{\{\overline{25200}\}} ~~~, \\
{~~}{\rm {Level}}-4 \,~~~~~~~ (2) \CMTB{\{120\}} \oplus \CMTB{\{\overline{126}\}} \oplus (2) \CMTB{\{320\}} \oplus (3) \CMTB{\{1728\}} \oplus (2) \CMTB{\{2970\}}  \\
~~~~~~~~~~~~~~~~~~~~~~~ \oplus (3) \CMTB{\{3696'\}} \oplus \CMTB{\{\overline{3696'}\}} \oplus (2) \CMTB{\{4312\}} \oplus \CMTB{\{4410\}} \oplus \CMTB{\{\overline{4950}\}}  \\
~~~~~~~~~~~~~~~~~~~~~~~ \oplus \CMTB{\{\overline{6930'}\}} \oplus \CMTB{\{10560\}} \oplus \CMTB{\{34398\}} \oplus (2) \CMTB{\{36750\}} \oplus \CMTB{\{\overline{48114}\}} ~~~,  \\
{~~}{\rm {Level}}-5 \,~~~~~~~ \CMTred{\{144\}} \oplus (3) \CMTred{\{560\}} \oplus (2) \CMTred{\{720\}} \oplus \CMTred{\{1200\}} \oplus (2) \CMTred{\{1440\}} \oplus \CMTred{\{2640\}}  \\
~~~~~~~~~~~~~~~~~~~~~~~ \oplus (3) \CMTred{\{3696\}} \oplus (2) \CMTred{\{5280\}} \oplus (2) \CMTred{\{8064\}} \oplus (4) \CMTred{\{8800\}} \oplus \CMTred{\{11088\}}  \\
~~~~~~~~~~~~~~~~~~~~~~~ \oplus (2) \CMTred{\{15120\}} \oplus \CMTred{\{25200\}} \oplus \CMTred{\{29568\}} \oplus (2) \CMTred{\{34992\}} \oplus \CMTred{\{38016\}} \\
~~~~~~~~~~~~~~~~~~~~~~~ \oplus \CMTred{\{43680\}} \oplus \CMTred{\{49280\}} \oplus \CMTred{\{144144\}} ~~~,  \\
{~~}{\rm {Level}}-6 \,~~~~~~~ \CMTgrn{\bf \{54\}} \oplus \CMTB{\{210\}} \oplus \CMTB{\{660\}} \oplus (2) \CMTB{\{770\}} \oplus (2) \CMTB{\{945\}} \oplus \CMTB{\{1050\}} \oplus (3) \CMTB{\{\overline{1050}\}}  \\
~~~~~~~~~~~~~~~~~~~~~~~ \oplus (2) \CMTB{\{1386\}} \oplus \CMTB{\{\overline{2772}\}} \oplus (2) \CMTB{\{4125\}} \oplus (3) \CMTB{\{5940\}} \oplus (2) \CMTB{\{\overline{6930}\}}  \\
~~~~~~~~~~~~~~~~~~~~~~~ \oplus (4) \CMTB{\{8085\}} \oplus \CMTB{\{8910\}} \oplus \CMTB{\{14784\}} \oplus \CMTB{\{16380\}} \oplus \CMTB{\{17325\}} \oplus \CMTB{\{\overline{17325}\}}  \\
~~~~~~~~~~~~~~~~~~~~~~~ \oplus (3) \CMTB{\{17920\}} \oplus  \CMTB{\{23040\}} \oplus (3) \CMTB{\{\overline{23040}\}}  \oplus \CMTB{\{\overline{50688}\}} \oplus (2) \CMTB{\{72765\}}  \\
~~~~~~~~~~~~~~~~~~~~~~~ \oplus \CMTB{\{73710\}} \oplus \CMTB{\{112320\}} \oplus \CMTB{\{\overline{128700}\}} \oplus \CMTB{\{143000\}} ~~~,  \\
{~~}{\rm {Level}}-7 \,~~~~~~~ \CMTred{\{\overline{144}\}} \oplus \CMTred{\{\overline{560}\}} \oplus \CMTred{\{\overline{672}\}} \oplus (2) \CMTred{\{\overline{720}\}} \oplus (2) \CMTred{\{\overline{1200}\}} \oplus \CMTred{\{\overline{1440}\}} \oplus (3) \CMTred{\{\overline{2640}\}} \\
~~~~~~~~~~~~~~~~~~~~~~~ \oplus (4) \CMTred{\{\overline{3696}\}} \oplus \CMTred{\{\overline{7920}\}} \oplus \CMTred{\{\overline{8064}\}} \oplus (4) \CMTred{\{\overline{8800}\}} \oplus (3) \CMTred{\{\overline{11088}\}}  \\
~~~~~~~~~~~~~~~~~~~~~~~ \oplus (3) \CMTred{\{\overline{15120}\}} \oplus \CMTred{\{\overline{17280}\}} \oplus \CMTred{\{\overline{23760}\}} \oplus (3) \CMTred{\{\overline{25200}\}} \oplus \CMTred{\{\overline{30800}\}}  \\
~~~~~~~~~~~~~~~~~~~~~~~ \oplus \CMTred{\{\overline{34992}\}} \oplus (3) \CMTred{\{\overline{38016}\}} \oplus \CMTred{\{\overline{43680}\}} \oplus \CMTred{\{\overline{48048}\}} \oplus (2) \CMTred{\{\overline{49280}\}} \\
~~~~~~~~~~~~~~~~~~~~~~~ \oplus \CMTred{\{\overline{55440}\}} \oplus \CMTred{\{\overline{124800}\}} \oplus \CMTred{\{\overline{144144}\}} \oplus \CMTred{\{\overline{196560}\}} \oplus \CMTred{\{\overline{205920}\}} ~~~,  \\
{~~}{\rm {Level}}-8 \,~~~~~~~ \CMTB{\{120\}} \oplus \CMTB{\{210'\}} \oplus \CMTB{\{320\}} \oplus (3) \CMTB{\{1728\}} \oplus (2) \CMTB{\{2970\}} \oplus (2) \CMTB{\{3696'\}} \\
~~~~~~~~~~~~~~~~~~~~~~~ \oplus (2) \CMTB{\{\overline{3696'}\}} \oplus (5) \CMTB{\{4312\}} \oplus \CMTB{\{4410\}} \oplus (2) \CMTB{\{4608\}} \oplus (2) \CMTB{\{4950\}} \\
~~~~~~~~~~~~~~~~~~~~~~~ \oplus (2) \CMTB{\{\overline{4950}\}} \oplus (3) \CMTB{\{10560\}} \oplus (2) \CMTB{\{27720\}} \oplus (3) \CMTB{\{28160\}} \oplus \CMTB{\{34398\}} \\
~~~~~~~~~~~~~~~~~~~~~~~ \oplus (4) \CMTB{\{36750\}} \oplus \CMTB{\{42120\}} \oplus (2) \CMTB{\{48114\}} \oplus (2) \CMTB{\{\overline{48114}\}} \oplus \CMTB{\{64680\}} \\
~~~~~~~~~~~~~~~~~~~~~~~ \oplus \CMTB{\{68640\}} \oplus \CMTB{\{70070'\}} \oplus \CMTB{\{90090\}} \oplus \CMTB{\{\overline{90090}\}} \oplus \CMTB{\{192192\}} \oplus \CMTB{\{299520\}} ~~~,  \\
{~~~~~~}  {~~~~} \vdots  {~~~~~~~~~\,~~~~~~} \vdots
\end{cases}
\end{equation}
\endgroup
where Level-9 to Level-16 are conjugate of Level-7 to Level-0 respectively. 

The superfield ${\cal V} \otimes \CMTB{\{120\}}$ can be interpreted as ${\cal V}_{\un{a}\un{b}\un{c}}$, a three form (three totally antisymmetric Lorentz indices). It provides four possibilities for the
embedding of traceless gravitons (at level-2, level-6, level-10, and level-14). Note that if one finds a $\CMTgrn{\bf\{54\}}$ in level-$n$, one finds a $\CMTred{\{\overline{144}\}}$ at level-$(n+1)$. This holds for all bosonic superfields listed in Table~\ref{Tab:tensoring_bosonic}. We can see this relation between graviton and gravitino clearly by considering the SUSY transformation law of the graviton in 10D $\mathcal{N} = 1$ theory. We know the undotted D-operator acting on the graviton represented by $h_{\un{a}\un{b}}$ gives a term proportional to the undotted gravitino in the on-shell case (in the off-shell case, there are several auxiliary fields showing up in the r.h.s. besides the undotted gravitino)
\begin{equation}
\label{equ:susytrans_graviton}
    {\rm D}_{\a} h_{\un{a}\un{b}} ~\propto~ (\sigma_{(\un{a}})_{\a\b} \, \psi_{\un{b})}{}^{\b}  ~~~.
\end{equation}
Note that the 
undotted spinor index on the gravitino is a superscript. According to our convention, $\psi_{\un{b}}{}^{\b}$ corresponds to the irrep $\CMTred{\{\overline{144}\}}$. 
Recall that acting the D-operator once will add one to the level. 
Thus, Equation~(\ref{equ:susytrans_graviton}) is exactly the mathematical expression of the statement that if you find a $\CMTgrn{\bf\{54\}}$ in level-$n$, you will find a $\CMTred{\{\overline{144}\}}$ in level-$(n+1)$. 

Based on Equation~(\ref{equ:Psitimes120}), we can see four possible embeddings for the graviton, ten possible embeddings for the gravitino, along with a number of auxiliary fields. The superfield ${\cal V}_{\un{a}\un{b}\un{c}}$ is also the simplest nontrivial bosonic superfield that contains traceless gravitons and gravitinos and can be used as the starting point to construct supergravity.   In fact, as long ago as 1982, Howe, Nicolai, and Van
Proeyen\cite{HNvP} had suggested this particular superfield might be an appropriate
point from which to construct a prepotential for 10D, $\cal N$ = 1 supergravity.

However, our more general study provides the basis for constructing possible viable
theories.  It also provides new insights into the origin of constraints in these systems.

Notice that  ${\cal V} \otimes \CMTB{\{660\}} $ has one and only one conformal graviton candidate at level-8. In recalling the structure of 4D, $\cal N$ = 1 supergravity theory, there it was seen the supergravity prepotential contained only a single field of the appropriate Lorentz representation to be identified with a conformal graviton and that was precisely at the ``middle level'' of the $\theta$-expansion. This immediately implies a question.  One can ask the question of how unique is this occurence?  To answer this, we created a search routine to explore other possible ``external tensor indices'' in addition to the above bosonic irreps. Table \ref{Tab:level8graviton} shows all the irreps up to dimension 73710, that when tensored on a scalar superfield, gives a bosonic superfield that have this property. These irreps all have congruency class (00).
\begingroup
\def\arraystretch{1.8}
\begin{table}[h!]
\centering
\begin{tabular}{|c|c|} \hline
    Dynkin Label & Irrep \\ \hline \hline
    $\CMTB{(40000)}$ & $\CMTB{\{660\}}$ \\ \hline
    $\CMTB{(60000)}$ & $\CMTB{\{4290\}}$ \\ \hline
    $\CMTB{(00022)}$ & $\CMTB{\{8910\}}$ \\ \hline
    $\CMTB{(41000)}$ & $\CMTB{\{12870\}}$ \\ \hline
    $\CMTB{(30100)}$ & $\CMTB{\{14784\}}$ \\ \hline
    $\CMTB{(10111)}$ & $\CMTB{\{72765\}}$ \\ \hline
    $\CMTB{(02011)}$ & $\CMTB{\{73710\}}$ \\ \hline
\end{tabular}
\caption{Summary of bosonic superfields that contain only one conformal graviton at level-8 \label{Tab:level8graviton}}
\end{table}
\endgroup

From this investigation it is obvious that the frequency of occurence of possessing a single graviton within a supermultiplet that occurs at the middle level in the $\theta$-expansion of superfield is not a rare event.

We know the ${\cal V} \otimes \CMTB{\{54\}} $ representation possesses a graviton condidate at Level-0 while the ${\cal V} \otimes \CMTB{\{660\}} $ representation possesses a graviton candidate at Level-8.  The interesting thing about the 
${\cal V} \otimes \CMTB{\{54\}} $ representation is that it may regarded as
part of the linearization of the full 10D, ${\cal N} = 1$ frame field ${\rm E}{}_{\un{a}}{}_{\un{b}}$.  This means one can write an equation of the form
\begin{equation}
\left[ \,{\cal V} \otimes \CMTB{\{54\}} \, \right]~=~ {\cal T}{}^{{\g}_1 \cdots{\g}_8}{}_{\CMTB{\{54\}} \, \CMTB{\{660\}}} \left[ \,
{\rm D}_{ {\g}_1} {\rm D}_{{\g}_2} {\rm D}_{{\g}_3}  {\rm D}_{{\g}_4} {\rm D}_{{\g}_5}  {\rm D}_{{\g}_6 } {\rm D}_{{\g}_7} {\rm D}_{{\g}_8} \, \right]
\left[ \, {\cal V} \otimes \CMTB{\{660\}} \, \right]  ~~~, 
\end{equation}
which has the effect of implying that the graviton condidate at Level-0 in ${\cal V} \otimes \CMTB{\{54\}} $ and the graviton candidate at Level-8 in the ${\cal V} \otimes \CMTB{\{660\}} $ representation are one and the same field.  In this
equation, the quantity ${\cal T}{}^{{\g}_1 \cdots{\g}_8}{}_{\CMTB{\{54\}} \, \CMTB{\{660\}}}$ are a set of quantities chosen so that the equation is consistent
with SO(1,9) Lorentz symmetry.

Without this equation, ${\cal V} \otimes \CMTB{\{54\}} $ is unconstrained.  However,
when this equation is imposed, it causes constraints that involve the ${\rm D}_{ {\a}}$ operator to be imposed on ${\cal V} \otimes \CMTB{\{54\}} $. This is the
origin of SG constraints for the perspective of adinkras.  Of course, the example
we chose is only one possibility.  One could make other choices in place of the
${\cal V} \otimes \CMTB{\{660\}} $ representation.  This would require using
other orders of the ${\rm D}_{ {\a}}$ operator and an appropriate replacement 
of the $\cal T$-tensor.  This is determined by where the candidate $\CMTgrn{\bf\{54\}}$ representation occurs in final superfield in the
equation.  Finally, though we only considered bosonic superfields above,
with further appropriate modifications the ${\cal V} \otimes \CMTB{\{660\}}$ can be replaced by fermionic superfields.

\newpage
\subsection{Fermionic Superfields}

Now we investigate fermionic superfields by attaching spinor indices onto the scalar superfield. Here are the spinorial irreps we will explicitly explore: $\CMTred{\{16\}}$, $\CMTred{\{\overline{16}\}}$, $\CMTred{\{144\}}$, $\CMTred{\{\overline{144}\}}$, $\CMTred{\{560\}}$, $\CMTred{\{\overline{560}\}}$, $\CMTred{\{672\}}$, $\CMTred{\{\overline{672}\}}$, $\CMTred{\{720\}}$, and $\CMTred{\{\overline{720}\}}$.  Before we present our results, some
words of motivation will likely serve the cause of explaining the purpose of
doing so.  

Ever since the work of \cite{SGSFN2}, it has been known that the prepotential superfield description of 4D, $\cal N$ = 2 supergravity  is described by a fundamental dynamical entity that is fermionic.  As a 4D, $\cal N$ = 2 supersymmetrical theory can be descended from a 6D minimally supersymmetrical one, this clearly points out the
possibility that higher dimensional theories can possess fermionic supergravity prepotentials.

There is a second reason why the study of fermionic superfields is suggested.
Some time ago \cite{Gates:1986is}, efforts were undetaken to investigate the structure of
1-form 10D, $\cal N$ = 1 gauge theories as theories involving superspace superconnections over fiber bundles.  This study reached a conclusion that {\it {all}} off-shell theories of this type {\it {must}} include a bosonic component field $\CMTblu{\{126\}}$. This observation was later confirmed by a number of subsequent studies \cite{X1,X2,X3,X4}.  This leads to a powerful restriction if the scanning process is applied to the 1-form 10D, $\cal N$ = 1 gauge theories.  At the same level in the expansion in Grassmann coordinates, both the $\CMTblu{\{10\}}$ and the $\CMTblu{\{126\}}$ SO(1,9) representations must be present.  In addition, there must
also occur the $\CMTR{\{16\}}$ at one level higher in the expansion to accommodate an accompanying gaugino. In a subsequent section we will discuss the relevance of the features uncovered to the case of the 1-form 10D, $\cal N$ = 1 gauge theory.  Furthermore, in order to reach the maximal level of simplication in our presentation, we will only consider the abelian case for the 1-form 10D, $\cal N$ = 1 gauge theory.

We will begin by only considering the feature revealed by our tensoring that are relevant to the case of 10D, $\cal N$ = 1 supergravity theory.  The summary of the expansions for the fermionic cases is listed in Table~\ref{Tab:tensoring_spinorial}. The third column shows which tensor product contributes $\CMTgrn{\bf\{54\}}$ and the decomposition results show that each one only contributes one $\CMTgrn{\bf\{54\}}$. 

\begingroup
\def\arraystretch{1.8}
\begin{table}[h!]
\centering
\begin{tabular}{|c|c|c|} \hline
    Dynkin Label & Irrep & Is there $\CMTgrn{\bf \{54\}}$? \\ \hline \hline
    $\CMTred{(00001)}$ & $\CMTred{\{16\}}$ & no \\ \hline
    $\CMTred{(00010)}$ & $\CMTred{\{\overline{16}\}}$ & no \\ \hline
    \multirow{2}{*}{ $\CMTred{(10010)}$ } & \multirow{2}{*}{ $\CMTred{\{144\}}$ } & level-1: $\CMTred{\{\overline{16}\}} \otimes \CMTred{\{144\}}$ \\
     & & level-5: $\CMTred{\{\overline{3696}\}} \otimes \CMTred{\{144\}}$ \\ \hline
    \multirow{2}{*}{ $\CMTred{(10001)}$ } &
    \multirow{2}{*}{ $\CMTred{\{\overline{144}\}}$ } & level-3: $\CMTred{\{560\}} \otimes \CMTred{\{\overline{144}\}}$ \\
     & & level-7: $\CMTred{\{2640\}} \otimes \CMTred{\{\overline{144}\}}$ \\ \hline
    $\CMTred{(01001)}$  & $\CMTred{\{560\}}$ & level-5: $\CMTred{\{\overline{3696}\}} \otimes \CMTred{\{560\}}$\\ \hline
     \multirow{2}{*}{ $\CMTred{(01010)}$ } & \multirow{2}{*}{$\CMTred{\{\overline{560}\}}$ }& level-3: $\CMTred{\{560\}} \otimes \CMTred{\{\overline{560}\}}$\\
    & & level-7: $\CMTred{\{8800\}} \otimes \CMTred{\{\overline{560}\}}$\\ \hline
    $\CMTred{(00030)}$& $\CMTred{\{672\}}$ & no \\ \hline
    $\CMTred{(00003)}$ & $\CMTred{\{\overline{672}\}}$ & no \\ \hline
     \multirow{2}{*}{ $\CMTred{(20001)}$ } & \multirow{2}{*}{ $\CMTred{\{720\}}$ } & level-1: $\CMTred{\{\overline{16}\}} \otimes \CMTred{\{720\}}$ \\
     & & level-5: $\CMTred{\{\overline{3696}\}} \otimes \CMTred{\{720\}}$ \\ \hline
     \multirow{2}{*}{ $\CMTred{(20010)}$ } & \multirow{2}{*}{ $\CMTred{\{\overline{720}\}}$ } & level-3: $\CMTred{\{560\}} \otimes \CMTred{\{\overline{720}\}}$ \\
     & & level-7: $\CMTred{\{2640\}} \otimes \CMTred{\{\overline{720}\}}$ \\ \hline
\end{tabular}
\caption{Summary of tensoring spinorial irreps onto the scalar superfield\label{Tab:tensoring_spinorial}}
\end{table}
\endgroup

Now, one can look at one of those fermionic superfields with traceless gravitons $\CMTgrn{\bf\{54\}}$ in some detail as an example.  The cases of other fermionic superfields with embeddings for traceless gravitons are given in Appendix~\ref{appen:spinorial_superfield}. Generally for $\mathcal{V}\otimes \CMTred{\{irrep\}}$, here we only list level-0 to level-8 results, and its level-9 to level-16 are conjugate of level-7 to level-0 of $\mathcal{V}\otimes \CMTred{\{\overline{irrep}\}}$. 

${\cal V} \otimes \CMTred{\{560\}}$ can be interpreted as ${\cal V}_{\un{a}\un{b}}{}^{\gamma}$, a fermionic superfield with two antisymmetric bosonic indices and one spinor index satisfying
\begin{equation}
    (\s^{\un a})_{\gamma\d}{\cal V}_{\un{a}\un{b}}{}^{\gamma} ~=~ 0~~~.
\end{equation}
It provides three possible embeddings for traceless gravitons in total (in level-5, level-9, and level-13). Note that if you find a $\CMTgrn{\bf\{54\}}$ in level-$n$, you can find a $\CMTred{\{\overline{144}\}}$ in level-$(n+1)$. This holds for all fermionic superfields listed in Table~\ref{Tab:tensoring_spinorial}, which is consistent with Equation~(\ref{equ:susytrans_graviton}).

There exists a work \cite{Gates:1986is} in the literature from 1987, where the proposal for the use of a fermionic 10D, $\cal N$ =
1 supergravity prepotential was first given.  However, the proposed prepotential was of the form ${\cal V}_{\un{a}}{}^{\gamma}$ which is equivalent to 
${\cal V} \otimes \CMTred{\{\overline{144}\}}$.

Based on Equation~(\ref{equ:Psitimes560}), we see three possible embeddings for gravitons, fifteen 
possible embeddings for gravitinos, and auxiliary fields. The superfield
${\cal V}_{\un{a}\un{b}}{}^{\gamma}$ is also the simplest nontrivial fermionic superfield that contains traceless gravitons and gravitinos and can be used to construct supergravity. 
\begingroup
\footnotesize
\begin{equation}
\label{equ:Psitimes560}
{\cal V} \otimes \CMTred{\{560\}} ~=~ \begin{cases}
{~~}{\rm {Level}}-0 \,~~~~~~~  \CMTred{\{560\}} ~~~,  \\
{~~}{\rm {Level}}-1 \,~~~~~~~\CMTB{\{45\}} \oplus \CMTB{\{210\}} \oplus \CMTB{\{770\}} \oplus \CMTB{\{945\}} \oplus \CMTB{\{\overline{1050}\}} \oplus \CMTB{\{5940\}}  ~~~,  \\
{~~}{\rm {Level}}-2 \,~~~~~~~ \CMTred{\{\overline{16}\}} \oplus (2) \CMTred{\{\overline{144}\}} \oplus (2) \CMTred{\{\overline{560}\}} \oplus \CMTred{\{\overline{672}\}} \oplus \CMTred{\{\overline{720}\}} \oplus (2) \CMTred{\{\overline{1200}\}} \oplus \CMTred{\{\overline{1440}\}}\\
~~~~~~~~~~~~~~~~~~~~~~~\oplus (2) \CMTred{\{\overline{3696}\}} \oplus \CMTred{\{\overline{8064}\}} \oplus \CMTred{\{\overline{8800}\}} \oplus \CMTred{\{\overline{11088}\}} \oplus \CMTred{\{\overline{25200}\}}
 ~~~, \\
{~~}{\rm {Level}}-3 \,~~~~~~~ \CMTB{\{10\}} \oplus (2) \CMTB{\{120\}} \oplus (2) \CMTB{\{\overline{126}\}} \oplus \CMTB{\{126\}} \oplus \CMTB{\{210'\}} \oplus (2) \CMTB{\{320\}} \oplus (4) \CMTB{\{1728\}}\\
~~~~~~~~~~~~~~~~~~~~~~~\oplus (3) \CMTB{\{2970\}} \oplus (3) \CMTB{\{\overline{3696'}\}} \oplus \CMTB{\{3696'\}} \oplus (2) \CMTB{\{4312\}} \oplus \CMTB{\{4410\}} \oplus \CMTB{\{4950\}}\\
~~~~~~~~~~~~~~~~~~~~~~~\oplus \CMTB{\{\overline{4950}\}} \oplus \CMTB{\{\overline{6930'}\}} \oplus (2) \CMTB{\{10560\}} \oplus \CMTB{\{27720\}} \oplus \CMTB{\{34398\}} \oplus (2) \CMTB{\{36750\}}\\
~~~~~~~~~~~~~~~~~~~~~~~\oplus \CMTB{\{\overline{46800}\}} \oplus \CMTB{\{\overline{48114}\}} ~~~, \\
{~~}{\rm {Level}}-4 \,~~~~~~~ \CMTred{\{16\}} \oplus (2) \CMTred{\{144\}} \oplus (4) \CMTred{\{560\}} \oplus (3) \CMTred{\{720\}} \oplus (3) \CMTred{\{1200\}} \oplus (3) \CMTred{\{1440\}} \oplus (2) \CMTred{\{2640\}}\\
~~~~~~~~~~~~~~~~~~~~~~~\oplus (4) \CMTred{\{3696\}} \oplus (2) \CMTred{\{5280\}} \oplus (3) \CMTred{\{8064\}} \oplus (6) \CMTred{\{8800\}} \oplus (2) \CMTred{\{11088\}}\\
~~~~~~~~~~~~~~~~~~~~~~~\oplus (2) \CMTred{\{15120\}} \oplus (2) \CMTred{\{25200\}} \oplus \CMTred{\{29568\}} \oplus \CMTred{\{30800\}} \oplus (3) \CMTred{\{34992\}} \oplus (2) \CMTred{\{38016\}}\\
~~~~~~~~~~~~~~~~~~~~~~~\oplus \CMTred{\{43680\}} \oplus \CMTred{\{49280\}} \oplus \CMTred{\{70560\}} \oplus \CMTred{\{102960\}} \oplus (2) \CMTred{\{144144\}} ~~~,  \\
{~~}{\rm {Level}}-5 \,~~~~~~~ \CMTB{\{45\}} \oplus \CMTgrn{\bf \{54\}} \oplus (2) \CMTB{\{210\}} \oplus \CMTB{\{660\}} \oplus (2) \CMTB{\{770\}} \oplus (5) \CMTB{\{945\}} \oplus (4) \CMTB{\{\overline{1050}\}}\\
~~~~~~~~~~~~~~~~~~~~~~~\oplus (2) \CMTB{\{1050\}} \oplus (3) \CMTB{\{1386\}} \oplus \CMTB{\{\overline{2772}\}} \oplus (3) \CMTB{\{4125\}} \oplus (6) \CMTB{\{5940\}} \oplus (4) \CMTB{\{\overline{6930}\}} \\
~~~~~~~~~~~~~~~~~~~~~~~\oplus \CMTB{\{6930\}} \oplus \CMTB{\{7644\}} \oplus (6) \CMTB{\{8085\}} \oplus \CMTB{\{8910\}} \oplus (2) \CMTB{\{14784\}} \oplus \CMTB{\{16380\}} \oplus \CMTB{\{17325\}}\\
~~~~~~~~~~~~~~~~~~~~~~~\oplus \CMTB{\{\overline{17325}\}} \oplus (5) \CMTB{\{17920\}} \oplus (5) \CMTB{\{\overline{23040}\}} \oplus (2) \CMTB{\{23040\}} \oplus (2) \CMTB{\{\overline{50688}\}} \oplus \CMTB{\{\overline{64350}\}}\\
~~~~~~~~~~~~~~~~~~~~~~~\oplus \CMTB{\{70070\}} \oplus (4) \CMTB{\{72765\}} \oplus (2) \CMTB{\{73710\}} \oplus \CMTB{\{112320\}} \oplus (2) \CMTB{\{\overline{128700}\}} \\
~~~~~~~~~~~~~~~~~~~~~~~\oplus (2) \CMTB{\{143000\}} \oplus \CMTB{\{174636\}} \oplus \CMTB{\{\overline{199017}\}} \oplus \CMTB{\{\overline{242550}\}} ~~~,  \\
{~~}{\rm {Level}}-6 \,~~~~~~~  (2) \CMTred{\{\overline{144}\}} \oplus (3) \CMTred{\{\overline{560}\}} \oplus (2) \CMTred{\{\overline{672}\}} \oplus (4) \CMTred{\{\overline{720}\}} \oplus (4) \CMTred{\{\overline{1200}\}} \oplus (2) \CMTred{\{\overline{1440}\}}\\
~~~~~~~~~~~~~~~~~~~~~~~\oplus (4) \CMTred{\{\overline{2640}\}} \oplus (7) \CMTred{\{\overline{3696}\}} \oplus \CMTred{\{\overline{5280}\}} \oplus \CMTred{\{\overline{7920}\}} \oplus (3) \CMTred{\{\overline{8064}\}} \oplus (7) \CMTred{\{\overline{8800}\}}\\
~~~~~~~~~~~~~~~~~~~~~~~\oplus (6) \CMTred{\{\overline{11088}\}} \oplus (5) \CMTred{\{\overline{15120}\}} \oplus (3) \CMTred{\{\overline{17280}\}} \oplus (2) \CMTred{\{\overline{23760}\}} \oplus (6) \CMTred{\{\overline{25200}\}} \\
~~~~~~~~~~~~~~~~~~~~~~~\oplus \CMTred{\{\overline{26400}\}} \oplus (2) \CMTred{\{\overline{30800}\}} \oplus (2) \CMTred{\{\overline{34992}\}} \oplus (6) \CMTred{\{\overline{38016}\}} \oplus (3) \CMTred{\{\overline{43680}\}}\\
~~~~~~~~~~~~~~~~~~~~~~~\oplus (2) \CMTred{\{\overline{48048}\}} \oplus (3) \CMTred{\{\overline{49280}\}} \oplus (2) \CMTred{\{\overline{55440}\}} \oplus \CMTred{\{\overline{124800}\}} \oplus (3) \CMTred{\{\overline{144144}\}}\\
~~~~~~~~~~~~~~~~~~~~~~~\oplus \CMTred{\{\overline{155232}\}} \oplus \CMTred{\{\overline{164736}\}} \oplus (2) \CMTred{\{\overline{196560}\}} \oplus \CMTred{\{\overline{196560'}\}} \oplus \CMTred{\{\overline{198000}\}}\\
~~~~~~~~~~~~~~~~~~~~~~~\oplus (3) \CMTred{\{\overline{205920}\}} \oplus \CMTred{\{\overline{258720}\}} \oplus \CMTred{\{\overline{529200}\}} ~~~,  \\
{~~}{\rm {Level}}-7 \,~~~~~~~\CMTB{\{120\}} \oplus \CMTB{\{126\}} \oplus \CMTB{\{\overline{126}\}} \oplus (2) \CMTB{\{210'\}} \oplus (3) \CMTB{\{320\}} \oplus (6) \CMTB{\{1728\}} \oplus (5) \CMTB{\{2970\}} \\
~~~~~~~~~~~~~~~~~~~~~~~\oplus (4) \CMTB{\{3696'\}} \oplus (4) \CMTB{\{\overline{3696'}\}} \oplus (7) \CMTB{\{4312\}} \oplus (3) \CMTB{\{4410\}} \oplus (3) \CMTB{\{4608\}} \oplus (5) \CMTB{\{\overline{4950}\}} \\
~~~~~~~~~~~~~~~~~~~~~~~\oplus (4) \CMTB{\{4950\}} \oplus (2) \CMTB{\{\overline{6930'}\}} \oplus \CMTB{\{6930'\}} \oplus (5) \CMTB{\{10560\}} \oplus \CMTB{\{\overline{20790}\}} \oplus (5) \CMTB{\{27720\}} \\
~~~~~~~~~~~~~~~~~~~~~~~\oplus (5) \CMTB{\{28160\}} \oplus (2) \CMTB{\{34398\}} \oplus (9) \CMTB{\{36750\}} \oplus \CMTB{\{42120\}} \oplus (2) \CMTB{\{\overline{46800}\}} \oplus \CMTB{\{46800\}}\\
~~~~~~~~~~~~~~~~~~~~~~~\oplus (5) \CMTB{\{\overline{48114}\}} \oplus (3) \CMTB{\{48114\}} \oplus \CMTB{\{48510\}} \oplus \CMTB{\{\overline{50050}\}} \oplus (2) \CMTB{\{64680\}} \oplus (4) \CMTB{\{68640\}}\\
~~~~~~~~~~~~~~~~~~~~~~~\oplus \CMTB{\{70070'\}} \oplus (3) \CMTB{\{\overline{90090}\}} \oplus (2) \CMTB{\{90090\}} \oplus \CMTB{\{\overline{144144'}\}} \oplus \CMTB{\{\overline{150150}\}} \oplus (3) \CMTB{\{192192\}} \\
~~~~~~~~~~~~~~~~~~~~~~~\oplus \CMTB{\{\overline{216216}\}} \oplus (3) \CMTB{\{299520\}} \oplus \CMTB{\{376320\}} \oplus \CMTB{\{380160\}} \oplus \CMTB{\{436590\}} \oplus \CMTB{\{\overline{705600}\}}  ~~~,  \\
{~~}{\rm {Level}}-8 \,~~~~~~~ (2) \CMTred{\{144\}} \oplus (3) \CMTred{\{560\}} \oplus (2) \CMTred{\{672\}} \oplus (4) \CMTred{\{720\}} \oplus (4) \CMTred{\{1200\}} \oplus (2) \CMTred{\{1440\}} \oplus (4) \CMTred{\{2640\}} \\
~~~~~~~~~~~~~~~~~~~~~~~\oplus (8) \CMTred{\{3696\}} \oplus (2) \CMTred{\{5280\}} \oplus (2) \CMTred{\{7920\}} \oplus (3) \CMTred{\{8064\}} \oplus (8) \CMTred{\{8800\}} \oplus (6) \CMTred{\{11088\}} \\
~~~~~~~~~~~~~~~~~~~~~~~\oplus (7) \CMTred{\{15120\}} \oplus (2) \CMTred{\{17280\}} \oplus \CMTred{\{23760\}} \oplus (6) \CMTred{\{25200\}} \oplus \CMTred{\{29568\}} \oplus (3) \CMTred{\{30800\}} \\
~~~~~~~~~~~~~~~~~~~~~~~\oplus (4) \CMTred{\{34992\}} \oplus (7) \CMTred{\{38016\}} \oplus (3) \CMTred{\{43680\}} \oplus (3) \CMTred{\{48048\}} \oplus (5) \CMTred{\{49280\}} \\
~~~~~~~~~~~~~~~~~~~~~~~\oplus (2) \CMTred{\{55440\}} \oplus \CMTred{\{80080\}} \oplus \CMTred{\{102960\}} \oplus (3) \CMTred{\{124800\}} \oplus \CMTred{\{129360\}} \\
~~~~~~~~~~~~~~~~~~~~~~~\oplus (5) \CMTred{\{144144\}} \oplus \CMTred{\{155232\}} \oplus \CMTred{\{164736\}} \oplus (2) \CMTred{\{196560\}} \oplus (3) \CMTred{\{205920\}} \\
~~~~~~~~~~~~~~~~~~~~~~~\oplus \CMTred{\{258720\}} \oplus (2) \CMTred{\{364000\}} \oplus \CMTred{\{465696\}} \oplus \CMTred{\{529200\}} \oplus \CMTred{\{769824\}} ~~~,  \\
{~~~~~~}  {~~~~} \vdots  {~~~~~~~~~\,~~~~~~} \vdots
\end{cases}
\end{equation}
\endgroup

\newpage
\subsection{10D, $\mathcal{N}=1$ Yang-Mills Supermultiplet}

The work previously described in this chapter on fermionic superfields also opens the doorway for using the techniques developed so far in this paper to the issue of scanning the component level description of a superspace connection $\G_{\un{A}}{} = (\G_{\a}, \G_{\un{a}})$ used to define a 1-form U(1) superspace covariant derivatives
\begin{equation}
    \nabla_{\un A} ~=~ {\rm D}_{\un A} ~+~ ig\Gamma_{\un A}\, {\rm t}  ~~~,
\end{equation}
where ${\rm t}$ denotes the U(1) generator and $g$ is a coupling constant. The commutators of two covariant derivatives take the forms \cite{Gates:1986is} 
\begin{align}
    &[ \nabla_{\a}, \nabla_{\b} \} ~=~ i(\s^{\un c})_{\a\b}\nabla_{\un c} ~+~ g\frac{1}{5!}(\s^{[5]})_{\a\b} f_{[5]} \, {\rm t}~~~,\\
    & [ \nabla_{\a}, \nabla_{\un b} \} ~=~ -ig\,\Big[ \lambda_{\a\un b} ~+~ i\frac{1}{\sqrt{2}}(\s_{\un b})_{\a\g}W^{\g} \Big] {\rm t}~~~,\\
    & (\s^{\un b})^{\a\b}\lambda_{\a\un b} ~=~ 0~~~,\\
    & [ \nabla_{\un a}, \nabla_{\un b} \} ~=~ ig F_{\un a\un b}\, {\rm t} ~~~.
\end{align}
The equations below are consistent with the solutions of Bianchi identities \cite{Gates:1986is} 
\begin{equation}
    i (\s^{\un{a}})_{\d\b} \nabla_{\un{a}} W^{\b} ~=~ - \frac{1}{2 \sqrt{2} \times 7!} \Big[~ i (\s_{\un{a}})^{\b\e} (\s^{[4]})_{\e}{}^{\g} (\s^{\un{a}\un{b}})_{\d}{}^{\l} \nabla_{\l} \nabla_{\b} \nabla_{\g} f_{\un{b}[4]} ~+~ 8 (\s^{[4]})_{\d}{}^{\b} \big\{ \nabla^{\un{b}}, \nabla_{\b} \big\} f_{\un{b}[4]}   ~\Big]  ~~~,
\end{equation}
and
\begin{equation}
\begin{split}
    \nabla^{\un b}F_{\un b\un c} ~=&~
    i\frac{1}{8\times5!}(\s^{\un b})^{\a\g}(\s^{[4]})_{\g}{}^{\b}\nabla_{\un b}\nabla_{\a}\nabla_{\b}f_{\un c[4]} \\
    & ~+~ \frac{1}{32\times 7!}\Big[ (\s_{\un c})^{\a\e}(\s^{\un a\un b})_{\e}{}^{\b}(\s_{\un a})^{\g\lambda}(\s^{[4]})_{\lambda}{}^{\d}\nabla_{\a}\nabla_{\b}\nabla_{\g}\nabla_{\d}f_{\un b[4]} \\
    & ~~~~~~~~~~~~~~~~-~i8(\s_{\un c})^{\a\g}(\s^{[4]})_{\g}{}^{\b}\nabla_{\a}\{ \nabla^{\un b},\nabla_{\b} \}f_{\un b[4]} \Big] ~~~.
\end{split}
\end{equation}
The results immediately above are the most important.  They show that if the component field variable $f_{[5]}$ is set to zero on the RHS of these equations, then the spinorial photino field $W^{\b}$ and the bosonic
Maxwell field $F_{\un{b}\un{c}}$ must both satisfy their equations of motion.  Stated another way, for these
two component fields to be off-shell, it requires the presence of the condition $f_{[5]}$ $\ne$ 0.

In the work of \cite{Gates:1986is}, the structures of the spinorial gauge connection superfield $\G_{\a}$ and its gauge parameter $K$ up to level-4 in $\theta$ are given by
\begin{align}
\begin{split}
    \G_{\a}{} ~=&~ \phi_{\a}{} ~+~ i \frac{1}{2} \q^{\b} \Big[~ (\s^{\un{c}})_{\a\b} A_{\un{c}}{} ~+~ (\s^{[3]})_{\a\b} \phi_{[3]}{} ~-~ (\s^{[5]})_{\a\b} f_{[5]}{}  ~\Big]  ~+~ i \q^{\b} \q^{\g} (\s^{[3]})_{\b\g} \S_{\a[3]}{}   \\
    &~+~ \q^{\b} \q^{\g} (\s^{[3]})_{\b\g} \q^{\d} \Big[~ (\s^{\un{c}})_{\a\d} B_{\un{c}[3]}{} ~+~ (\s^{[3']})_{\a\d} B_{[3'][3]}{} ~+~ (\s^{[5]})_{\a\d} B_{[5][3]}{}   ~\Big]  \\
    &~+~ \q^{\b} \q^{\g} (\s^{[3]})_{\b\g} \q^{\d} \q^{\e} (\s^{[3']})_{\d\e} \S_{\a[3][3']}{} ~+~ \mathcal{O}(\q^{5}) ~~~, \\
    &{~}
\end{split}  \label{equ:connection} \\
\begin{split}
    K ~=&~ k~+~ \theta^{\a}\pi_{\a}{}~+~ i\theta^{\b}\theta^{\g}(\s^{[3]})_{\b\g}H_{[3]}{} ~+~ \theta^{\b}\theta^{\g}(\s^{[3]})_{\b\g}\theta^{\a}\pi_{\a[3]}{} \\
    & ~+~ \theta^{\a}\theta^{\b}(\s^{[3]})_{\a\b}\theta^{\g}\theta^{\d}(\s^{[3']})_{\g\d}H_{[3][3']}{} ~+~ \mathcal{O}(\theta^5) ~~~,
\end{split}
\end{align}
respectively.  The irreducible component fields in Figure~\ref{Fig:10DTypeI} and Figure~\ref{Fig:10DTypeI_dynkin} are contained in
$K$ in the following way.  The component fields in $K$ at levels 0, 1, 2, and 3 in the $\theta$-expansion correspond to
the nodes at the same height assignments in the adinkra.  The component field $H_{[3][3']}$ at level-4 in the $\theta$-expansion of  $K$ is to be expanded {\it {solely}} over the the irreducible representations shown at the fourth
level of the adinkra. The {\em component} content of the connection superfield $\G_{\a}$ is encoded by tensoring the components of the scalar superfield with a lower spinor index, i.e. ${\cal V} \otimes \CMTred{\{\overline{16}\}}$. 
\begingroup
\small
\begin{equation}
\G_{\a} ~=~ \begin{cases}
{~~}{\rm {Level}}-0 \,~~~~~~~ \CMTred {\{\overline{16}\}} ~~~,  \\
{~~}{\rm {Level}}-1 \,~~~~~~~ \CMTB {\{10\}} \oplus \CMTB {\{120\}} \oplus \CMTB {\{126\}} ~~~,  \\
{~~}{\rm {Level}}-2 \,~~~~~~~ \CMTred {\{16\}} \oplus \CMTred {\{144\}} \oplus \CMTred {\{560\}} \oplus \CMTred {\{1200\}} ~~~, \\
{~~}{\rm {Level}}-3 \,~~~~~~~ \CMTB {\{45\}} \oplus \CMTB {\{210\}} \oplus \CMTB {\{770\}} \oplus \CMTB {\{945\}} \oplus \CMTB {\{\overline{1050}\}} \oplus \CMTB {\{5940\}} ~~~, \\
{~~}{\rm {Level}}-4 \,~~~~~~~ \CMTred {\{\overline{144}\}} \oplus \CMTred {\{\overline{560}\}} \oplus \CMTred {\{\overline{672}\}} \oplus \CMTred {\{\overline{1200}\}} \oplus (2) \CMTred {\{\overline{3696}\}} \oplus \CMTred {\{\overline{8064}\}} \oplus \CMTred {\{\overline{11088}\}} ~~~,  \\
{~~}{\rm {Level}}-5 \,~~~~~~~ \CMTB {\{\overline{126}\}} \oplus \CMTB {\{320\}} \oplus \CMTB {\{1728\}} \oplus \CMTB {\{2970\}} \oplus (2) \CMTB {\{\overline{3696'}\}} \oplus \CMTB {\{4312\}} \oplus \CMTB {\{4410\}}  \\
~~~~~~~~~~~~~~~~~~~~~~~ \oplus \CMTB {\{\overline{4950}\}} \oplus \CMTB {\{\overline{6930'}\}} \oplus \CMTB {\{36750\}} ~~~,  \\
{~~}{\rm {Level}}-6 \,~~~~~~~ \CMTred {\{560\}} \oplus \CMTred {\{720\}} \oplus \CMTred {\{1440\}} \oplus \CMTred {\{2640\}} \oplus \CMTred {\{3696\}} \oplus \CMTred {\{5280\}} \oplus \CMTred {\{8064\}}  \\
~~~~~~~~~~~~~~~~~~~~~~~ \oplus (2) \CMTred {\{8800\}} \oplus \CMTred {\{15120\}} \oplus \CMTred {\{34992\}} \oplus \CMTred {\{38016\}} ~~~,  \\
{~~}{\rm {Level}}-7 \,~~~~~~~ \CMTB {\{660\}} \oplus \CMTB {\{945\}} \oplus \CMTB {\{\overline{1050}\}} \oplus \CMTB {\{1386\}} \oplus \CMTB {\{4125\}} \oplus \CMTB {\{5940\}} \oplus \CMTB {\{\overline{6930}\}}  \\
~~~~~~~~~~~~~~~~~~~~~~~ \oplus (2) \CMTB {\{8085\}} \oplus \CMTB {\{14784\}} \oplus \CMTB {\{17325\}} \oplus \CMTB {\{17920\}} \oplus \CMTB {\{\overline{23040}\}} \oplus \CMTB {\{72765\}} ~~~,  \\
{~~}{\rm {Level}}-8 \,~~~~~~~ \CMTred {\{\overline{720}\}} \oplus \CMTred {\{\overline{1200}\}} \oplus (2) \CMTred {\{\overline{2640}\}} \oplus \CMTred {\{\overline{3696}\}} \oplus \CMTred {\{\overline{7920}\}} \oplus (2) \CMTred {\{\overline{8800}\}}  \\
~~~~~~~~~~~~~~~~~~~~~~~ \oplus \CMTred {\{\overline{11088}\}} \oplus \CMTred {\{\overline{15120}\}} \oplus \CMTred {\{\overline{25200}\}} \oplus \CMTred {\{\overline{30800}\}} \oplus \CMTred {\{\overline{38016}\}} \oplus \CMTred {\{\overline{49280}\}}  ~~~,  \\
{~~}{\rm {Level}}-9 \,~~~~~~~ \CMTB {\{210'\}} \oplus \CMTB {\{1728\}} \oplus \CMTB {\{2970\}} \oplus \CMTB {\{3696'\}} \oplus (2) \CMTB {\{4312\}} \oplus \CMTB {\{4608\}} \oplus \CMTB {\{4950\}}  \\
~~~~~~~~~~~~~~~~~~~~~~~ \oplus \CMTB {\{\overline{4950}\}} \oplus \CMTB {\{10560\}} \oplus \CMTB {\{27720\}} \oplus \CMTB {\{28160\}} \oplus \CMTB {\{36750\}} \oplus \CMTB {\{48114\}} ~~~,  \\
{~~}{\rm {Level}}-10 ~~~~~~~~~ \CMTred {\{672\}} \oplus \CMTred {\{720\}} \oplus \CMTred {\{1200\}} \oplus \CMTred {\{2640\}} \oplus (2) \CMTred {\{3696\}} \oplus \CMTred {\{8800\}} \oplus \CMTred {\{11088\}}  \\
~~~~~~~~~~~~~~~~~~~~~~~ \oplus \CMTred {\{15120\}} \oplus \CMTred {\{17280\}} \oplus \CMTred {\{25200\}} \oplus \CMTred {\{38016\}} ~~~,  \\
{~~}{\rm {Level}}-11 ~~~~~~~~~ \CMTB {\{770\}} \oplus \CMTB {\{945\}} \oplus (2) \CMTB {\{1050\}} \oplus \CMTB {\{1386\}} \oplus \CMTB {\{2772\}} \oplus \CMTB {\{5940\}} \oplus \CMTB {\{6930\}}  \\
~~~~~~~~~~~~~~~~~~~~~~~ \oplus \CMTB {\{8085\}} \oplus \CMTB {\{17920\}} \oplus \CMTB {\{23040\}} ~~~,  \\
{~~}{\rm {Level}}-12 ~~~~~~~~~ (2) \CMTred {\{\overline{560}\}} \oplus \CMTred {\{\overline{720}\}} \oplus \CMTred {\{\overline{1440}\}} \oplus \CMTred {\{\overline{3696}\}} \oplus \CMTred {\{\overline{5280}\}} \oplus \CMTred {\{\overline{8064}\}} \oplus \CMTred {\{\overline{8800}\}} ~~~,  \\
{~~}{\rm {Level}}-13 ~~~~~~~~~ \CMTB {\{120\}} \oplus \CMTB {\{126\}} \oplus \CMTB {\{320\}} \oplus \CMTB {\{1728\}} \oplus \CMTB {\{2970\}} \oplus \CMTB {\{3696'\}} ~~~,  \\
{~~}{\rm {Level}}-14 ~~~~~~~~~ \CMTred {\{16\}} \oplus \CMTred {\{144\}} \oplus \CMTred {\{560\}} \oplus \CMTred {\{1200\}} ~~~,  \\
{~~}{\rm {Level}}-15 ~~~~~~~~~ \CMTB {\{1\}} \oplus \CMTB {\{45\}} \oplus \CMTB {\{210\}} ~~~,  \\
{~~}{\rm {Level}}-16 ~~~~~~~~~ \CMTred {\{\overline{16}\}} ~~~.  \\
\end{cases} \label{equ:connectiondecomp}
\end{equation}
\endgroup
By comparing this with Equation (\ref{equ:connection}), we read that in level-0 $\phi_{\a}$ corresponds to $\CMTred{\{\overline{16}\}}$; level-1 $A_{\un{c}}$, $\phi_{[3]}$ and $f_{[5]}$ correspond to $\CMTB{\{10\}}$, $\CMTB{\{120\}}$ and $\CMTB{\{126\}}$; level-2 $\S_{\a[3]}$ corresponds to $\CMTB{\{120\}} \otimes \CMTred{\{\overline{16}\}} ~=~ \overline{\ell_{2}} \otimes \CMTred{\{\overline{16}\}}$ which is exactly the decomposition at level-2 of (\ref{equ:connectiondecomp}), where $\ell_{2}$ is the level-2 $\theta-$monomial of the scalar superfield. One point to note is that $f_{[5]}$ is $\CMTB{\{126\}}$ but not $\CMTB{\{\overline{126}\}}$. This is because it is contracted with $(\s^{[5]})_{\a\b}$ which a self-dual 5-form by Equation (\ref{equ:5formdual}). The conditions for embedding the component fields into $\G_{\a}$ requires at some level-$n$ (here $n=1$) there must occur $\CMTblu{\{{10}\}}$ and the $\CMTblu{\{126\}}$ while at the level-$(n + 1)$ there must occur a $\CMTR{\{{16}\}}$, as we noted earlier.

Now, if we want to put the gauge fields to the middle level and apply Wess-Zumino gauge, we look for superfields that have gauge fields $\CMTB{\{10\}}$ and $\CMTB{\{126\}}$ at level-8, and the gaugino $\CMTred{\{16\}}$ at the next level, i.e. level-9. And we want the triplet ($\CMTB{\{10\}}$, $\CMTB{\{126\}}$, and $\CMTred{\{16\}}$) to show up at level-8 and 9 only. Table \ref{Tab:YM-WZ} shows all the irreps (up to dimension 73710) that when tensored into the scalar superfield, would give us the superfields with that property. It also shows the numbers of gauge 1-form(s), gauge 5-form(s) and gaugino(s) in level-8 and 9 that appear in each superfield listed. We denote them by $( b_{\CMTB{\{10\}}}, b_{\CMTB{\{126\}}}, b_{\CMTred{\{16\}}} )$. All these irreps have conjugacy class (02).
\begingroup
\def\arraystretch{1.8}
\begin{table}[h!]
\centering
\begin{tabular}{|c|c|c|} \hline
    Dynkin Label & Irrep & $( b_{\CMTB{\{10\}}}, b_{\CMTB{\{126\}}}, b_{\CMTred{\{16\}}} )$ \\ \hline \hline
    $\CMTB{(30000)}$ & $\CMTB{\{210'\}}$ & (1,1,1)  \\ \hline 
    $\CMTB{(20100)}$ & $\CMTB{\{4312\}}$ & (1,2,2)  \\ \hline 
    $\CMTB{(31000)}$ & $\CMTB{\{4608\}}$ & (1,1,1)  \\ \hline 
    $\CMTB{(20002)}$ & $\CMTB{\{\overline{4950}\}}$ & (1,2,1)  \\ \hline 
    $\CMTB{(00111)}$ & $\CMTB{\{10560\}}$ & (1,2,1)  \\ \hline 
    $\CMTB{(10200)}$ & $\CMTB{\{27720\}}$ & (1,2,1)  \\ \hline 
    $\CMTB{(30011)}$ & $\CMTB{\{28160\}}$ & (1,2,1)  \\ \hline 
    $\CMTB{(11011)}$ & $\CMTB{\{36750\}}$ & (1,3,1)  \\ \hline 
\end{tabular}
\caption{Summary of bosonic superfields that contain gauge 1-form(s) and 5-form(s) at level-8, and gaugino(s) at level-9 only \label{Tab:YM-WZ}}
\end{table}
\endgroup

All the remarks that were made at the end of the discussion regarding the possibility of the conformal supergravity theory apply with minor modifications here.

\newpage
\section{10D, $\mathcal{N}=2$A Scalar Superfield Decomposition and Superconformal Multiplet}

\subsection{Methodology for 10D, $\mathcal{N}=2$A Scalar Superfield Construction}
\label{subsec:2Amethod}

One can introduce two sets of 10D spinor coordinates denoted by $\theta^{\alpha}$ and $\theta^{\dot{\alpha}}$ so that a 10D, $\mathcal{N} = (1,1)$ superfield can be expressed in the form ${\cal V} (x^{\un{a}}, \, \theta^{\a}, \, \theta^{\dot{\a}})$.  It is possible to organize the expansion so that it takes the form
\begin{equation}
    {\cal V} (x^{\un{a}}, \, \theta^{\a}, \, \theta^{\dot{\a}}) ~=~ {\cal V}^{(0)} (x^{\un{a}}, \, \theta^{\a}) ~+~ \theta^{\dot{\a}} \, {\cal V}^{(1)}_{\dot{\a}} (x^{\un{a}}, \, \theta^{\a}) ~+~ \theta^{\dot{\a}} \theta^{\dot{\b}} \, {\cal V}^{(2)}_{\dot{\a}\dot{\b}}  (x^{\un{a}}, \, \theta^{\a}) ~+~ \dots   ~~~,
\end{equation}
and the point is that each of the expansion coefficients ${\cal V}^{(0)} (x^{\un{a}}, \, \theta^{\a})$, ${\cal V}^{(1)}_{\dot{\a}} (x^{\un{a}}, \, \theta^{\a})$, ${\cal V}^{(2)}_{\dot{\a}\dot{\b}}  (x^{\un{a}}, \, \theta^{\a}), \dots$ is a 10D, $\mathcal{N} = 1$ superfield. More explicitly, from Equation (\ref{equ:10DIsuperfield}) one can write
\begin{equation}
\begin{split}
    {\cal V} (x^{\un{a}}, \, \theta^{\a}, \, \theta^{\dot{\a}}) ~=&~ {\cal V}^{(0)} (x^{\un{a}}, \, \theta^{\a}) ~+~ \theta^{\dot{\a}} \, {\cal V}^{(1)}_{\dot{\a}} (x^{\un{a}}, \, \theta^{\a}) ~+~ \theta^{\dot{\a}} \theta^{\dot{\b}} \, {\cal V}^{(2)}_{\dot{\a}\dot{\b}}  (x^{\un{a}}, \, \theta^{\a}) ~+~ \dots   \\
    ~=&~ \big[~ \varphi^{(0)} (x^{\un{a}}) ~+~ \theta^{\a} \, \varphi^{(1)}_{\a} (x^{\un{a}}) ~+~ \theta^{\a} \theta^{\b} \, \varphi^{(2)}_{\a\b}  (x^{\un{a}}) ~+~ \dots ~\big]   \\
    &~+~ \theta^{\dot{\a}} \, \big[~ \varphi^{(0)}_{\dot{\a}} (x^{\un{a}}) ~+~ \theta^{\a} \, \varphi^{(1)}_{\a\dot{\a}} (x^{\un{a}}) ~+~ \theta^{\a} \theta^{\b} \, \varphi^{(2)}_{\a\b\dot{\a}}  (x^{\un{a}}) ~+~ \dots ~\big]  \\
    &~+~ \theta^{\dot{\a}} \theta^{\dot{\b}} \, \big[~ \varphi^{(0)}_{\dot{\a}\dot{\b}} (x^{\un{a}}) ~+~ \theta^{\a} \, \varphi^{(1)}_{\a\dot{\a}\dot{\b}} (x^{\un{a}}) ~+~ \theta^{\a} \theta^{\b} \, \varphi^{(2)}_{\a\b\dot{\a}\dot{\b}}  (x^{\un{a}}) ~+~ \dots ~\big]  \\
    &~+~ \dots  \\
    ~=&~ \varphi^{(0)} (x^{\un{a}})   \\
    &~+~ \theta^{\a} \, \varphi^{(1)}_{\a} (x^{\un{a}}) ~+~ \theta^{\dot{\a}} \, \varphi^{(0)}_{\dot{\a}} (x^{\un{a}})  \\
    &~+~ \theta^{\a} \theta^{\b} \, \varphi^{(2)}_{\a\b}  (x^{\un{a}})  ~+~ \theta^{\dot{\a}} \theta^{\a} \, \varphi^{(1)}_{\a\dot{\a}} (x^{\un{a}}) ~+~ \theta^{\dot{\a}} \theta^{\dot{\b}} \, \varphi^{(0)}_{\dot{\a}\dot{\b}} (x^{\un{a}})  \\
    &~+~ \theta^{\a} \theta^{\b} \theta^{\g} \, \varphi^{(3)}_{\a\b\g} (x^{\un{a}}) ~+~ \theta^{\dot{\a}} \theta^{\a} \theta^{\b} \, \varphi^{(2)}_{\a\b\dot{\a}}  (x^{\un{a}}) ~+~ \theta^{\dot{\a}} \theta^{\dot{\b}} \theta^{\a} \, \varphi^{(1)}_{\a\dot{\a}\dot{\b}} (x^{\un{a}}) ~+~ \theta^{\dot{\a}} \theta^{\dot{\b}} \theta^{\dot{\g}} \, \varphi^{(0)}_{\dot{\a}\dot{\b}\dot{\g}} (x^{\un{a}}) \\
    &~+~ \dots  ~~~.
\end{split}
\end{equation}
Attaching $n$ totally antisymmetric dotted $\theta$'s to an undotted $\theta-$monomial corresponds to tensoring the $\CMTred{\{\overline{16}\}}^{\wedge n}$ representation on it, i.e. tensoring the conjugate of the $n$th level of the $\theta-$monomial of the 10D, $\mathcal{N}=1$ scalar superfield on it. Therefore, if we denote level-$n$ $\theta-$monomial of 10D, $\mathcal{N}=1$ scalar superfield as $\ell_{n}$, we can obtain the 10D, $\mathcal{N}=2$A scalar superfield $\theta-$monomial decomposition by
\begin{equation}
\label{equ:2A_formula}
{\cal V} ~=~ \begin{cases}
{~~}{\rm {Level}}-0 \,~~~~~~~~~~ \ell_{0} \otimes \overline{\ell_{0}} ~~~,  \\
{~~}{\rm {Level}}-1 \,~~~~~~~~~~ \ell_{1} \otimes \overline{\ell_{0}} \, \oplus \, \ell_{0} \otimes \overline{\ell_{1}} ~~~,  \\
{~~}{\rm {Level}}-2 \,~~~~~~~~~~ \ell_{2} \otimes \overline{\ell_{0}} \, \oplus \, \ell_{1} \otimes \overline{\ell_{1}} \, \oplus \, \ell_{0} \otimes \overline{\ell_{2}} ~~~,  \\
{~~}{\rm {Level}}-3 \,~~~~~~~~~~ \ell_{3} \otimes \overline{\ell_{0}} \, \oplus \, \ell_{2} \otimes \overline{\ell_{1}} \, \oplus \, \ell_{1} \otimes \overline{\ell_{2}} \, \oplus \, \ell_{0} \otimes \overline{\ell_{3}} ~~~,  \\
{~~~~~~}  {~~~~} \vdots  {~~~~~~~~~\,~~~~~~} \vdots \\
{~~}{\rm {Level}}-n \,~~~~~~~~~~ \ell_{n} \otimes \overline{\ell_{0}} \, \oplus \, \ell_{n-1} \otimes \overline{\ell_{1}} \, \oplus \, \dots \, \oplus \, \ell_{1} \otimes \overline{\ell_{n-1}} \, \oplus \, \ell_{0} \otimes \overline{\ell_{n}} ~~~,  \\
{~~~~~~}  {~~~~} \vdots  {~~~~~~~~~\,~~~~~~} \vdots \\
{~~}{\rm {Level}}-16 ~~~~~~~~~ \ell_{16} \otimes \overline{\ell_{0}} \, \oplus \, \ell_{15} \otimes \overline{\ell_{1}} \, \oplus \, \dots \, \oplus \, \ell_{1} \otimes \overline{\ell_{15}} \, \oplus \, \ell_{0} \otimes \overline{\ell_{16}} ~~~,  \\
{~~~~~~}  {~~~~} \vdots  {~~~~~~~~~\,~~~~~~} \vdots 
\end{cases}
\end{equation}
and level-17 to 32 are the conjugates of level-15 to 0 respectively. From Equation~(\ref{equ:2A_formula}), it's clear that each level is self-conjugate and therefore level-17 to 32 are the same as level-15 to 0. Moreover, the decompositions of component fields are the same as that of the $\theta-$monomials.

\subsection{10D, $\mathcal{N} = 2$A Scalar Superfield Decomposition Results and Superconformal Multiplet}
\label{sec:2A}

Based on Equations~(\ref{equ:10DTypeI}) and (\ref{equ:2A_formula}), one can directly obtain the scalar superfield decomposition in 10D, $\mathcal{N} = 2$A. The results from level-0 to level-16 are listed below. We use green color to highlight the irrep $\CMTgrn{\bf \{54\}}$ which corresponds to the traceless graviton in 10D. Unlike in 10D, $\mathcal{N} = 1$ theory, one can find possible embeddings for gravitons and gravitinos in the $\theta-$expansion of the scalar 10D, $\mathcal{N} = 2$A superfield. 
As discussed in chapter~\ref{sec:4D}, we can translate irreps into component fields and see 72 graviton embeddings, 280 gravitino embeddings, and associated 
hosts of auxiliary fields. The scalar superfield is the simplest superfield that contains traceless gravitons and gravitinos and can be used to construct supergravity. 

Note that if you find a $\CMTgrn{\bf\{54\}}$ in level-$n$, you can find $\CMTred{\{144\}}$ and $\CMTred{\{\overline{144}\}}$ in level-$(n+1)$. This is also consistent with SUSY transformation laws of the graviton $h_{\un{a}\un{b}}$ in 10D, $\mathcal{N} = 2$A theory. Acting the undotted D-operator on the graviton gives a term proportional to the undotted gravitino in the on-shell case (in the off-shell case, there are several auxiliary fields showing up in the r.h.s. besides the undotted gravitino)
\begin{equation}
    {\rm D}_{\a} h_{\un{a}\un{b}} ~\propto~ (\sigma_{(\un{a}})_{\a\b} \, \psi_{\un{b})}{}^{\b} ~~~.
\end{equation}
Note that the gravitino here has an undotted superscript spinor index and hence corresponds to irrep $\CMTred{\{\overline{144}\}}$ in our convention. Meanwhile the dotted D-operator acting on the graviton satisfies
\begin{equation}
    {\rm D}_{\dot{\a}} h_{\un{a}\un{b}} ~\propto~ (\sigma_{(\un{a}})_{\dot{\a}\dot{\b}} \, \psi_{\un{b})}{}^{\dot{\b}} ~~~,
\end{equation}
which gives a term proportional to the dotted gravitino in the on-shell case. The superscript dotted spinor index on the gravitino indicates that it corresponds to irrep $\CMTred{\{144\}}$ in our convention.

\begin{itemize}
\item Level-0: $\CMTB{\{1\}}$
\item Level-1: $\CMTred{\{16\}} \oplus \CMTred{\{\overline{16}\}}$
\item Level-2: $\CMTB{\{1\}} \oplus \CMTB{\{45\}} \oplus (2)\CMTB{\{120\}} \oplus \CMTB{\{210\}}$
\item Level-3: $\CMTred{\{16\}} \oplus \CMTred{\{\overline{16}\}} \oplus (2)\CMTred{\{560\}} \oplus (2)\CMTred{\{\overline{560}\}} \oplus \CMTred{\{144\}} \oplus \CMTred{\{\overline{144}\}} \oplus \CMTred{\{1200\}} \oplus \CMTred{\{\overline{1200}\}}$
\item Level-4: $ \CMTB{\{1\}} \oplus \CMTB{\{45\}} \oplus \CMTgrn{\bf \{54\} } \oplus (2) \CMTB{\{120\}} \oplus \CMTB{\{126\}} \oplus \CMTB{\{\overline{126}\}} \oplus (2) \CMTB{\{210\}} \oplus (2) \CMTB{\{320\}} \oplus (3) \CMTB{\{770\}} \oplus \CMTB{\{945\}} \oplus (2) \CMTB{\{1050\}} \oplus (2) \CMTB{\{\overline{1050}\}} \oplus (2) \CMTB{\{1728\}} \oplus (2) \CMTB{\{2970\}} \oplus \CMTB{\{3696'\}} \oplus \CMTB{\{\overline{3696'}\}} \oplus \CMTB{\{4125\}} \oplus \CMTB{\{5940\}}$
\item Level-5: $ \CMTred{\{16\}} \oplus \CMTred{\{\overline{16}\}} \oplus (2) \CMTred{\{144\}} \oplus (2) \CMTred{\{\overline{144}\}} \oplus (4) \CMTred{\{560\}} \oplus (4) \CMTred{\{\overline{560}\}} \oplus (2) \CMTred{\{672\}} \oplus (2) \CMTred{\{\overline{672}\}} \oplus (2) \CMTred{\{720\}} \oplus (2) \CMTred{\{\overline{720}\}} \oplus (2) \CMTred{\{1200\}} \oplus (2) \CMTred{\{\overline{1200}\}} \oplus (2) \CMTred{\{1440\}} \oplus (2) \CMTred{\{\overline{1440}\}} \oplus (4) \CMTred{\{3696\}} \oplus (4) \CMTred{\{\overline{3696}\}} \oplus \CMTred{\{5280\}} \oplus \CMTred{\{\overline{5280}\}} \oplus (2) \CMTred{\{8064\}} \oplus (2) \CMTred{\{\overline{8064}\}} \oplus (2) \CMTred{\{8800\}} \oplus (2) \CMTred{\{\overline{8800}\}} \oplus \CMTred{\{11088\}} \oplus \CMTred{\{\overline{11088}\}} \oplus \CMTred{\{25200\}} \oplus \CMTred{\{\overline{25200}\}}$
\item Level-6: $ \CMTB{\{1\}} \oplus (2) \CMTB{\{45\}} \oplus \CMTgrn{\bf \{54\} } \oplus (4) \CMTB{\{120\}} \oplus \CMTB{\{126\}} \oplus \CMTB{\{\overline{126}\}} \oplus (3) \CMTB{\{210\}} \oplus (4) \CMTB{\{320\}} \oplus (4) \CMTB{\{770\}} \oplus (5) \CMTB{\{945\}} \oplus (4) \CMTB{\{1050\}} \oplus (4) \CMTB{\{\overline{1050}\}} \oplus (3) \CMTB{\{1386\}} \oplus (6) \CMTB{\{1728\}} \oplus \CMTB{\{2772\}} \oplus \CMTB{\{\overline{2772}\}} \oplus (4) \CMTB{\{2970\}} \oplus (5) \CMTB{\{3696'\}} \oplus (5) \CMTB{\{\overline{3696'}\}} \oplus \CMTB{\{4125\}} \oplus (6) \CMTB{\{4312\}} \oplus (2) \CMTB{\{4410\}} \oplus \CMTB{\{4950\}} \oplus \CMTB{\{\overline{4950}\}} \oplus (6) \CMTB{\{5940\}} \oplus (2) \CMTB{\{6930\}} \oplus (2) \CMTB{\{\overline{6930}\}} \oplus \CMTB{\{6930'\}} \oplus \CMTB{\{\overline{6930'}\}} \oplus \CMTB{\{7644\}} \oplus (4) \CMTB{\{8085\}} \oplus \CMTB{\{8910\}} \oplus (2) \CMTB{\{10560\}} \oplus (4) \CMTB{\{17920\}} \oplus (2) \CMTB{\{23040\}} \oplus (2) \CMTB{\{\overline{23040}\}} \oplus (2) \CMTB{\{34398\}} \oplus (4) \CMTB{\{36750\}} \oplus \CMTB{\{48114\}} \oplus \CMTB{\{\overline{48114}\}} \oplus \CMTB{\{72765\}} \oplus \CMTB{\{73710\}}$
\item Level-7: $(2) \CMTred{\{16\}} \oplus (2) \CMTred{\{\overline{16}\}} \oplus (4) \CMTred{\{144\}} \oplus (4) \CMTred{\{\overline{144}\}} \oplus (7) \CMTred{\{560\}} \oplus (7) \
\CMTred{\{\overline{560}\}} \oplus (2) \CMTred{\{672\}} \oplus (2) \CMTred{\{\overline{672}\}} \oplus (5) \CMTred{\{720\}} \oplus (5) \CMTred{\{\overline{720}\}} \oplus (6) \CMTred{\{1200\}} \oplus (6) \CMTred{\{\overline{1200}\}} \oplus (4) \CMTred{\{1440\}} \oplus (4) \CMTred{\{\overline{1440}\}} \oplus (4) \CMTred{\{2640\}} \oplus (4) \CMTred{\{\overline{2640}\}} \oplus (10) \CMTred{\{3696\}} \oplus (10) \CMTred{\{\overline{3696}\}} \oplus (2) \CMTred{\{5280\}} \oplus (2) \CMTred{\{\overline{5280}\}} \oplus (5) \CMTred{\{8064\}} \oplus (5) \CMTred{\{\overline{8064}\}} \oplus (10) \CMTred{\{8800\}} \oplus (10) \CMTred{\{\overline{8800}\}} \oplus (6) \CMTred{\{11088\}} \oplus (6) \CMTred{\{\overline{11088}\}} \oplus (5) \CMTred{\{15120\}} \oplus (5) \CMTred{\{\overline{15120}\}} \oplus (2) \CMTred{\{17280\}} \oplus (2) \CMTred{\{\overline{17280}\}} \oplus (5) \CMTred{\{25200\}} \oplus (5) \CMTred{\{\overline{25200}\}} \oplus \CMTred{\{29568\}} \oplus \CMTred{\{\overline{29568}\}} \oplus (4) \CMTred{\{34992\}} \oplus (4) \CMTred{\{\overline{34992}\}} \oplus (4) \CMTred{\{38016\}} \oplus (4) \CMTred{\{\overline{38016}\}} \oplus (3) \CMTred{\{43680\}} \oplus (3) \CMTred{\{\overline{43680}\}} \oplus (2) \CMTred{\{49280\}} \oplus (2) \CMTred{\{\overline{49280}\}} \oplus \CMTred{\{55440\}} \oplus \CMTred{\{\overline{55440}\}} \oplus \CMTred{\{70560\}} \oplus \CMTred{\{\overline{70560}\}} \oplus (2) \CMTred{\{144144\}} \oplus (2) \CMTred{\{\overline{144144}\}} \oplus \CMTred{\{205920\}} \oplus \CMTred{\{\overline{205920}\}}$
\item Level-8: $(2) \CMTB{\{1\}} \oplus (2) \CMTB{\{10\}} \oplus (3) \CMTB{\{45\}} \oplus (4) \CMTgrn{\bf \{54\} } \oplus (6) \CMTB{\{120\}} \oplus (3) \CMTB{\{126\}} \oplus (3) \CMTB{\{\overline{126}\}} \oplus (7) \CMTB{\{210\}} \oplus (4) \CMTB{\{210'\}} \oplus (6) \CMTB{\{320\}} \oplus (5) \CMTB{\{660\}} \oplus (8) \CMTB{\{770\}} \oplus (9) \CMTB{\{945\}} \oplus (8) \CMTB{\{1050\}} \oplus (8) \CMTB{\{\overline{1050}\}} \oplus (6) \CMTB{\{1386\}} \oplus (14) \CMTB{\{1728\}} \oplus \CMTB{\{2772\}} \oplus \CMTB{\{\overline{2772}\}} \oplus (12) \CMTB{\{2970\}} \oplus (8) \CMTB{\{3696'\}} \oplus (8) \CMTB{\{\overline{3696'}\}} \oplus (10) \CMTB{\{4125\}} \oplus (12) \CMTB{\{4312\}} \oplus (6) \CMTB{\{4410\}} \oplus (4) \CMTB{\{4608\}} \oplus (7) \CMTB{\{4950\}} \oplus (7) \CMTB{\{\overline{4950}\}} \oplus (14) \CMTB{\{5940\}} \oplus (4) \CMTB{\{6930\}} \oplus (4) \CMTB{\{\overline{6930}\}} \oplus (2) \CMTB{\{6930'\}} \oplus (2) \CMTB{\{\overline{6930'}\}} \oplus (2) \CMTB{\{7644\}} \oplus (16) \CMTB{\{8085\}} \oplus (5) \CMTB{\{8910\}} \oplus (10) \CMTB{\{10560\}} \oplus (3) \CMTB{\{14784\}} \oplus (4) \CMTB{\{16380\}} \oplus (3) \CMTB{\{17325\}} \oplus (3) \CMTB{\{\overline{17325}\}} \oplus (12) \CMTB{\{17920\}} \oplus (8) \CMTB{\{23040\}} \oplus (8) \CMTB{\{\overline{23040}\}} \oplus (6) \CMTB{\{27720\}} \oplus (6) \CMTB{\{28160\}} \oplus (4) \CMTB{\{34398\}} \oplus (16) \CMTB{\{36750\}} \oplus (2) \CMTB{\{37632\}} \oplus (3) \CMTB{\{46800\}} \oplus (3) \CMTB{\{\overline{46800}\}} \oplus (5) \CMTB{\{48114\}} \oplus (5) \CMTB{\{\overline{48114}\}} \oplus (2) \CMTB{\{50688\}} \oplus (2) \CMTB{\{\overline{50688}\}} \oplus \CMTB{\{52920\}} \oplus (4) \CMTB{\{64680\}} \oplus (4) \CMTB{\{68640\}} \oplus (9) \CMTB{\{72765\}} \oplus (6) \CMTB{\{73710\}} \oplus (2) \CMTB{\{90090\}} \oplus (2) \CMTB{\{\overline{90090}\}} \oplus (3) \CMTB{\{112320\}} \oplus (2) \CMTB{\{128700\}} \oplus (2) \CMTB{\{\overline{128700}\}} \oplus (5) \CMTB{\{143000\}} \oplus \CMTB{\{150150\}} \oplus \CMTB{\{\overline{150150}\}} \oplus \CMTB{\{174636\}} \oplus \CMTB{\{189189\}} \oplus (2) \CMTB{\{192192\}} \oplus \CMTB{\{242550\}} \oplus \CMTB{\{\overline{242550}\}} \oplus \CMTB{\{274560\}} \oplus (2) \CMTB{\{299520\}} \oplus (2) \CMTB{\{380160\}}$
\item Level-9: $(4) \CMTred{\{16\}} \oplus (4) \CMTred{\{\overline{16}\}} \oplus (8) \CMTred{\{144\}} \oplus (8) \CMTred{\{\overline{144}\}} \oplus (12) \CMTred{\{560\}} \oplus (12) \CMTred{\{\overline{560}\}} \oplus (3) \CMTred{\{672\}} \oplus (3) \CMTred{\{\overline{672}\}} \oplus (11) \CMTred{\{720\}} \oplus (11) \CMTred{\{\overline{720}\}} \oplus (13) \CMTred{\{1200\}} \oplus (13) \CMTred{\{\overline{1200}\}} \oplus (9) \CMTred{\{1440\}} \oplus (9) \CMTred{\{\overline{1440}\}} \oplus (12) \CMTred{\{2640\}} \oplus (12) \CMTred{\{\overline{2640}\}} \oplus (20) \CMTred{\{3696\}} \oplus (20) \CMTred{\{\overline{3696}\}} \oplus (4) \CMTred{\{5280\}} \oplus (4) \CMTred{\{\overline{5280}\}} \oplus (4) \CMTred{\{7920\}} \oplus (4) \CMTred{\{\overline{7920}\}} \oplus (12) \CMTred{\{8064\}} \oplus (12) \CMTred{\{\overline{8064}\}} \oplus (24) \CMTred{\{8800\}} \oplus (24) \CMTred{\{\overline{8800}\}} \oplus (16) \CMTred{\{11088\}} \oplus (16) \CMTred{\{\overline{11088}\}} \oplus (15) \CMTred{\{15120\}} \oplus (15) \CMTred{\{\overline{15120}\}} \oplus (3) \CMTred{\{17280\}} \oplus (3) \CMTred{\{\overline{17280}\}} \oplus (3) \CMTred{\{23760\}} \oplus (3) \CMTred{\{\overline{23760}\}} \oplus (16) \CMTred{\{25200\}} \oplus (16) \CMTred{\{\overline{25200}\}} \oplus \CMTred{\{29568\}} \oplus \CMTred{\{\overline{29568}\}} \oplus (6) \CMTred{\{30800\}} \oplus (6) \CMTred{\{\overline{30800}\}} \oplus (11) \CMTred{\{34992\}} \oplus (11) \CMTred{\{\overline{34992}\}} \oplus (15) \CMTred{\{38016\}} \oplus (15) \CMTred{\{\overline{38016}\}} \oplus (2) \CMTred{\{39600\}} \oplus (2) \CMTred{\{\overline{39600}\}} \oplus (10) \CMTred{\{43680\}} \oplus (10) \CMTred{\{\overline{43680}\}} \oplus (5) \CMTred{\{48048\}} \oplus (5) \CMTred{\{\overline{48048}\}} \oplus (12) \CMTred{\{49280\}} \oplus (12) \CMTred{\{\overline{49280}\}} \oplus (5) \CMTred{\{55440\}} \oplus (5) \CMTred{\{\overline{55440}\}} \oplus (3) \CMTred{\{70560\}} \oplus (3) \CMTred{\{\overline{70560}\}} \oplus (2) \CMTred{\{102960\}} \oplus (2) \CMTred{\{\overline{102960}\}} \oplus (4) \CMTred{\{124800\}} \oplus (4) \CMTred{\{\overline{124800}\}} \oplus (10) \CMTred{\{144144\}} \oplus (10) \CMTred{\{\overline{144144}\}} \oplus (3) \CMTred{\{155232\}} \oplus (3) \CMTred{\{\overline{155232}\}} \oplus (2) \CMTred{\{164736\}} \oplus (2) \CMTred{\{\overline{164736}\}} \oplus (3) \CMTred{\{196560\}} \oplus (3) \CMTred{\{\overline{196560}\}} \oplus \CMTred{\{198000\}} \oplus \CMTred{\{\overline{198000}\}} \oplus (8) \CMTred{\{205920\}} \oplus (8) \CMTred{\{\overline{205920}\}} \oplus \CMTred{\{258720\}} \oplus \CMTred{\{\overline{258720}\}} \oplus \CMTred{\{274560'\}} \oplus \CMTred{\{\overline{274560'}\}} \oplus \CMTred{\{332640\}} \oplus \CMTred{\{\overline{332640}\}} \oplus (2) \CMTred{\{364000\}} \oplus (2) \CMTred{\{\overline{364000}\}} \oplus (2) \CMTred{\{388080\}} \oplus (2) \CMTred{\{\overline{388080}\}} \oplus (2) \CMTred{\{529200\}} \oplus (2) \CMTred{\{\overline{529200}\}} \oplus \CMTred{\{769824\}} \oplus \CMTred{\{\overline{769824}\}}$
\item Level-10: $(2) \CMTB{\{1\}} \oplus (2) \CMTB{\{10\}} \oplus (6) \CMTB{\{45\}} \oplus (4) \CMTgrn{\bf \{54\} } \oplus (12) \CMTB{\{120\}} \oplus (4) \CMTB{\{126\}} \oplus (4) \CMTB{\{\overline{126}\}} \oplus (11) \CMTB{\{210\}} \oplus (4) \CMTB{\{210'\}} \oplus (12) \CMTB{\{320\}} \oplus (5) \CMTB{\{660\}} \oplus (11) \CMTB{\{770\}} \oplus (21) \CMTB{\{945\}} \oplus (14) \CMTB{\{1050\}} \oplus (14) \CMTB{\{\overline{1050}\}} \oplus (15) \CMTB{\{1386\}} \oplus (26) \CMTB{\{1728\}} \oplus \CMTB{\{2772\}} \oplus \CMTB{\{\overline{2772}\}} \oplus (20) \CMTB{\{2970\}} \oplus (19) \CMTB{\{3696'\}} \oplus (19) \CMTB{\{\overline{3696'}\}} \oplus (14) \CMTB{\{4125\}} \oplus (30) \CMTB{\{4312\}} \oplus (12) \CMTB{\{4410\}} \oplus (10) \CMTB{\{4608\}} \oplus (13) \CMTB{\{4950\}} \oplus (13) \CMTB{\{\overline{4950}\}} \oplus (30) \CMTB{\{5940\}} \oplus (12) \CMTB{\{6930\}} \oplus (12) \CMTB{\{\overline{6930}\}} \oplus (4) \CMTB{\{6930'\}} \oplus (4) \CMTB{\{\overline{6930'}\}} \oplus (6) \CMTB{\{7644\}} \oplus (31) \CMTB{\{8085\}} \oplus (9) \CMTB{\{8910\}} \oplus (22) \CMTB{\{10560\}} \oplus (3) \CMTB{\{12870\}} \oplus (15) \CMTB{\{14784\}} \oplus (8) \CMTB{\{16380\}} \oplus (9) \CMTB{\{17325\}} \oplus (9) \CMTB{\{\overline{17325}\}} \oplus (28) \CMTB{\{17920\}} \oplus (20) \CMTB{\{23040\}} \oplus (20) \CMTB{\{\overline{23040}\}} \oplus (14) \CMTB{\{27720\}} \oplus (18) \CMTB{\{28160\}} \oplus (14) \CMTB{\{34398\}} \oplus (38) \CMTB{\{36750\}} \oplus (4) \CMTB{\{37632\}} \oplus (6) \CMTB{\{42120\}} \oplus (6) \CMTB{\{46800\}} \oplus (6) \CMTB{\{\overline{46800}\}} \oplus (16) \CMTB{\{48114\}} \oplus (16) \CMTB{\{\overline{48114}\}} \oplus (2) \CMTB{\{48510\}} \oplus \CMTB{\{50050\}} \oplus \CMTB{\{\overline{50050}\}} \oplus (6) \CMTB{\{50688\}} \oplus (6) \CMTB{\{\overline{50688}\}} \oplus \CMTB{\{52920\}} \oplus (12) \CMTB{\{64680\}} \oplus (14) \CMTB{\{68640\}} \oplus (7) \CMTB{\{70070\}} \oplus (4) \CMTB{\{70070'\}} \oplus (28) \CMTB{\{72765\}} \oplus (15) \CMTB{\{73710\}} \oplus (4) \CMTB{\{81081\}} \oplus (11) \CMTB{\{90090\}} \oplus (11) \CMTB{\{\overline{90090}\}} \oplus (7) \CMTB{\{112320\}} \oplus \CMTB{\{124740\}} \oplus (6) \CMTB{\{128700\}} \oplus (6) \CMTB{\{\overline{128700}\}} \oplus (20) \CMTB{\{143000\}} \oplus (3) \CMTB{\{150150\}} \oplus (3) \CMTB{\{\overline{150150}\}} \oplus \CMTB{\{165165\}} \oplus (6) \CMTB{\{174636\}} \oplus (6) \CMTB{\{189189\}} \oplus (12) \CMTB{\{192192\}} \oplus (6) \CMTB{\{199017\}} \oplus (6) \CMTB{\{\overline{199017}\}} \oplus (4) \CMTB{\{207360\}} \oplus (3) \CMTB{\{210210\}} \oplus (2) \CMTB{\{216216\}} \oplus (2) \CMTB{\{\overline{216216}\}} \oplus (4) \CMTB{\{242550\}} \oplus (4) \CMTB{\{\overline{242550}\}} \oplus (2) \CMTB{\{270270\}} \oplus (5) \CMTB{\{274560\}} \oplus (2) \CMTB{\{275968\}} \oplus (2) \CMTB{\{\overline{275968}\}} \oplus (12) \CMTB{\{299520\}} \oplus \CMTB{\{371250\}} \oplus \CMTB{\{\overline{371250}\}} \oplus (2) \CMTB{\{376320\}} \oplus (8) \CMTB{\{380160\}} \oplus (4) \CMTB{\{436590\}} \oplus \CMTB{\{577500\}} \oplus (2) \CMTB{\{590490\}} \oplus \CMTB{\{630630\}} \oplus \CMTB{\{\overline{630630}\}} \oplus (2) \CMTB{\{705600\}} \oplus (2) \CMTB{\{\overline{705600}\}} \oplus \CMTB{\{848925\}} \oplus \CMTB{\{\overline{848925}\}} \oplus \CMTB{\{945945\}} \oplus (4) \CMTB{\{1048576\}} \oplus (2) \CMTB{\{1064448\}} \oplus \CMTB{\{1299078\}}$
\item Level-11: $(4) \CMTred{\{16\}} \oplus (4) \CMTred{\{\overline{16}\}} \oplus (12) \CMTred{\{144\}} 
\oplus (12) \CMTred{\{\overline{144}\}} \oplus (22) \CMTred{\{560\}} \oplus (22) 
\CMTred{\{\overline{560}\}} \oplus (8) \CMTred{\{672\}} \oplus (8) 
\CMTred{\{\overline{672}\}} \oplus (19) \CMTred{\{720\}} \oplus (19) 
\CMTred{\{\overline{720}\}} \oplus (21) \CMTred{\{1200\}} \oplus (21) 
\CMTred{\{\overline{1200}\}} \oplus (16) \CMTred{\{1440\}} \oplus (16) 
\CMTred{\{\overline{1440}\}} \oplus (18) \CMTred{\{2640\}} \oplus (18) 
\CMTred{\{\overline{2640}\}} \oplus (40) \CMTred{\{3696\}} \oplus (40) 
\CMTred{\{\overline{3696}\}} \oplus (11) \CMTred{\{5280\}} \oplus (11) 
\CMTred{\{\overline{5280}\}} \oplus (8) \CMTred{\{7920\}} \oplus (8) 
\CMTred{\{\overline{7920}\}} \oplus (23) \CMTred{\{8064\}} \oplus (23) 
\CMTred{\{\overline{8064}\}} \oplus (43) \CMTred{\{8800\}} \oplus (43) 
\CMTred{\{\overline{8800}\}} \oplus (30) \CMTred{\{11088\}} \oplus (30) 
\CMTred{\{\overline{11088}\}} \oplus (33) \CMTred{\{15120\}} \oplus (33) 
\CMTred{\{\overline{15120}\}} \oplus (9) \CMTred{\{17280\}} \oplus (9) 
\CMTred{\{\overline{17280}\}} \oplus (2) \CMTred{\{20592\}} \oplus (2) 
\CMTred{\{\overline{20592}\}} \oplus (8) \CMTred{\{23760\}} \oplus (8) 
\CMTred{\{\overline{23760}\}} \oplus (32) \CMTred{\{25200\}} \oplus (32) 
\CMTred{\{\overline{25200}\}} \oplus (4) \CMTred{\{29568\}} \oplus (4) 
\CMTred{\{\overline{29568}\}} \oplus (12) \CMTred{\{30800\}} \oplus (12) 
\CMTred{\{\overline{30800}\}} \oplus (26) \CMTred{\{34992\}} \oplus (26) 
\CMTred{\{\overline{34992}\}} \oplus (35) \CMTred{\{38016\}} \oplus (35) 
\CMTred{\{\overline{38016}\}} \oplus (3) \CMTred{\{39600\}} \oplus (3) 
\CMTred{\{\overline{39600}\}} \oplus (22) \CMTred{\{43680\}} \oplus (22) 
\CMTred{\{\overline{43680}\}} \oplus (16) \CMTred{\{48048\}} \oplus (16) 
\CMTred{\{\overline{48048}\}} \oplus (25) \CMTred{\{49280\}} \oplus (25) 
\CMTred{\{\overline{49280}\}} \oplus (13) \CMTred{\{55440\}} \oplus (13) 
\CMTred{\{\overline{55440}\}} \oplus (6) \CMTred{\{70560\}} \oplus (6) 
\CMTred{\{\overline{70560}\}} \oplus (3) \CMTred{\{80080\}} \oplus (3) 
\CMTred{\{\overline{80080}\}} \oplus (6) \CMTred{\{102960\}} \oplus (6) 
\CMTred{\{\overline{102960}\}} \oplus (15) \CMTred{\{124800\}} \oplus (15) 
\CMTred{\{\overline{124800}\}} \oplus (4) \CMTred{\{129360\}} \oplus (4) 
\CMTred{\{\overline{129360}\}} \oplus (29) \CMTred{\{144144\}} \oplus (29) 
\CMTred{\{\overline{144144}\}} \oplus (9) \CMTred{\{155232\}} \oplus (9) 
\CMTred{\{\overline{155232}\}} \oplus (9) \CMTred{\{164736\}} \oplus (9) 
\CMTred{\{\overline{164736}\}} \oplus (11) \CMTred{\{196560\}} \oplus (11) 
\CMTred{\{\overline{196560}\}} \oplus (2) \CMTred{\{196560'\}} \oplus (2) 
\CMTred{\{\overline{196560'}\}} \oplus \CMTred{\{198000\}} \oplus 
\CMTred{\{\overline{198000}\}} \oplus (22) \CMTred{\{205920\}} \oplus (22) 
\CMTred{\{\overline{205920}\}} \oplus \CMTred{\{224224\}} \oplus 
\CMTred{\{\overline{224224}\}} \oplus (7) \CMTred{\{258720\}} \oplus (7) 
\CMTred{\{\overline{258720}\}} \oplus (3) \CMTred{\{274560'\}} \oplus (3) 
\CMTred{\{\overline{274560'}\}} \oplus (2) \CMTred{\{332640\}} \oplus (2) 
\CMTred{\{\overline{332640}\}} \oplus (2) \CMTred{\{343200\}} \oplus (2) 
\CMTred{\{\overline{343200}\}} \oplus (10) \CMTred{\{364000\}} \oplus (10) 
\CMTred{\{\overline{364000}\}} \oplus (5) \CMTred{\{388080\}} \oplus (5) 
\CMTred{\{\overline{388080}\}} \oplus (2) \CMTred{\{388080'\}} \oplus (2) 
\CMTred{\{\overline{388080'}\}} \oplus (2) \CMTred{\{443520\}} \oplus (2) 
\CMTred{\{\overline{443520}\}} \oplus \CMTred{\{458640\}} \oplus 
\CMTred{\{\overline{458640}\}} \oplus (3) \CMTred{\{465696\}} \oplus (3) 
\CMTred{\{\overline{465696}\}} \oplus (10) \CMTred{\{529200\}} \oplus (10) 
\CMTred{\{\overline{529200}\}} \oplus \CMTred{\{758912\}} \oplus 
\CMTred{\{\overline{758912}\}} \oplus \CMTred{\{764400\}} \oplus 
\CMTred{\{\overline{764400}\}} \oplus (7) \CMTred{\{769824\}} \oplus (7) 
\CMTred{\{\overline{769824}\}} \oplus \CMTred{\{784784\}} \oplus 
\CMTred{\{\overline{784784}\}} \oplus (2) \CMTred{\{905520\}} \oplus (2) 
\CMTred{\{\overline{905520}\}} \oplus (2) \CMTred{\{1260000\}} \oplus (2) 
\CMTred{\{\overline{1260000}\}} \oplus (2) \CMTred{\{1441440\}} \oplus (2) 
\CMTred{\{\overline{1441440}\}} \oplus \CMTred{\{1544400\}} \oplus 
\CMTred{\{\overline{1544400}\}} \oplus \CMTred{\{1924560\}} \oplus 
\CMTred{\{\overline{1924560}\}}$
\item Level-12: $(2) \CMTB{\{1\}} \oplus (2) \CMTB{\{10\}} \oplus (6) \CMTB{\{45\}} \oplus (9) \CMTgrn{\bf \{54\} } 
\oplus (12) \CMTB{\{120\}} \oplus (9) \CMTB{\{126\}} \oplus (9) \CMTB{\{\overline{126}\}} \oplus (16) \CMTB{\{210\}} \oplus (8) \CMTB{\{210'\}} \oplus (22) 
\CMTB{\{320\}} \oplus (8) \CMTB{\{660\}} \oplus (27) \CMTB{\{770\}} \oplus (29) \CMTB{\{945\}} 
\oplus (27) \CMTB{\{1050\}} \oplus (27) \CMTB{\{\overline{1050}\}} \oplus (23) 
\CMTB{\{1386\}} \oplus (42) \CMTB{\{1728\}} \oplus (2) \CMTB{\{1782\}} \oplus (5) \CMTB{\{2772\}} 
\oplus (5) \CMTB{\{\overline{2772}\}} \oplus (38) \CMTB{\{2970\}} \oplus (30) 
\CMTB{\{3696'\}} \oplus (30) \CMTB{\{\overline{3696'}\}} \oplus 
(25) \CMTB{\{4125\}} \oplus \CMTB{\{4290\}} \oplus (40) \CMTB{\{4312\}} \oplus (26) \CMTB{\{4410\}} \oplus (16) \CMTB{\{4608\}} \oplus (27) \CMTB{\{4950\}} \oplus (27) 
\CMTB{\{\overline{4950}\}} \oplus (48) \CMTB{\{5940\}} \oplus (19) \CMTB{\{6930\}} 
\oplus (19) \CMTB{\{\overline{6930}\}} \oplus (12) \CMTB{\{6930'\}} 
\oplus (12) \CMTB{\{\overline{6930'}\}} \oplus (7) \CMTB{\{7644\}} 
\oplus (51) \CMTB{\{8085\}} \oplus (18) \CMTB{\{8910\}} \oplus (34) \CMTB{\{10560\}} 
\oplus (7) \CMTB{\{12870\}} \oplus (22) \CMTB{\{14784\}} \oplus (20) \CMTB{\{16380\}} 
\oplus (19) \CMTB{\{17325\}} \oplus (19) \CMTB{\{\overline{17325}\}} \oplus 
(52) \CMTB{\{17920\}} \oplus (3) \CMTB{\{20790\}} \oplus (3) \CMTB{\{\overline{20790}\}} \oplus (38) \CMTB{\{23040\}} \oplus (38) \CMTB{\{\overline{23040}\}} \oplus 
(30) \CMTB{\{27720\}} \oplus (32) \CMTB{\{28160\}} \oplus (2) \CMTB{\{31680\}} \oplus (22) 
\CMTB{\{34398\}} \oplus (70) \CMTB{\{36750\}} \oplus (6) \CMTB{\{37632\}} \oplus (10) 
\CMTB{\{42120\}} \oplus (16) \CMTB{\{46800\}} \oplus (16) \CMTB{\{\overline{46800}\}} 
\oplus (29) \CMTB{\{48114\}} \oplus (29) \CMTB{\{\overline{48114}\}} \oplus (8)
\CMTB{\{48510\}} \oplus (7) \CMTB{\{50050\}} \oplus (7) \CMTB{\{\overline{50050}\}} 
\oplus (14) \CMTB{\{50688\}} \oplus (14) \CMTB{\{\overline{50688}\}} \oplus (2) 
\CMTB{\{52920\}} \oplus (22) \CMTB{\{64680\}} \oplus (34) \CMTB{\{68640\}} \oplus (13) 
\CMTB{\{70070\}} \oplus (4) \CMTB{\{70070'\}} \oplus (51) \CMTB{\{72765\}} \oplus 
(32) \CMTB{\{73710\}} \oplus (12) \CMTB{\{81081\}} \oplus (23) \CMTB{\{90090\}} \oplus 
(23) \CMTB{\{\overline{90090}\}} \oplus (2) \CMTB{\{90090'\}} \oplus 
(2) \CMTB{\{\overline{90090'}\}} \oplus \CMTB{\{105105\}} \oplus (23) 
\CMTB{\{112320\}} \oplus (3) \CMTB{\{123750\}} \oplus \CMTB{\{124740\}} \oplus (2) 
\CMTB{\{126126\}} \oplus (2) \CMTB{\{\overline{126126}\}} \oplus (17) 
\CMTB{\{128700\}} \oplus (17) \CMTB{\{\overline{128700}\}} \oplus (43) \CMTB{\{143000\}} \oplus (4) \CMTB{\{144144'\}} \oplus (4) \CMTB{\{\overline{144144'}\}} \oplus (7) \CMTB{\{150150\}} \oplus (7) \CMTB{\{\overline{150150}\}} 
\oplus \CMTB{\{165165\}} \oplus (14) \CMTB{\{174636\}} \oplus (11) \CMTB{\{189189\}} 
\oplus (28) \CMTB{\{192192\}} \oplus (14) \CMTB{\{199017\}} \oplus (14) 
\CMTB{\{\overline{199017}\}} \oplus (2) \CMTB{\{203840\}} \oplus (12) 
\CMTB{\{207360\}} \oplus (7) \CMTB{\{210210\}} \oplus (7) \CMTB{\{216216\}} \oplus (7) 
\CMTB{\{\overline{216216}\}} \oplus (2) \CMTB{\{219648\}} \oplus (2) 
\CMTB{\{\overline{219648}\}} \oplus (10) \CMTB{\{242550\}} \oplus (10) 
\CMTB{\{\overline{242550}\}} \oplus (2) \CMTB{\{258720'\}} \oplus (4) 
\CMTB{\{270270\}} \oplus (13) \CMTB{\{274560\}} \oplus (8) \CMTB{\{275968\}} \oplus (8)
\CMTB{\{\overline{275968}\}} \oplus \CMTB{\{294294\}} \oplus 
\CMTB{\{\overline{294294}\}} \oplus (30) \CMTB{\{299520\}} \oplus (4) 
\CMTB{\{351000\}} \oplus \CMTB{\{371250\}} \oplus \CMTB{\{\overline{371250}\}} \oplus 
(10) \CMTB{\{376320\}} \oplus (20) \CMTB{\{380160\}} \oplus (14) \CMTB{\{436590\}} \oplus 
(2) \CMTB{\{536250\}} \oplus (2) \CMTB{\{577500\}} \oplus (4) \CMTB{\{590490\}} \oplus (3) 
\CMTB{\{620928\}} \oplus (4) \CMTB{\{630630\}} \oplus (4) \CMTB{\{\overline{630630}\}} 
\oplus (2) \CMTB{\{698880\}} \oplus (2) \CMTB{\{\overline{698880}\}} \oplus (8) 
\CMTB{\{705600\}} \oplus (8) \CMTB{\{\overline{705600}\}} \oplus \CMTB{\{720720\}} 
\oplus \CMTB{\{\overline{720720}\}} \oplus (2) \CMTB{\{831600\}} \oplus (2) 
\CMTB{\{\overline{831600}\}} \oplus (3) \CMTB{\{848925\}} \oplus (3) 
\CMTB{\{\overline{848925}\}} \oplus (2) \CMTB{\{882882\}} \oplus (2) 
\CMTB{\{\overline{882882}\}} \oplus (2) \CMTB{\{884520\}} \oplus \CMTB{\{928125\}} 
\oplus (6) \CMTB{\{945945\}} \oplus (16) \CMTB{\{1048576\}} \oplus (6) \CMTB{\{1064448\}} 
\oplus \CMTB{\{1137500\}} \oplus (2) \CMTB{\{1281280\}} \oplus (5) \CMTB{\{1299078\}} 
\oplus \CMTB{\{1316250\}} \oplus \CMTB{\{1387386\}} \oplus \CMTB{\{\overline{1387386}\}} \oplus \CMTB{\{1698840\}} \oplus (2) \CMTB{\{1774080\}} \oplus \CMTB{\{1897280\}} 
\oplus \CMTB{\{2502500\}} \oplus \CMTB{\{\overline{2502500}\}} \oplus (2) 
\CMTB{\{3706560\}}$
\item Level-13: $(4) \CMTred{\{16\}} \oplus (4) \CMTred{\{\overline{16}\}} \oplus (18) \CMTred{\{144\}} 
\oplus (18) \CMTred{\{\overline{144}\}} \oplus (34) \CMTred{\{560\}} \oplus (34) 
\CMTred{\{\overline{560}\}} \oplus (12) \CMTred{\{672\}} \oplus (12) 
\CMTred{\{\overline{672}\}} \oplus (28) \CMTred{\{720\}} \oplus (28) 
\CMTred{\{\overline{720}\}} \oplus (31) \CMTred{\{1200\}} \oplus (31) 
\CMTred{\{\overline{1200}\}} \oplus (24) \CMTred{\{1440\}} \oplus (24) 
\CMTred{\{\overline{1440}\}} \oplus (25) \CMTred{\{2640\}} \oplus (25) 
\CMTred{\{\overline{2640}\}} \oplus (60) \CMTred{\{3696\}} \oplus (60) 
\CMTred{\{\overline{3696}\}} \oplus (19) \CMTred{\{5280\}} \oplus (19) 
\CMTred{\{\overline{5280}\}} \oplus (13) \CMTred{\{7920\}} \oplus (13) 
\CMTred{\{\overline{7920}\}} \oplus (39) \CMTred{\{8064\}} \oplus (39) 
\CMTred{\{\overline{8064}\}} \oplus (64) \CMTred{\{8800\}} \oplus (64) 
\CMTred{\{\overline{8800}\}} \oplus (48) \CMTred{\{11088\}} \oplus (48) 
\CMTred{\{\overline{11088}\}} \oplus (51) \CMTred{\{15120\}} \oplus (51) 
\CMTred{\{\overline{15120}\}} \oplus (16) \CMTred{\{17280\}} \oplus (16) 
\CMTred{\{\overline{17280}\}} \oplus (5) \CMTred{\{20592\}} \oplus (5) 
\CMTred{\{\overline{20592}\}} \oplus (15) \CMTred{\{23760\}} \oplus (15) 
\CMTred{\{\overline{23760}\}} \oplus (52) \CMTred{\{25200\}} \oplus (52) 
\CMTred{\{\overline{25200}\}} \oplus (3) \CMTred{\{26400\}} \oplus (3) 
\CMTred{\{\overline{26400}\}} \oplus (10) \CMTred{\{29568\}} \oplus (10) 
\CMTred{\{\overline{29568}\}} \oplus (20) \CMTred{\{30800\}} \oplus (20) 
\CMTred{\{\overline{30800}\}} \oplus (41) \CMTred{\{34992\}} \oplus (41) 
\CMTred{\{\overline{34992}\}} \oplus (56) \CMTred{\{38016\}} \oplus (56) 
\CMTred{\{\overline{38016}\}} \oplus (6) \CMTred{\{39600\}} \oplus (6) 
\CMTred{\{\overline{39600}\}} \oplus (39) \CMTred{\{43680\}} \oplus (39) 
\CMTred{\{\overline{43680}\}} \oplus (28) \CMTred{\{48048\}} \oplus (28) 
\CMTred{\{\overline{48048}\}} \oplus \CMTred{\{48048'\}} \oplus 
\CMTred{\{\overline{48048'}\}} \oplus (42) \CMTred{\{49280\}} \oplus (42) 
\CMTred{\{\overline{49280}\}} \oplus (21) \CMTred{\{55440\}} \oplus (21) 
\CMTred{\{\overline{55440}\}} \oplus (10) \CMTred{\{70560\}} \oplus (10) 
\CMTred{\{\overline{70560}\}} \oplus (7) \CMTred{\{80080\}} \oplus (7) 
\CMTred{\{\overline{80080}\}} \oplus (13) \CMTred{\{102960\}} \oplus (13) 
\CMTred{\{\overline{102960}\}} \oplus (28) \CMTred{\{124800\}} \oplus (28) 
\CMTred{\{\overline{124800}\}} \oplus (10) \CMTred{\{129360\}} \oplus (10) 
\CMTred{\{\overline{129360}\}} \oplus (54) \CMTred{\{144144\}} \oplus (54) 
\CMTred{\{\overline{144144}\}} \oplus (19) \CMTred{\{155232\}} \oplus (19) 
\CMTred{\{\overline{155232}\}} \oplus (19) \CMTred{\{164736\}} \oplus (19) 
\CMTred{\{\overline{164736}\}} \oplus (2) \CMTred{\{185328\}} \oplus (2) 
\CMTred{\{\overline{185328}\}} \oplus (22) \CMTred{\{196560\}} \oplus (22)
\CMTred{\{\overline{196560}\}} \oplus (7) \CMTred{\{196560'\}} \oplus (7)
\CMTred{\{\overline{196560'}\}} \oplus (4) \CMTred{\{198000\}} \oplus (4) 
\CMTred{\{\overline{198000}\}} \oplus (40) \CMTred{\{205920\}} \oplus (40)
\CMTred{\{\overline{205920}\}} \oplus (3) \CMTred{\{224224\}} \oplus (3) 
\CMTred{\{\overline{224224}\}} \oplus (15) \CMTred{\{258720\}} \oplus (15) 
\CMTred{\{\overline{258720}\}} \oplus (5) \CMTred{\{274560'\}} \oplus (5) 
\CMTred{\{\overline{274560'}\}} \oplus (2) \CMTred{\{308880\}} \oplus (2) 
\CMTred{\{\overline{308880}\}} \oplus (4) \CMTred{\{332640\}} \oplus (4) 
\CMTred{\{\overline{332640}\}} \oplus (7) \CMTred{\{343200\}} \oplus (7) 
\CMTred{\{\overline{343200}\}} \oplus (19) \CMTred{\{364000\}} \oplus (19) 
\CMTred{\{\overline{364000}\}} \oplus (12) \CMTred{\{388080\}} \oplus (12) 
\CMTred{\{\overline{388080}\}} \oplus (5) \CMTred{\{388080'\}} \oplus (5) 
\CMTred{\{\overline{388080'}\}} \oplus (5) \CMTred{\{443520\}} \oplus (5) 
\CMTred{\{\overline{443520}\}} \oplus (5) \CMTred{\{458640\}} \oplus (5) 
\CMTred{\{\overline{458640}\}} \oplus (7) \CMTred{\{465696\}} \oplus (7) 
\CMTred{\{\overline{465696}\}} \oplus (2) \CMTred{\{498960\}} \oplus (2) 
\CMTred{\{\overline{498960}\}} \oplus \CMTred{\{524160\}} \oplus 
\CMTred{\{\overline{524160}\}} \oplus (24) \CMTred{\{529200\}} \oplus (24) 
\CMTred{\{\overline{529200}\}} \oplus \CMTred{\{758912\}} \oplus 
\CMTred{\{\overline{758912}\}} \oplus (7) \CMTred{\{764400\}} \oplus (7) 
\CMTred{\{\overline{764400}\}} \oplus \CMTred{\{764400'\}} \oplus 
\CMTred{\{\overline{764400'}\}} \oplus (16) \CMTred{\{769824\}} \oplus (16) 
\CMTred{\{\overline{769824}\}} \oplus (3) \CMTred{\{784784\}} \oplus (3) 
\CMTred{\{\overline{784784}\}} \oplus \CMTred{\{831600'\}} \oplus 
\CMTred{\{\overline{831600'}\}} \oplus (6) \CMTred{\{905520\}} \oplus (6) 
\CMTred{\{\overline{905520}\}} \oplus \CMTred{\{917280\}} \oplus 
\CMTred{\{\overline{917280}\}} \oplus \CMTred{\{1123200\}} \oplus 
\CMTred{\{\overline{1123200}\}} \oplus \CMTred{\{1153152\}} \oplus 
\CMTred{\{\overline{1153152}\}} \oplus (6) \CMTred{\{1260000\}} \oplus (6)
\CMTred{\{\overline{1260000}\}} \oplus (6) \CMTred{\{1441440\}} \oplus (6) 
\CMTred{\{\overline{1441440}\}} \oplus (4) \CMTred{\{1544400\}} \oplus (4) 
\CMTred{\{\overline{1544400}\}} \oplus (3) \CMTred{\{1924560\}} \oplus (3) 
\CMTred{\{\overline{1924560}\}} \oplus (2) \CMTred{\{2274480\}} \oplus (2) \CMTred{\{\overline{2274480}\}} \oplus \CMTred{\{2310000\}} \oplus \CMTred{\{\overline{2310000}\}} \oplus (2) \CMTred{\{2402400\}} \oplus (2) \CMTred{\{\overline{2402400}\}} \oplus \CMTred{\{4756752\}} \oplus \CMTred{\{\overline{4756752}\}}$
\item Level-14: $(2) \CMTB{\{1\}} \oplus (2) \CMTB{\{10\}} \oplus (13) \CMTB{\{45\}} \oplus (9) \CMTgrn{\bf \{54\} } \oplus (26) \CMTB{\{120\}} \oplus (9) \CMTB{\{126\}} \oplus (9) \CMTB{\{\overline{126}\}} \oplus (23) \CMTB{\{210\}} \oplus (8) \CMTB{\{210'\}} \oplus (28) \CMTB{\{320\}} \oplus (8) \CMTB{\{660\}} \oplus (27) \CMTB{\{770\}} \oplus (46) \CMTB{\{945\}} \oplus (32) \CMTB{\{1050\}} \oplus (32) \CMTB{\{\overline{1050}\}} \oplus (31) \CMTB{\{1386\}} \oplus (58) \CMTB{\{1728\}} \oplus (2) \CMTB{\{1782\}} \oplus (5) \CMTB{\{2772\}} \oplus (5) \CMTB{\{\overline{2772}\}} \oplus (48) \CMTB{\{2970\}} \oplus (43) \CMTB{\{3696'\}} \oplus (43) \CMTB{\{\overline{3696'}\}} \oplus (30) \CMTB{\{4125\}} \oplus \CMTB{\{4290\}} \oplus (60) \CMTB{\{4312\}} \oplus (32) \CMTB{\{4410\}} \oplus (22) \CMTB{\{4608\}} \oplus (32) \CMTB{\{4950\}} \oplus (32) \CMTB{\{\overline{4950}\}} \oplus (71) \CMTB{\{5940\}} \oplus (30) \CMTB{\{6930\}} \oplus (30) \CMTB{\{\overline{6930}\}} \oplus (15) \CMTB{\{6930'\}} \oplus (15) \CMTB{\{\overline{6930'}\}} \oplus (16) \CMTB{\{7644\}} \oplus (68) \CMTB{\{8085\}} \oplus (21) \CMTB{\{8910\}} \oplus (50) \CMTB{\{10560\}} \oplus (11) \CMTB{\{12870\}} \oplus (37) \CMTB{\{14784\}} \oplus (24) \CMTB{\{16380\}} \oplus (24) \CMTB{\{17325\}} \oplus (24) \CMTB{\{\overline{17325}\}} \oplus (72) \CMTB{\{17920\}} \oplus (5) \CMTB{\{20790\}} \oplus (5) \CMTB{\{\overline{20790}\}} \oplus (54) \CMTB{\{23040\}} \oplus (54) \CMTB{\{\overline{23040}\}} \oplus (40) \CMTB{\{27720\}} \oplus (46) \CMTB{\{28160\}} \oplus (4) \CMTB{\{31680\}} \oplus (40) \CMTB{\{34398\}} \oplus (100) \CMTB{\{36750\}} \oplus (12) \CMTB{\{37632\}} \oplus (20) \CMTB{\{42120\}} \oplus (22) \CMTB{\{46800\}} \oplus (22) \CMTB{\{\overline{46800}\}} \oplus (46) \CMTB{\{48114\}} \oplus (46) \CMTB{\{\overline{48114}\}} \oplus (10) \CMTB{\{48510\}} \oplus (9) \CMTB{\{50050\}} \oplus (9) \CMTB{\{\overline{50050}\}} \oplus (22) \CMTB{\{50688\}} \oplus (22) \CMTB{\{\overline{50688}\}} \oplus (2) \CMTB{\{52920\}} \oplus (4) \CMTB{\{64350\}} \oplus (4) \CMTB{\{\overline{64350}\}} \oplus (30) \CMTB{\{64680\}} \oplus (48) \CMTB{\{68640\}} \oplus (25) \CMTB{\{70070\}} \oplus (10) \CMTB{\{70070'\}} \oplus \CMTB{\{70785\}} \oplus (75) \CMTB{\{72765\}} \oplus (47) \CMTB{\{73710\}} \oplus (20) \CMTB{\{81081\}} \oplus (37) \CMTB{\{90090\}} \oplus (37) \CMTB{\{\overline{90090}\}} \oplus (3) \CMTB{\{90090'\}} \oplus (3) \CMTB{\{\overline{90090'}\}} \oplus (2) \CMTB{\{102960'\}} \oplus (2) \CMTB{\{\overline{102960'}\}} \oplus (6) \CMTB{\{105105\}} \oplus (30) \CMTB{\{112320\}} \oplus (3) \CMTB{\{123750\}} \oplus (4) \CMTB{\{124740\}} \oplus (3) \CMTB{\{126126\}} \oplus (3) \CMTB{\{\overline{126126}\}} \oplus (25) \CMTB{\{128700\}} \oplus (25) \CMTB{\{\overline{128700}\}} \oplus (67) \CMTB{\{143000\}} \oplus (5) \CMTB{\{144144'\}} \oplus (5) \CMTB{\{\overline{144144'}\}} \oplus (14) \CMTB{\{150150\}} \oplus (14) \CMTB{\{\overline{150150}\}} \oplus \CMTB{\{165165\}} \oplus (26) \CMTB{\{174636\}} \oplus (18) \CMTB{\{189189\}} \oplus (42) \CMTB{\{192192\}} \oplus (27) \CMTB{\{199017\}} \oplus (27) \CMTB{\{\overline{199017}\}} \oplus (6) \CMTB{\{203840\}} \oplus (20) \CMTB{\{207360\}} \oplus (11) \CMTB{\{210210\}} \oplus (12) \CMTB{\{216216\}} \oplus (12) \CMTB{\{\overline{216216}\}} \oplus (4) \CMTB{\{219648\}} \oplus (4) \CMTB{\{\overline{219648}\}} \oplus (2) \CMTB{\{237160\}} \oplus (18) \CMTB{\{242550\}} \oplus (18) \CMTB{\{\overline{242550}\}} \oplus (2) \CMTB{\{258720'\}} \oplus (6) \CMTB{\{270270\}} \oplus (18) \CMTB{\{274560\}} \oplus (14) \CMTB{\{275968\}} \oplus (14) \CMTB{\{\overline{275968}\}} \oplus \CMTB{\{294294\}} \oplus \CMTB{\{\overline{294294}\}} \oplus (48) \CMTB{\{299520\}} \oplus (8) \CMTB{\{351000\}} \oplus (2) \CMTB{\{369600\}} \oplus (2) \CMTB{\{\overline{369600}\}} \oplus (4) \CMTB{\{371250\}} \oplus (4) \CMTB{\{\overline{371250}\}} \oplus (18) \CMTB{\{376320\}} \oplus (34) \CMTB{\{380160\}} \oplus (4) \CMTB{\{420420\}} \oplus (26) \CMTB{\{436590\}} \oplus \CMTB{\{462462\}} \oplus (4) \CMTB{\{536250\}} \oplus (5) \CMTB{\{577500\}} \oplus (10) \CMTB{\{590490\}} \oplus (3) \CMTB{\{620928\}} \oplus (10) \CMTB{\{630630\}} \oplus (10) \CMTB{\{\overline{630630}\}} \oplus \CMTB{\{660660\}} \oplus \CMTB{\{\overline{660660}\}} \oplus (4) \CMTB{\{698880\}} \oplus (4) \CMTB{\{\overline{698880}\}} \oplus (16) \CMTB{\{705600\}} \oplus (16) \CMTB{\{\overline{705600}\}} \oplus (3) \CMTB{\{720720\}} \oplus (3) \CMTB{\{\overline{720720}\}} \oplus (3) \CMTB{\{831600\}} \oplus (3) \CMTB{\{\overline{831600}\}} \oplus (7) \CMTB{\{848925\}} \oplus (7) \CMTB{\{\overline{848925}\}} \oplus (4) \CMTB{\{882882\}} \oplus (4) \CMTB{\{\overline{882882}\}} \oplus (4) \CMTB{\{884520\}} \oplus (2) \CMTB{\{928125\}} \oplus (15) \CMTB{\{945945\}} \oplus (28) \CMTB{\{1048576\}} \oplus (10) \CMTB{\{1064448\}} \oplus (5) \CMTB{\{1137500\}} \oplus (2) \CMTB{\{1202850\}} \oplus (2) \CMTB{\{\overline{1202850}\}} \oplus (6) \CMTB{\{1281280\}} \oplus (9) \CMTB{\{1299078\}} \oplus (5) \CMTB{\{1316250\}} \oplus (2) \CMTB{\{1387386\}} \oplus (2) \CMTB{\{\overline{1387386}\}} \oplus (2) \CMTB{\{1559250\}} \oplus \CMTB{\{1683990\}} \oplus \CMTB{\{\overline{1683990}\}} \oplus (2) \CMTB{\{1698840\}} \oplus (4) \CMTB{\{1774080\}} \oplus \CMTB{\{1897280\}} \oplus \CMTB{\{1990170\}} \oplus \CMTB{\{\overline{1990170}\}} \oplus (3) \CMTB{\{2502500\}} \oplus (3) \CMTB{\{\overline{2502500}\}} \oplus (2) \CMTB{\{2520000\}} \oplus \CMTB{\{2866500\}} \oplus \CMTB{\{3071250\}} \oplus \CMTB{\{\overline{3071250}\}} \oplus (6) \CMTB{\{3706560\}} \oplus \CMTB{\{3838185\}} \oplus \CMTB{\{4802490\}}$
\item Level-15: $(12) \CMTred{\{16\}} \oplus (12) \CMTred{\{\overline{16}\}} \oplus (25) \CMTred{\{144\}} \oplus (25) \CMTred{\{\overline{144}\}} \oplus (40) \CMTred{\{560\}} \oplus (40) \CMTred{\{\overline{560}\}} \oplus (12) \CMTred{\{672\}} \oplus (12) \CMTred{\{\overline{672}\}} \oplus (34) \CMTred{\{720\}} \oplus (34) \CMTred{\{\overline{720}\}} \oplus (43) \CMTred{\{1200\}} \oplus (43) \CMTred{\{\overline{1200}\}} \oplus (30) \CMTred{\{1440\}} \oplus (30) \CMTred{\{\overline{1440}\}} \oplus (32) \CMTred{\{2640\}} \oplus (32) \CMTred{\{\overline{2640}\}} \oplus (69) \CMTred{\{3696\}} \oplus (69) \CMTred{\{\overline{3696}\}} \oplus (21) \CMTred{\{5280\}} \oplus (21) \CMTred{\{\overline{5280}\}} \oplus (18) \CMTred{\{7920\}} \oplus (18) \CMTred{\{\overline{7920}\}} \oplus (47) \CMTred{\{8064\}} \oplus (47) \CMTred{\{\overline{8064}\}} \oplus (82) \CMTred{\{8800\}} \oplus (82) \CMTred{\{\overline{8800}\}} \oplus \CMTred{\{9504\}} \oplus \CMTred{\{\overline{9504}\}} \oplus (59) \CMTred{\{11088\}} \oplus (59) \CMTred{\{\overline{11088}\}} \oplus (62) \CMTred{\{15120\}} \oplus (62) \CMTred{\{\overline{15120}\}} \oplus (20) \CMTred{\{17280\}} \oplus (20) \CMTred{\{\overline{17280}\}} \oplus (8) \CMTred{\{20592\}} \oplus (8) \CMTred{\{\overline{20592}\}} \oplus (19) \CMTred{\{23760\}} \oplus (19) \CMTred{\{\overline{23760}\}} \oplus (68) \CMTred{\{25200\}} \oplus (68) \CMTred{\{\overline{25200}\}} \oplus (4) \CMTred{\{26400\}} \oplus (4) \CMTred{\{\overline{26400}\}} \oplus (12) \CMTred{\{29568\}} \oplus (12) \CMTred{\{\overline{29568}\}} \oplus (28) \CMTred{\{30800\}} \oplus (28) \CMTred{\{\overline{30800}\}} \oplus (53) \CMTred{\{34992\}} \oplus (53) \CMTred{\{\overline{34992}\}} \oplus (72) \CMTred{\{38016\}} \oplus (72) \CMTred{\{\overline{38016}\}} \oplus (7) \CMTred{\{39600\}} \oplus (7) \CMTred{\{\overline{39600}\}} \oplus (51) \CMTred{\{43680\}} \oplus (51) \CMTred{\{\overline{43680}\}} \oplus (35) \CMTred{\{48048\}} \oplus (35) \CMTred{\{\overline{48048}\}} \oplus (3) \CMTred{\{48048'\}} \oplus (3) \CMTred{\{\overline{48048'}\}} \oplus (53) \CMTred{\{49280\}} \oplus (53) \CMTred{\{\overline{49280}\}} \oplus (27) \CMTred{\{55440\}} \oplus (27) \CMTred{\{\overline{55440}\}} \oplus (2) \CMTred{\{68640'\}} \oplus (2) \CMTred{\{\overline{68640'}\}} \oplus (18) \CMTred{\{70560\}} \oplus (18) \CMTred{\{\overline{70560}\}} \oplus (9) \CMTred{\{80080\}} \oplus (9) \CMTred{\{\overline{80080}\}} \oplus (21) \CMTred{\{102960\}} \oplus (21) \CMTred{\{\overline{102960}\}} \oplus \CMTred{\{1029'\}} \oplus \CMTred{\{\overline{10296'}\}} \oplus (39) \CMTred{\{124800\}} \oplus (39) \CMTred{\{\overline{124800}\}} \oplus (13) \CMTred{\{129360\}} \oplus (13) \CMTred{\{\overline{129360}\}} \oplus (70) \CMTred{\{144144\}} \oplus (70) \CMTred{\{\overline{144144}\}} \oplus (26) \CMTred{\{155232\}} \oplus (26) \CMTred{\{\overline{155232}\}} \oplus (26) \CMTred{\{164736\}} \oplus (26) \CMTred{\{\overline{164736}\}} \oplus (4) \CMTred{\{185328\}} \oplus (4) \CMTred{\{\overline{185328}\}} \oplus (30) \CMTred{\{196560\}} \oplus (30) \CMTred{\{\overline{196560}\}} \oplus (11) \CMTred{\{196560'\}} \oplus (11) \CMTred{\{\overline{196560'}\}} \oplus (8) \CMTred{\{198000\}} \oplus (8) \CMTred{\{\overline{198000}\}} \oplus \CMTred{\{203280\}} \oplus \CMTred{\{\overline{203280}\}} \oplus (54) \CMTred{\{205920\}} \oplus (54) \CMTred{\{\overline{205920}\}} \oplus (3) \CMTred{\{224224\}} \oplus (3) \CMTred{\{\overline{224224}\}} \oplus (21) \CMTred{\{258720\}} \oplus (21) \CMTred{\{\overline{258720}\}} \oplus (7) \CMTred{\{274560'\}} \oplus (7) \CMTred{\{\overline{274560'}\}} \oplus (3) \CMTred{\{308880\}} \oplus (3) \CMTred{\{\overline{308880}\}} \oplus (8) \CMTred{\{332640\}} \oplus (8) \CMTred{\{\overline{332640}\}} \oplus (13) \CMTred{\{343200\}} \oplus (13) \CMTred{\{\overline{343200}\}} \oplus (26) \CMTred{\{364000\}} \oplus (26) \CMTred{\{\overline{364000}\}} \oplus (16) \CMTred{\{388080\}} \oplus (16) \CMTred{\{\overline{388080}\}} \oplus (9) \CMTred{\{388080'\}} \oplus (9) \CMTred{\{\overline{388080'}\}} \oplus (2) \CMTred{\{428064\}} \oplus (2) \CMTred{\{\overline{428064}\}} \oplus (7) \CMTred{\{443520\}} \oplus (7) \CMTred{\{\overline{443520}\}} \oplus (8) \CMTred{\{458640\}} \oplus (8) \CMTred{\{\overline{458640}\}} \oplus (11) \CMTred{\{465696\}} \oplus (11) \CMTred{\{\overline{465696}\}} \oplus (4) \CMTred{\{498960\}} \oplus (4) \CMTred{\{\overline{498960}\}} \oplus \CMTred{\{522720\}} \oplus \CMTred{\{\overline{522720}\}} \oplus \CMTred{\{524160\}} \oplus \CMTred{\{\overline{524160}\}} \oplus (34) \CMTred{\{529200\}} \oplus (34) \CMTred{\{\overline{529200}\}} \oplus (2) \CMTred{\{758912\}} \oplus (2) \CMTred{\{\overline{758912}\}} \oplus (12) \CMTred{\{764400\}} \oplus (12) \CMTred{\{\overline{764400}\}} \oplus \CMTred{\{764400'\}} \oplus \CMTred{\{\overline{764400'}\}} \oplus (24) \CMTred{\{769824\}} \oplus (24) \CMTred{\{\overline{769824}\}} \oplus (5) \CMTred{\{784784\}} \oplus (5) \CMTred{\{\overline{784784}\}} \oplus (3) \CMTred{\{831600'\}} \oplus (3) \CMTred{\{\overline{831600'}\}} \oplus (8) \CMTred{\{905520\}} \oplus (8) \CMTred{\{\overline{905520}\}} \oplus \CMTred{\{917280\}} \oplus \CMTred{\{\overline{917280}\}} \oplus (2) \CMTred{\{990000\}} \oplus (2) \CMTred{\{\overline{990000}\}} \oplus \CMTred{\{1121120\}} \oplus \CMTred{\{\overline{1121120}\}} \oplus \CMTred{\{1123200\}} \oplus \CMTred{\{\overline{1123200}\}} \oplus \CMTred{\{1153152\}} \oplus \CMTred{\{\overline{1153152}\}} \oplus (9) \CMTred{\{1260000\}} \oplus (9) \CMTred{\{\overline{1260000}\}} \oplus (10) \CMTred{\{1441440\}} \oplus (10) \CMTred{\{\overline{1441440}\}} \oplus (8) \CMTred{\{1544400\}} \oplus (8) \CMTred{\{\overline{1544400}\}} \oplus (5) \CMTred{\{1924560\}} \oplus (5) \CMTred{\{\overline{1924560}\}} \oplus (4) \CMTred{\{2274480\}} \oplus (4) \CMTred{\{\overline{2274480}\}} \oplus (3) \CMTred{\{2310000\}} \oplus (3) \CMTred{\{\overline{2310000}\}} \oplus (4) \CMTred{\{2402400\}} \oplus (4) \CMTred{\{\overline{2402400}\}} \oplus \CMTred{\{2469600\}} \oplus \CMTred{\{\overline{2469600}\}} \oplus \CMTred{\{2642640\}} \oplus \CMTred{\{\overline{2642640}\}} \oplus \CMTred{\{4410000\}} \oplus \CMTred{\{\overline{4410000}\}} \oplus (2) \CMTred{\{4756752\}} \oplus (2) \CMTred{\{\overline{4756752}\}}$
\item Level-16: $(11) \CMTB{\{1\}} \oplus (10) \CMTB{\{10\}} \oplus (13) \CMTB{\{45\}} \oplus (16) \CMTgrn{\bf \{54\} } \oplus (26) \CMTB{\{120\}} \oplus (16) \CMTB{\{126\}} \oplus (16) \CMTB{\{\overline{126}\}} \oplus (30) \CMTB{\{210\}} \oplus (14) \CMTB{\{210'\}} \oplus (28) \CMTB{\{320\}} \oplus (14) \CMTB{\{660\}} \oplus (32) \CMTB{\{770\}} \oplus (46) \CMTB{\{945\}} \oplus (38) \CMTB{\{1050\}} \oplus (38) \CMTB{\{\overline{1050}\}} \oplus (31) \CMTB{\{1386\}} \oplus (64) \CMTB{\{1728\}} \oplus (6) \CMTB{\{1782\}} \oplus (6) \CMTB{\{2772\}} \oplus (6) \CMTB{\{\overline{2772}\}} \oplus (58) \CMTB{\{2970\}} \oplus (43) \CMTB{\{3696'\}} \oplus (43) \CMTB{\{\overline{3696'}\}} \oplus (45) \CMTB{\{4125\}} \oplus (4) \CMTB{\{4290\}} \oplus (60) \CMTB{\{4312\}} \oplus (36) \CMTB{\{4410\}} \oplus (24) \CMTB{\{4608\}} \oplus (40) \CMTB{\{4950\}} \oplus (40) \CMTB{\{\overline{4950}\}} \oplus (76) \CMTB{\{5940\}} \oplus (30) \CMTB{\{6930\}} \oplus (30) \CMTB{\{\overline{6930}\}} \oplus (16) \CMTB{\{6930'\}} \oplus (16) \CMTB{\{\overline{6930'}\}} \oplus (16) \CMTB{\{7644\}} \oplus (76) \CMTB{\{8085\}} \oplus (27) \CMTB{\{8910\}} \oplus (2) \CMTB{\{9438\}} \oplus (60) \CMTB{\{10560\}} \oplus (11) \CMTB{\{12870\}} \oplus (38) \CMTB{\{14784\}} \oplus (30) \CMTB{\{16380\}} \oplus (30) \CMTB{\{17325\}} \oplus (30) \CMTB{\{\overline{17325}\}} \oplus (80) \CMTB{\{17920\}} \oplus \CMTB{\{19305\}} \oplus (7) \CMTB{\{20790\}} \oplus (7) \CMTB{\{\overline{20790}\}} \oplus (60) \CMTB{\{23040\}} \oplus (60) \CMTB{\{\overline{23040}\}} \oplus (52) \CMTB{\{27720\}} \oplus (52) \CMTB{\{28160\}} \oplus \CMTB{\{28314\}} \oplus \CMTB{\{\overline{28314}\}} \oplus (4) \CMTB{\{31680\}} \oplus (40) \CMTB{\{34398\}} \oplus (110) \CMTB{\{36750\}} \oplus (16) \CMTB{\{37632\}} \oplus (20) \CMTB{\{42120\}} \oplus (31) \CMTB{\{46800\}} \oplus (31) \CMTB{\{\overline{46800}\}} \oplus (49) \CMTB{\{48114\}} \oplus (49) \CMTB{\{\overline{48114}\}} \oplus (12) \CMTB{\{48510\}} \oplus (13) \CMTB{\{50050\}} \oplus (13) \CMTB{\{\overline{50050}\}} \oplus (24) \CMTB{\{50688\}} \oplus (24) \CMTB{\{\overline{50688}\}} \oplus (7) \CMTB{\{52920\}} \oplus (5) \CMTB{\{64350\}} \oplus (5) \CMTB{\{\overline{64350}\}} \oplus (36) \CMTB{\{64680\}} \oplus (56) \CMTB{\{68640\}} \oplus (25) \CMTB{\{70070\}} \oplus (10) \CMTB{\{70070'\}} \oplus \CMTB{\{70785\}} \oplus (86) \CMTB{\{72765\}} \oplus (56) \CMTB{\{73710\}} \oplus (24) \CMTB{\{81081\}} \oplus \CMTB{\{84942\}} \oplus \CMTB{\{\overline{84942}\}} \oplus (41) \CMTB{\{90090\}} \oplus (41) \CMTB{\{\overline{90090}\}} \oplus (6) \CMTB{\{90090'\}} \oplus (6) \CMTB{\{\overline{90090'}\}} \oplus (2) \CMTB{\{102960'\}} \oplus (2) \CMTB{\{\overline{102960'}\}} \oplus (6) \CMTB{\{105105\}} \oplus (40) \CMTB{\{112320\}} \oplus (4) \CMTB{\{123750\}} \oplus (4) \CMTB{\{124740\}} \oplus (5) \CMTB{\{126126\}} \oplus (5) \CMTB{\{\overline{126126}\}} \oplus (31) \CMTB{\{128700\}} \oplus (31) \CMTB{\{\overline{128700}\}} \oplus \CMTB{\{141570\}} \oplus \CMTB{\{\overline{141570}\}} \oplus (74) \CMTB{\{143000\}} \oplus (8) \CMTB{\{144144'\}} \oplus (8) \CMTB{\{\overline{144144'}\}} \oplus (15) \CMTB{\{150150\}} \oplus (15) \CMTB{\{\overline{150150}\}} \oplus (2) \CMTB{\{165165\}} \oplus (29) \CMTB{\{174636\}} \oplus (19) \CMTB{\{189189\}} \oplus (48) \CMTB{\{192192\}} \oplus (29) \CMTB{\{199017\}} \oplus (29) \CMTB{\{\overline{199017}\}} \oplus (8) \CMTB{\{203840\}} \oplus (24) \CMTB{\{207360\}} \oplus (14) \CMTB{\{210210\}} \oplus (15) \CMTB{\{216216\}} \oplus (15) \CMTB{\{\overline{216216}\}} \oplus (4) \CMTB{\{219648\}} \oplus (4) \CMTB{\{\overline{219648}\}} \oplus (2) \CMTB{\{237160\}} \oplus (24) \CMTB{\{242550\}} \oplus (24) \CMTB{\{\overline{242550}\}} \oplus (6) \CMTB{\{258720'\}} \oplus (6) \CMTB{\{270270\}} \oplus (24) \CMTB{\{274560\}} \oplus (16) \CMTB{\{275968\}} \oplus (16) \CMTB{\{\overline{275968}\}} \oplus \CMTB{\{286650\}} \oplus \CMTB{\{\overline{286650}\}} \oplus \CMTB{\{294294\}} \oplus \CMTB{\{\overline{294294}\}} \oplus (56) \CMTB{\{299520\}} \oplus (14) \CMTB{\{351000\}} \oplus (4) \CMTB{\{369600\}} \oplus (4) \CMTB{\{\overline{369600}\}} \oplus (4) \CMTB{\{371250\}} \oplus (4) \CMTB{\{\overline{371250}\}} \oplus (20) \CMTB{\{376320\}} \oplus (40) \CMTB{\{380160\}} \oplus (5) \CMTB{\{420420\}} \oplus (30) \CMTB{\{436590\}} \oplus (2) \CMTB{\{462462\}} \oplus (6) \CMTB{\{536250\}} \oplus (5) \CMTB{\{577500\}} \oplus (10) \CMTB{\{590490\}} \oplus (6) \CMTB{\{620928\}} \oplus (11) \CMTB{\{630630\}} \oplus (11) \CMTB{\{\overline{630630}\}} \oplus \CMTB{\{660660\}} \oplus \CMTB{\{\overline{660660}\}} \oplus \CMTB{\{680625\}} \oplus \CMTB{\{\overline{680625}\}} \oplus (4) \CMTB{\{698880\}} \oplus (4) \CMTB{\{\overline{698880}\}} \oplus (20) \CMTB{\{705600\}} \oplus (20) \CMTB{\{\overline{705600}\}} \oplus (3) \CMTB{\{720720\}} \oplus (3) \CMTB{\{\overline{720720}\}} \oplus \CMTB{\{749112\}} \oplus (6) \CMTB{\{831600\}} \oplus (6) \CMTB{\{\overline{831600}\}} \oplus (9) \CMTB{\{848925\}} \oplus (9) \CMTB{\{\overline{848925}\}} \oplus (6) \CMTB{\{882882\}} \oplus (6) \CMTB{\{\overline{882882}\}} \oplus (6) \CMTB{\{884520\}} \oplus (2) \CMTB{\{917280'\}} \oplus (5) \CMTB{\{928125\}} \oplus (20) \CMTB{\{945945\}} \oplus (32) \CMTB{\{1048576\}} \oplus (12) \CMTB{\{1064448\}} \oplus (6) \CMTB{\{1137500\}} \oplus (2) \CMTB{\{1202850\}} \oplus (2) \CMTB{\{\overline{1202850}\}} \oplus (8) \CMTB{\{1281280\}} \oplus (12) \CMTB{\{1299078\}} \oplus (5) \CMTB{\{1316250\}} \oplus (2) \CMTB{\{1387386\}} \oplus (2) \CMTB{\{\overline{1387386}\}} \oplus \CMTB{\{1524600\}} \oplus \CMTB{\{\overline{1524600}\}} \oplus (2) \CMTB{\{1559250\}} \oplus \CMTB{\{1683990\}} \oplus \CMTB{\{\overline{1683990}\}} \oplus (2) \CMTB{\{1698840\}} \oplus (4) \CMTB{\{1774080\}} \oplus (2) \CMTB{\{1897280\}} \oplus \CMTB{\{1990170\}} \oplus \CMTB{\{\overline{1990170}\}} \oplus \CMTB{\{2388750\}} \oplus (5) \CMTB{\{2502500\}} \oplus (5) \CMTB{\{\overline{2502500}\}} \oplus (4) \CMTB{\{2520000\}} \oplus \CMTB{\{2866500\}} \oplus \CMTB{\{3071250\}} \oplus \CMTB{\{\overline{3071250}\}} \oplus \CMTB{\{3320240\}} \oplus \CMTB{\{\overline{3320240}\}} \oplus (8) \CMTB{\{3706560\}} \oplus \CMTB{\{3838185\}} \oplus (2) \CMTB{\{4802490\}}$
\end{itemize}

\newpage
\subsection{Adinkra Diagram for 10D, $\mathcal{N} = 2$A Superfield}

In this section, we present the ten dimensional $\mathcal{N} = 2$A adinkra diagram up to the cubic level. Based on the results listed in Section \ref{sec:2A}, we can draw the complete adinkra diagram in principle: use open nodes to denote bosonic component fields and put their corresponding irreps in the center. For fermionic component fields, use closed nodes. The number of level represents the height assignment. Black edges connect nodes in the adjacent levels, meaning SUSY transformations. Due to the limited space of the paper, we only draw the adinkra up to the cubic level. 

\begin{figure}[htp!]
\centering
\includegraphics[width=0.8\textwidth]{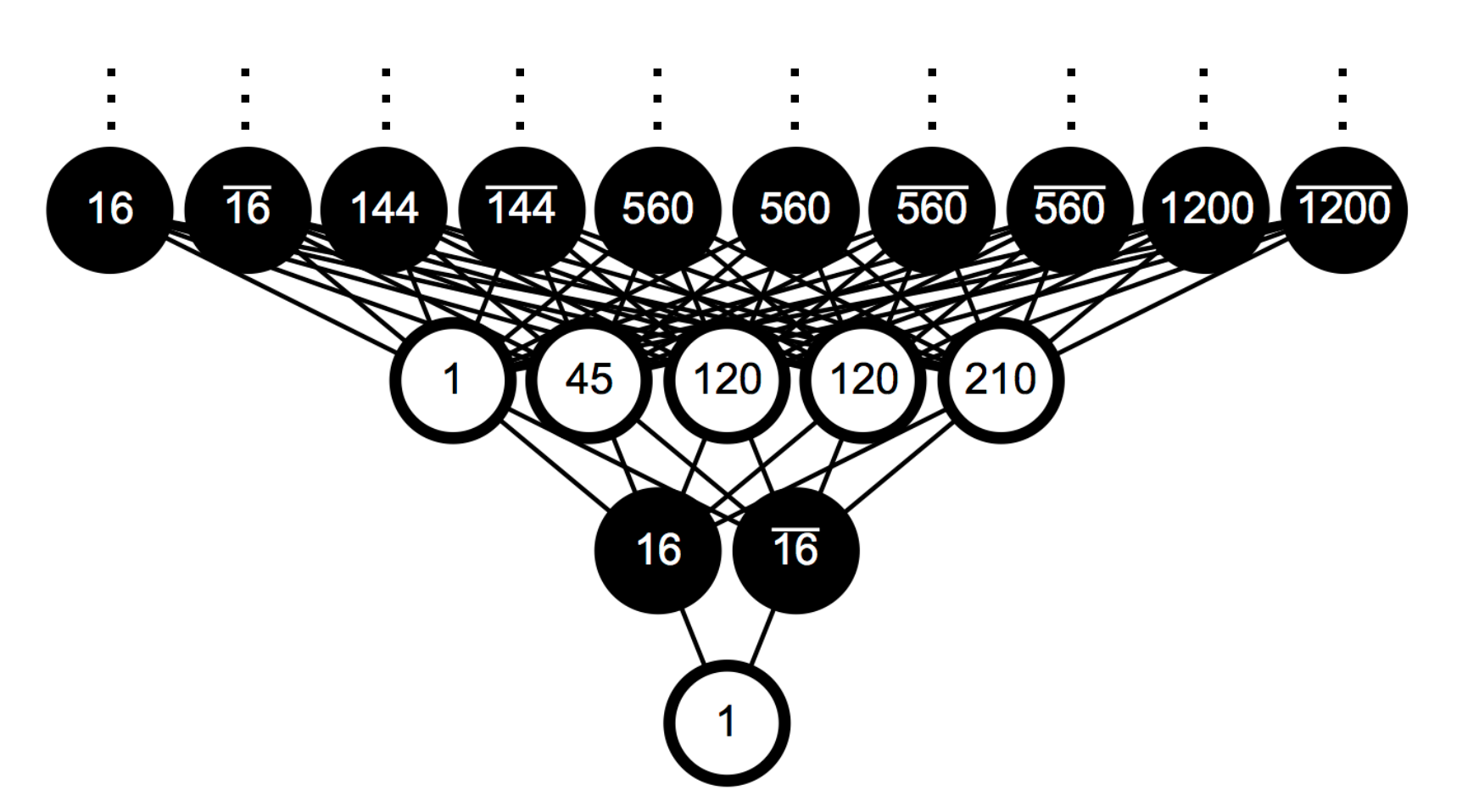}
\caption{Adinkra Diagram for 10D, $\mathcal{N} = 2$A At Low Orders\label{Fig:10DTypeIIA}}
\end{figure}

\newpage
\section{10D, $\mathcal{N}=2$B Scalar Superfield Decomposition and Superconformal Multiplet}

\subsection{Methodology for 10D, $\mathcal{N}=2$B Scalar Superfield Construction}
\label{subsec:2Bmethod}

The Type IIB theory means $\mathcal{N} = (2,0)$, i.e. we have two sets of spinor coordinates denoted by $\theta^{\a^{1}}$ and $\theta^{\a^{2}}$. Following the same logic as in Type IIA, one can organize the $\mathcal{N}=2$B scalar superfield as
\begin{equation}
    {\cal V} (x^{\un{a}}, \, \theta^{\a^{1}}, \, \theta^{\a^{2}}) ~=~ {\cal V}^{(0)} (x^{\un{a}}, \, \theta^{\a^{1}}) ~+~ \theta^{\a^{2}} \, {\cal V}^{(1)}_{\a^{2}} (x^{\un{a}}, \, \theta^{\a^{1}}) ~+~ \theta^{\a^{2}} \theta^{\b^{2}} \, {\cal V}^{(2)}_{\a^{2}\b^{2}}  (x^{\un{a}}, \, \theta^{\a^{1}}) ~+~ \dots   ~~~,
\end{equation}
where the expansion coefficients ${\cal V}^{(0)} (x^{\un{a}}, \, \theta^{\a^{1}})$, ${\cal V}^{(1)}_{\a^{2}} (x^{\un{a}}, \, \theta^{\a^{1}})$, ${\cal V}^{(2)}_{\a^{2}\b^{2}}  (x^{\un{a}}, \, \theta^{\a^{1}}), \dots$ are 10D, $\mathcal{N} = 1$ superfields. 
Attaching another copy of $n$ totally antisymmetric undotted $\theta$'s to an undotted $\theta-$monomial corresponds to tensoring the $\CMTred{\{16\}}^{\wedge n}$ representation on it, i.e. the $n$th level of the $\theta-$monomial of the 10D, $\mathcal{N}=1$ superfield itself, without any conjugate. Therefore, the $\mathcal{N}=2$B scalar superfield $\theta-$monomial decomposition can be summarized as
\begin{equation}
\label{equ:2B_formula}
{\cal V} ~=~ \begin{cases}
{~~}{\rm {Level}}-0 \,~~~~~~~~~~ \ell_{0} \otimes \ell_{0} ~~~,  \\
{~~}{\rm {Level}}-1 \,~~~~~~~~~~ \ell_{1} \otimes \ell_{0} \, \oplus \, \ell_{0} \otimes \ell_{1} ~~~,  \\
{~~}{\rm {Level}}-2 \,~~~~~~~~~~ \ell_{2} \otimes \ell_{0} \, \oplus \, \ell_{1} \otimes \ell_{1} \, \oplus \, \ell_{0} \otimes \ell_{2} ~~~,  \\
{~~}{\rm {Level}}-3 \,~~~~~~~~~~ \ell_{3} \otimes \ell_{0} \, \oplus \, \ell_{2} \otimes \ell_{1} \, \oplus \, \ell_{1} \otimes \ell_{2} \, \oplus \, \ell_{0} \otimes \ell_{3} ~~~,  \\
{~~~~~~}  {~~~~} \vdots  {~~~~~~~~~\,~~~~~~} \vdots \\
{~~}{\rm {Level}}-n \,~~~~~~~~~~ \ell_{n} \otimes \ell_{0} \, \oplus \, \ell_{n-1} \otimes \ell_{1} \, \oplus \, \dots \, \oplus \, \ell_{1} \otimes \ell_{n-1} \, \oplus \, \ell_{0} \otimes \ell_{n} ~~~,  \\
{~~~~~~}  {~~~~} \vdots  {~~~~~~~~~\,~~~~~~} \vdots \\
{~~}{\rm {Level}}-16 ~~~~~~~~~ \ell_{16} \otimes \ell_{0} \, \oplus \, \ell_{15} \otimes \ell_{1} \, \oplus \, \dots \, \oplus \, \ell_{1} \otimes \ell_{15} \, \oplus \, \ell_{0} \otimes \ell_{16} ~~~,  \\
{~~~~~~}  {~~~~} \vdots  {~~~~~~~~~\,~~~~~~} \vdots 
\end{cases}
\end{equation}
and level-17 to 32 are the conjugates of level-15 to 0 respectively. The irreps corresponding to the component fields are the conjugates of the $\theta-$monomials at each level.  

\subsection{10D, $\mathcal{N} = 2$B Scalar Superfield Decomposition Results and Superconformal Multiplet}
\label{sec:2B}

Based on Equations~(\ref{equ:10DTypeI}) and (\ref{equ:2B_formula}), we can directly obtain the scalar superfield decomposition in 10D, $\mathcal{N} = 2$B. The results for $\theta-$monomials from level-0 to level-16 are listed below.  Same as in the Type IIA case, we can find graviton embeddings and gravitino embeddings in the $\theta-$expansion of the scalar superfield.  We use green color to highlight the irrep $\CMTgrn{\bf \{54\}}$ which corresponds to the traceless graviton in 10D as well. 
We can translate irreps into component fields and see 72 graviton embeddings, 280 gravitino embeddings, and accompanying auxiliary fields. The scalar superfield is the simplest superfield that contains traceless graviton and gravitino embeddings that can be used as a starting point to construct supergravity. 
The numbers of gravitons and gravitinos are the same as those in Type IIA case. 

Note that if one find a $\CMTgrn{\bf\{54\}}$ in level-$n$, one can find $\CMTred{\{144\}}$ in level-$(n+1)$, which can be interpreted by SUSY transformation laws of the graviton $h_{\un{a}\un{b}}$ in 10D, $\mathcal{N} = 2$B theory. Since in Type IIB theory we have two copies of $\theta^{\a}$ rather than $\theta^{\a}$ and $\theta^{\Dot{\a}}$, we have D-operators ${\rm D}_{\a^i}$ where $i=1,2$. 
Acting the D-operators on the graviton gives a term proportional to the gravitino $\psi_{\un{b}}{}^{\b^i}$ in the on-shell case (in the off-shell case, there are several auxiliary fields showing up in the r.h.s. besides the undotted gravitino)
\begin{equation}
    {\rm D}_{\a^i} h_{\un{a}\un{b}} ~\propto~ (\sigma_{(\un{a}})_{\a^i\b^i} \, \psi_{\un{b})}{}^{\b^i} ~~~.
\end{equation}
Note that the spinor indices on the gravitinos are superscript. It suggests that the gravitinos correspond to irrep $\CMTred{\{\overline{144}\}}$ in our convention for both $i=1$ and $2$. In the context of $\theta-$monomials, the conjugate of $\CMTgrn{\bf\{54\}}$ is still $\CMTgrn{\bf\{54\}}$, and the conjugate of $\CMTred{\{\overline{144}\}}$ is $\CMTred{\{144\}}$.  

\begin{itemize}
\item Level-0: $\CMTB{\{1\}}$
\item Level-1: $\CMTred{\{16\}}~\oplus~\CMTred{\{16\}}$
\item Level-2: $\CMTB{\{10\}}~\oplus~(3)\CMTB{\{120\}}~\oplus~\CMTB{\{\overline{126}\}}$
\item Level-3: $(2) \CMTred{\{\overline{16}\}} \oplus (2) \CMTred{\{\overline{144}\}} \oplus \
(4) \CMTred{\{\overline{560}\}} \oplus (2) \CMTred{\{\overline{1200}\}}$
\item Level-4: $\CMTB{\{1\}} \oplus (3) \CMTB{\{45\}} \oplus \CMTgrn{\bf \{54\}} \oplus (4) \CMTB{\{210\}} \oplus (5) \
\CMTB{\{770\}} \oplus (3) \CMTB{\{945\}} \oplus (5) \CMTB{\{1050\}} \oplus \
\CMTB{\{\overline{1050}\}} \oplus \CMTB{\{4125\}} \oplus (3) \CMTB{\{5940\}}$
\item Level-5: $(2) \CMTred{\{16\}} \oplus (6) \CMTred{\{144\}} \oplus (6) \CMTred{\{560\}} \oplus (6) \CMTred{\{672\}} \
\oplus (2) \CMTred{\{720\}} \oplus (6) \CMTred{\{1200\}} \oplus (2) \CMTred{\{1440\}} \oplus \
(10) \CMTred{\{3696\}} \oplus (4) \CMTred{\{8064\}} \oplus (2) \CMTred{\{8800\}} \oplus (4) \
\CMTred{\{11088\}} \oplus (2) \CMTred{\{25200\}}$
\item Level-6: $\CMTB{\{10\}} \oplus (6) \CMTB{\{120\}} \oplus (6) \CMTB{\{126\}} \oplus \CMTB{\{\overline{126}\}} \oplus \CMTB{\{210'\}} \oplus (8) \CMTB{\{320\}} \oplus (12) \CMTB{\{1728\}} \
\oplus (9) \CMTB{\{2970\}} \oplus (15) \CMTB{\{3696'\}} \oplus (3) \
\CMTB{\{\overline{3696'}\}} \oplus (10) \CMTB{\{4312\}} \oplus (5) \
\CMTB{\{4410\}} \oplus (5) \CMTB{\{4950\}} \oplus \CMTB{\{\overline{4950}\}} \oplus \
(5) \CMTB{\{6930'\}} \oplus (4) \CMTB{\{10560\}} \oplus \CMTB{\{27720\}} \oplus \
(3) \CMTB{\{34398\}} \oplus (8) \CMTB{\{36750\}} \oplus \CMTB{\{46800\}} \oplus (3) \
\CMTB{\{48114\}}$
\item Level-7: $(2) \CMTred{\{\overline{16}\}} \oplus (6) \CMTred{\{\overline{144}\}} \oplus \
(16) \CMTred{\{\overline{560}\}} \oplus (12) \CMTred{\{\overline{720}\}} \oplus \
(8) \CMTred{\{\overline{1200}\}} \oplus (12) \CMTred{\{\overline{1440}\}} \
\oplus (10) \CMTred{\{\overline{2640}\}} \oplus (16) \CMTred{\{\overline{3696}\
\}} \oplus (10) \CMTred{\{\overline{5280}\}} \oplus (12) \CMTred{\{\overline{8064}\}} \oplus (26) \CMTred{\{\overline{8800}\}} \oplus (6) \
\CMTred{\{\overline{11088}\}} \oplus (10) \CMTred{\{\overline{15120}\}} \oplus \
(6) \CMTred{\{\overline{25200}\}} \oplus (4) \CMTred{\{\overline{29568}\}} \
\oplus (2) \CMTred{\{\overline{30800}\}} \oplus (12) \CMTred{\{\overline{34992}\}} \oplus (8) \CMTred{\{\overline{38016}\}} \oplus (4) \
\CMTred{\{\overline{43680}\}} \oplus (4) \CMTred{\{\overline{49280}\}} \oplus \
(2) \CMTred{\{\overline{70560}\}} \oplus (2) \CMTred{\{\overline{102960}\}} \
\oplus (6) \CMTred{\{\overline{144144}\}}$
\item Level-8: $\CMTB{\{1\}} \oplus (3) \CMTB{\{45\}} \oplus (6) \CMTgrn{\bf \{54\}} \oplus (9) \CMTB{\{210\}} \oplus \
(10) \CMTB{\{660\}} \oplus (11) \CMTB{\{770\}} \oplus (21) \CMTB{\{945\}} \oplus (21) \
\CMTB{\{1050\}} \oplus (9) \CMTB{\{\overline{1050}\}} \oplus (15) \CMTB{\{1386\}} \
\oplus (5) \CMTB{\{2772\}} \oplus (19) \CMTB{\{4125\}} \oplus (27) \CMTB{\{5940\}} \oplus \
(18) \CMTB{\{6930\}} \oplus (3) \CMTB{\{\overline{6930}\}} \oplus (3) \CMTB{\{7644\}} \
\oplus (33) \CMTB{\{8085\}} \oplus (6) \CMTB{\{8910\}} \oplus (10) \CMTB{\{14784\}} \oplus \
(5) \CMTB{\{16380\}} \oplus (5) \CMTB{\{17325\}} \oplus (8) \CMTB{\{\overline{17325}\}} \oplus (24) \CMTB{\{17920\}} \oplus (24) \CMTB{\{23040\}} \oplus (8) \
\CMTB{\{\overline{23040}\}} \oplus (8) \CMTB{\{50688\}} \oplus \CMTB{\{52920\}} \oplus \
(3) \CMTB{\{64350\}} \oplus (3) \CMTB{\{70070\}} \oplus (19) \CMTB{\{72765\}} \oplus (9) \
\CMTB{\{73710\}} \oplus \CMTB{\{90090'\}} \oplus (6) \CMTB{\{112320\}} \oplus (9) \
\CMTB{\{128700\}} \oplus (8) \CMTB{\{143000\}} \oplus (3) \CMTB{\{174636\}} \oplus (3) \
\CMTB{\{199017\}} \oplus (4) \CMTB{\{242550\}}$
\item Level-9: $(2) \CMTred{\{16\}} \oplus (12) \CMTred{\{144\}} \oplus (16) \CMTred{\{560\}} \oplus (10) \CMTred{\{672\
\}} \oplus (24) \CMTred{\{720\}} \oplus (26) \CMTred{\{1200\}} \oplus (12) \CMTred{\{1440\}} \
\oplus (30) \CMTred{\{2640\}} \oplus (42) \CMTred{\{3696\}} \oplus (4) \CMTred{\{5280\}} \oplus \
(10) \CMTred{\{7920\}} \oplus (16) \CMTred{\{8064\}} \oplus (48) \CMTred{\{8800\}} \oplus (36) \
\CMTred{\{11088\}} \oplus (30) \CMTred{\{15120\}} \oplus (16) \CMTred{\{17280\}} \oplus (12) \
\CMTred{\{23760\}} \oplus (38) \CMTred{\{25200\}} \oplus (4) \CMTred{\{26400\}} \oplus (14) \
\CMTred{\{30800\}} \oplus (12) \CMTred{\{34992\}} \oplus (36) \CMTred{\{38016\}} \oplus (18) \
\CMTred{\{43680\}} \oplus (10) \CMTred{\{48048\}} \oplus (20) \CMTred{\{49280\}} \oplus (12) \
\CMTred{\{55440\}} \oplus (2) \CMTred{\{68640'\}} \oplus (2) \CMTred{\{70560\}} \oplus \
(8) \CMTred{\{124800\}} \oplus (16) \CMTred{\{144144\}} \oplus (4) \CMTred{\{155232\}} \oplus \
(4) \CMTred{\{164736\}} \oplus (12) \CMTred{\{196560\}} \oplus (6) \CMTred{\{196560'\}} \
\oplus (6) \CMTred{\{198000\}} \oplus (18) \CMTred{\{205920\}} \oplus (6) \CMTred{\{258720\}} \
\oplus (2) \CMTred{\{332640\}} \oplus (2) \CMTred{\{388080'\}} \oplus (6) \
\CMTred{\{529200\}}$
\item Level-10: $\CMTB{\{10\}} \oplus (13) \CMTB{\{120\}} \oplus (6) \CMTB{\{126\}} \oplus (6) \
\CMTB{\{\overline{126}\}} \oplus (15) \CMTB{\{210'\}} \oplus (20) \CMTB{\{320\
\}} \oplus (48) \CMTB{\{1728\}} \oplus \CMTB{\{1782\}} \oplus (39) \CMTB{\{2970\}} \oplus \
(33) \CMTB{\{3696'\}} \oplus (34) \CMTB{\{\overline{3696'}\}} \
\oplus (63) \CMTB{\{4312\}} \oplus (21) \CMTB{\{4410\}} \oplus (24) \CMTB{\{4608\}} \oplus \
(36) \CMTB{\{4950\}} \oplus (30) \CMTB{\{\overline{4950}\}} \oplus (13) \CMTB{\{6930'\}} \oplus (5) \CMTB{\{\overline{6930'}\}} \oplus (44) \
\CMTB{\{10560\}} \oplus (9) \CMTB{\{20790\}} \oplus (39) \CMTB{\{27720\}} \oplus (40) \
\CMTB{\{28160\}} \oplus \CMTB{\{28314\}} \oplus (21) \CMTB{\{34398\}} \oplus (72) \
\CMTB{\{36750\}} \oplus (4) \CMTB{\{37632\}} \oplus (10) \CMTB{\{42120\}} \oplus (18) \
\CMTB{\{46800\}} \oplus (6) \CMTB{\{\overline{46800}\}} \oplus (42) \CMTB{\{48114\}} \
\oplus (27) \CMTB{\{\overline{48114}\}} \oplus (5) \CMTB{\{48510\}} \oplus (5) \
\CMTB{\{50050\}} \oplus \CMTB{\{\overline{50050}\}} \oplus (16) \CMTB{\{64680\}} \
\oplus (29) \CMTB{\{68640\}} \oplus (10) \CMTB{\{70070'\}} \oplus (27) \
\CMTB{\{90090\}} \oplus (15) \CMTB{\{\overline{90090}\}} \oplus (3) \CMTB{\{102960'\}} \oplus (6) \CMTB{\{144144'\}} \oplus (11) \CMTB{\{150150\}} \
\oplus (24) \CMTB{\{192192\}} \oplus (9) \CMTB{\{216216\}} \oplus \CMTB{\{258720'\
\}} \oplus (24) \CMTB{\{299520\}} \oplus \CMTB{\{351000\}} \oplus (4) \CMTB{\{369600\}} \
\oplus (3) \CMTB{\{371250\}} \oplus (8) \CMTB{\{376320\}} \oplus (12) \CMTB{\{380160\}} \
\oplus (8) \CMTB{\{436590\}} \oplus (3) \CMTB{\{590490\}} \oplus (3) \CMTB{\{630630\}} \
\oplus (12) \CMTB{\{705600\}} \oplus \CMTB{\{831600\}}$
\item Level-11: $(2) \CMTred{\{\overline{16}\}} \oplus (20) \CMTred{\{\overline{144}\}} \oplus \
(30) \CMTred{\{\overline{560}\}} \oplus (22) \CMTred{\{\overline{672}\}} \oplus \
(40) \CMTred{\{\overline{720}\}} \oplus (42) \CMTred{\{\overline{1200}\}} \
\oplus (24) \CMTred{\{\overline{1440}\}} \oplus (42) \CMTred{\{\overline{2640}\
\}} \oplus (82) \CMTred{\{\overline{3696}\}} \oplus (22) \CMTred{\{\overline{5280}\}} \oplus (20) \CMTred{\{\overline{7920}\}} \oplus (36) \
\CMTred{\{\overline{8064}\}} \oplus (86) \CMTred{\{\overline{8800}\}} \oplus \
(2) \CMTred{\{\overline{9504}\}} \oplus (60) \CMTred{\{\overline{11088}\}} \
\oplus (72) \CMTred{\{\overline{15120}\}} \oplus (20) \CMTred{\{\overline{17280}\}} \oplus (2) \CMTred{\{\overline{20592}\}} \oplus (12) \
\CMTred{\{\overline{23760}\}} \oplus (66) \CMTred{\{\overline{25200}\}} \oplus \
(16) \CMTred{\{\overline{29568}\}} \oplus (32) \CMTred{\{\overline{30800}\}} \
\oplus (48) \CMTred{\{\overline{34992}\}} \oplus (72) \CMTred{\{\overline{38016}\}} \oplus (4) \CMTred{\{\overline{39600}\}} \oplus (36) \
\CMTred{\{\overline{43680}\}} \oplus (30) \CMTred{\{\overline{48048}\}} \oplus \
(56) \CMTred{\{\overline{49280}\}} \oplus (20) \CMTred{\{\overline{55440}\}} \
\oplus (8) \CMTred{\{\overline{70560}\}} \oplus (12) \CMTred{\{\overline{80080}\}} \oplus (24) \CMTred{\{\overline{102960}\}} \oplus (36) \
\CMTred{\{\overline{124800}\}} \oplus (10) \CMTred{\{\overline{129360}\}} \
\oplus (64) \CMTred{\{\overline{144144}\}} \oplus (18) \
\CMTred{\{\overline{155232}\}} \oplus (12) \CMTred{\{\overline{164736}\}} \
\oplus (6) \CMTred{\{\overline{185328}\}} \oplus (24) \CMTred{\{\overline{196560}\}} \oplus (2) \CMTred{\{\overline{203280}\}} \oplus (32) \
\CMTred{\{\overline{205920}\}} \oplus (12) \CMTred{\{\overline{258720}\}} \
\oplus (2) \CMTred{\{\overline{332640}\}} \oplus (2) \CMTred{\{\overline{343200}\}} \oplus (24) \CMTred{\{\overline{364000}\}} \oplus (10) \
\CMTred{\{\overline{388080}\}} \oplus (4) \CMTred{\{\overline{443520}\}} \oplus \
(4) \CMTred{\{\overline{458640}\}} \oplus (12) \CMTred{\{\overline{465696}\}} \
\oplus (6) \CMTred{\{\overline{498960}\}} \oplus (16) \CMTred{\{\overline{529200}\}} \oplus (6) \CMTred{\{\overline{764400}\}} \oplus (18) \
\CMTred{\{\overline{769824}\}} \oplus (6) \CMTred{\{\overline{784784}\}} \oplus \
(2) \CMTred{\{\overline{990000}\}} \oplus (6) \CMTred{\{\overline{1260000}\}} \
\oplus (2) \CMTred{\{\overline{1544400}\}}$
\item Level-12: $\CMTB{\{1\}} \oplus (10) \CMTB{\{45\}} \oplus (15) \CMTgrn{\bf \{54\}} \oplus (25) \CMTB{\{210\}} \
\oplus (15) \CMTB{\{660\}} \oplus (45) \CMTB{\{770\}} \oplus (60) \CMTB{\{945\}} \oplus \
(35) \CMTB{\{1050\}} \oplus (66) \CMTB{\{\overline{1050}\}} \oplus (50) \
\CMTB{\{1386\}} \oplus (6) \CMTB{\{2772\}} \oplus (16) \CMTB{\{\overline{2772}\}} \
\oplus (46) \CMTB{\{4125\}} \oplus \CMTB{\{4290\}} \oplus (93) \CMTB{\{5940\}} \oplus (33) \
\CMTB{\{6930\}} \oplus (50) \CMTB{\{\overline{6930}\}} \oplus (13) \CMTB{\{7644\}} \
\oplus (105) \CMTB{\{8085\}} \oplus (30) \CMTB{\{8910\}} \oplus (15) \CMTB{\{12870\}} \
\oplus (52) \CMTB{\{14784\}} \oplus (36) \CMTB{\{16380\}} \oplus (46) \CMTB{\{17325\}} \
\oplus (30) \CMTB{\{\overline{17325}\}} \oplus (104) \CMTB{\{17920\}} \oplus \
(72) \CMTB{\{23040\}} \oplus (80) \CMTB{\{\overline{23040}\}} \oplus (32) \
\CMTB{\{50688\}} \oplus (24) \CMTB{\{\overline{50688}\}} \oplus \CMTB{\{52920\}} \
\oplus (7) \CMTB{\{64350\}} \oplus (33) \CMTB{\{70070\}} \oplus (104) \CMTB{\{72765\}} \
\oplus (57) \CMTB{\{73710\}} \oplus (24) \CMTB{\{81081\}} \oplus \CMTB{\{84942\}} \oplus \
(9) \CMTB{\{90090'\}} \oplus (3) \CMTB{\{105105\}} \oplus (44) \CMTB{\{112320\}} \
\oplus (5) \CMTB{\{123750\}} \oplus (3) \CMTB{\{124740\}} \oplus (5) \CMTB{\{126126\}} \
\oplus \CMTB{\{\overline{126126}\}} \oplus (38) \CMTB{\{128700\}} \oplus (30) \
\CMTB{\{\overline{128700}\}} \oplus (87) \CMTB{\{143000\}} \oplus (27) \CMTB{\{174636\
\}} \oplus (18) \CMTB{\{189189\}} \oplus (45) \CMTB{\{199017\}} \oplus (18) \
\CMTB{\{\overline{199017}\}} \oplus (24) \CMTB{\{207360\}} \oplus (16) \CMTB{\{210210\
\}} \oplus (8) \CMTB{\{219648\}} \oplus (21) \CMTB{\{242550\}} \oplus (16) \
\CMTB{\{\overline{242550}\}} \oplus (22) \CMTB{\{274560\}} \oplus (24) \CMTB{\{275968\
\}} \oplus (8) \CMTB{\{\overline{275968}\}} \oplus (3) \CMTB{\{420420\}} \oplus \
(3) \CMTB{\{577500\}} \oplus (6) \CMTB{\{620928\}} \oplus (3) \CMTB{\{660660\}} \oplus \
\CMTB{\{680625\}} \oplus (8) \CMTB{\{698880\}} \oplus (12) \CMTB{\{848925\}} \oplus (9) \
\CMTB{\{882882\}} \oplus \CMTB{\{928125\}} \oplus (12) \CMTB{\{945945\}} \oplus (32) \
\CMTB{\{1048576\}} \oplus (3) \CMTB{\{1137500\}} \oplus (3) \CMTB{\{1202850\}} \oplus (9) \
\CMTB{\{1299078\}} \oplus (3) \CMTB{\{1316250\}} \oplus (4) \CMTB{\{2502500\}}$
\item Level-13: $(10) \CMTred{\{16\}} \oplus (30) \CMTred{\{144\}} \oplus (80) \CMTred{\{560\}} \oplus (10) \CMTred{\{672\
\}} \oplus (60) \CMTred{\{720\}} \oplus (50) \CMTred{\{1200\}} \oplus (60) \CMTred{\{1440\}} \
\oplus (50) \CMTred{\{2640\}} \oplus (110) \CMTred{\{3696\}} \oplus (50) \CMTred{\{5280\}} \
\oplus (24) \CMTred{\{7920\}} \oplus (84) \CMTred{\{8064\}} \oplus (138) \CMTred{\{8800\}} \
\oplus (84) \CMTred{\{11088\}} \oplus (100) \CMTred{\{15120\}} \oplus (24) \CMTred{\{17280\}} \
\oplus (12) \CMTred{\{20592\}} \oplus (36) \CMTred{\{23760\}} \oplus (96) \CMTred{\{25200\}} \
\oplus (6) \CMTred{\{26400\}} \oplus (20) \CMTred{\{29568\}} \oplus (40) \CMTred{\{30800\}} \
\oplus (90) \CMTred{\{34992\}} \oplus (116) \CMTred{\{38016\}} \oplus (12) \CMTred{\{39600\}} \
\oplus (78) \CMTred{\{43680\}} \oplus (62) \CMTred{\{48048\}} \oplus (2) \CMTred{\{48048'\}} \oplus (80) \CMTred{\{49280\}} \oplus (42) \CMTred{\{55440\}} \oplus (2) \
\CMTred{\{68640'\}} \oplus (22) \CMTred{\{70560\}} \oplus (4) \CMTred{\{80080\}} \oplus \
(20) \CMTred{\{102960\}} \oplus (48) \CMTred{\{124800\}} \oplus (16) \CMTred{\{129360\}} \oplus \
(106) \CMTred{\{144144\}} \oplus (36) \CMTred{\{155232\}} \oplus (48) \CMTred{\{164736\}} \oplus \
(48) \CMTred{\{196560\}} \oplus (22) \CMTred{\{196560'\}} \oplus (8) \CMTred{\{198000\}} \
\oplus (84) \CMTred{\{205920\}} \oplus (10) \CMTred{\{224224\}} \oplus (28) \CMTred{\{258720\}} \
\oplus (12) \CMTred{\{274560'\}} \oplus (4) \CMTred{\{308880\}} \oplus (6) \
\CMTred{\{332640\}} \oplus (18) \CMTred{\{343200\}} \oplus (32) \CMTred{\{364000\}} \oplus (24) \
\CMTred{\{388080\}} \oplus (18) \CMTred{\{388080'\}} \oplus (2) \CMTred{\{428064\}} \
\oplus (12) \CMTred{\{443520\}} \oplus (6) \CMTred{\{458640\}} \oplus (12) \CMTred{\{465696\}} \
\oplus (2) \CMTred{\{522720\}} \oplus (4) \CMTred{\{524160\}} \oplus (54) \CMTred{\{529200\}} \
\oplus (10) \CMTred{\{764400\}} \oplus (4) \CMTred{\{764400'\}} \oplus (24) \
\CMTred{\{769824\}} \oplus (2) \CMTred{\{831600'\}} \oplus (16) \CMTred{\{905520\}} \
\oplus (12) \CMTred{\{1260000\}} \oplus (18) \CMTred{\{1441440\}} \oplus (6) \
\CMTred{\{1544400\}} \oplus (6) \CMTred{\{1924560\}} \oplus (6) \CMTred{\{2274480\}} \oplus (2) \
\CMTred{\{2310000\}} \oplus (6) \CMTred{\{2402400\}} \oplus (2) \CMTred{\{2642640\}}$
\item Level-14: $(10) \CMTB{\{10\}} \oplus (55) \CMTB{\{120\}} \oplus (15) \CMTB{\{126\}} \oplus (35) \
\CMTB{\{\overline{126}\}} \oplus (15) \CMTB{\{210'\}} \oplus (60) \CMTB{\{320\
\}} \oplus (120) \CMTB{\{1728\}} \oplus (6) \CMTB{\{1782\}} \oplus (105) \CMTB{\{2970\}} \
\oplus (65) \CMTB{\{3696'\}} \oplus (105) \CMTB{\{\overline{3696'}\}} \oplus (105) \CMTB{\{4312\}} \oplus (70) \CMTB{\{4410\}} \oplus (40) \
\CMTB{\{4608\}} \oplus (60) \CMTB{\{4950\}} \oplus (68) \CMTB{\{\overline{4950}\}} \
\oplus (21) \CMTB{\{6930'\}} \oplus (40) \CMTB{\{\overline{6930'}\}} \oplus \CMTB{\{9438\}} \oplus (100) \CMTB{\{10560\}} \oplus (9) \
\CMTB{\{20790\}} \oplus (5) \CMTB{\{\overline{20790}\}} \oplus (76) \CMTB{\{27720\}} \
\oplus (88) \CMTB{\{28160\}} \oplus (8) \CMTB{\{31680\}} \oplus (78) \CMTB{\{34398\}} \
\oplus (203) \CMTB{\{36750\}} \oplus (28) \CMTB{\{37632\}} \oplus (36) \CMTB{\{42120\}} \
\oplus (45) \CMTB{\{46800\}} \oplus (55) \CMTB{\{\overline{46800}\}} \oplus \
(81) \CMTB{\{48114\}} \oplus (90) \CMTB{\{\overline{48114}\}} \oplus (21) \
\CMTB{\{48510\}} \oplus (27) \CMTB{\{50050\}} \oplus (14) \CMTB{\{\overline{50050}\}} \
\oplus (70) \CMTB{\{64680\}} \oplus (99) \CMTB{\{68640\}} \oplus (13) \CMTB{\{70070'\}} \oplus (82) \CMTB{\{90090\}} \oplus (63) \CMTB{\{\overline{90090}\}} \
\oplus (3) \CMTB{\{102960'\}} \oplus \CMTB{\{141570\}} \oplus (14) \
\CMTB{\{144144'\}} \oplus (6) \CMTB{\{\overline{144144'}\}} \
\oplus (27) \CMTB{\{150150\}} \oplus (27) \CMTB{\{\overline{150150}\}} \oplus \
(84) \CMTB{\{192192\}} \oplus (12) \CMTB{\{203840\}} \oplus (32) \CMTB{\{216216\}} \oplus \
(16) \CMTB{\{\overline{216216}\}} \oplus (3) \CMTB{\{237160\}} \oplus (6) \
\CMTB{\{258720'\}} \oplus (15) \CMTB{\{270270\}} \oplus \
\CMTB{\{\overline{286650}\}} \oplus (5) \CMTB{\{294294\}} \oplus (96) \
\CMTB{\{299520\}} \oplus (18) \CMTB{\{351000\}} \oplus (4) \CMTB{\{369600\}} \oplus (3) \
\CMTB{\{371250\}} \oplus (10) \CMTB{\{\overline{371250}\}} \oplus (32) \CMTB{\{376320\
\}} \oplus (72) \CMTB{\{380160\}} \oplus (51) \CMTB{\{436590\}} \oplus (9) \
\CMTB{\{536250\}} \oplus (18) \CMTB{\{590490\}} \oplus (27) \CMTB{\{630630\}} \oplus (11) \
\CMTB{\{\overline{630630}\}} \oplus (36) \CMTB{\{705600\}} \oplus (24) \
\CMTB{\{\overline{705600}\}} \oplus (8) \CMTB{\{720720\}} \oplus (3) \
\CMTB{\{\overline{720720}\}} \oplus (9) \CMTB{\{831600\}} \oplus (6) \
\CMTB{\{\overline{831600}\}} \oplus (9) \CMTB{\{884520\}} \oplus \CMTB{\{917280'\}} \oplus (24) \CMTB{\{1064448\}} \oplus (12) \CMTB{\{1281280\}} \oplus (8) \
\CMTB{\{1387386\}} \oplus \CMTB{\{1524600\}} \oplus (3) \CMTB{\{1559250\}} \oplus (8) \
\CMTB{\{1774080\}} \oplus (3) \CMTB{\{1990170\}} \oplus (4) \CMTB{\{2520000\}} \oplus (3) \
\CMTB{\{3071250\}} \oplus \CMTB{\{3320240\}} \oplus (12) \CMTB{\{3706560\}}$
\item Level-15: $(36) \CMTred{\{\overline{16}\}} \oplus (64) \CMTred{\{\overline{144}\}} \oplus \
(90) \CMTred{\{\overline{560}\}} \oplus (30) \CMTred{\{\overline{672}\}} \oplus \
(60) \CMTred{\{\overline{720}\}} \oplus (100) \CMTred{\{\overline{1200}\}} \
\oplus (60) \CMTred{\{\overline{1440}\}} \oplus (50) \CMTred{\{\overline{2640}\
\}} \oplus (146) \CMTred{\{\overline{3696}\}} \oplus (30) \
\CMTred{\{\overline{5280}\}} \oplus (32) \CMTred{\{\overline{7920}\}} \oplus \
(104) \CMTred{\{\overline{8064}\}} \oplus (150) \CMTred{\{\overline{8800}\}} \
\oplus (130) \CMTred{\{\overline{11088}\}} \oplus (120) \
\CMTred{\{\overline{15120}\}} \oplus (40) \CMTred{\{\overline{17280}\}} \oplus \
(16) \CMTred{\{\overline{20592}\}} \oplus (30) \CMTred{\{\overline{23760}\}} \
\oplus (140) \CMTred{\{\overline{25200}\}} \oplus (4) \CMTred{\{\overline{26400}\}} \oplus (16) \CMTred{\{\overline{29568}\}} \oplus (44) \
\CMTred{\{\overline{30800}\}} \oplus (108) \CMTred{\{\overline{34992}\}} \oplus \
(132) \CMTred{\{\overline{38016}\}} \oplus (20) \CMTred{\{\overline{39600}\}} \
\oplus (114) \CMTred{\{\overline{43680}\}} \oplus (66) \CMTred{\{\overline{48048}\}} \oplus (6) \CMTred{\{\overline{48048'}\}} \oplus (108) \
\CMTred{\{\overline{49280}\}} \oplus (60) \CMTred{\{\overline{55440}\}} \oplus \
(42) \CMTred{\{\overline{70560}\}} \oplus (22) \CMTred{\{\overline{80080}\}} \
\oplus (38) \CMTred{\{\overline{102960}\}} \oplus (2) \CMTred{\{\overline{102960'}\}} \oplus (80) \CMTred{\{\overline{124800}\}} \oplus \
(28) \CMTred{\{\overline{129360}\}} \oplus (138) \CMTred{\{\overline{144144}\}} \
\oplus (56) \CMTred{\{\overline{155232}\}} \oplus (48) \
\CMTred{\{\overline{164736}\}} \oplus (6) \CMTred{\{\overline{185328}\}} \oplus \
(48) \CMTred{\{\overline{196560}\}} \oplus (12) \CMTred{\{\overline{196560'}\}} \oplus (14) \CMTred{\{\overline{198000}\}} \oplus (116) \
\CMTred{\{\overline{205920}\}} \oplus (4) \CMTred{\{\overline{224224}\}} \oplus \
(42) \CMTred{\{\overline{258720}\}} \oplus (20) \CMTred{\{\overline{274560'}\}} \oplus (6) \CMTred{\{\overline{308880}\}} \oplus (20) \
\CMTred{\{\overline{332640}\}} \oplus (24) \CMTred{\{\overline{343200}\}} \
\oplus (58) \CMTred{\{\overline{364000}\}} \oplus (36) \
\CMTred{\{\overline{388080}\}} \oplus (12) \CMTred{\{\overline{388080'}\}} \oplus (2) \CMTred{\{\overline{428064}\}} \oplus (12) \
\CMTred{\{\overline{443520}\}} \oplus (18) \CMTred{\{\overline{458640}\}} \
\oplus (18) \CMTred{\{\overline{465696}\}} \oplus (6) \CMTred{\{\overline{498960}\}} \oplus (64) \CMTred{\{\overline{529200}\}} \oplus (8) \
\CMTred{\{\overline{758912}\}} \oplus (24) \CMTred{\{\overline{764400}\}} \
\oplus (54) \CMTred{\{\overline{769824}\}} \oplus (12) \
\CMTred{\{\overline{784784}\}} \oplus (6) \CMTred{\{\overline{831600'}\
\}} \oplus (16) \CMTred{\{\overline{905520}\}} \oplus (4) \
\CMTred{\{\overline{917280}\}} \oplus (2) \CMTred{\{\overline{990000}\}} \oplus \
(2) \CMTred{\{\overline{1121120}\}} \oplus (4) \CMTred{\{\overline{1123200}\}} \
\oplus (4) \CMTred{\{\overline{1153152}\}} \oplus (16) \
\CMTred{\{\overline{1260000}\}} \oplus (18) \CMTred{\{\overline{1441440}\}} \
\oplus (18) \CMTred{\{\overline{1544400}\}} \oplus (12) \
\CMTred{\{\overline{1924560}\}} \oplus (6) \CMTred{\{\overline{2274480}\}} \
\oplus (6) \CMTred{\{\overline{2310000}\}} \oplus (6) \
\CMTred{\{\overline{2402400}\}} \oplus (2) \CMTred{\{\overline{2469600}\}} \
\oplus (2) \CMTred{\{\overline{4410000}\}} \oplus (6) \
\CMTred{\{\overline{4756752}\}}$
\item Level-16: $(27) \CMTB{\{1\}} \oplus (45) \CMTB{\{45\}} \oplus (28) \CMTgrn{\bf \{54\}} \oplus (80) \CMTB{\{210\}} \
\oplus (16) \CMTB{\{660\}} \oplus (70) \CMTB{\{770\}} \oplus (100) \CMTB{\{945\}} \oplus \
(75) \CMTB{\{1050\}} \oplus (75) \CMTB{\{\overline{1050}\}} \oplus (57) \
\CMTB{\{1386\}} \oplus (5) \CMTB{\{2772\}} \oplus (5) \CMTB{\{\overline{2772}\}} \
\oplus (75) \CMTB{\{4125\}} \oplus (6) \CMTB{\{4290\}} \oplus (170) \CMTB{\{5940\}} \oplus \
(60) \CMTB{\{6930\}} \oplus (60) \CMTB{\{\overline{6930}\}} \oplus (48) \
\CMTB{\{7644\}} \oplus (140) \CMTB{\{8085\}} \oplus (63) \CMTB{\{8910\}} \oplus (23) \
\CMTB{\{12870\}} \oplus (68) \CMTB{\{14784\}} \oplus (60) \CMTB{\{16380\}} \oplus (51) \
\CMTB{\{17325\}} \oplus (51) \CMTB{\{\overline{17325}\}} \oplus (160) \CMTB{\{17920\}} \
\oplus \CMTB{\{19305\}} \oplus (120) \CMTB{\{23040\}} \oplus (120) \
\CMTB{\{\overline{23040}\}} \oplus (48) \CMTB{\{50688\}} \oplus (48) \
\CMTB{\{\overline{50688}\}} \oplus (15) \CMTB{\{52920\}} \oplus (3) \CMTB{\{64350\}} \
\oplus (3) \CMTB{\{\overline{64350}\}} \oplus (43) \CMTB{\{70070\}} \oplus (3) \
\CMTB{\{70785\}} \oplus (168) \CMTB{\{72765\}} \oplus (126) \CMTB{\{73710\}} \oplus (48) \
\CMTB{\{81081\}} \oplus (6) \CMTB{\{90090'\}} \oplus (6) \
\CMTB{\{\overline{90090'}\}} \oplus (14) \CMTB{\{105105\}} \oplus (66) \
\CMTB{\{112320\}} \oplus (6) \CMTB{\{123750\}} \oplus (10) \CMTB{\{124740\}} \oplus (9) \
\CMTB{\{126126\}} \oplus (9) \CMTB{\{\overline{126126}\}} \oplus (54) \
\CMTB{\{128700\}} \oplus (54) \CMTB{\{\overline{128700}\}} \oplus (154) \
\CMTB{\{143000\}} \oplus (8) \CMTB{\{165165\}} \oplus (63) \CMTB{\{174636\}} \oplus (55) \
\CMTB{\{189189\}} \oplus (57) \CMTB{\{199017\}} \oplus (57) \CMTB{\{\overline{199017}\
\}} \oplus (48) \CMTB{\{207360\}} \oplus (24) \CMTB{\{210210\}} \oplus (8) \
\CMTB{\{219648\}} \oplus (8) \CMTB{\{\overline{219648}\}} \oplus (49) \
\CMTB{\{242550\}} \oplus (49) \CMTB{\{\overline{242550}\}} \oplus (54) \CMTB{\{274560\
\}} \oplus (32) \CMTB{\{275968\}} \oplus (32) \CMTB{\{\overline{275968}\}} \
\oplus (7) \CMTB{\{420420\}} \oplus (4) \CMTB{\{462462\}} \oplus (15) \CMTB{\{577500\}} \
\oplus (6) \CMTB{\{620928\}} \oplus (8) \CMTB{\{698880\}} \oplus (8) \
\CMTB{\{\overline{698880}\}} \oplus \CMTB{\{749112\}} \oplus (19) \CMTB{\{848925\}} \
\oplus (19) \CMTB{\{\overline{848925}\}} \oplus (9) \CMTB{\{882882\}} \oplus \
(9) \CMTB{\{\overline{882882}\}} \oplus (9) \CMTB{\{928125\}} \oplus (40) \
\CMTB{\{945945\}} \oplus (64) \CMTB{\{1048576\}} \oplus (12) \CMTB{\{1137500\}} \oplus (3) \
\CMTB{\{1202850\}} \oplus (3) \CMTB{\{\overline{1202850}\}} \oplus (24) \
\CMTB{\{1299078\}} \oplus (11) \CMTB{\{1316250\}} \oplus (3) \CMTB{\{1683990\}} \oplus (3) \
\CMTB{\{\overline{1683990}\}} \oplus (8) \CMTB{\{1698840\}} \oplus (6) \
\CMTB{\{1897280\}} \oplus \CMTB{\{2388750\}} \oplus (9) \CMTB{\{2502500\}} \oplus (9) \
\CMTB{\{\overline{2502500}\}} \oplus (3) \CMTB{\{2866500\}} \oplus (3) \
\CMTB{\{3838185\}} \oplus (4) \CMTB{\{4802490\}}$
\end{itemize}

\subsection{Adinkra Diagram for 10D, $\mathcal{N} = 2$B Superfield}

Here we draw the ten dimensional $\mathcal{N} = 2$B adinkra diagram up to the cubic level.

\begin{figure}[htp!]
\centering
\includegraphics[width=0.8\textwidth]{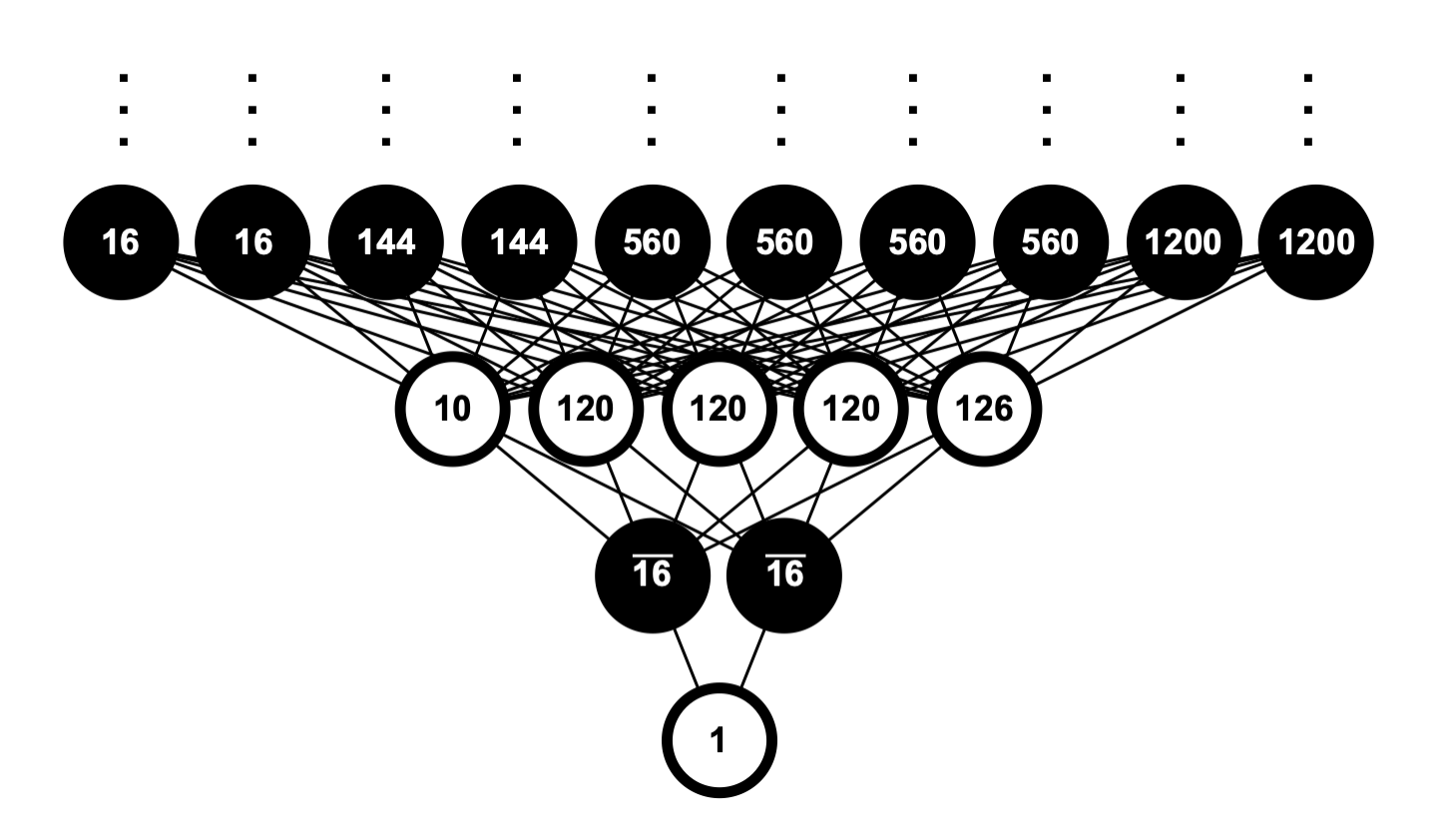}
\caption{Adinkra Diagram for 10D, $\mathcal{N} = 2$B At Low Orders\label{Fig:10DTypeIIB}}
\end{figure}

\newpage
\section{Conclusion}

This work gives a new group-theoretic free-of-dynamics method for the complete decompositions of scalar superfields to irreducible component field representations of the 10D Lorentz group, and a proposal for identifying the corresponding superconformal multiplets by applying the Breitenlohner approach in $\mathcal{N} = 1$, $\mathcal{N} = 2$A, and $\mathcal{N} = 2$B superspaces. The new results here provide a foundation for future extensions.  Our efforts also mark a new beginning for the search for irreducible off-shell formulations of the 10D
Yang-Mills supermultiplet derived from superfields. 

We believe it is important to comment on the graphs shown in Figure~\ref{TetMin} in comparisons to those shown in Figure~\ref{Fig:10DTypeI},  Figure~\ref{Fig:10DTypeI_dynkin}, Figure~\ref{Fig:10DTypeIIA}, and Figure~\ref{Fig:10DTypeIIB}.
While the latter two of these are incomplete\footnote{The only reason for their incompleteness is convenience of presentation.}, they share an attribute with those in Figure~\ref{Fig:10DTypeI} and Figure~\ref{Fig:10DTypeI_dynkin}.  The graphs in Figure~\ref{TetMin} were constructed by starting from an off-shell {\it {component}} formulation.
In contrast, the graphs in Figure~\ref{Fig:10DTypeI},  Figure~\ref{Fig:10DTypeI_dynkin}, Figure~\ref{Fig:10DTypeIIA}, and Figure~\ref{Fig:10DTypeIIB} were 
constructed {\it {without}} {\it {any}} information from a {\it {component}} formulation.  The presentations of the latter four graphs provide demonstrations that with this work a new level of completeness has been achieved about the structure of superfields in high dimensions.  
 
There is a certain tension in the path we are pursuing with the use of adinkras and superfields in comparison with the well established results in the literature.  Many years ago, Nahm \cite{Nahm} pointed out the absence of a superconformal current above six dimensions.  This most certainly
 suggests an obstruction may exist.  
 
On the other hand, in works going back decades\cite{10DScLR,Y2,Y3,Y4,Y5}, there have been increasing explorations of the concepts of conformal symmetry within the context of 10D superspaces.  Our scans suggest the possibilities of a number of superfields for embedding the component-level conformal gravitons into 10D superfields.  This supports the idea of the eventual success of these efforts.    Though we do not understand how this tension will be resolved... if it can... we would point the reader to what may prove to be a similar situation.
 
In the works of \cite{GRanaV,GatRodg}, the Witt algebra (i.e. the ``centerless Virasoro algebra'') was investigated.  It was found that the form of the  Witt algebra undergoes a radical change dependent on the number of supercharges under investigation.  When the number of supercharges is four or less, the form of the Witt algebra is simple and uniform.  However, when the number of supercharges exceeds four, the form of the Witt algebra changes dramatically with the appearance of new generators that are not present for the cases when the number of supercharges is less than four.  It seems likely to us this phenomenon may provide a route by which these embeddings that are obvious from the supergeometrical side could lead to conformal supergravity theories in higher dimensions.

In a future work, we will also dive far more deeply into the relations between analytical expressions of the irreducible $\theta-$monomials, Young Tableaux, and Dynkin labels.  Along a different direction of future activities lies the extension of our results to the case of 11D, $\cal N$
= 1 supergravity.  The results in this work regarding the case of the 10D, $\cal N$ = 2A supergravity
theory already contain a lot of information about the eleven dimensional theory as it is
equivalent upon toroidal compactification to the 10D, $\cal N$ = 2A supergravity theory.  In principle, it is straightforward to construct 
the component level contents from our approach that uses branching rules.  However, in practice this is considerably more computationally challenging than the equivalent work in the 10D, $\cal N$ = 2A supergravity theory. Currently the work on this is underway.

We end by concisely summarizing a surprising and heretofore unknown result
discovered in this work and cast it into the form of a conjecture.

Consider a Lorentz superspace of signature SO(1, D - 1) and let $d$ denote the dimension of
the smallest spinor representation consistent with this signature.  Furthermore, let $b_{ \CMTB
{\{  {\cal R}  \}  } }$,  and $b_{ \CMTR {\{  {\cal R}  \}  } }$ denote a set of non-negative integers. 
Finally, let  $d_{ \CMTB {\{  {\cal R}  \}  } } $,  and $d_{ \CMTR {\{  {\cal R}  \}  } }$ respectively
denote the dimensionality of some bosonic and fermionic representations of the SO(1, D - 1)
Lorentz algebra.   For a scalar superfield $\cal V$ in this superspace, the number of bosonic
degrees of freedom at the $n$-th level of the corresponding adinkra is given by $\frac {d!}{n! (d - n)!}
$ for even values of $n$ and the number of fermionic
degrees of freedom at the $n$-th level of the corresponding adinkra is given by $\frac {d!}{n! (d - n)!}
$ for odd values of $n$.

 $~~~~$ {\it{Conjecture}}:
\newline $~~~~~~~~~~$ {\it{Let}} $\cal V$ {\it{denote a scalar superfield in
a Lorentz superspace of signature}} SO(1, D - 1),  \newline
$~~~~~~~~~~$ {\it{then at each even level of the superfield the equation}}
 \newline $~~~~~~~~~~$
\begin{equation} \label{equ:BIN2}
\frac {d!}{n! (d - n)!} ~=~
\sum_{ \CMTB {\cal R}} \, b_{ \CMTB {\{  {\cal R}  \}  } }    \,  d_{ \CMTB {\{  {\cal R}  \}  } }   
\end{equation}
$~~~~~~~~~~$   {\it{and at each odd level of the superfield the equation}}
\begin{equation} \label{equ:BIN2}
\frac {d!}{n! (d - n)!} ~=~
\sum_{ \CMTR {\cal R}} \, b_{ \CMTR {\{  {\cal R}  \}  } }    \,  d_{ \CMTR {\{  {\cal R}  \}  } }   
\end{equation}
$~~~~~~~~~~$   {\it{are both determined by the
branching rules of the totally antisymmetric representa-}} 
\newline \noindent
$~~~~~~~~~~~$   {\it{tions of ${\cal A}_{d-1}$ series of the Cartan classification of compact Lie algebras
under the}}
\newline \noindent
$~~~~~~~~~~~$   {\it{projection to its}} SO(1, D - 1) {\it{subalgebra.}}

\vspace{.05in}
 \begin{center}
\parbox{4in}{{\it ``Our knowledge can only be finite, while our ignorance \\ $~~$
must necessarily be infinite.'' \\ ${~}$
 \\ ${~}$
\\ ${~}$ }\,\,-\,\, Karl Popper}
 \parbox{4in}{
 $~~$} 
 \end{center}

$$~~~~{\,}~~~~~
$$
\noindent
{\bf Acknowledgements}\\[.1in] \indent
This research of S.\ J.\ G., Y.\ Hu, and S.-N.\ Mak is supported in part by the
endowment of the Ford Foundation Professorship of Physics at Brown University.
S.\ J.\ G.\ makes an additional acknowledgment to the National Science
Foundation grant PHY-1620074 and all the authors gratefully acknowledge
the support of the Brown Theoretical Physics Center.  Additionally, we acknowledge
the referee whose critique spurred expanded historical and content additions
to the final version of this work.

\newpage
\appendix

\section{Chiral and Vector Supermultiplets from the 4D, $\mathcal{N}=1$ Unconstrained Scalar \\ Superfield}
\label{appen:4DCSVS}

In this appendix, we present an expanded discussion of the mathematical
structures that support the validity of the ``splitting'' of the adinikras presented in Figure \ref{Fig:4DtoVSCS}. The vector supermultiplet shown in the figure is well familiar, but this is not so for the chiral supermultiplet shown.  We will begin with the vector supermultiplet.  The notational conventions used in the following discussion are those of {\it {Superspace}} \cite{SpRSp8c}.

The vector supermultiplet contains a 1-form gauge potential $A_{\un{a}}$
that is part of a super 1-form with different sectors given by
\begin{equation}
\begin{gathered}
    A_{\a} ~=~ i {\rm D}_{\a}{ V}   ~~~,~~~ 
    A_{\Dot {\a}} ~=~-i {\overline {\rm D}}{}_{\Dot {\a}}V ~~~,~~~
    A_{\un{a}} ~=~ [ \overline{\rm D}{}_{\Dot{\a}} , {\rm D}{}_{\a} ] {V}  ~~~.
\end{gathered}
\end{equation}
The vector supermultiplet field strength superfield $W_{\alpha}$
is determined in terms of an unconstrained real scalar superfield $V$ by 
\begin{equation}
    W_{\alpha} ~=~ i\overline{\rm D}^2{\rm D}_{\alpha} V ~~~,~~~ \overline{W}_{\Dot{\alpha}} ~=~ -i{\rm D}^2\overline{\rm D}_{\Dot{\alpha}} V ~~~,~~~ V ~=~ \overline{V} ~~~.
\end{equation}
The definition of $W^{\alpha}$ is invariant under gauge transformations with a chiral parameter $\Lambda$ that
takes the form
\begin{equation}
    V' ~=~ V ~+~ i(\overline{\Lambda}-\Lambda)~~,~~ \overline{\rm D}_{\Dot{\alpha}}\Lambda~=~{\rm D}_{\alpha}\overline{\Lambda}~=~0~~.
\end{equation}
The components of the prepotential superfield $V$ can be
defined by the projection method
\begin{equation}
\begin{gathered}
    C ~=~ V \,| ~~~,~~~ 
    \chi_{\a} ~=~ i{\rm D}_{\alpha}V\,| ~~~,~~~
    \overline{\chi}_{\Dot{\alpha}} ~=~ -i \overline{\rm D}_{\Dot{\alpha}} V \,| ~~~, \\
    M ~=~ {\rm D}^2 V \,| ~~~,~~~ 
    \overline{M} ~=~ \overline{\rm D}^2 V\,| ~~~,~~~
    A_{\a\Dot\a} ~=~ \frac{1}{2}[\overline{\rm D}_{\Dot{\alpha}}~,~ {\rm D}_{\alpha}] V \,| ~~~,\\
    \lambda_{\a} ~=~ i \overline{\rm D}^2{\rm D}_{\alpha} V \,| ~~~,~~~ 
    \overline{\lambda}_{\Dot{\alpha}} ~=~ -i{\rm D}^2 \overline{\rm D}_{\Dot{\alpha}}V \,| ~~~,~~~ 
    D' ~=~ \frac{1}{2}{\rm D}^{\alpha}\overline{\rm D}^2 {\rm D}_{\alpha} V \,| ~~~.
\end{gathered}
\end{equation}
All the components of $V$ can be gauged away by nonderivative gauge transformations except for $A_{\a\Dot\a}$, $\lambda_{\a}$, $\overline{\lambda}_{\Dot\a}$, and $D'$, where the vector and spinor are the physical components and $D'$ is auxiliary component. 
We find the action to be
\begin{equation}
    S ~=~ \int d^4x d^2\theta\, W^2 ~=~ \frac{1}{2} \int d^4x d^4\theta ~ V \, {\rm D}^{\alpha}\overline{\rm D}^2 {\rm D}_{\alpha} V ~~~.
\end{equation}

However, a superfield with the same structure as $V$ can also be used to describe an entirely different supermultiplet.  To distinguish this second supermultiplet
from the first, we will use the symbol $\cal V$ to
denote its prepotential.  This second supermultiplet
contains a 3-form gauge potential $A_{\un{a}\un{b}\un{c}}$
that is part of a super 3-form with different sectors given by
\begin{equation}
\label{equ:super3form_chiral}
\begin{gathered}
    A_{\a\b\g} ~=~ A_{\a\b\Dot{\g}} ~=~ A_{\a\b\un{c}} ~=~ 0   ~~~,~~~ A_{\a\Dot{\b}\un{c}} ~=~ i C_{\a \g}
    C_{\Dot{\b}\Dot{\g}} {\cal V}  ~~~,  \\
    A_{\a\un{b}\un{c}} ~=~ - C_{\Dot{\b}\Dot{\g}} C_{\a(\b} {\rm D}_{\g)}{\cal V}   ~~~,~~~  A_{\un{a}\un{b}\un{c}} ~=~ \e_{\un{d}\un{a}\un{b}\un{c}} [ \overline{\rm D}^{\Dot{\d}} , {\rm D}^{\d} ] {\cal V}  ~~~.
\end{gathered}
\end{equation}
This super 3-form theory has been discussed previously in the works of \cite{SpRSp8c,SpRpF} and more recently in \cite{CmpLx} in terms of component fields.

The chiral supermultiplet that is contained in a real unconstrained scalar superfield has a gauge invariant superfield field strength of the form
\begin{equation}
    \overline{\Pi} ~=~ {\rm D}^{2} {\cal V} ~~,~~
    {\Pi } ~=~ {\overline {\rm D}}{}^2 {\cal V} ~~,~~
    {\cal V} ~=~ \overline{\cal V}
    ~~~. 
\end{equation}
where the prepotential $\cal V$ has gauge transformations
\begin{equation}
    {\cal V}{}^{\prime} ~=~ {\cal V} - \frac{1}{2} \big( {\rm D}^{\a} \w_{\a} + \overline{\rm D}^{\Dot{\a}} \overline{\w}_{\Dot{\a}} \big)   ~~~,~~~   {\rm D}_{\a} \overline{\w}_{\Dot{\a}} ~=~ 0    ~~~.
\end{equation}
The physical component fields of this gauge 3-form multiplet are
\begin{equation}
\label{equ:comp_chiral_mul}
\begin{gathered}
    \overline{\phi} ~=~ \overline{\Pi} | ~=~ {\rm D}^{2} {\cal V} |   ~~~,~~~  \phi ~=~ \Pi | ~=~ \overline{\rm D}^{2} {\cal V} |  ~~~, \\
    \psi_{\a} ~=~ {\rm D}_{\a} \Pi | ~=~ {\rm D}_{\a} \overline{\rm D}^{2} {\cal V} |  ~~~,~~~  \overline{\psi}_{\Dot{\a}} ~=~ \overline{\rm D}_{\Dot\a} \overline{\Pi} | ~=~ \overline{\rm D}_{\Dot\a} {\rm D}^{2} {\cal V} |  ~~~, \\
    h ~=~ \big( {\rm D}^{2} \Pi + \overline{\rm D}^{2} \overline{\Pi} \big) | ~=~ \{ {\rm D}^{2} , \overline{\rm D}^{2} \} {\cal V} |  ~~~,  \\
    f ~=~ - i \big( {\rm D}^{2} \Pi - \overline{\rm D}^{2} \overline{\Pi} \big) |  ~=~  \frac{1}{2 \cdot 3!} \e_{{\un a}{\un b} {\un c} {\un d} } \pa^{\un{a}} A {}^{{\un b} {\un c} {\un d}} |  ~=~  \frac{1}{2} \pa^{\a\Dot{\a}} [ \overline{\rm D}_{\Dot{\a}} , {\rm D}_{\a} ] {\cal V} |   ~~~.
\end{gathered}
\end{equation}
The quantity $f$ is the field strength of the component gauge 3-form $A_{\un{a}\un{b}\un{c}}$, so that the field strength $f$ is invariant. Interpreted the other way, the chiral supermultiplet is that with component content $(\overline{\phi}, \phi, \psi_{\a}, \overline{\psi}_{\Dot\a}, h, f)$, and the supermultiplet corresponding to the bottom right diagram of Figure \ref{Fig:4DtoVSCS} is one of the Hodge-dual variants\footnote{There are three of them. Two of them can correspond to Figure \ref{Fig:4DtoVSCS} with one of the auxiliary fields replaced by its Hodge-dual - see equations (3.6) and (3.7) in \cite{CmpLx}. The third one is obtained by replacing both of the auxiliary fields to their Hodge-duals, as shown in (3.8) in the reference.} of the chiral supermultiplet with component content $(\overline{\phi}, \phi, \psi_{\a}, \overline{\psi}_{\Dot\a}, h, A_{\un{a}\un{b}\un{c}})$. This is done by taking one of the auxiliary fields $f$ to its Hodge-dual gauge 3-form $A_{\un{a}\un{b}\un{c}}$ as shown in the last line of (\ref{equ:comp_chiral_mul}).

The action for this supermultiplet is given by
\begin{equation}
    S ~=~ \int d^4x d^2\theta\, {\overline \Pi} \,
    {\Pi }~=~ \frac{1}{2} \int d^4x \, d^4\theta\, 
    d^4 {\overline \theta} ~ {\cal V} \,
    \big\{ {\rm D}^{2} \overline{\rm D}^{2} ~+~  \overline{\rm D}^{2} {\rm D}^{2}  \big\}{\cal V} ~~~.
\end{equation}
and when this is evaluated at the level of component
fields, this action contains the d'Alembertian for
the complex component scalar field $\phi$, the Dirac
operator for the component spinor $\psi{}_{\a}$,
the square of $f$ - the kinetic term for the component
3-form and the square of $h$ - real spin-0 auxiliary
field.

The first result in (\ref{equ:comp_chiral_mul}) shows that the propagating
spin-0 states associate with $\phi$ occur at the quadratic $\theta$ level of $\cal V$.  However, the final
equation in (\ref{equ:super3form_chiral}) shows the component 3-form is at
the same level.

\newpage
\section{SO(10) Irreducible Representations}
\label{appen:so10}

Here we list the SO(10) irreducible representations by Dynkin labels and dimensions \cite{so11table}.

\begin{table}[h!]
\centering
\begin{tabular}{|c|c|c|}
\hline
 Dynkin label & Dimension\\
 \hline
 (00000) & $\CMTB{1}$\\
 \hline
 (10000) & $\CMTB{10}$\\
 \hline
 (00001) & $\CMTred{16}$\\
 \hline
 (00010) & $\CMTred{\overline{16}}$\\
 \hline
 (01000) & $\CMTB{45}$ \\
 \hline
 (20000) & $\CMTB{54}$ \\
 \hline
 (00100) & $\CMTB{120}$\\
 \hline
 (00020) & $\CMTB{126}$ \\
 \hline
 (00002) & $\CMTB{\overline{126}}$ \\
 \hline
 (10010) & $\CMTred{144}$ \\
 \hline
 (10001) & $\CMTred{\overline{144}}$ \\
 \hline
 (00011) & $\CMTB{210}$ \\
 \hline
 (30000) & $\CMTB{210'}$\\
 \hline
 (11000) & $\CMTB{320}$ \\
 \hline
 (01001) & $\CMTred{560}$ \\
 \hline
 (01010) & $\CMTred{\overline{560}}$ \\
 \hline
 (40000) & $\CMTB{660}$ \\
 \hline
 (00030) & $\CMTred{672}$ \\
 \hline
 (00003) & $\CMTred{\overline{672}}$ \\
 \hline
 (20001) & $\CMTred{720}$ \\
 \hline
 (20010) & $\CMTred{\overline{720}}$ \\
 \hline
 (02000) & $\CMTB{770}$ \\
 \hline
 (10100) & $\CMTB{945}$ \\
 \hline
 (10020) & $\CMTB{1,050}$\\
 \hline
 (10002) & $\CMTB{\overline{1,050}}$ \\
 \hline
 (00110) & $\CMTred{1,200}$ \\
 \hline
 (00101) & $\CMTred{\overline{1,200}}$ \\
 \hline
 (21000) & $\CMTB{1,386}$ \\
 \hline
 (00012) & $\CMTred{1,440}$\\
 \hline
 (00021) & $\CMTred{\overline{1,440}}$\\
 \hline
 (10011) & $\CMTB{1,728}$ \\
 \hline
 (50000) & $\CMTB{1,782}$ \\
 \hline
 (30010) & $\CMTred{2,640}$ \\
 \hline
 (30001) & $\CMTred{\overline{2,640}}$ \\
 \hline
 (00040) & $\CMTB{2,772}$ \\
 \hline
 (00004) & $\CMTB{\overline{2,772}}$ \\
 \hline
\end{tabular}
\end{table}

\newpage
\begin{table}[h!]
\centering
\begin{tabular}{|c|c|c|}
\hline
 Dynkin label & Dimension  \\
 \hline
 (01100) & $\CMTB{2,970}$ \\
 \hline
 (11010) & $\CMTred{3,696}$\\
 \hline
 (11001) & $\CMTred{\overline{3,696}}$ \\
 \hline
 (01020) & $\CMTB{3,696'}$ \\
 \hline
 (01002) & $\CMTB{\overline{3,696'}}$ \\
 \hline
 (00200) & $\CMTB{4,125}$ \\
 \hline
 (60000) & $\CMTB{4,290}$ \\
 \hline
 (20100) & $\CMTB{4,312}$ \\
 \hline
 (12000) & $\CMTB{4,410}$ \\
 \hline
 (31000) & $\CMTB{4,608}$ \\
 \hline
 (20020) & $\CMTB{4,950}$ \\
 \hline
 (20002) & $\CMTB{\overline{4,950}}$\\
 \hline
 (10003) & $\CMTred{5,280}$ \\
 \hline
 (10030) & $\CMTred{\overline{5,280}}$ \\
 \hline
 (01011) & $\CMTB{5,940}$ \\
 \hline
 (00120) & $\CMTB{6,930}$ \\
 \hline
 (00102) & $\CMTB{\overline{6,930}}$\\
 \hline
 (00031) & $\CMTB{6,930'}$ \\
 \hline
 (00013) & $\CMTB{\overline{6,930'}}$\\
 \hline
 (03000) & $\CMTB{7,644}$\\
 \hline
 (40001) & $\CMTred{7,920}$ \\
 \hline
 (40010) & $\CMTred{\overline{7,920}}$ \\
 \hline
 (02001) & $\CMTred{8,064}$\\
 \hline
 (02010) & $\CMTred{\overline{8,064}}$ \\
 \hline
 (20011) & $\CMTB{8,085}$ \\
 \hline
 (10101) & $\CMTred{8,800}$ \\
 \hline
 (10110) & $\CMTred{\overline{8,800}}$ \\
 \hline
 (00022) & $\CMTB{8,910}$ \\
 \hline
 (70000) & $\CMTB{9,438}$ \\
 \hline
 (00005) & $\CMTred{9,504}$ \\
 \hline
 (00050) & $\CMTred{\overline{9,504}}$ \\
 \hline
 (00111) & $\CMTB{10,560}$ \\
 \hline
 \end{tabular}
\caption{SO(10) irreducible representations \protect\cite{so11table}}
\end{table}

\newpage
\section{Bosonic Superfield Decompositions}
\label{appen:bosonic_superfield}

In this appendix, we list a few bosonic superfields that contain the traceless graviton $\CMTgrn{\bf \{54\}}$ $h_{\un a\un b}$ and the traceless gravitino $\CMTred{\{\overline{144}\}}$ $\psi_{\un a}{}^{\g}$ and $\CMTred{\{144\}}$ $\psi_{\un a\g}$. Here every irrep corresponds to a {\em component} field.

\begingroup
\footnotesize
\begin{equation}
{\cal V} \otimes \CMTB{\{54\}} ~=~ \begin{cases}
{~~}{\rm {Level}}-0 \,~~~~~~~ \CMTgrn{\bf \{54\}} ~~~,  \\
{~~}{\rm {Level}}-1 \,~~~~~~~ \CMTred{\{\overline{144}\}} \oplus \CMTred{\{\overline{720}\}} ~~~,  \\
{~~}{\rm {Level}}-2 \,~~~~~~~ \CMTB{\{120\}} \oplus \CMTB{\{320\}} \oplus \CMTB{\{1728\}} \oplus \CMTB{\{4312\}}  ~~~, \\
{~~}{\rm {Level}}-3 \,~~~~~~~ \CMTred{\{144\}} \oplus \CMTred{\{560\}} \oplus \CMTred{\{720\}} \oplus \CMTred{\{1200\}} \oplus \CMTred{\{3696\}} \\
~~~~~~~~~~~~~~~~~~~~~~~ \oplus \CMTred{\{8800\}} \oplus \CMTred{\{15120\}}   ~~~, \\
{~~}{\rm {Level}}-4 \,~~~~~~~ \CMTgrn{\bf \{54\}} \oplus \CMTB{\{210\}} \oplus \CMTB{\{770\}} \oplus \CMTB{\{945\}} \oplus \CMTB{\{1050\}} \oplus \CMTB{\{\overline{1050}\}} \\
~~~~~~~~~~~~~~~~~~~~~~~ \oplus \CMTB{\{1386\}} \oplus \CMTB{\{4125\}} \oplus \CMTB{\{5940\}} \oplus \CMTB{\{8085\}} \oplus \CMTB{\{16380\}} \\
~~~~~~~~~~~~~~~~~~~~~~~ \oplus \CMTB{\{\overline{17325}\}} \oplus \CMTB{\{17920\}} \oplus \CMTB{\{\overline{23040}\}} ~~~,  \\
{~~}{\rm {Level}}-5 \,~~~~~~~ \CMTred{\{\overline{144}\}} \oplus \CMTred{\{\overline{560}\}} \oplus \CMTred{\{\overline{720}\}} \oplus \CMTred{\{\overline{1200}\}} \oplus \CMTred{\{\overline{1440}\}} \oplus \CMTred{\{\overline{2640}\}} \\
~~~~~~~~~~~~~~~~~~~~~~~ \oplus (2) \CMTred{\{\overline{3696}\}} \oplus \CMTred{\{\overline{8064}\}} \oplus \CMTred{\{\overline{8800}\}} \oplus \CMTred{\{\overline{11088}\}} \oplus \CMTred{\{\overline{15120}\}} \\
~~~~~~~~~~~~~~~~~~~~~~~ \oplus \CMTred{\{\overline{23760}\}} \oplus \CMTred{\{\overline{25200}\}} \oplus \CMTred{\{\overline{38016}\}} \oplus \CMTred{\{\overline{43680}\}} \oplus \CMTred{\{\overline{48048}\}} ~~~,  \\
{~~}{\rm {Level}}-6 \,~~~~~~~ \CMTB{\{120\}} \oplus \CMTB{\{320\}} \oplus (2) \CMTB{\{1728\}} \oplus \CMTB{\{2970\}} \oplus \CMTB{\{3696'\}}  \\
~~~~~~~~~~~~~~~~~~~~~~~ \oplus \CMTB{\{\overline{3696'}\}} \oplus (2) \CMTB{\{4312\}} \oplus \CMTB{\{4410\}} \oplus \CMTB{\{4608\}} \oplus \CMTB{\{\overline{4950}\}}  \\
~~~~~~~~~~~~~~~~~~~~~~~ \oplus \CMTB{\{10560\}} \oplus \CMTB{\{28160\}} \oplus \CMTB{\{34398\}} \oplus (2) \CMTB{\{36750\}}  \\
~~~~~~~~~~~~~~~~~~~~~~~ \oplus \CMTB{\{42120\}} \oplus \CMTB{\{\overline{48114}\}} \oplus \CMTB{\{68640\}} \oplus \CMTB{\{\overline{90090}\}} ~~~,  \\
{~~}{\rm {Level}}-7 \,~~~~~~~ \CMTred{\{144\}} \oplus \CMTred{\{560\}} \oplus \CMTred{\{720\}} \oplus \CMTred{\{1200\}} \oplus \CMTred{\{1440\}} \oplus \CMTred{\{2640\}}  \\
~~~~~~~~~~~~~~~~~~~~~~~ \oplus (2) \CMTred{\{3696\}} \oplus \CMTred{\{7920\}} \oplus \CMTred{\{8064\}} \oplus (2) \CMTred{\{8800\}} \oplus \CMTred{\{11088\}}  \\
~~~~~~~~~~~~~~~~~~~~~~~ \oplus (2) \CMTred{\{15120\}} \oplus \CMTred{\{20592\}} \oplus \CMTred{\{25200\}} \oplus \CMTred{\{34992\}} \oplus \CMTred{\{38016\}}  \\
~~~~~~~~~~~~~~~~~~~~~~~ \oplus \CMTred{\{43680\}} \oplus \CMTred{\{48048\}} \oplus \CMTred{\{49280\}} \oplus \CMTred{\{124800\}} \oplus \CMTred{\{144144\}} ~~~,  \\
{~~}{\rm {Level}}-8 \,~~~~~~~ \CMTgrn{\bf \{54\}} \oplus \CMTB{\{210\}} \oplus \CMTB{\{660\}} \oplus \CMTB{\{770\}} \oplus \CMTB{\{945\}} \oplus \CMTB{\{1050\}} \\
~~~~~~~~~~~~~~~~~~~~~~~  \oplus \CMTB{\{\overline{1050}\}} \oplus \CMTB{\{1386\}} \oplus \CMTB{\{4125\}} \oplus \CMTB{\{4290\}} \oplus (2) \CMTB{\{5940\}} \\
~~~~~~~~~~~~~~~~~~~~~~~  \oplus (2) \CMTB{\{8085\}} \oplus \CMTB{\{8910\}} \oplus \CMTB{\{12870\}} \oplus \CMTB{\{14784\}} \oplus \CMTB{\{16380\}} \\
~~~~~~~~~~~~~~~~~~~~~~~  \oplus \CMTB{\{17325\}} \oplus \CMTB{\{\overline{17325}\}} \oplus (2) \CMTB{\{17920\}} \oplus \CMTB{\{23040\}} \oplus \CMTB{\{\overline{23040}\}} \\
~~~~~~~~~~~~~~~~~~~~~~~  \oplus \CMTB{\{72765\}} \oplus \CMTB{\{73710\}} \oplus \CMTB{\{81081\}} \oplus \CMTB{\{112320\}} \oplus \CMTB{\{143000\}} ~~~,  \\
{~~~~~~}  {~~~~} \vdots  {~~~~~~~~~\,~~~~~~} \vdots
\end{cases}
{~~~~~~~~~~~~~~~}
\end{equation}
\endgroup

\begingroup
\footnotesize
\begin{equation}
{\cal V} \otimes \CMTB{\{210\}} ~=~ \begin{cases}
{~~}{\rm {Level}}-0 \,~~~~~~~ \CMTB{\{210\}} ~~~,  \\
{~~}{\rm {Level}}-1 \,~~~~~~~ \CMTred{\{\overline{16}\}} \oplus \CMTred{\{\overline{144}\}} \oplus \CMTred{\{\overline{560}\}} \oplus \CMTred{\{\overline{1200}\}} \oplus \CMTred{\{\overline{1440}\}}  ~~~, \\
{~~}{\rm {Level}}-2 \,~~~~~~~  \CMTB{\{10\}} \oplus (2) \CMTB{\{120\}} \oplus \CMTB{\{126\}} \oplus \CMTB{\{\overline{126}\}} \oplus \CMTB{\{320\}} \oplus (2) \CMTB{\{1728\}}\\
~~~~~~~~~~~~~~~~~~~~~~~\oplus \CMTB{\{2970\}} \oplus \CMTB{\{3696'\}} \oplus \CMTB{\{\overline{3696'}\}} \oplus \CMTB{\{10560\}}  ~~~, \\
{~~}{\rm {Level}}-3 \,~~~~~~~ \CMTred{\{16\}} \oplus (2) \CMTred{\{144\}} \oplus (3) \CMTred{\{560\}} \oplus \CMTred{\{720\}} \oplus (2) \CMTred{\{1200\}} \oplus (2) \CMTred{\{1440\}}\\
~~~~~~~~~~~~~~~~~~~~~~~\oplus (2) \CMTred{\{3696\}} \oplus \CMTred{\{5280\}} \oplus \CMTred{\{8064\}} \oplus (2) \CMTred{\{8800\}} \oplus \CMTred{\{11088\}}\\
~~~~~~~~~~~~~~~~~~~~~~~\oplus \CMTred{\{25200\}} \oplus \CMTred{\{34992\}}
 ~~~, \\
{~~}{\rm {Level}}-4 \,~~~~~~~\CMTB{\{45\}} \oplus \CMTgrn{\bf \{54\}} \oplus (2) \CMTB{\{210\}} \oplus (2) \CMTB{\{770\}} \oplus (3) \CMTB{\{945\}} \oplus (3) \CMTB{\{\overline{1050}\}} \\
~~~~~~~~~~~~~~~~~~~~~~~\oplus \CMTB{\{1050\}} \oplus \CMTB{\{1386\}} \oplus \CMTB{\{\overline{2772}\}} \oplus \CMTB{\{4125\}} \oplus (4) \CMTB{\{5940\}} \oplus (2) \CMTB{\{\overline{6930}\}} \\
~~~~~~~~~~~~~~~~~~~~~~~\oplus (2) \CMTB{\{8085\}} \oplus \CMTB{\{8910\}} \oplus (2) \CMTB{\{17920\}} \oplus (2) \CMTB{\{\overline{23040}\}} \oplus \CMTB{\{23040\}}\\
~~~~~~~~~~~~~~~~~~~~~~~\oplus \CMTB{\{\overline{50688}\}} \oplus \CMTB{\{72765\}} \oplus \CMTB{\{73710\}}  ~~~,  \\
{~~}{\rm {Level}}-5 \,~~~~~~~ (2) \CMTred{\{\overline{144}\}} \oplus (2) \CMTred{\{\overline{560}\}} \oplus (2) \CMTred{\{\overline{672}\}} \oplus (2) \CMTred{\{\overline{720}\}} \oplus (3) \CMTred{\{\overline{1200}\}} \oplus \CMTred{\{\overline{1440}\}}\\
~~~~~~~~~~~~~~~~~~~~~~~\oplus \CMTred{\{\overline{2640}\}} \oplus (5) \CMTred{\{\overline{3696}\}} \oplus (2) \CMTred{\{\overline{8064}\}} \oplus (3) \CMTred{\{\overline{8800}\}} \oplus (4) \CMTred{\{\overline{11088}\}}\\
~~~~~~~~~~~~~~~~~~~~~~~\oplus (2) \CMTred{\{\overline{15120}\}} \oplus (2) \CMTred{\{\overline{17280}\}} \oplus \CMTred{\{\overline{23760}\}} \oplus (3) \CMTred{\{\overline{25200}\}} \oplus \CMTred{\{\overline{26400}\}}\\
~~~~~~~~~~~~~~~~~~~~~~~\oplus \CMTred{\{\overline{34992}\}} \oplus (2) \CMTred{\{\overline{38016}\}} \oplus \CMTred{\{\overline{43680}\}} \oplus \CMTred{\{\overline{49280}\}} \oplus \CMTred{\{\overline{55440}\}}\\
~~~~~~~~~~~~~~~~~~~~~~~\oplus \CMTred{\{\overline{144144}\}} \oplus \CMTred{\{\overline{205920}\}} ~~~,  \\
{~~}{\rm {Level}}-6 \,~~~~~~~ \CMTB{\{120\}} \oplus \CMTB{\{\overline{126}\}} \oplus \CMTB{\{210'\}} \oplus (2) \CMTB{\{320\}} \oplus (4) \CMTB{\{1728\}} \oplus (3) \CMTB{\{2970\}}\\
~~~~~~~~~~~~~~~~~~~~~~~\oplus (4) \CMTB{\{\overline{3696'}\}} \oplus \CMTB{\{3696'\}} \oplus (4) \CMTB{\{4312\}} \oplus (2) \CMTB{\{4410\}} \oplus \CMTB{\{4608\}}\\
~~~~~~~~~~~~~~~~~~~~~~~\oplus (3) \CMTB{\{\overline{4950}\}} \oplus (2) \CMTB{\{4950\}} \oplus (2) \CMTB{\{\overline{6930'}\}} \oplus (3) \CMTB{\{10560\}} \oplus \CMTB{\{\overline{20790}\}}\\
~~~~~~~~~~~~~~~~~~~~~~~\oplus (2) \CMTB{\{27720\}} \oplus (2) \CMTB{\{28160\}} \oplus \CMTB{\{34398\}} \oplus (5) \CMTB{\{36750\}} \oplus \CMTB{\{\overline{46800}\}}\\
~~~~~~~~~~~~~~~~~~~~~~~\oplus (3) \CMTB{\{\overline{48114}\}} \oplus \CMTB{\{48114\}} \oplus \CMTB{\{64680\}} \oplus \CMTB{\{68640\}} \oplus \CMTB{\{90090\}}\\
~~~~~~~~~~~~~~~~~~~~~~~\oplus \CMTB{\{\overline{90090}\}} \oplus \CMTB{\{\overline{150150}\}} \oplus \CMTB{\{192192\}} \oplus \CMTB{\{299520\}} ~~~,  \\
{~~}{\rm {Level}}-7 \,~~~~~~~ \CMTred{\{144\}} \oplus (2) \CMTred{\{560\}} \oplus (3) \CMTred{\{720\}} \oplus (2) \CMTred{\{1200\}} \oplus (2) \CMTred{\{1440\}} \oplus (3) \CMTred{\{2640\}} \\
~~~~~~~~~~~~~~~~~~~~~~~ \oplus (4) \CMTred{\{3696\}} \oplus (2) \CMTred{\{5280\}} \oplus \CMTred{\{7920\}} \oplus (2) \CMTred{\{8064\}} \oplus (6) \CMTred{\{8800\}}\\
~~~~~~~~~~~~~~~~~~~~~~~ \oplus (3) \CMTred{\{11088\}} \oplus (4) \CMTred{\{15120\}} \oplus \CMTred{\{23760\}} \oplus (3) \CMTred{\{25200\}} \oplus \CMTred{\{29568\}}\\
~~~~~~~~~~~~~~~~~~~~~~~ \oplus (2) \CMTred{\{30800\}} \oplus (3) \CMTred{\{34992\}} \oplus (4) \CMTred{\{38016\}} \oplus \CMTred{\{43680\}} \oplus \CMTred{\{48048\}}\\
~~~~~~~~~~~~~~~~~~~~~~~ \oplus (3) \CMTred{\{49280\}} \oplus \CMTred{\{55440\}} \oplus \CMTred{\{102960\}} \oplus \CMTred{\{124800\}} \oplus (2) \CMTred{\{144144\}}\\
~~~~~~~~~~~~~~~~~~~~~~~ \oplus \CMTred{\{164736\}} \oplus \CMTred{\{196560\}} \oplus \CMTred{\{205920\}} \oplus \CMTred{\{364000\}} ~~~,  \\
{~~}{\rm {Level}}-8 \,~~~~~~~ \CMTgrn{\bf \{54\}} \oplus \CMTB{\{210\}} \oplus \CMTB{\{660\}} \oplus \CMTB{\{770\}} \oplus (3) \CMTB{\{945\}} \oplus (2) \CMTB{\{1050\}} \oplus (2) \CMTB{\{\overline{1050}\}}\\
~~~~~~~~~~~~~~~~~~~~~~~ \oplus (3) \CMTB{\{1386\}} \oplus (3) \CMTB{\{4125\}} \oplus (4) \CMTB{\{5940\}} \oplus (2) \CMTB{\{6930\}} \oplus (2) \CMTB{\{\overline{6930}\}}\\
~~~~~~~~~~~~~~~~~~~~~~~ \oplus (6) \CMTB{\{8085\}} \oplus \CMTB{\{8910\}} \oplus (3) \CMTB{\{14784\}} \oplus \CMTB{\{16380\}} \oplus (2) \CMTB{\{17325\}}\\
~~~~~~~~~~~~~~~~~~~~~~~ \oplus (2) \CMTB{\{\overline{17325}\}} \oplus (4) \CMTB{\{17920\}} \oplus (3) \CMTB{\{23040\}} \oplus (3) \CMTB{\{\overline{23040}\}} \oplus \CMTB{\{50688\}}\\
~~~~~~~~~~~~~~~~~~~~~~~ \oplus \CMTB{\{\overline{50688}\}} \oplus \CMTB{\{70070\}} \oplus (5) \CMTB{\{72765\}} \oplus \CMTB{\{73710\}} \oplus \CMTB{\{81081\}}\\
~~~~~~~~~~~~~~~~~~~~~~~ \oplus \CMTB{\{112320\}} \oplus \CMTB{\{128700\}} \oplus \CMTB{\{\overline{128700}\}} \oplus (2) \CMTB{\{143000\}} \oplus \CMTB{\{199017\}}\\
~~~~~~~~~~~~~~~~~~~~~~~ \oplus \CMTB{\{\overline{199017}\}} \oplus \CMTB{\{210210\}} \oplus \CMTB{\{274560\}}  ~~~,  \\
{~~~~~~}  {~~~~} \vdots  {~~~~~~~~~\,~~~~~~} \vdots
\end{cases}
{~~~~~~~~~~~~~~~}
\end{equation}
\endgroup

\begingroup
\footnotesize
\begin{equation}
{\cal V} \otimes \CMTB{\{320\}} ~=~ \begin{cases}
{~~}{\rm {Level}}-0 \,~~~~~~~ \CMTB{\{320\}} ~~~,  \\
{~~}{\rm {Level}}-1 \,~~~~~~~ \CMTred{\{144\}} \oplus \CMTred{\{560\}} \oplus \CMTred{\{720\}} \oplus \CMTred{\{3696\}} ~~~, \\
{~~}{\rm {Level}}-2 \,~~~~~~~ \CMTB{\{45\}} \oplus \CMTgrn{\bf \{54\}} \oplus \CMTB{\{210\}} \oplus \CMTB{\{770\}} \oplus (2) \CMTB{\{945\}} \oplus \CMTB{\{1050\}} \oplus \CMTB{\{\overline{1050}\}} \oplus \CMTB{\{1386\}}\\
~~~~~~~~~~~~~~~~~~~~~~~\oplus \CMTB{\{5940\}} \oplus \CMTB{\{8085\}} \oplus \CMTB{\{17920\}} ~~~, \\
{~~}{\rm {Level}}-3 \,~~~~~~~ \CMTred{\{\overline{16}\}} \oplus (2) \CMTred{\{\overline{144}\}} \oplus (2) \CMTred{\{\overline{560}\}} \oplus (2) \CMTred{\{\overline{720}\}} \oplus (2) \CMTred{\{\overline{1200}\}} \oplus \CMTred{\{\overline{1440}\}} \oplus \CMTred{\{\overline{2640}\}}\\
~~~~~~~~~~~~~~~~~~~~~~~\oplus (3) \CMTred{\{\overline{3696}\}} \oplus \CMTred{\{\overline{8064}\}} \oplus (2) \CMTred{\{\overline{8800}\}} \oplus \CMTred{\{\overline{11088}\}} \oplus \CMTred{\{\overline{15120}\}} \oplus \CMTred{\{\overline{25200}\}}\\
~~~~~~~~~~~~~~~~~~~~~~~\oplus \CMTred{\{\overline{38016}\}} \oplus \CMTred{\{\overline{43680}\}} ~~~, \\
{~~}{\rm {Level}}-4 \,~~~~~~~ \CMTB{\{10\}} \oplus (2) \CMTB{\{120\}} \oplus \CMTB{\{126\}} \oplus \CMTB{\{\overline{126}\}} \oplus \CMTB{\{210'\}} \oplus (2) \CMTB{\{320\}} \oplus (4) \CMTB{\{1728\}}\\
~~~~~~~~~~~~~~~~~~~~~~~\oplus (3) \CMTB{\{2970\}} \oplus (2) \CMTB{\{\overline{3696'}\}}  \oplus \CMTB{\{3696'\}} \oplus (3) \CMTB{\{4312\}} \oplus (2) \CMTB{\{4410\}} \oplus \CMTB{\{4608\}}\\
~~~~~~~~~~~~~~~~~~~~~~~\oplus (2) \CMTB{\{\overline{4950}\}} \oplus \CMTB{\{4950\}} \oplus (2) \CMTB{\{10560\}} \oplus \CMTB{\{27720\}}  \oplus \CMTB{\{28160\}} \oplus \CMTB{\{34398\}} \\
~~~~~~~~~~~~~~~~~~~~~~~\oplus (3) \CMTB{\{36750\}} \oplus \CMTB{\{37632\}}  \oplus \CMTB{\{\overline{46800}\}} \oplus \CMTB{\{\overline{48114}\}} \oplus \CMTB{\{68640\}} \oplus \CMTB{\{\overline{90090}\}} ~~~,  \\
{~~}{\rm {Level}}-5 \,~~~~~~~ \CMTred{\{16\}} \oplus (2) \CMTred{\{144\}} \oplus (3) \CMTred{\{560\}} \oplus (3) \CMTred{\{720\}} \oplus (3) \CMTred{\{1200\}} \oplus (2) \CMTred{\{1440\}}\\
~~~~~~~~~~~~~~~~~~~~~~~\oplus (2) \CMTred{\{2640\}} \oplus (4) \CMTred{\{3696\}} \oplus \CMTred{\{5280\}} \oplus \CMTred{\{7920\}} \oplus (3) \CMTred{\{8064\}}\\
~~~~~~~~~~~~~~~~~~~~~~~\oplus (5) \CMTred{\{8800\}} \oplus (2) \CMTred{\{11088\}} \oplus (4) \CMTred{\{15120\}} \oplus (2) \CMTred{\{25200\}} \oplus \CMTred{\{30800\}}\\
~~~~~~~~~~~~~~~~~~~~~~~\oplus (2) \CMTred{\{34992\}} \oplus (2) \CMTred{\{38016\}} \oplus (2) \CMTred{\{43680\}} \oplus \CMTred{\{48048\}} \oplus (2) \CMTred{\{49280\}}\\
~~~~~~~~~~~~~~~~~~~~~~~\oplus \CMTred{\{70560\}} \oplus \CMTred{\{102960\}} \oplus \CMTred{\{124800\}} \oplus (2) \CMTred{\{144144\}} \oplus \CMTred{\{155232\}} ~~~,  \\
{~~}{\rm {Level}}-6 \,~~~~~~~ \CMTB{\{45\}} \oplus \CMTgrn{\bf \{54\}} \oplus (2) \CMTB{\{210\}} \oplus \CMTB{\{660\}} \oplus (2) \CMTB{\{770\}} \oplus (4) \CMTB{\{945\}}\\
~~~~~~~~~~~~~~~~~~~~~~~\oplus (3) \CMTB{\{\overline{1050}\}} \oplus (2) \CMTB{\{1050\}} \oplus (3) \CMTB{\{1386\}} \oplus (2) \CMTB{\{4125\}} \oplus (5) \CMTB{\{5940\}}\\
~~~~~~~~~~~~~~~~~~~~~~~\oplus (2) \CMTB{\{\overline{6930}\}} \oplus \CMTB{\{6930\}} \oplus \CMTB{\{7644\}} \oplus (5) \CMTB{\{8085\}} \oplus \CMTB{\{8910\}} \oplus \CMTB{\{12870\}}\\
~~~~~~~~~~~~~~~~~~~~~~~\oplus (3) \CMTB{\{14784\}} \oplus (2) \CMTB{\{16380\}} \oplus (2) \CMTB{\{\overline{17325}\}} \oplus \CMTB{\{17325\}} \oplus (5) \CMTB{\{17920\}}\\
~~~~~~~~~~~~~~~~~~~~~~~\oplus (4) \CMTB{\{\overline{23040}\}} \oplus (2) \CMTB{\{23040\}} \oplus \CMTB{\{\overline{50688}\}} \oplus \CMTB{\{70070\}} \oplus (3) \CMTB{\{72765\}}\\
~~~~~~~~~~~~~~~~~~~~~~~\oplus (2) \CMTB{\{73710\}} \oplus \CMTB{\{81081\}} \oplus \CMTB{\{112320\}} \oplus \CMTB{\{\overline{128700}\}} \oplus (3) \CMTB{\{143000\}}\\
~~~~~~~~~~~~~~~~~~~~~~~\oplus \CMTB{\{174636\}} \oplus \CMTB{\{\overline{199017}\}} \oplus \CMTB{\{207360\}} \oplus \CMTB{\{\overline{242550}\}} ~~~,  \\
{~~}{\rm {Level}}-7 \,~~~~~~~ (2) \CMTred{\{\overline{144}\}} \oplus (3) \CMTred{\{\overline{560}\}} \oplus \CMTred{\{\overline{672}\}} \oplus (3) \CMTred{\{\overline{720}\}} \oplus (3) \CMTred{\{\overline{1200}\}} \oplus (2) \CMTred{\{\overline{1440}\}}\\
~~~~~~~~~~~~~~~~~~~~~~~\oplus (3) \CMTred{\{\overline{2640}\}} \oplus (6) \CMTred{\{\overline{3696}\}} \oplus \CMTred{\{\overline{5280}\}} \oplus (2) \CMTred{\{\overline{7920}\}} \oplus (3) \CMTred{\{\overline{8064}\}}\\
~~~~~~~~~~~~~~~~~~~~~~~\oplus (5) \CMTred{\{\overline{8800}\}} \oplus (4) \CMTred{\{\overline{11088}\}} \oplus (5) \CMTred{\{\overline{15120}\}} \oplus \CMTred{\{\overline{17280}\}} \oplus \CMTred{\{\overline{20592}\}}\\
~~~~~~~~~~~~~~~~~~~~~~~\oplus \CMTred{\{\overline{23760}\}} \oplus (4) \CMTred{\{\overline{25200}\}} \oplus \CMTred{\{\overline{30800}\}} \oplus (2) \CMTred{\{\overline{34992}\}} \oplus (5) \CMTred{\{\overline{38016}\}}\\
~~~~~~~~~~~~~~~~~~~~~~~\oplus (3) \CMTred{\{\overline{43680}\}} \oplus (3) \CMTred{\{\overline{48048}\}} \oplus (2) \CMTred{\{\overline{49280}\}} \oplus \CMTred{\{\overline{55440}\}}\\
~~~~~~~~~~~~~~~~~~~~~~~\oplus (2) \CMTred{\{\overline{124800}\}} \oplus \CMTred{\{\overline{129360}\}} \oplus (3) \CMTred{\{\overline{144144}\}} \oplus \CMTred{\{\overline{155232}\}} \oplus \CMTred{\{\overline{164736}\}}\\
~~~~~~~~~~~~~~~~~~~~~~~\oplus \CMTred{\{\overline{196560}\}} \oplus (2) \CMTred{\{\overline{205920}\}} \oplus \CMTred{\{\overline{258720}\}} \oplus \CMTred{\{\overline{529200}\}} ~~~,  \\
{~~}{\rm {Level}}-8 \,~~~~~~~\CMTB{\{120\}} \oplus \CMTB{\{126\}} \oplus \CMTB{\{\overline{126}\}} \oplus \CMTB{\{210'\}} \oplus (3) \CMTB{\{320\}} \oplus (5) \CMTB{\{1728\}} \oplus \CMTB{\{1782\}}\\
~~~~~~~~~~~~~~~~~~~~~~~\oplus (4) \CMTB{\{2970\}} \oplus (3) \CMTB{\{3696'\}} \oplus (3) \CMTB{\{\overline{3696'}\}} \oplus (5) \CMTB{\{4312\}} \oplus (3) \CMTB{\{4410\}}\\
~~~~~~~~~~~~~~~~~~~~~~~\oplus (3) \CMTB{\{4608\}} \oplus (3) \CMTB{\{4950\}} \oplus (3) \CMTB{\{\overline{4950}\}} \oplus \CMTB{\{6930'\}} \oplus \CMTB{\{\overline{6930'}\}} \\
~~~~~~~~~~~~~~~~~~~~~~~\oplus (3) \CMTB{\{10560\}} \oplus (3) \CMTB{\{27720\}} \oplus (4) \CMTB{\{28160\}} \oplus \CMTB{\{31680\}} \oplus (2) \CMTB{\{34398\}}\\
~~~~~~~~~~~~~~~~~~~~~~~\oplus (7) \CMTB{\{36750\}} \oplus (2) \CMTB{\{42120\}} \oplus \CMTB{\{46800\}} \oplus \CMTB{\{\overline{46800}\}} \oplus (2) \CMTB{\{48114\}}\\
~~~~~~~~~~~~~~~~~~~~~~~\oplus (2) \CMTB{\{\overline{48114}\}} \oplus \CMTB{\{48510\}} \oplus \CMTB{\{50050\}} \oplus \CMTB{\{\overline{50050}\}} \oplus \CMTB{\{64680\}} \oplus (4) \CMTB{\{68640\}}\\
~~~~~~~~~~~~~~~~~~~~~~~\oplus (2) \CMTB{\{90090\}} \oplus (2) \CMTB{\{\overline{90090}\}} \oplus (2) \CMTB{\{192192\}} \oplus (2) \CMTB{\{299520\}}\\
~~~~~~~~~~~~~~~~~~~~~~~\oplus \CMTB{\{376320\}} \oplus \CMTB{\{380160\}} \oplus \CMTB{\{436590\}}
   ~~~,  \\
{~~~~~~}  {~~~~} \vdots  {~~~~~~~~~\,~~~~~~} \vdots
\end{cases}
{~~~~~~~~~~~~~~~}
\end{equation}
\endgroup

\begingroup
\footnotesize
\begin{equation}
{\cal V} \otimes \CMTB{\{660\}} ~=~ \begin{cases}
{~~}{\rm {Level}}-0 \,~~~~~~~ \CMTB{\{660\}} ~~~,  \\
{~~}{\rm {Level}}-1 \,~~~~~~~ \CMTred{\{\overline{2640}\}} \oplus \CMTred{\{\overline{7920}\}} ~~~, \\
{~~}{\rm {Level}}-2 \,~~~~~~~\CMTB{\{4312\}} \oplus \CMTB{\{4608\}} \oplus \CMTB{\{28160\}} \oplus \CMTB{\{42120\}}  ~~~, \\
{~~}{\rm {Level}}-3 \,~~~~~~~\CMTred{\{2640\}} \oplus \CMTred{\{3696\}} \oplus \CMTred{\{7920\}} \oplus \CMTred{\{15120\}} \oplus \CMTred{\{38016\}} \oplus \CMTred{\{48048\}} \oplus \CMTred{\{124800\}}\\
~~~~~~~~~~~~~~~~~~~~~~~\oplus \CMTred{\{129360\}}  ~~~, \\
{~~}{\rm {Level}}-4 \,~~~~~~~ \CMTB{\{660\}} \oplus \CMTB{\{770\}} \oplus \CMTB{\{1050\}} \oplus \CMTB{\{1386\}} \oplus \CMTB{\{8085\}} \oplus \CMTB{\{12870\}} \oplus \CMTB{\{14784\}}\\
~~~~~~~~~~~~~~~~~~~~~~~\oplus \CMTB{\{16380\}} \oplus \CMTB{\{17325\}} \oplus \CMTB{\{\overline{17325}\}} \oplus \CMTB{\{17920\}} \oplus \CMTB{\{23040\}} \oplus \CMTB{\{81081\}}\\
~~~~~~~~~~~~~~~~~~~~~~~\oplus \CMTB{\{112320\}} \oplus \CMTB{\{123750\}} \oplus \CMTB{\{\overline{126126}\}} \oplus \CMTB{\{143000\}} \oplus \CMTB{\{207360\}} \oplus \CMTB{\{\overline{275968}\}} ~~~,  \\
{~~}{\rm {Level}}-5 \,~~~~~~~ \CMTred{\{\overline{560}\}} \oplus \CMTred{\{\overline{720}\}} \oplus \CMTred{\{\overline{2640}\}} \oplus \CMTred{\{\overline{3696}\}} \oplus \CMTred{\{\overline{5280}\}} \oplus \CMTred{\{\overline{7920}\}} \oplus \CMTred{\{\overline{8064}\}} \oplus \CMTred{\{\overline{8800}\}}\\
~~~~~~~~~~~~~~~~~~~~~~~\oplus (2) \CMTred{\{\overline{15120}\}} \oplus \CMTred{\{\overline{20592}\}} \oplus \CMTred{\{\overline{38016}\}} \oplus \CMTred{\{\overline{43680}\}} \oplus (2) \CMTred{\{\overline{48048}\}} \oplus \CMTred{\{\overline{49280}\}}\\
~~~~~~~~~~~~~~~~~~~~~~~\oplus \CMTred{\{\overline{124800}\}} \oplus \CMTred{\{\overline{129360}\}} \oplus \CMTred{\{\overline{144144}\}} \oplus \CMTred{\{\overline{155232}\}} \oplus \CMTred{\{\overline{164736}\}} \oplus \CMTred{\{\overline{224224}\}}\\
~~~~~~~~~~~~~~~~~~~~~~~\oplus \CMTred{\{\overline{308880}\}} \oplus \CMTred{\{\overline{343200}\}} \oplus \CMTred{\{\overline{443520}\}} \oplus \CMTred{\{\overline{529200}\}} ~~~,  \\
{~~}{\rm {Level}}-6 \,~~~~~~~ \CMTB{\{120\}} \oplus \CMTB{\{320\}} \oplus \CMTB{\{1728\}} \oplus \CMTB{\{2970\}} \oplus \CMTB{\{3696'\}} \oplus (2) \CMTB{\{4312\}} \oplus \CMTB{\{4410\}}\\
~~~~~~~~~~~~~~~~~~~~~~~\oplus \CMTB{\{4608\}} \oplus \CMTB{\{4950\}} \oplus (2) \CMTB{\{28160\}} \oplus \CMTB{\{31680\}} \oplus \CMTB{\{34398\}} \oplus (2) \CMTB{\{36750\}}\\
~~~~~~~~~~~~~~~~~~~~~~~\oplus \CMTB{\{37632\}} \oplus (2) \CMTB{\{42120\}} \oplus \CMTB{\{48114\}} \oplus \CMTB{\{48510\}} \oplus \CMTB{\{\overline{50050}\}} \oplus (2) \CMTB{\{68640\}} \\
~~~~~~~~~~~~~~~~~~~~~~~\oplus \CMTB{\{90090\}} \oplus \CMTB{\{\overline{90090}\}} \oplus \CMTB{\{203840\}} \oplus \CMTB{\{237160\}} \oplus \CMTB{\{299520\}} \oplus \CMTB{\{380160\}}\\
~~~~~~~~~~~~~~~~~~~~~~~\oplus (2) \CMTB{\{436590\}} \oplus \CMTB{\{536250\}} \oplus \CMTB{\{590490\}} \oplus \CMTB{\{\overline{630630}\}} \oplus \CMTB{\{\overline{720720}\}} ~~~,  \\
{~~}{\rm {Level}}-7 \,~~~~~~~ \CMTred{\{16\}} \oplus \CMTred{\{144\}} \oplus \CMTred{\{560\}} \oplus \CMTred{\{720\}} \oplus \CMTred{\{1200\}} \oplus \CMTred{\{2640\}} \oplus (2) \CMTred{\{3696\}} \oplus \CMTred{\{7920\}}\\
~~~~~~~~~~~~~~~~~~~~~~~\oplus \CMTred{\{8064\}} \oplus \CMTred{\{8800\}} \oplus \CMTred{\{11088\}} \oplus (2) \CMTred{\{15120\}} \oplus \CMTred{\{20592\}} \oplus \CMTred{\{25200\}}\\
~~~~~~~~~~~~~~~~~~~~~~~\oplus (2) \CMTred{\{38016\}} \oplus (2) \CMTred{\{43680\}} \oplus (2) \CMTred{\{48048\}} \oplus \CMTred{\{48048'\}} \oplus \CMTred{\{49280\}} \\
~~~~~~~~~~~~~~~~~~~~~~~\oplus \CMTred{\{70560\}} \oplus \CMTred{\{102960''\}} \oplus (2) \CMTred{\{124800\}} \oplus (2) \CMTred{\{129360\}} \oplus \CMTred{\{144144\}}\\
~~~~~~~~~~~~~~~~~~~~~~~\oplus (2) \CMTred{\{155232\}} \oplus \CMTred{\{164736\}} \oplus \CMTred{\{205920\}} \oplus \CMTred{\{308880\}} \oplus \CMTred{\{332640\}}\oplus \CMTred{\{343200\}} \\
~~~~~~~~~~~~~~~~~~~~~~~\oplus \CMTred{\{443520\}} \oplus \CMTred{\{458640\}} \oplus \CMTred{\{529200\}} \oplus \CMTred{\{769824\}} \oplus \CMTred{\{831600'\}} \oplus \CMTred{\{1544400\}} ~~~,  \\
{~~}{\rm {Level}}-8 \,~~~~~~~ \CMTB{\{1\}} \oplus \CMTB{\{45\}} \oplus \CMTgrn{\bf \{54\}} \oplus \CMTB{\{210\}} \oplus \CMTB{\{660\}} \oplus \CMTB{\{770\}} \oplus \CMTB{\{945\}} \oplus \CMTB{\{1050\}} \oplus \CMTB{\{\overline{1050}\}}\\
~~~~~~~~~~~~~~~~~~~~~~~\oplus \CMTB{\{1386\}} \oplus \CMTB{\{4125\}} \oplus \CMTB{\{4290\}} \oplus \CMTB{\{5940\}} \oplus \CMTB{\{7644\}} \oplus (2) \CMTB{\{8085\}} \oplus \CMTB{\{12870\}}\\
~~~~~~~~~~~~~~~~~~~~~~~\oplus \CMTB{\{14784\}} \oplus (2) \CMTB{\{16380\}} \oplus \CMTB{\{17325\}} \oplus \CMTB{\{\overline{17325}\}} \oplus (2) \CMTB{\{17920\}} \oplus \CMTB{\{19305\}}\\
~~~~~~~~~~~~~~~~~~~~~~~\oplus \CMTB{\{23040\}} \oplus \CMTB{\{\overline{23040}\}} \oplus \CMTB{\{52920\}} \oplus \CMTB{\{70785\}} \oplus \CMTB{\{72765\}} \oplus \CMTB{\{73710\}}\\
~~~~~~~~~~~~~~~~~~~~~~~\oplus (2) \CMTB{\{81081\}} \oplus \CMTB{\{105105\}} \oplus \CMTB{\{112320\}} \oplus \CMTB{\{123750\}} \oplus \CMTB{\{124740\}} \oplus \CMTB{\{126126\}}\\
~~~~~~~~~~~~~~~~~~~~~~~\oplus \CMTB{\{\overline{126126}\}} \oplus (3) \CMTB{\{143000\}} \oplus \CMTB{\{174636\}} \oplus (2) \CMTB{\{207360\}} \oplus \CMTB{\{242550\}}\\
~~~~~~~~~~~~~~~~~~~~~~~\oplus \CMTB{\{\overline{242550}\}} \oplus \CMTB{\{274560\}} \oplus \CMTB{\{275968\}} \oplus \CMTB{\{\overline{275968}\}} \oplus \CMTB{\{462462\}} \oplus \CMTB{\{928125\}}\\
~~~~~~~~~~~~~~~~~~~~~~~\oplus \CMTB{\{945945\}} \oplus \CMTB{\{1137500\}} \oplus \CMTB{\{1299078\}} ~~~,  \\
{~~~~~~}  {~~~~} \vdots  {~~~~~~~~~\,~~~~~~} \vdots
\end{cases}
{~~~~~~~~~~~~~~~}
\end{equation}
\endgroup

\begingroup
\scriptsize
\begin{equation}
{\cal V} \otimes \CMTB{\{770\}} ~=~ \begin{cases}
{~~}{\rm {Level}}-0 \,~~~~~~~ \CMTB{\{770\}} ~~~,  \\
{~~}{\rm {Level}}-1 \,~~~~~~~ \CMTred{\{\overline{560}\}} \oplus \CMTred{\{\overline{3696}\}} \oplus \CMTred{\{\overline{8064}\}} ~~~, \\
{~~}{\rm {Level}}-2 \,~~~~~~~ \CMTB{\{120\}} \oplus \CMTB{\{320\}} \oplus \CMTB{\{1728\}} \oplus \CMTB{\{2970\}} \oplus \CMTB{\{3696'\}} \oplus \CMTB{\{\overline{3696'}\}} \oplus \CMTB{\{4312\}}\\
~~~~~~~~~~~~~~~~~~~~~~~\oplus \CMTB{\{4410\}} \oplus \CMTB{\{34398\}} \oplus \CMTB{\{36750\}} ~~~, \\
{~~}{\rm {Level}}-3 \,~~~~~~~ \CMTred{\{16\}} \oplus \CMTred{\{144\}} \oplus (2) \CMTred{\{560\}} \oplus \CMTred{\{720\}} \oplus \CMTred{\{1200\}} \oplus \CMTred{\{1440\}} \oplus \CMTred{\{2640\}}\\
~~~~~~~~~~~~~~~~~~~~~~~\oplus (2) \CMTred{\{3696\}} \oplus (2) \CMTred{\{8064\}} \oplus (2) \CMTred{\{8800\}} \oplus \CMTred{\{11088\}} \oplus \CMTred{\{15120\}} \oplus \CMTred{\{25200\}} \\
~~~~~~~~~~~~~~~~~~~~~~~\oplus \CMTred{\{34992\}} \oplus \CMTred{\{38016\}} \oplus \CMTred{\{43680\}} \oplus \CMTred{\{70560\}} \oplus \CMTred{\{144144\}} ~~~, \\
{~~}{\rm {Level}}-4 \,~~~~~~~ \CMTB{\{1\}} \oplus \CMTB{\{45\}} \oplus \CMTgrn{\bf \{54\}} \oplus (2) \CMTB{\{210\}} \oplus \CMTB{\{660\}} \oplus (2) \CMTB{\{770\}} \oplus (2) \CMTB{\{945\}} \oplus (2) \CMTB{\{\overline{1050}\}}\\
~~~~~~~~~~~~~~~~~~~~~~~\oplus \CMTB{\{1050\}} \oplus \CMTB{\{1386\}} \oplus (2) \CMTB{\{4125\}} \oplus (3) \CMTB{\{5940\}} \oplus \CMTB{\{\overline{6930}\}} \oplus \CMTB{\{7644\}} \\
~~~~~~~~~~~~~~~~~~~~~~~\oplus (3) \CMTB{\{8085\}} \oplus \CMTB{\{8910\}} \oplus \CMTB{\{14784\}} \oplus \CMTB{\{16380\}} \oplus \CMTB{\{17325\}} \oplus (3) \CMTB{\{17920\}}\\
~~~~~~~~~~~~~~~~~~~~~~~\oplus (2) \CMTB{\{\overline{23040}\}} \oplus \CMTB{\{23040\}} \oplus \CMTB{\{52920\}} \oplus (2) \CMTB{\{72765\}} \oplus (2) \CMTB{\{73710\}}\\
~~~~~~~~~~~~~~~~~~~~~~~\oplus \CMTB{\{112320\}} \oplus \CMTB{\{\overline{128700}\}} \oplus \CMTB{\{143000\}} \oplus \CMTB{\{174636\}} \oplus \CMTB{\{\overline{242550}\}} ~~~,  \\
{~~}{\rm {Level}}-5 \,~~~~~~~ \CMTred{\{\overline{16}\}} \oplus (2) \CMTred{\{\overline{144}\}} \oplus (2) \CMTred{\{\overline{560}\}} \oplus \CMTred{\{\overline{672}\}} \oplus (2) \CMTred{\{\overline{720}\}} \oplus (3) \CMTred{\{\overline{1200}\}} \oplus \CMTred{\{\overline{1440}\}} \\
~~~~~~~~~~~~~~~~~~~~~~~\oplus (2) \CMTred{\{\overline{2640}\}} \oplus (4) \CMTred{\{\overline{3696}\}} \oplus \CMTred{\{\overline{7920}\}} \oplus (2) \CMTred{\{\overline{8064}\}} \oplus (4) \CMTred{\{\overline{8800}\}} \oplus (3) \CMTred{\{\overline{11088}\}}\\
~~~~~~~~~~~~~~~~~~~~~~~\oplus (3) \CMTred{\{\overline{15120}\}} \oplus \CMTred{\{\overline{17280}\}} \oplus (4) \CMTred{\{\overline{25200}\}} \oplus \CMTred{\{\overline{30800}\}} \oplus \CMTred{\{\overline{34992}\}} \oplus (3) \CMTred{\{\overline{38016}\}}\\
~~~~~~~~~~~~~~~~~~~~~~~\oplus (3) \CMTred{\{\overline{43680}\}} \oplus \CMTred{\{\overline{48048}\}} \oplus (2) \CMTred{\{\overline{49280}\}} \oplus \CMTred{\{\overline{55440}\}} \oplus \CMTred{\{\overline{70560}\}} \oplus \CMTred{\{\overline{124800}\}}\\
~~~~~~~~~~~~~~~~~~~~~~~\oplus (2) \CMTred{\{\overline{144144}\}} \oplus \CMTred{\{\overline{155232}\}} \oplus \CMTred{\{\overline{196560}\}} \oplus \CMTred{\{\overline{198000}\}} \oplus (2) \CMTred{\{\overline{205920}\}}\\
~~~~~~~~~~~~~~~~~~~~~~~\oplus \CMTred{\{\overline{258720}\}} \oplus \CMTred{\{\overline{332640}\}} \oplus \CMTred{\{\overline{529200}\}} ~~~,  \\
{~~}{\rm {Level}}-6 \,~~~~~~~(2) \CMTB{\{120\}} \oplus \CMTB{\{\overline{126}\}} \oplus \CMTB{\{210'\}} \oplus (2) \CMTB{\{320\}} \oplus (4) \CMTB{\{1728\}} \oplus (3) \CMTB{\{2970\}} \oplus (4) \CMTB{\{\overline{3696'}\}}\\
~~~~~~~~~~~~~~~~~~~~~~~\oplus (2) \CMTB{\{3696'\}} \oplus (5) \CMTB{\{4312\}} \oplus (2) \CMTB{\{4410\}} \oplus (2) \CMTB{\{4608\}} \oplus (2) \CMTB{\{4950\}} \oplus (2) \CMTB{\{\overline{4950}\}}\\
~~~~~~~~~~~~~~~~~~~~~~~\oplus \CMTB{\{\overline{6930'}\}} \oplus (3) \CMTB{\{10560\}} \oplus (3) \CMTB{\{27720\}} \oplus (3) \CMTB{\{28160\}} \oplus (3) \CMTB{\{34398\}}\\
~~~~~~~~~~~~~~~~~~~~~~~\oplus (6) \CMTB{\{36750\}} \oplus \CMTB{\{37632\}} \oplus \CMTB{\{42120\}} \oplus (2) \CMTB{\{\overline{46800}\}} \oplus (3) \CMTB{\{\overline{48114}\}} \\
~~~~~~~~~~~~~~~~~~~~~~~\oplus (2) \CMTB{\{48114\}} \oplus \CMTB{\{48510\}} \oplus \CMTB{\{64680\}} \oplus (3) \CMTB{\{68640\}} \oplus \CMTB{\{70070'\}} \oplus (2) \CMTB{\{90090\}}\\
~~~~~~~~~~~~~~~~~~~~~~~\oplus (2) \CMTB{\{\overline{90090}\}} \oplus \CMTB{\{\overline{150150}\}} \oplus (2) \CMTB{\{192192\}} \oplus (2) \CMTB{\{299520\}} \oplus \CMTB{\{\overline{371250}\}}\\
~~~~~~~~~~~~~~~~~~~~~~~\oplus \CMTB{\{376320\}} \oplus (2) \CMTB{\{380160\}} \oplus \CMTB{\{436590\}} \oplus \CMTB{\{590490\}} \oplus \CMTB{\{\overline{705600}\}}  ~~~,  \\
{~~}{\rm {Level}}-7 \,~~~~~~~ \CMTred{\{144\}} \oplus (3) \CMTred{\{560\}} \oplus \CMTred{\{672\}} \oplus (3) \CMTred{\{720\}} \oplus (2) \CMTred{\{1200\}} \oplus (2) \CMTred{\{1440\}} \oplus (3) \CMTred{\{2640\}} \\
~~~~~~~~~~~~~~~~~~~~~~~\oplus (5) \CMTred{\{3696\}} \oplus (2) \CMTred{\{5280\}} \oplus \CMTred{\{7920\}} \oplus (3) \CMTred{\{8064\}} \oplus (6) \CMTred{\{8800\}} \oplus (3) \CMTred{\{11088\}} \\
~~~~~~~~~~~~~~~~~~~~~~~\oplus (5) \CMTred{\{15120\}} \oplus \CMTred{\{17280\}} \oplus \CMTred{\{23760\}} \oplus (4) \CMTred{\{25200\}} \oplus \CMTred{\{29568\}} \oplus (2) \CMTred{\{30800\}} \\
~~~~~~~~~~~~~~~~~~~~~~~\oplus (3) \CMTred{\{34992\}} \oplus (5) \CMTred{\{38016\}} \oplus (3) \CMTred{\{43680\}} \oplus (3) \CMTred{\{48048\}} \oplus (3) \CMTred{\{49280\}}\\
~~~~~~~~~~~~~~~~~~~~~~~\oplus \CMTred{\{55440\}} \oplus \CMTred{\{70560\}} \oplus \CMTred{\{102960\}} \oplus (2) \CMTred{\{124800\}} \oplus \CMTred{\{129360\}} \oplus (5) \CMTred{\{144144\}}\\
~~~~~~~~~~~~~~~~~~~~~~~\oplus (2) \CMTred{\{155232\}} \oplus \CMTred{\{164736\}} \oplus (2) \CMTred{\{196560\}} \oplus (2) \CMTred{\{205920\}} \oplus \CMTred{\{258720\}}\oplus \CMTred{\{364000\}}\\
~~~~~~~~~~~~~~~~~~~~~~~\oplus \CMTred{\{388080\}} \oplus \CMTred{\{443520\}} \oplus \CMTred{\{465696\}} \oplus (2) \CMTred{\{529200\}} \oplus \CMTred{\{769824\}} \oplus \CMTred{\{1260000\}}  ~~~,  \\
{~~}{\rm {Level}}-8 \,~~~~~~~\CMTgrn{\bf \{54\}} \oplus \CMTB{\{210\}} \oplus \CMTB{\{660\}} \oplus (3) \CMTB{\{770\}} \oplus (3) \CMTB{\{945\}} \oplus (3) \CMTB{\{1050\}} \oplus (3) \CMTB{\{\overline{1050}\}}\\
~~~~~~~~~~~~~~~~~~~~~~~\oplus (3) \CMTB{\{1386\}} \oplus \CMTB{\{2772\}} \oplus \CMTB{\{\overline{2772}\}} \oplus (3) \CMTB{\{4125\}} \oplus (4) \CMTB{\{5940\}} \oplus (2) \CMTB{\{6930\}}\\
~~~~~~~~~~~~~~~~~~~~~~~\oplus (2) \CMTB{\{\overline{6930}\}}\oplus (6) \CMTB{\{8085\}} \oplus \CMTB{\{8910\}} \oplus \CMTB{\{12870\}} \oplus (3) \CMTB{\{14784\}} \oplus (3) \CMTB{\{16380\}} \\
~~~~~~~~~~~~~~~~~~~~~~~\oplus (2) \CMTB{\{17325\}} \oplus (2) \CMTB{\{\overline{17325}\}} \oplus (6) \CMTB{\{17920\}} \oplus (4) \CMTB{\{23040\}} \oplus (4) \CMTB{\{\overline{23040}\}} \\
~~~~~~~~~~~~~~~~~~~~~~~\oplus \CMTB{\{50688\}} \oplus \CMTB{\{\overline{50688}\}} \oplus (2) \CMTB{\{70070\}} \oplus (5) \CMTB{\{72765\}} \oplus (3) \CMTB{\{73710\}} \oplus \CMTB{\{81081\}}\\
~~~~~~~~~~~~~~~~~~~~~~~\oplus (3) \CMTB{\{112320\}} \oplus \CMTB{\{123750\}} \oplus (2) \CMTB{\{128700\}} \oplus (2) \CMTB{\{\overline{128700}\}} \oplus (5) \CMTB{\{143000\}} \\
~~~~~~~~~~~~~~~~~~~~~~~\oplus (2) \CMTB{\{174636\}} \oplus \CMTB{\{199017\}} \oplus \CMTB{\{\overline{199017}\}} \oplus (2) \CMTB{\{207360\}} \oplus \CMTB{\{210210\}} \\
~~~~~~~~~~~~~~~~~~~~~~~\oplus \CMTB{\{242550\}}\oplus \CMTB{\{\overline{242550}\}}\oplus \CMTB{\{274560\}} \oplus \CMTB{\{275968\}} \oplus \CMTB{\{\overline{275968}\}} \oplus \CMTB{\{620928\}}\\
~~~~~~~~~~~~~~~~~~~~~~~\oplus (2) \CMTB{\{1048576\}} \oplus \CMTB{\{1299078\}}  ~~~,  \\
{~~~~~~}  {~~~~} \vdots  {~~~~~~~~~\,~~~~~~} \vdots
\end{cases}
{~~~~~~~~~~~~~~~}
\end{equation}
\endgroup

\newpage
\section{Fermionic Superfield Decompositions}
\label{appen:spinorial_superfield}

In this appendix, we list a few fermionic superfields that contain the traceless graviton $\CMTgrn{\bf \{54\}}$ $h_{\un a\un b}$ and the traceless gravitino $\CMTred{\{\overline{144}\}}$ $\psi_{\un a}{}^{\g}$ and $\CMTred{\{144\}}$ $\psi_{\un a\g}$. Here every irrep corresponds to a {\em component} field.

\begingroup
\footnotesize
\begin{equation}
{\cal V} \otimes \CMTred{\{144\}} ~=~ \begin{cases}
{~~}{\rm {Level}}-0 \,~~~~~~~ \CMTred{\{144\}}
 ~~~,  \\
{~~}{\rm {Level}}-1 \,~~~~~~~ \CMTB{\{45\}} \oplus \CMTgrn{\bf \{54\}} \oplus \CMTB{\{210\}} \oplus \CMTB{\{945\}} \oplus \CMTB{\{1050\}} ~~~,  \\
{~~}{\rm {Level}}-2 \,~~~~~~~ \CMTred{\{\overline{16}\}} \oplus (2) \CMTred{\{\overline{144}\}} \oplus (2) \CMTred{\{\overline{560}\}} \oplus \CMTred{\{\overline{720}\}} \oplus \CMTred{\{\overline{1200}\}} \oplus \CMTred{\{\overline{1440}\}} \oplus \CMTred{\{\overline{3696}\}} \oplus \CMTred{\{\overline{8800}\}} ~~~, \\
{~~}{\rm {Level}}-3 \,~~~~~~~\CMTB{\{10\}} \oplus (2) \CMTB{\{120\}} \oplus \CMTB{\{126\}} \oplus \CMTB{\{\overline{126}\}} \oplus (2) \CMTB{\{320\}} \oplus (3) \CMTB{\{1728\}} \oplus (2) \CMTB{\{2970\}} \\
~~~~~~~~~~~~~~~~~~~~~~~\oplus \CMTB{\{3696'\}} \oplus \CMTB{\{\overline{3696'}\}} \oplus \CMTB{\{4312\}} \oplus \CMTB{\{4410\}} \oplus \CMTB{\{\overline{4950}\}} \oplus \CMTB{\{10560\}} \oplus \CMTB{\{36750\}}  ~~~, \\
{~~}{\rm {Level}}-4 \,~~~~~~~ \CMTred{\{16\}} \oplus (2) \CMTred{\{144\}} \oplus (3) \CMTred{\{560\}} \oplus (2) \CMTred{\{720\}} \oplus (2) \CMTred{\{1200\}} \oplus (2) \CMTred{\{1440\}} \oplus (3) \CMTred{\{3696\}}\\
~~~~~~~~~~~~~~~~~~~~~~~\oplus \CMTred{\{5280\}} \oplus (2) \CMTred{\{8064\}} \oplus (3) \CMTred{\{8800\}} \oplus \CMTred{\{11088\}} \oplus (2) \CMTred{\{15120\}} \oplus \CMTred{\{25200\}} \\
~~~~~~~~~~~~~~~~~~~~~~~\oplus \CMTred{\{34992\}} \oplus \CMTred{\{43680\}} \oplus \CMTred{\{49280\}} ~~~,  \\
{~~}{\rm {Level}}-5 \,~~~~~~~ \CMTB{\{45\}} \oplus \CMTgrn{\bf \{54\}} \oplus (2) \CMTB{\{210\}} \oplus (2) \CMTB{\{770\}} \oplus (3) \CMTB{\{945\}} \oplus (3) \CMTB{\{\overline{1050}\}} \oplus \CMTB{\{1050\}} \\
~~~~~~~~~~~~~~~~~~~~~~~\oplus (2) \CMTB{\{1386\}} \oplus \CMTB{\overline{4125}\}} \oplus (4) \CMTB{\{5940\}} \oplus (2) \CMTB{\{\overline{6930}\}} \oplus \CMTB{\{7644\}} \oplus (3) \CMTB{\{8085\}}\\
~~~~~~~~~~~~~~~~~~~~~~~\oplus \CMTB{\{8910\}} \oplus \CMTB{\{14784\}} \oplus \CMTB{\{16380\}} \oplus \CMTB{\{\overline{17325}\}} \oplus (3) \CMTB{\{17920\}} \oplus (3) \CMTB{\{23040\}} \\
~~~~~~~~~~~~~~~~~~~~~~~\oplus \CMTB{\{\overline{23040}\}} \oplus \CMTB{\{\overline{50688}\}} \oplus \CMTB{\{72765\}} \oplus \CMTB{\{73710\}} \oplus \CMTB{\{143000\}} ~~~,  \\
{~~}{\rm {Level}}-6 \,~~~~~~~  (2) \CMTred{\{\overline{144}\}} \oplus (2) \CMTred{\{\overline{560}\}} \oplus \CMTred{\{\overline{672}\}} \oplus (2) \CMTred{\{\overline{720}\}} \oplus (3) \CMTred{\{\overline{1200}\}} \oplus \CMTred{\{\overline{1440}\}} \oplus (2) \CMTred{\{\overline{2640}\}}\\
~~~~~~~~~~~~~~~~~~~~~~~\oplus (5) \CMTred{\{\overline{3696}\}} \oplus \CMTred{\{\overline{7920}\}} \oplus (2) \CMTred{\{\overline{8064}\}} \oplus (3) \CMTred{\{\overline{8800}\}} \oplus (4) \CMTred{\{\overline{11088}\}} \oplus (3) \CMTred{\{\overline{15120}\}}\\
~~~~~~~~~~~~~~~~~~~~~~~\oplus \CMTred{\{\overline{17280}\}} \oplus \CMTred{\{\overline{23760}\}} \oplus (3) \CMTred{\{\overline{25200}\}} \oplus \CMTred{\{\overline{34992}\}} \oplus (3) \CMTred{\{\overline{38016}\}} \oplus (2) \CMTred{\{\overline{43680}\}}\\
~~~~~~~~~~~~~~~~~~~~~~~\oplus \CMTred{\{\overline{48048}\}} \oplus \CMTred{\{\overline{49280}\}} \oplus \CMTred{\{\overline{55440}\}} \oplus \CMTred{\{\overline{124800}\}} \oplus \CMTred{\{\overline{144144}\}} \oplus \CMTred{\{\overline{205920}\}} ~~~,  \\
{~~}{\rm {Level}}-7 \,~~~~~~~ \CMTB{\{120\}} \oplus \CMTB{\{\overline{126}\}} \oplus \CMTB{\{210'\}} \oplus (2) \CMTB{\{320\}} \oplus (4) \CMTB{\{1728\}} \oplus \CMTB{\{1782\}} \oplus (3) \CMTB{\{2970\}}\\
~~~~~~~~~~~~~~~~~~~~~~~\oplus (3) \CMTB{\{\overline{3696'}\}} \oplus \CMTB{\{3696'\}} \oplus (4) \CMTB{\{4312\}} \oplus (2) \CMTB{\{4410\}} \oplus (2) \CMTB{\{4608\}} \oplus (3) \CMTB{\{\overline{4950}\}}\\
~~~~~~~~~~~~~~~~~~~~~~~\oplus (2) \CMTB{\{4950\}} \oplus \CMTB{\{\overline{6930'}\}} \oplus (3) \CMTB{\{10560\}} \oplus (2) \CMTB{\{27720\}} \oplus (3) \CMTB{\{28160\}} \oplus \CMTB{\{34398\}} \\
~~~~~~~~~~~~~~~~~~~~~~~\oplus (5) \CMTB{\{36750\}} \oplus \CMTB{\{42120\}} \oplus \CMTB{\{\overline{46800}\}} \oplus (2) \CMTB{\{\overline{48114}\}} \oplus \CMTB{\{48114\}} \oplus \CMTB{\{50050\}} \\
~~~~~~~~~~~~~~~~~~~~~~~\oplus \CMTB{\{64680\}} \oplus (2\overline{)} \CMTB{\{68640\}} \oplus \CMTB{\{90090\}} \oplus \CMTB{\{90090\}} \oplus \CMTB{\{192192\}} \oplus \CMTB{\{299520\}} ~~~,  \\
{~~}{\rm {Level}}-8 \,~~~~~~~ \CMTred{\{144\}} \oplus (2) \CMTred{\{560\}} \oplus (3) \CMTred{\{720\}} \oplus (2) \CMTred{\{1200\}} \oplus (2) \CMTred{\{1440\}} \oplus (3) \CMTred{\{2640\}}\\
~~~~~~~~~~~~~~~~~~~~~~~\oplus (4) \CMTred{\{3696\}} \oplus \CMTred{\{5280\}} \oplus (2) \CMTred{\{7920\}} \oplus (2) \CMTred{\{8064\}} \oplus (5) \CMTred{\{8800\}} \oplus (3) \CMTred{\{11088\}}\\
~~~~~~~~~~~~~~~~~~~~~~~\oplus (4) \CMTred{\{15120\}} \oplus \CMTred{\{20592\}} \oplus \CMTred{\{23760\}} \oplus (3) \CMTred{\{25200\}} \oplus \CMTred{\{30800\}} \oplus (2) \CMTred{\{34992\}} \\
~~~~~~~~~~~~~~~~~~~~~~~\oplus (4) \CMTred{\{38016\}} \oplus \CMTred{\{43680\}} \oplus (2) \CMTred{\{48048\}} \oplus (2) \CMTred{\{49280\}} \oplus \CMTred{\{55440\}} \oplus \CMTred{\{124800\}}\\
~~~~~~~~~~~~~~~~~~~~~~~\oplus (2) \CMTred{\{144144\}} \oplus \CMTred{\{164736\}} \oplus \CMTred{\{196560\}} \oplus \CMTred{\{205920\}} ~~~,  \\
{~~~~~~}  {~~~~} \vdots  {~~~~~~~~~\,~~~~~~} \vdots
\end{cases}
{~~~~~~~~~~~~~~~}
\end{equation}
\endgroup

\begingroup
\footnotesize
\begin{equation}
\label{equ:Psitimes144}
{\cal V} \otimes \CMTred{\{\overline{144}\}} ~=~ \begin{cases}
{~~}{\rm {Level}}-0 \,~~~~~~~\CMTred{\{\overline{144}\}}  ~~~,  \\
{~~}{\rm {Level}}-1 \,~~~~~~~ \CMTB{\{10\}} \oplus \CMTB{\{120\}} \oplus \CMTB{\{\overline{126}\}} \oplus \CMTB{\{320\}} \oplus \CMTB{\{1728\}}
 ~~~,  \\
{~~}{\rm {Level}}-2 \,~~~~~~~\CMTred{\{16\}} \oplus (2) \CMTred{\{144\}} \oplus (2) \CMTred{\{560\}} \oplus \CMTred{\{720\}} \oplus \CMTred{\{1200\}} \oplus \CMTred{\{1440\}} \oplus \CMTred{\{3696\}} \oplus \CMTred{\{8800\}}  ~~~, \\
{~~}{\rm {Level}}-3 \,~~~~~~~ \CMTB{\{45\}} \oplus \CMTgrn{\bf \{54\}} \oplus (2) \CMTB{\{210\}} \oplus \CMTB{\{770\}} \oplus (3) \CMTB{\{945\}} \oplus (2) \CMTB{\{\overline{1050}\}} \oplus \CMTB{\{1050\}} \oplus \CMTB{\{1386\}} \\
~~~~~~~~~~~~~~~~~~~~~~~\oplus \CMTB{\{4125\}} \oplus (2) \CMTB{\{5940\}} \oplus \CMTB{\{\overline{6930}\}} \oplus \CMTB{\{8085\}} \oplus \CMTB{\{17920\}} \oplus \CMTB{\{\overline{23040}\}} ~~~, \\
{~~}{\rm {Level}}-4 \,~~~~~~~(2) \CMTred{\{\overline{144}\}} \oplus (2) \CMTred{\{\overline{560}\}} \oplus \CMTred{\{\overline{672}\}} \oplus (2) \CMTred{\{\overline{720}\}} \oplus (3) \CMTred{\{\overline{1200}\}} \oplus \CMTred{\{\overline{1440}\}} \oplus \CMTred{\{\overline{2640}\}} \\
~~~~~~~~~~~~~~~~~~~~~~~\oplus (4) \CMTred{\{\overline{3696}\}} \oplus \CMTred{\{\overline{8064}\}} \oplus (2) \CMTred{\{\overline{8800}\}} \oplus (2) \CMTred{\{\overline{11088}\}} \oplus \CMTred{\{\overline{15120}\}} \oplus \CMTred{\{\overline{17280}\}}\\
~~~~~~~~~~~~~~~~~~~~~~~\oplus \CMTred{\{\overline{23760}\}} \oplus (2) \CMTred{\{\overline{25200}\}} \oplus \CMTred{\{\overline{38016}\}} \oplus \CMTred{\{\overline{43680}\}}  ~~~,  \\
{~~}{\rm {Level}}-5 \,~~~~~~~ \CMTB{\{120\}} \oplus \CMTB{\{\overline{126}\}} \oplus \CMTB{\{210'\}} \oplus (2) \CMTB{\{320\}} \oplus (4) \CMTB{\{1728\}} \oplus (3) \CMTB{\{2970\}} \oplus (3) \CMTB{\{\overline{3696'}\}}\\
~~~~~~~~~~~~~~~~~~~~~~~\oplus \CMTB{\{3696'\}} \oplus (3) \CMTB{\{4312\}} \oplus (2) \CMTB{\{4410\}} \oplus \CMTB{\{4608\}} \oplus (3) \CMTB{\{\overline{4950}\}} \oplus \CMTB{\{4950\}} \\
~~~~~~~~~~~~~~~~~~~~~~~\oplus \CMTB{\{\overline{6930'}\}} \oplus (2) \CMTB{\{10560\}} \oplus \CMTB{\{\overline{20790}\}} \oplus \CMTB{\{27720\}} \oplus \CMTB{\{28160\}} \oplus \CMTB{\{34398\}}\\
~~~~~~~~~~~~~~~~~~~~~~~\oplus (3) \CMTB{\{36750\}} \oplus \CMTB{\{\overline{46800}\}} \oplus (2) \CMTB{\{\overline{48114}\}} \oplus \CMTB{\{68640\}} \oplus \CMTB{\{\overline{90090}\}} ~~~,  \\
{~~}{\rm {Level}}-6 \,~~~~~~~ \CMTred{\{144\}} \oplus (2) \CMTred{\{560\}} \oplus (3) \CMTred{\{720\}} \oplus (2) \CMTred{\{1200\}} \oplus (2) \CMTred{\{1440\}} \oplus (2) \CMTred{\{2640\}} \\
~~~~~~~~~~~~~~~~~~~~~~~\oplus (4) \CMTred{\{3696\}} \oplus (2) \CMTred{\{5280\}} \oplus \CMTred{\{7920\}} \oplus (2) \CMTred{\{8064\}} \oplus (5) \CMTred{\{8800\}} \oplus (2) \CMTred{\{11088\}}\\
~~~~~~~~~~~~~~~~~~~~~~~\oplus (4) \CMTred{\{15120\}} \oplus (2) \CMTred{\{25200\}} \oplus \CMTred{\{29568\}} \oplus \CMTred{\{30800\}} \oplus (2) \CMTred{\{34992\}} \\
~~~~~~~~~~~~~~~~~~~~~~~\oplus (2) \CMTred{\{38016\}} \oplus \CMTred{\{43680\}} \oplus \CMTred{\{48048\}} \oplus (2) \CMTred{\{49280\}} \oplus \CMTred{\{102960\}} \oplus \CMTred{\{124800\}} \\
~~~~~~~~~~~~~~~~~~~~~~~\oplus (2) \CMTred{\{144144\}} ~~~,  \\
{~~}{\rm {Level}}-7 \,~~~~~~~ \CMTgrn{\bf \{54\}} \oplus \CMTB{\{210\}} \oplus \CMTB{\{660\}} \oplus \CMTB{\{770\}} \oplus (3) \CMTB{\{945\}} \oplus (2) \CMTB{\{1050\}} \oplus (2) \CMTB{\{\overline{1050}\}} \\
~~~~~~~~~~~~~~~~~~~~~~~\oplus (3) \CMTB{\{1386\}} \oplus (2) \CMTB{\{4125\}} \oplus (4) \CMTB{\{5940\}} \oplus (2) \CMTB{\{\overline{6930}\}} \oplus \CMTB{\{6930\}} \oplus (5) \CMTB{\{8085\}}\\
~~~~~~~~~~~~~~~~~~~~~~~\oplus \CMTB{\{8910\}} \oplus \CMTB{\{12870\}} \oplus (3) \CMTB{\{14784\}} \oplus \CMTB{\{16380\}} \oplus (2) \CMTB{\{\overline{17325}\}} \oplus \CMTB{\{17325\}} \\
~~~~~~~~~~~~~~~~~~~~~~~\oplus (4) \CMTB{\{17920\}} \oplus (3) \CMTB{\{\overline{23040}\}} \oplus (2) \CMTB{\{23040\}} \oplus \CMTB{\{\overline{50688}\}} \oplus \CMTB{\{70070\}} \oplus (3) \CMTB{\{72765\}} \\
~~~~~~~~~~~~~~~~~~~~~~~\oplus \CMTB{\{73710\}} \oplus \CMTB{\{81081\}} \oplus \CMTB{\{112320\}} \oplus \CMTB{\{\overline{128700}\}} \oplus (2) \CMTB{\{143000\}} \oplus \CMTB{\{\overline{199017}\}} ~~~,  \\
{~~}{\rm {Level}}-8 \,~~~~~~~  \CMTred{\{\overline{144}\}} \oplus (2) \CMTred{\{\overline{560}\}} \oplus (3) \CMTred{\{\overline{720}\}} \oplus (2) \CMTred{\{\overline{1200}\}} \oplus (2) \CMTred{\{\overline{1440}\}} \oplus (3) \CMTred{\{\overline{2640}\}} \\
~~~~~~~~~~~~~~~~~~~~~~~\oplus (4) \CMTred{\{\overline{3696}\}} \oplus \CMTred{\{\overline{5280}\}} \oplus (2) \CMTred{\{\overline{7920}\}} \oplus (2) \CMTred{\{\overline{8064}\}} \oplus (5) \CMTred{\{\overline{8800}\}} \oplus (3) \CMTred{\{\overline{11088}\}}\\
~~~~~~~~~~~~~~~~~~~~~~~\oplus (4) \CMTred{\{\overline{15120}\}} \oplus \CMTred{\{\overline{20592}\}} \oplus \CMTred{\{\overline{23760}\}} \oplus (3) \CMTred{\{\overline{25200}\}} \oplus \CMTred{\{\overline{30800}\}} \oplus (2) \CMTred{\{\overline{34992}\}}\\
~~~~~~~~~~~~~~~~~~~~~~~\oplus (4) \CMTred{\{\overline{38016}\}} \oplus \CMTred{\{\overline{43680}\}} \oplus (2) \CMTred{\{\overline{48048}\}} \oplus (2) \CMTred{\{\overline{49280}\}} \oplus \CMTred{\{\overline{55440}\}} \oplus \CMTred{\{\overline{124800}\}} \\
~~~~~~~~~~~~~~~~~~~~~~~\oplus (2) \CMTred{\{\overline{144144}\}} \oplus \CMTred{\{\overline{164736}\}} \oplus \CMTred{\{\overline{196560}\}} \oplus \CMTred{\{\overline{205920}\}}
 ~~~,  \\
{~~~~~~}  {~~~~} \vdots  {~~~~~~~~~\,~~~~~~} \vdots
\end{cases}
{~~~~~~~~~~~~~~~}
\end{equation}
\endgroup

\begingroup
\scriptsize
\begin{equation}
\label{equ:Psitimes560bar}
{\cal V} \otimes \CMTred{\{\overline{560}\}} ~=~ \begin{cases}
{~~}{\rm {Level}}-0 \,~~~~~~~ \CMTred{\{\overline{560}\}}  ~~~,  \\
{~~}{\rm {Level}}-1 \,~~~~~~~ \CMTB{\{120\}} \oplus \CMTB{\{126\}} \oplus \CMTB{\{320\}} \oplus \CMTB{\{1728\}} \oplus \CMTB{\{2970\}} \oplus \CMTB{\{3696'\}} ~~~,  \\
{~~}{\rm {Level}}-2 \,~~~~~~~ \CMTred{\{16\}} \oplus (2) \CMTred{\{144\}} \oplus (2) \CMTred{\{560\}} \oplus \CMTred{\{672\}} \oplus \CMTred{\{720\}} \oplus (2) \CMTred{\{1200\}} \oplus \CMTred{\{1440\}}\\
~~~~~~~~~~~~~~~~~~~~~~~\oplus (2) \CMTred{\{3696\}} \oplus \CMTred{\{8064\}} \oplus \CMTred{\{8800\}} \oplus \CMTred{\{11088\}} \oplus \CMTred{\{25200\}}  ~~~, \\
{~~}{\rm {Level}}-3 \,~~~~~~~ \CMTB{\{1\}} \oplus (2) \CMTB{\{45\}} \oplus \CMTgrn{\bf \{54\}} \oplus (3) \CMTB{\{210\}} \oplus (2) \CMTB{\{770\}} \oplus (3) \CMTB{\{945\}} \oplus (2) \CMTB{\{1050\}}\\
~~~~~~~~~~~~~~~~~~~~~~~\oplus (2) \CMTB{\{\overline{1050}\}} \oplus \CMTB{\{1386\}} \oplus \CMTB{\{4125\}} \oplus (4) \CMTB{\{5940\}} \oplus \CMTB{\{6930\}} \oplus \CMTB{\{\overline{6930}\}} \oplus \CMTB{\{7644\}} \\
~~~~~~~~~~~~~~~~~~~~~~~\oplus (2) \CMTB{\{8085\}} \oplus \CMTB{\{8910\}} \oplus (2) \CMTB{\{17920\}} \oplus \CMTB{\{23040\}} \oplus \CMTB{\{\overline{23040}\}} \\
~~~~~~~~~~~~~~~~~~~~~~~\oplus \CMTB{\{72765\}} \oplus \CMTB{\{73710\}} ~~~, \\
{~~}{\rm {Level}}-4 \,~~~~~~~(2) \CMTred{\{\overline{16}\}} \oplus (3) \CMTred{\{\overline{144}\}} \oplus (4) \CMTred{\{\overline{560}\}} \oplus \CMTred{\{\overline{672}\}} \oplus (2) \CMTred{\{\overline{720}\}} \oplus (4) \CMTred{\{\overline{1200}\}} \oplus (2) \CMTred{\{\overline{1440}\}} \\
~~~~~~~~~~~~~~~~~~~~~~~\oplus \CMTred{\{\overline{2640}\}} \oplus (5) \CMTred{\{\overline{3696}\}} \oplus (3) \CMTred{\{\overline{8064}\}} \oplus (4) \CMTred{\{\overline{8800}\}} \oplus (4) \CMTred{\{\overline{11088}\}} \oplus (2) \CMTred{\{\overline{15120}\}}\\
~~~~~~~~~~~~~~~~~~~~~~~\oplus \CMTred{\{\overline{17280}\}} \oplus (3) \CMTred{\{\overline{25200}\}} \oplus (2) \CMTred{\{\overline{34992}\}} \oplus (2) \CMTred{\{\overline{38016}\}} \oplus (2) \CMTred{\{\overline{43680}\}} \\
~~~~~~~~~~~~~~~~~~~~~~~\oplus \CMTred{\{\overline{49280}\}} \oplus \CMTred{\{\overline{55440}\}} \oplus \CMTred{\{\overline{70560}\}} \oplus \CMTred{\{\overline{144144}\}} \oplus \CMTred{\{\overline{205920}\}}  ~~~,  \\
{~~}{\rm {Level}}-5 \,~~~~~~~ \CMTB{\{10\}} \oplus (3) \CMTB{\{120\}} \oplus (2) \CMTB{\{\overline{126}\}} \oplus \CMTB{\{126\}} \oplus \CMTB{\{210'\}} \oplus (3) \CMTB{\{320\}} \oplus (6) \CMTB{\{1728\}} \\
~~~~~~~~~~~~~~~~~~~~~~~\oplus (5) \CMTB{\{2970\}} \oplus (5) \CMTB{\{\overline{3696'}\}} \oplus (2) \CMTB{\{3696'\}} \oplus (4) \CMTB{\{4312\}} \oplus (3) \CMTB{\{4410\}} \oplus \CMTB{\{4608\}}\\
~~~~~~~~~~~~~~~~~~~~~~~\oplus (3) \CMTB{\{\overline{4950}\}} \oplus (2) \CMTB{\{4950\}} \oplus (2) \CMTB{\{\overline{6930'}\}} \oplus (4) \CMTB{\{10560\}} \oplus (2) \CMTB{\{27720\}} \\
~~~~~~~~~~~~~~~~~~~~~~~\oplus (2) \CMTB{\{28160\}} \oplus (2) \CMTB{\{34398\}} \oplus (7) \CMTB{\{36750\}} \oplus \CMTB{\{37632\}} \oplus (2) \CMTB{\{\overline{46800}\}}\\
~~~~~~~~~~~~~~~~~~~~~~~\oplus \CMTB{\{46800\}} \oplus (3) \CMTB{\{\overline{48114}\}} \oplus \CMTB{\{48114\}} \oplus (2) \CMTB{\{64680\}} \oplus (2) \CMTB{\{68640\}} \oplus \CMTB{\{90090\}} \\
~~~~~~~~~~~~~~~~~~~~~~~\oplus \CMTB{\{\overline{90090}\}} \oplus \CMTB{\{\overline{150150}\}} \oplus \CMTB{\{192192\}} \oplus \CMTB{\{299520\}} \oplus \CMTB{\{380160\}} ~~~,  \\
{~~}{\rm {Level}}-6 \,~~~~~~~\CMTred{\{16\}} \oplus (2) \CMTred{\{144\}} \oplus (5) \CMTred{\{560\}} \oplus (4) \CMTred{\{720\}} \oplus (3) \CMTred{\{1200\}} \oplus (4) \CMTred{\{1440\}} \oplus (3) \CMTred{\{2640\}} \\
~~~~~~~~~~~~~~~~~~~~~~~\oplus (6) \CMTred{\{3696\}} \oplus (3) \CMTred{\{5280\}} \oplus \CMTred{\{7920\}} \oplus (5) \CMTred{\{8064\}} \oplus (8) \CMTred{\{8800\}} \oplus (4) \CMTred{\{11088\}} \\
~~~~~~~~~~~~~~~~~~~~~~~\oplus (5) \CMTred{\{15120\}} \oplus \CMTred{\{23760\}} \oplus (4) \CMTred{\{25200\}} \oplus \CMTred{\{29568\}} \oplus (2) \CMTred{\{30800\}} \oplus (5) \CMTred{\{34992\}}\\
~~~~~~~~~~~~~~~~~~~~~~~\oplus (5) \CMTred{\{38016\}} \oplus \CMTred{\{39600\}} \oplus (3) \CMTred{\{43680\}} \oplus (2) \CMTred{\{48048\}} \oplus (4) \CMTred{\{49280\}} \oplus \CMTred{\{55440\}}\\
~~~~~~~~~~~~~~~~~~~~~~~\oplus \CMTred{\{70560\}} \oplus \CMTred{\{102960\}} \oplus \CMTred{\{124800\}} \oplus (4) \CMTred{\{144144\}} \oplus \CMTred{\{155232\}} \oplus \CMTred{\{164736\}}\\
~~~~~~~~~~~~~~~~~~~~~~~\oplus \CMTred{\{196560\}} \oplus (2) \CMTred{\{205920\}} \oplus \CMTred{\{364000\}} \oplus \CMTred{\{388080\}} \oplus \CMTred{\{529200\}}  ~~~,  \\
{~~}{\rm {Level}}-7 \,~~~~~~~ \CMTB{\{45\}} \oplus \CMTgrn{\bf \{54\}} \oplus (2) \CMTB{\{210\}} \oplus \CMTB{\{660\}} \oplus (3) \CMTB{\{770\}} \oplus (5) \CMTB{\{945\}} \oplus (5) \CMTB{\{\overline{1050}\}}\\
~~~~~~~~~~~~~~~~~~~~~~~\oplus (2) \CMTB{\{1050\}} \oplus (4) \CMTB{\{1386\}} \oplus \CMTB{\{\overline{2772}\}} \oplus (3) \CMTB{\{4125\}} \oplus (7) \CMTB{\{5940\}} \oplus (4) \CMTB{\{\overline{6930}\}} \\
~~~~~~~~~~~~~~~~~~~~~~~\oplus (2) \CMTB{\{6930\}} \oplus \CMTB{\{7644\}} \oplus (8) \CMTB{\{8085\}} \oplus (2) \CMTB{\{8910\}} \oplus \CMTB{\{12870\}} \oplus (4) \CMTB{\{14784\}} \\
~~~~~~~~~~~~~~~~~~~~~~~\oplus (2) \CMTB{\{16380\}} \oplus (2) \CMTB{\{\overline{17325}\}} \oplus (3) \CMTB{\{17325\}} \oplus (7) \CMTB{\{17920\}} \oplus (6) \CMTB{\{\overline{23040}\}} \\
~~~~~~~~~~~~~~~~~~~~~~~\oplus (4) \CMTB{\{23040\}} \oplus (2) \CMTB{\{\overline{50688}\}} \oplus \CMTB{\{50688\}} \oplus (2) \CMTB{\{70070\}} \oplus (7) \CMTB{\{72765\}} \\
~~~~~~~~~~~~~~~~~~~~~~~\oplus (3) \CMTB{\{73710\}} \oplus \CMTB{\{81081\}} \oplus (2) \CMTB{\{112320\}} \oplus (2) \CMTB{\{\overline{128700}\}} \oplus \CMTB{\{128700\}}\\
~~~~~~~~~~~~~~~~~~~~~~~\oplus (5) \CMTB{\{143000\}} \oplus \CMTB{\{174636\}} \oplus \CMTB{\{189189\}} \oplus \CMTB{\{\overline{199017}\}} \oplus (2) \CMTB{\{199017\}}\\
~~~~~~~~~~~~~~~~~~~~~~~\oplus \CMTB{\{207360\}} \oplus \CMTB{\{210210\}} \oplus \CMTB{\{\overline{242550}\}} \oplus \CMTB{\{274560\}} \oplus \CMTB{\{275968\}} \oplus \CMTB{\{1048576\}} ~~~,  \\
{~~}{\rm {Level}}-8 \,~~~~~~~ (2) \CMTred{\{\overline{144}\}} \oplus (3) \CMTred{\{\overline{560}\}} \oplus (2) \CMTred{\{\overline{672}\}} \oplus (4) \CMTred{\{\overline{720}\}} \oplus (4) \CMTred{\{\overline{1200}\}} \oplus (2) \CMTred{\{\overline{1440}\}}\\
~~~~~~~~~~~~~~~~~~~~~~~\oplus (4) \CMTred{\{\overline{2640}\}} \oplus (8) \CMTred{\{\overline{3696}\}} \oplus (2) \CMTred{\{\overline{5280}\}} \oplus (2) \CMTred{\{\overline{7920}\}} \oplus (3) \CMTred{\{\overline{8064}\}} \oplus (8) \CMTred{\{\overline{8800}\}}\\
~~~~~~~~~~~~~~~~~~~~~~~\oplus (6) \CMTred{\{\overline{11088}\}} \oplus (7) \CMTred{\{\overline{15120}\}} \oplus (2) \CMTred{\{\overline{17280}\}} \oplus \CMTred{\{\overline{23760}\}} \oplus (6) \CMTred{\{\overline{25200}\}} \oplus \CMTred{\{\overline{29568}\}}\\
~~~~~~~~~~~~~~~~~~~~~~~\oplus (3) \CMTred{\{\overline{30800}\}} \oplus (4) \CMTred{\{\overline{34992}\}} \oplus (7) \CMTred{\{\overline{38016}\}} \oplus (3) \CMTred{\{\overline{43680}\}} \oplus (3) \CMTred{\{\overline{48048}\}}\\
~~~~~~~~~~~~~~~~~~~~~~~\oplus (5) \CMTred{\{\overline{49280}\}} \oplus (2) \CMTred{\{\overline{55440}\}} \oplus \CMTred{\{\overline{80080}\}} \oplus \CMTred{\{\overline{102960}\}} \oplus (3) \CMTred{\{\overline{124800}\}} \oplus \CMTred{\{\overline{129360}\}}\\
~~~~~~~~~~~~~~~~~~~~~~~\oplus (5) \CMTred{\{\overline{144144}\}} \oplus \CMTred{\{\overline{155232}\}} \oplus \CMTred{\{\overline{164736}\}} \oplus (2) \CMTred{\{\overline{196560}\}} \oplus (3) \CMTred{\{\overline{205920}\}} \\
~~~~~~~~~~~~~~~~~~~~~~~\oplus \CMTred{\{\overline{258720}\}} \oplus (2) \CMTred{\{\overline{364000}\}} \oplus \CMTred{\{\overline{465696}\}} \oplus \CMTred{\{\overline{529200}\}} \oplus \CMTred{\{\overline{769824}\}} ~~~,  \\
{~~~~~~}  {~~~~} \vdots  {~~~~~~~~~\,~~~~~~} \vdots
\end{cases}
{~~~~~~~~~~~~~~~}
\end{equation}
\endgroup

\begingroup
\scriptsize
\begin{equation}
{\cal V} \otimes \CMTred{\{720\}} ~=~ \begin{cases}
{~~}{\rm {Level}}-0 \,~~~~~~~  \CMTred{\{720\}} ~~~,  \\
{~~}{\rm {Level}}-1 \,~~~~~~~\CMTgrn{\bf \{54\}} \oplus \CMTB{\{945\}} \oplus \CMTB{\{\overline{1050}\}} \oplus \CMTB{\{1386\}} \oplus \CMTB{\{8085\}}  ~~~,  \\
{~~}{\rm {Level}}-2 \,~~~~~~~\CMTred{\{\overline{144}\}} \oplus \CMTred{\{\overline{560}\}} \oplus (2) \CMTred{\{\overline{720}\}} \oplus \CMTred{\{\overline{1200}\}} \oplus \CMTred{\{\overline{2640}\}} \oplus (2) \CMTred{\{\overline{3696}\}} \oplus \CMTred{\{\overline{8800}\}}\\
~~~~~~~~~~~~~~~~~~~~~~~\oplus \CMTred{\{\overline{11088}\}} \oplus \CMTred{\{\overline{15120}\}} \oplus \CMTred{\{\overline{38016}\}}  ~~~, \\
{~~}{\rm {Level}}-3 \,~~~~~~~\CMTB{\{120\}} \oplus \CMTB{\{126\}} \oplus \CMTB{\{210'\}} \oplus (2) \CMTB{\{320\}} \oplus (3) \CMTB{\{1728\}} \oplus (2) \CMTB{\{2970\}} \oplus \CMTB{\{3696'\}}\\
~~~~~~~~~~~~~~~~~~~~~~~\oplus \CMTB{\{\overline{3696'}\}}\oplus (3) \CMTB{\{4312\}} \oplus \CMTB{\{4410\}} \oplus \CMTB{\{4608\}} \oplus (2) \CMTB{\{\overline{4950}\}} \oplus \CMTB{\{4950\}} \oplus \CMTB{\{10560\}} \\
~~~~~~~~~~~~~~~~~~~~~~~\oplus \CMTB{\{27720\}} \oplus \CMTB{\{28160\}} \oplus (2) \CMTB{\{36750\}} \oplus \CMTB{\{\overline{48114}\}} \oplus \CMTB{\{68640\}} \oplus \CMTB{\{\overline{90090}\}}  ~~~, \\
{~~}{\rm {Level}}-4 \,~~~~~~~ (2) \CMTred{\{144\}} \oplus (2) \CMTred{\{560\}} \oplus \CMTred{\{672\}} \oplus (3) \CMTred{\{720\}} \oplus (3) \CMTred{\{1200\}} \oplus \CMTred{\{1440\}} \oplus (2) \CMTred{\{2640\}} \\
~~~~~~~~~~~~~~~~~~~~~~~\oplus (4) \CMTred{\{3696\}} \oplus \CMTred{\{5280\}} \oplus \CMTred{\{7920\}} \oplus (2) \CMTred{\{8064\}} \oplus (4) \CMTred{\{8800\}} \oplus (2) \CMTred{\{11088\}}\\
~~~~~~~~~~~~~~~~~~~~~~~\oplus (4) \CMTred{\{15120\}} \oplus (2) \CMTred{\{25200\}} \oplus \CMTred{\{30800\}} \oplus \CMTred{\{34992\}} \oplus (2) \CMTred{\{38016\}} \oplus \CMTred{\{43680\}}\\
~~~~~~~~~~~~~~~~~~~~~~~\oplus \CMTred{\{48048\}} \oplus (2) \CMTred{\{49280\}} \oplus \CMTred{\{80080\}} \oplus \CMTred{\{102960\}} \oplus \CMTred{\{124800\}} \oplus (2) \CMTred{\{144144\}} \\
~~~~~~~~~~~~~~~~~~~~~~~\oplus \CMTred{\{155232\}} ~~~,  \\
{~~}{\rm {Level}}-5 \,~~~~~~~\CMTB{\{45\}} \oplus \CMTgrn{\bf \{54\}} \oplus (2) \CMTB{\{210\}} \oplus \CMTB{\{660\}} \oplus (2) \CMTB{\{770\}} \oplus (4) \CMTB{\{945\}} \oplus (2) \CMTB{\{\overline{1050}\}} \\
~~~~~~~~~~~~~~~~~~~~~~~\oplus (3) \CMTB{\{1050\}} \oplus (3) \CMTB{\{1386\}} \oplus (2) \CMTB{\{4125\}} \oplus (5) \CMTB{\{5940\}} \oplus \CMTB{\{\overline{6930}\}} \oplus (2) \CMTB{\{6930\}} \\
~~~~~~~~~~~~~~~~~~~~~~~\oplus \CMTB{\{7644\}} \oplus (5) \CMTB{\{8085\}} \oplus \CMTB{\{8910\}} \oplus \CMTB{\{12870\}} \oplus (3) \CMTB{\{14784\}} \oplus (2) \CMTB{\{16380\}}\\
~~~~~~~~~~~~~~~~~~~~~~~\oplus (3) \CMTB{\{\overline{17325}\}} \oplus \CMTB{\{17325\}} \oplus (5) \CMTB{\{17920\}} \oplus (4) \CMTB{\{\overline{23040}\}} \oplus (2) \CMTB{\{23040\}} \oplus \CMTB{\{\overline{50688}\}} \\
~~~~~~~~~~~~~~~~~~~~~~~\oplus \CMTB{\{70070\}} \oplus (3) \CMTB{\{72765\}} \oplus (2) \CMTB{\{73710\}} \oplus \CMTB{\{81081\}} \oplus \CMTB{\{\overline{90090'}\}} \oplus \CMTB{\{112320\}}\\
~~~~~~~~~~~~~~~~~~~~~~~\oplus \CMTB{\{\overline{128700}\}} \oplus (3) \CMTB{\{143000\}} \oplus \CMTB{\{174636\}} \oplus (2) \CMTB{\{\overline{199017}\}} \oplus \CMTB{\{207360\}}\\
~~~~~~~~~~~~~~~~~~~~~~~\oplus \CMTB{\{\overline{242550}\}} \oplus \CMTB{\{\overline{275968}\}}  ~~~,  \\
{~~}{\rm {Level}}-6 \,~~~~~~~ \CMTred{\{\overline{16}\}} \oplus (2) \CMTred{\{\overline{144}\}} \oplus (4) \CMTred{\{\overline{560}\}} \oplus (3) \CMTred{\{\overline{720}\}} \oplus (3) \CMTred{\{\overline{1200}\}} \oplus (3) \CMTred{\{\overline{1440}\}} \oplus (3) \CMTred{\{\overline{2640}\}} \\
~~~~~~~~~~~~~~~~~~~~~~~\oplus (6) \CMTred{\{\overline{3696}\}} \oplus \CMTred{\{\overline{5280}\}} \oplus (2) \CMTred{\{\overline{7920}\}} \oplus (4) \CMTred{\{\overline{8064}\}} \oplus (6) \CMTred{\{\overline{8800}\}} \oplus (4) \CMTred{\{\overline{11088}\}}\\
~~~~~~~~~~~~~~~~~~~~~~~\oplus (5) \CMTred{\{\overline{15120}\}} \oplus \CMTred{\{\overline{17280}\}} \oplus \CMTred{\{\overline{20592}\}} \oplus (2) \CMTred{\{\overline{23760}\}} \oplus (4) \CMTred{\{\overline{25200}\}} \oplus \CMTred{\{\overline{30800}\}}\\
~~~~~~~~~~~~~~~~~~~~~~~\oplus (3) \CMTred{\{\overline{34992}\}} \oplus (6) \CMTred{\{\overline{38016}\}} \oplus (4) \CMTred{\{\overline{43680}\}} \oplus (4) \CMTred{\{\overline{48048}\}} \oplus (2) \CMTred{\{\overline{49280}\}} \\
~~~~~~~~~~~~~~~~~~~~~~~\oplus \CMTred{\{\overline{55440}\}} \oplus \CMTred{\{\overline{70560}\}} \oplus (2) \CMTred{\{\overline{124800}\}} \oplus \CMTred{\{\overline{129360}\}} \oplus (3) \CMTred{\{\overline{144144}\}} \oplus \CMTred{\{\overline{155232}\}}\\
~~~~~~~~~~~~~~~~~~~~~~~\oplus (2) \CMTred{\{\overline{164736}\}} \oplus \CMTred{\{\overline{196560}\}} \oplus \CMTred{\{\overline{196560'}\}} \oplus (3) \CMTred{\{\overline{205920}\}} \oplus \CMTred{\{\overline{258720}\}} \\
~~~~~~~~~~~~~~~~~~~~~~~\oplus \CMTred{\{\overline{343200}\}} \oplus \CMTred{\{\overline{388080'}\}} \oplus (2) \CMTred{\{\overline{529200}\}}  ~~~,  \\
{~~}{\rm {Level}}-7 \,~~~~~~~ \CMTB{\{10\}} \oplus (2) \CMTB{\{120\}} \oplus \CMTB{\{\overline{126}\}} \oplus (2) \CMTB{\{126\}} \oplus \CMTB{\{210'\}} \oplus (3) \CMTB{\{320\}} \oplus (6) \CMTB{\{1728\}} \\
~~~~~~~~~~~~~~~~~~~~~~~\oplus \CMTB{\{1782\}} \oplus (5) \CMTB{\{2970\}} \oplus (3) \CMTB{\{\overline{3696'}\}} \oplus (4) \CMTB{\{3696'\}} \oplus (5) \CMTB{\{4312\}} \oplus (4) \CMTB{\{4410\}}\\
~~~~~~~~~~~~~~~~~~~~~~~\oplus (3) \CMTB{\{4608\}} \oplus (3) \CMTB{\{\overline{4950}\}} \oplus (3) \CMTB{\{4950\}} \oplus \CMTB{\{6930'\}} \oplus \CMTB{\{\overline{6930'}\}} \oplus (4) \CMTB{\{10560\}}\\
~~~~~~~~~~~~~~~~~~~~~~~\oplus (3) \CMTB{\{27720\}} \oplus (5) \CMTB{\{28160\}} \oplus \CMTB{\{31680\}} \oplus (3) \CMTB{\{34398\}} \oplus (9) \CMTB{\{36750\}} \oplus \CMTB{\{37632\}}\\
~~~~~~~~~~~~~~~~~~~~~~~\oplus (3) \CMTB{\{42120\}} \oplus (2) \CMTB{\{46800\}} \oplus (2) \CMTB{\{\overline{46800}\}} \oplus (3) \CMTB{\{\overline{48114}\}} \oplus (2) \CMTB{\{48114\}}\\
~~~~~~~~~~~~~~~~~~~~~~~\oplus \CMTB{\{48510\}} \oplus (2) \CMTB{\{\overline{50050}\}} \oplus \CMTB{\{50050\}} \oplus (2) \CMTB{\{64680\}} \oplus (5) \CMTB{\{68640\}} \oplus (4) \CMTB{\{\overline{90090}\}}\\
~~~~~~~~~~~~~~~~~~~~~~~\oplus (2) \CMTB{\{90090\}} \oplus \CMTB{\{\overline{150150}\}} \oplus (2) \CMTB{\{192192\}} \oplus \CMTB{\{203840\}} \oplus \CMTB{\{\overline{216216}\}} \\
~~~~~~~~~~~~~~~~~~~~~~~\oplus (3) \CMTB{\{299520\}} \oplus \CMTB{\{351000\}} \oplus \CMTB{\{376320\}} \oplus (2) \CMTB{\{380160\}} \oplus (2) \CMTB{\{436590\}}\\
~~~~~~~~~~~~~~~~~~~~~~~\oplus \CMTB{\{\overline{630630}\}} \oplus \CMTB{\{\overline{705600}\}} ~~~,  \\
{~~}{\rm {Level}}-8 \,~~~~~~~\CMTred{\{16\}} \oplus (3) \CMTred{\{144\}} \oplus (4) \CMTred{\{560\}} \oplus \CMTred{\{672\}} \oplus (3) \CMTred{\{720\}} \oplus (4) \CMTred{\{1200\}} \oplus (3) \CMTred{\{1440\}} \\
~~~~~~~~~~~~~~~~~~~~~~~\oplus (3) \CMTred{\{2640\}} \oplus (7) \CMTred{\{3696\}} \oplus \CMTred{\{5280\}} \oplus (3) \CMTred{\{7920\}} \oplus (5) \CMTred{\{8064\}} \oplus (6) \CMTred{\{8800\}} \\
~~~~~~~~~~~~~~~~~~~~~~~\oplus (5) \CMTred{\{11088\}} \oplus (6) \CMTred{\{15120\}} \oplus \CMTred{\{17280\}} \oplus (2) \CMTred{\{20592\}} \oplus \CMTred{\{23760\}} \oplus (5) \CMTred{\{25200\}}\\
~~~~~~~~~~~~~~~~~~~~~~~\oplus \CMTred{\{30800\}} \oplus (4) \CMTred{\{34992\}} \oplus (6) \CMTred{\{38016\}} \oplus \CMTred{\{39600\}} \oplus (5) \CMTred{\{43680\}} \oplus (4) \CMTred{\{48048\}} \\
~~~~~~~~~~~~~~~~~~~~~~~\oplus \CMTred{\{48048'\}} \oplus (4) \CMTred{\{49280\}} \oplus \CMTred{\{55440\}} \oplus \CMTred{\{70560\}} \oplus \CMTred{\{80080\}} \oplus \CMTred{\{102960\}}\\
~~~~~~~~~~~~~~~~~~~~~~~\oplus (4) \CMTred{\{124800\}} \oplus (2) \CMTred{\{129360\}} \oplus (5) \CMTred{\{144144\}} \oplus (2) \CMTred{\{155232\}} \oplus (2) \CMTred{\{164736\}}\\
~~~~~~~~~~~~~~~~~~~~~~~\oplus \CMTred{\{196560\}} \oplus (3) \CMTred{\{205920\}} \oplus \CMTred{\{258720\}} \oplus \CMTred{\{343200\}} \oplus \CMTred{\{364000\}} \oplus \CMTred{\{388080\}}\\
~~~~~~~~~~~~~~~~~~~~~~~\oplus \CMTred{\{458640\}} \oplus (2) \CMTred{\{529200\}} \oplus \CMTred{\{764400\}} \oplus \CMTred{\{769824\}}  ~~~,  \\
{~~~~~~}  {~~~~} \vdots  {~~~~~~~~~\,~~~~~~} \vdots
\end{cases}
{~~~~~~~~~~~~~~~}
\end{equation}
\endgroup

\begingroup
\scriptsize
\begin{equation}
{\cal V} \otimes \CMTred{\{\overline{720}\}} ~=~ \begin{cases}
{~~}{\rm {Level}}-0 \,~~~~~~~  \CMTred{\{\overline{720}\}} ~~~,  \\
{~~}{\rm {Level}}-1 \,~~~~~~~\CMTB{\{210'\}} \oplus \CMTB{\{320\}} \oplus \CMTB{\{1728\}} \oplus \CMTB{\{4312\}} \oplus \CMTB{\{4950\}}  ~~~,  \\
{~~}{\rm {Level}}-2 \,~~~~~~~  \CMTred{\{144\}} \oplus \CMTred{\{560\}} \oplus (2) \CMTred{\{720\}} \oplus \CMTred{\{1200\}} \oplus \CMTred{\{2640\}} \oplus (2) \CMTred{\{3696\}} \oplus \CMTred{\{8800\}} \\
~~~~~~~~~~~~~~~~~~~~~~~\oplus \CMTred{\{11088\}} \oplus \CMTred{\{15120\}} \oplus \CMTred{\{38016\}} ~~~, \\
{~~}{\rm {Level}}-3 \,~~~~~~~  \CMTB{\{45\}} \oplus \CMTgrn{\bf \{54\}} \oplus \CMTB{\{210\}} \oplus \CMTB{\{770\}} \oplus (3) \CMTB{\{945\}} \oplus \CMTB{\{\overline{1050}\}} \oplus (2) \CMTB{\{1050\}} \oplus (2) \CMTB{\{1386\}} \\
~~~~~~~~~~~~~~~~~~~~~~~\oplus \CMTB{\{4125\}} \oplus (2) \CMTB{\{5940\}} \oplus \CMTB{\{6930\}} \oplus (3) \CMTB{\{8085\}} \oplus \CMTB{\{14784\}} \oplus \CMTB{\{16380\}}\\
~~~~~~~~~~~~~~~~~~~~~~~\oplus \CMTB{\{\overline{17325}\}} \oplus (2) \CMTB{\{17920\}} \oplus \CMTB{\{23040\}} \oplus \CMTB{\{\overline{23040}\}} \oplus \CMTB{\{72765\}} \oplus \CMTB{\{143000\}} ~~~, \\
{~~}{\rm {Level}}-4 \,~~~~~~~  \CMTred{\{\overline{16}\}} \oplus (2) \CMTred{\{\overline{144}\}} \oplus (3) \CMTred{\{\overline{560}\}} \oplus (3) \CMTred{\{\overline{720}\}} \oplus (2) \CMTred{\{\overline{1200}\}} \oplus (2) \CMTred{\{\overline{1440}\}} \oplus (2) \CMTred{\{\overline{2640}\}}\\
~~~~~~~~~~~~~~~~~~~~~~~\oplus (4) \CMTred{\{\overline{3696}\}} \oplus \CMTred{\{\overline{5280}\}} \oplus (2) \CMTred{\{\overline{8064}\}} \oplus (4) \CMTred{\{\overline{8800}\}} \oplus (2) \CMTred{\{\overline{11088}\}} \oplus (3) \CMTred{\{\overline{15120}\}}\\
~~~~~~~~~~~~~~~~~~~~~~~\oplus \CMTred{\{\overline{23760}\}} \oplus (2) \CMTred{\{\overline{25200}\}} \oplus \CMTred{\{\overline{30800}\}} \oplus \CMTred{\{\overline{34992}\}} \oplus (3) \CMTred{\{\overline{38016}\}} \oplus (2) \CMTred{\{\overline{43680}\}}\\
~~~~~~~~~~~~~~~~~~~~~~~\oplus (2) \CMTred{\{\overline{48048}\}} \oplus \CMTred{\{\overline{49280}\}} \oplus \CMTred{\{\overline{144144}\}} \oplus \CMTred{\{\overline{155232}\}} \oplus \CMTred{\{\overline{164736}\}} \oplus \CMTred{\{\overline{205920}\}} ~~~,  \\
{~~}{\rm {Level}}-5 \,~~~~~~~\CMTB{\{10\}} \oplus (2) \CMTB{\{120\}} \oplus \CMTB{\{\overline{126}\}} \oplus (2) \CMTB{\{126\}} \oplus \CMTB{\{210'\}} \oplus (3) \CMTB{\{320\}} \oplus (5) \CMTB{\{1728\}} \\
~~~~~~~~~~~~~~~~~~~~~~~\oplus (4) \CMTB{\{2970\}} \oplus (2) \CMTB{\{\overline{3696'}\}} \oplus (3) \CMTB{\{3696'\}} \oplus (4) \CMTB{\{4312\}} \oplus (3) \CMTB{\{4410\}} \oplus (2) \CMTB{\{4608\}} \\
~~~~~~~~~~~~~~~~~~~~~~~\oplus (3) \CMTB{\{\overline{4950}\}} \oplus (2) \CMTB{\{4950\}} \oplus \CMTB{\{6930'\}} \oplus (3) \CMTB{\{10560\}} \oplus (2) \CMTB{\{27720\}} \oplus (3) \CMTB{\{28160\}}\\
~~~~~~~~~~~~~~~~~~~~~~~\oplus (2) \CMTB{\{34398\}} \oplus (6) \CMTB{\{36750\}} \oplus \CMTB{\{37632\}} \oplus \CMTB{\{42120\}} \oplus \CMTB{\{46800\}} \oplus \CMTB{\{\overline{46800}\}}\\
~~~~~~~~~~~~~~~~~~~~~~~\oplus (2) \CMTB{\{\overline{48114}\}} \oplus \CMTB{\{48114\}} \oplus \CMTB{\{48510\}} \oplus \CMTB{\{\overline{50050}\}} \oplus \CMTB{\{64680\}} \oplus (3) \CMTB{\{68640\}}\\
~~~~~~~~~~~~~~~~~~~~~~~\oplus (3) \CMTB{\{\overline{90090}\}} \oplus \CMTB{\{90090\}} \oplus \CMTB{\{192192\}} \oplus \CMTB{\{\overline{216216}\}} \oplus \CMTB{\{299520\}} \oplus \CMTB{\{380160\}}\\
~~~~~~~~~~~~~~~~~~~~~~~\oplus \CMTB{\{436590\}}  ~~~,  \\
{~~}{\rm {Level}}-6 \,~~~~~~~  \CMTred{\{16\}} \oplus (3) \CMTred{\{144\}} \oplus (4) \CMTred{\{560\}} \oplus \CMTred{\{672\}} \oplus (3) \CMTred{\{720\}} \oplus (4) \CMTred{\{1200\}} \oplus (2) \CMTred{\{1440\}}\\
~~~~~~~~~~~~~~~~~~~~~~~\oplus (2) \CMTred{\{2640\}} \oplus (6) \CMTred{\{3696\}} \oplus \CMTred{\{5280\}} \oplus (2) \CMTred{\{7920\}} \oplus (4) \CMTred{\{8064\}} \oplus (6) \CMTred{\{8800\}} \\
~~~~~~~~~~~~~~~~~~~~~~~\oplus (4) \CMTred{\{11088\}} \oplus (6) \CMTred{\{15120\}} \oplus \CMTred{\{17280\}} \oplus \CMTred{\{20592\}} \oplus (4) \CMTred{\{25200\}} \oplus \CMTred{\{30800\}}\\
~~~~~~~~~~~~~~~~~~~~~~~\oplus (3) \CMTred{\{34992\}} \oplus (4) \CMTred{\{38016\}} \oplus (4) \CMTred{\{43680\}} \oplus (3) \CMTred{\{48048\}} \oplus (4) \CMTred{\{49280\}}\\
~~~~~~~~~~~~~~~~~~~~~~~\oplus \CMTred{\{55440\}} \oplus \CMTred{\{70560\}} \oplus \CMTred{\{80080\}} \oplus \CMTred{\{102960\}} \oplus (3) \CMTred{\{124800\}} \oplus \CMTred{\{129360\}}\\
~~~~~~~~~~~~~~~~~~~~~~~\oplus (4) \CMTred{\{144144\}} \oplus (2) \CMTred{\{155232\}} \oplus \CMTred{\{164736\}} \oplus (2) \CMTred{\{205920\}} \oplus \CMTred{\{258720\}} \\
~~~~~~~~~~~~~~~~~~~~~~~\oplus \CMTred{\{343200\}} \oplus \CMTred{\{364000\}} \oplus \CMTred{\{529200\}} \oplus \CMTred{\{769824\}} ~~~,  \\
{~~}{\rm {Level}}-7 \,~~~~~~~ (2) \CMTB{\{45\}} \oplus \CMTgrn{\bf \{54\}} \oplus (3) \CMTB{\{210\}} \oplus \CMTB{\{660\}} \oplus (3) \CMTB{\{770\}} \oplus (5) \CMTB{\{945\}} \oplus (3) \CMTB{\{1050\}}\\
~~~~~~~~~~~~~~~~~~~~~~~\oplus (3) \CMTB{\{\overline{1050}\}} \oplus (3) \CMTB{\{1386\}} \oplus (2) \CMTB{\{4125\}} \oplus \CMTB{\{4290\}} \oplus (7) \CMTB{\{5940\}} \oplus (2) \CMTB{\{6930\}}\\
~~~~~~~~~~~~~~~~~~~~~~~\oplus (2) \CMTB{\{\overline{6930}\}} \oplus (2) \CMTB{\{7644\}} \oplus (6) \CMTB{\{8085\}} \oplus (2) \CMTB{\{8910\}} \oplus (2) \CMTB{\{12870\}} \\
~~~~~~~~~~~~~~~~~~~~~~~\oplus (4) \CMTB{\{14784\}} \oplus (3) \CMTB{\{16380\}} \oplus (3) \CMTB{\{\overline{17325}\}} \oplus (2) \CMTB{\{17325\}} \oplus (7) \CMTB{\{17920\}}\\
~~~~~~~~~~~~~~~~~~~~~~~\oplus (5) \CMTB{\{\overline{23040}\}} \oplus (4) \CMTB{\{23040\}} \oplus \CMTB{\{50688\}} \oplus \CMTB{\{\overline{50688}\}} \oplus \CMTB{\{70070\}} \oplus (5) \CMTB{\{72765\}}\\
~~~~~~~~~~~~~~~~~~~~~~~\oplus (4) \CMTB{\{73710\}} \oplus (3) \CMTB{\{81081\}} \oplus \CMTB{\{105105\}} \oplus (2) \CMTB{\{112320\}} \oplus \CMTB{\{126126\}} \\
~~~~~~~~~~~~~~~~~~~~~~~\oplus \CMTB{\{128700\}} \oplus \CMTB{\{\overline{128700}\}} \oplus (6) \CMTB{\{143000\}} \oplus (2) \CMTB{\{174636\}} \oplus \CMTB{\{189189\}} \\
~~~~~~~~~~~~~~~~~~~~~~~\oplus (2) \CMTB{\{\overline{199017}\}} \oplus \CMTB{\{199017\}} \oplus (2) \CMTB{\{207360\}} \oplus \CMTB{\{242550\}} \oplus \CMTB{\{\overline{242550}\}}\\
~~~~~~~~~~~~~~~~~~~~~~~\oplus \CMTB{\{274560\}}\oplus \CMTB{\{\overline{275968}\}} \oplus \CMTB{\{275968\}} \oplus \CMTB{\{945945\}} \oplus \CMTB{\{1048576\}} ~~~,  \\
{~~}{\rm {Level}}-8 \,~~~~~~~ \CMTred{\{\overline{16}\}} \oplus (3) \CMTred{\{\overline{144}\}} \oplus (4) \CMTred{\{\overline{560}\}} \oplus \CMTred{\{\overline{672}\}} \oplus (3) \CMTred{\{\overline{720}\}} \oplus (4) \CMTred{\{\overline{1200}\}} \oplus (3) \CMTred{\{\overline{1440}\}} \\
~~~~~~~~~~~~~~~~~~~~~~~\oplus (3) \CMTred{\{\overline{2640}\}} \oplus (7) \CMTred{\{\overline{3696}\}} \oplus \CMTred{\{\overline{5280}\}} \oplus (3) \CMTred{\{\overline{7920}\}} \oplus (5) \CMTred{\{\overline{8064}\}} \oplus (6) \CMTred{\{\overline{8800}\}}\\
~~~~~~~~~~~~~~~~~~~~~~~\oplus (5) \CMTred{\{\overline{11088}\}} \oplus (6) \CMTred{\{\overline{15120}\}} \oplus \CMTred{\{\overline{17280}\}} \oplus (2) \CMTred{\{\overline{20592}\}} \oplus \CMTred{\{\overline{23760}\}} \oplus (5) \CMTred{\{\overline{25200}\}}\\
~~~~~~~~~~~~~~~~~~~~~~~\oplus \CMTred{\{\overline{30800}\}} \oplus (4) \CMTred{\{\overline{34992}\}} \oplus (6) \CMTred{\{\overline{38016}\}} \oplus \CMTred{\{\overline{39600}\}} \oplus (5) \CMTred{\{\overline{43680}\}} \oplus (4) \CMTred{\{\overline{48048}\}}\\
~~~~~~~~~~~~~~~~~~~~~~~\oplus \CMTred{\{\overline{48048'}\}} \oplus (4) \CMTred{\{\overline{49280}\}} \oplus \CMTred{\{\overline{55440}\}} \oplus \CMTred{\{\overline{70560}\}} \oplus \CMTred{\{\overline{80080}\}} \oplus \CMTred{\{\overline{102960}\}} \\
~~~~~~~~~~~~~~~~~~~~~~~\oplus (4) \CMTred{\{\overline{124800}\}} \oplus (2) \CMTred{\{\overline{129360}\}} \oplus (5) \CMTred{\{\overline{144144}\}} \oplus (2) \CMTred{\{\overline{155232}\}} \oplus (2) \CMTred{\{\overline{164736}\}}\\
~~~~~~~~~~~~~~~~~~~~~~~\oplus \CMTred{\{\overline{196560}\}} \oplus (3) \CMTred{\{\overline{205920}\}} \oplus \CMTred{\{\overline{258720}\}} \oplus \CMTred{\{\overline{343200}\}} \oplus \CMTred{\{\overline{364000}\}} \oplus \CMTred{\{\overline{388080}\}}\\
~~~~~~~~~~~~~~~~~~~~~~~\oplus \CMTred{\{\overline{458640}\}} \oplus (2) \CMTred{\{\overline{529200}\}} \oplus \CMTred{\{\overline{764400}\}} \oplus \CMTred{\{\overline{769824}\}} ~~~,  \\
{~~~~~~}  {~~~~} \vdots  {~~~~~~~~~\,~~~~~~} \vdots
\end{cases}
{~~~~~~~~~~~~~~~}
\end{equation}
\endgroup

\newpage
$$~~$$

\end{document}